\documentclass{aastex631}    
\usepackage{graphicx}	
\usepackage{amsmath}	
\usepackage{amssymb}	

\usepackage{import}						
\usepackage{cleveref}					 
\crefformat{section}{\S#2#1#3}	        
\usepackage{makecell}					
\usepackage{enumerate}
\usepackage{threeparttable}
\usepackage{bm}
\usepackage{nicematrix}
\usepackage{booktabs}
\usepackage{tabularx}
\usepackage[T1]{fontenc}

\hypersetup{
	colorlinks=true,
	linkcolor=black,			
	filecolor=black,      
	urlcolor=blue,
	citecolor=black
}

\definecolor{lavendergray}{rgb}{0.77, 0.76, 0.82}
\definecolor{cadetgray}{rgb}{0.57, 0.64, 0.69}
\definecolor{gray}{rgb}{0.5, 0.5, 0.5}
\definecolor{debianred}{rgb}{0.84, 0.04, 0.33}
\definecolor{ballblue}{rgb}{0.13, 0.67, 0.8}
\definecolor{frenchblue}{rgb}{0.0, 0.45, 0.73}
\definecolor{indigo}{rgb}{0.0, 0.25, 0.42}
\definecolor{teal}{rgb}{0.0, 0.5, 0.5}
\definecolor{lavenderpurple}{rgb}{0.59, 0.48, 0.71}
\definecolor{darklavender}{rgb}{0.45, 0.31, 0.59}
\definecolor{darkgray}{rgb}{0.66, 0.66, 0.66}
\definecolor{gray}{rgb}{0.5, 0.5, 0.5}
\definecolor{cadetblue}{rgb}{0.37, 0.62, 0.63}
\definecolor{lightseagreen}{rgb}{0.13, 0.7, 0.67}
\definecolor{palerobineggblue}{rgb}{0.59, 0.87, 0.82}
\definecolor{cadetblue}{rgb}{0.37, 0.62, 0.63}

\definecolor{snow}{rgb}{1.0, 0.98, 0.98}
\colorlet{Lightpalerobineggblue}{snow!30!palerobineggblue}
\colorlet{Lightcadetblue}{snow!60!cadetblue}
\colorlet{extraLightcadetblue}{snow!70!cadetblue}

\definecolor{blagn}{rgb}{0.0, 0.55, 0.55}
\definecolor{nlagn}{rgb}{0.0, 0.48, 0.65}
\definecolor{galaxy}{rgb}{0.58, 0.77, 0.45}
\definecolor{GClstr}{rgb}{0.0, 0.18, 0.39}
\colorlet{GClstrlight}{snow!30!GClstr}
\definecolor{unclassified}{rgb}{0.78, 0.64, 0.78}
\definecolor{galactic}{rgb}{1.0, 0.6, 0.6}
\definecolor{tealblue}{rgb}{0.21, 0.46, 0.53}
\definecolor{Fcontinuum}{rgb}{0.59, 0.87, 0.82}
\colorlet{Fcontinuumdark}{tealblue!40!Fcontinuum}

\defcitealias{lansbury2017_cat}{L17}

\newcommand{\nustar}{{\it NuSTAR} }

\newcommand{\xray}{X-ray }
\newcommand{\wise}{{\it WISE} }
\newcommand{\xmm}{{\it XMM-Newton} }
\newcommand{\chandra}{{\it Chandra} }
\newcommand{\xrt}{{\it Swift}-XRT }

\shorttitle{The NSS80 catalog}
\shortauthors{Greenwell, Klindt, et al.}
\graphicspath{{./}{}}

\begin{document}

\title{The \textit{NuSTAR} Serendipitous Survey: the 80-month catalog and source properties of the high-energy emitting AGN and quasar population}

\correspondingauthor{Claire L. Greenwell}
\email{claire.l.greenwell@durham.ac.uk}

\author[0000-0002-7719-5809]{Claire L. Greenwell}
\affiliation{Centre for Extragalactic Astronomy, Department of Physics, Durham University, Durham, DH1 3LE, UK}

\author[0000-0001-9307-9026]{Lizelke Klindt}
\affiliation{Centre for Extragalactic Astronomy, Department of Physics, Durham University, Durham, DH1 3LE, UK}

\author[0000-0002-5328-9827]{George B. Lansbury}
\affiliation{European Southern Observatory, Karl-Schwarzschild str. 2, D-85748 Garching bei München, Germany}

\author[0000-0002-0001-3587]{David J. Rosario}
\affiliation{School of Mathematics, Statistics and Physics, Herschel Building, Newcastle University, Newcastle upon Tyne, NE1 7RU, UK}
\affiliation{Centre for Extragalactic Astronomy, Department of Physics, Durham University, Durham, DH1 3LE, UK}

\author[0000-0002-5896-6313]{David M. Alexander}
\affiliation{Centre for Extragalactic Astronomy, Department of Physics, Durham University, Durham, DH1 3LE, UK}

\author[0000-0003-1908-8463]{James Aird}
\affiliation{Institute for Astronomy, University of Edinburgh, Royal Observatory, Edinburgh EH9 3HJ, UK}
\affiliation{School of Physics \& Astronomy, University of Leicester, University Road, Leicester LE1 7RJ, UK}

\author[0000-0003-2686-9241]{Daniel Stern}
\affiliation{Jet Propulsion Laboratory, California Institute of Technology, 4800 Oak Grove Drive, Mail Stop 169-221, Pasadena, CA 91109, USA}

\author{Karl Forster}
\affiliation{Cahill Center for Astrophysics, 1216 East California Boulevard, California Institute of Technology, Pasadena, CA 91125, USA}

\author[0000-0002-7998-9581]{Michael J. Koss}
\affiliation{Eureka Scientific, 2452 Delmer Street, Suite 100, Oakland, CA 94602-3017, USA}
\affiliation{Space Science Institute, 4750 Walnut Street, Suite 205, Boulder, CO 80301, USA}

\author[0000-0002-8686-8737]{Franz E. Bauer}
\affiliation{Instituto de Astrof{\'{\i}}sica and Centro de Astroingenier{\'{\i}}a, Facultad de F{\'{i}}sica, Pontificia Universidad Cat{\'{o}}lica de Chile, Casilla 306, Santiago 22, Chile} 
\affiliation{Millennium Institute of Astrophysics, Nuncio Monse{\~{n}}or S{\'{o}}tero Sanz 100, Of 104, Providencia, Santiago, Chile} 

\author{Claudio Ricci}
\affiliation{Instituto de Estudios Astrof\'isicos, Facultad de Ingenier\'ia y Ciencias, Universidad Diego Portales, Av. Ej\'ercito Libertador 441, Santiago, Chile}
\affiliation{Kavli Institute for Astronomy and Astrophysics, Peking University, Beijing 100871, China}

\author[0000-0001-5506-9855]{John Tomsick}
\affiliation{Space Sciences Laboratory, 7 Gauss Way, University of California, Berkeley, CA 94720-7450, USA}

\author[0000-0002-0167-2453]{William N. Brandt}
\affiliation{Department of Astronomy and Astrophysics, The Pennsylvania State University, 525 Davey Lab, University Park, PA 16802, USA}
\affiliation{Institute for Gravitation and the Cosmos, The Pennsylvania State University, University Park, PA 16802, USA}
\affiliation{Department of Physics, The Pennsylvania State University, University Park, PA 16802, USA}

\author[0000-0002-7898-7664]{Thomas Connor}
\affiliation{Jet Propulsion Laboratory, California Institute of Technology, 4800 Oak Grove Drive, Mail Stop 169-221, Pasadena, CA 91109, USA}

\author[0000-0001-9379-4716]{Peter G. Boorman}
\affiliation{Cahill Center for Astrophysics, 1216 East California Boulevard, California Institute of Technology, Pasadena, CA 91125, USA}
\affiliation{Astronomical Institute, Academy of Sciences, Bo\u{c}n\'{i}' II 1401, CZ-14131 Prague, Czech Republic}

\author{Adlyka Annuar}
\affiliation{Department of Applied Physics, Faculty of Science and Technology, Universiti Kebangsaan Malaysia, 43600 UKM Bangi, Selangor, Malaysia}

\author[0000-0001-8128-6976]{David R. Ballantyne}
\affiliation{Center for Relativistic Astrophysics, School of Physics, Georgia Institute of Technology, 837 State Street, Atlanta, GA 30332-0430, USA}

\author[0000-0002-4945-5079]{Chien-Ting Chen}
\affiliation{Science and Technology Institute, Universities Space Research Association, Huntsville, AL 35805, USA}
\affiliation{Astrophysics Office, NASA Marshall Space Flight Center, ZP12, Huntsville, AL 35812, USA}

\author[0000-0002-2115-1137]{Francesca Civano}
\affiliation{NASA Goddard Space Flight Center, Greenbelt, MD, USA}

\author[0000-0003-3451-9970]{Andrea Comastri}
\affiliation{INAF - Osservatorio di Astrofisica e Scienza dello Spazio di Bologna, Via Piero Gobetti, 93/3, 40129, Bologna, Italy}

\author[0000-0003-1251-532X]{Victoria A. Fawcett}
\affiliation{School of Mathematics, Statistics and Physics, Herschel Building, Newcastle University, Newcastle upon Tyne, NE1 7RU, UK}

\author[0000-0002-9286-9963]{Francesca M. Fornasini}
\affiliation{Stonehill College, 320 Washington Street, Easton, MA 02357, USA}
\affiliation{Center for Astrophysics $\vert$ Harvard \& Smithsonian, 60 Garden Street, Cambridge, MA 02138, USA}

\author[0000-0003-3105-2615]{Poshak Gandhi}
\affiliation{School of Physics \& Astronomy, University of Southampton, Southampton SO17 1BJ, UK}

\author{Fiona Harrison}
\affiliation{Cahill Center for Astrophysics, 1216 East California Boulevard, California Institute of Technology, Pasadena, CA 91125, USA}

\author[0000-0002-1082-7496]{Marianne Heida}
\affiliation{European Southern Observatory, Karl-Schwarzschild str. 2, D-85748 Garching bei München, Germany}

\author{Ryan Hickox}
\affiliation{Department of Physics and Astronomy, Dartmouth College, 6127 Wilder Laboratory, Hanover, NH 03755, USA}

\author[0000-0002-0273-218X]{Elias S. Kammoun}
\affiliation{Dipartimento di Matematica e Fisica, Universit\`{a} Roma Tre, via della Vasca Navale 84, I-00146 Rome, Italy}
\affiliation{IRAP, Universit\'e de Toulouse, CNRS, UPS, CNES 9, Avenue du Colonel Roche, BP 44346, F-31028, Toulouse Cedex 4, France}
\affiliation{INAF -- Osservatorio Astrofisico di Arcetri, Largo Enrico Fermi 5, I-50125 Firenze, Italy}

\author[0000-0002-3249-8224]{Lauranne Lanz}
\affiliation{Department of Physics, The College of New Jersey, 2000 Pennington Road, Ewing, NJ 08628, USA}

\author[0000-0001-5544-0749]{Stefano Marchesi}
\affiliation{Dipartimento di Fisica e Astronomia (DIFA) Augusto Righi, Università di Bologna, via Gobetti 93/2, I-40129 Bologna, Italy}
\affiliation{INAF - Osservatorio di Astrofisica e Scienza dello Spazio di Bologna, Via Piero Gobetti, 93/3, 40129, Bologna, Italy}
\affiliation{Department of Physics and Astronomy, Clemson University, Kinard Lab of Physics, Clemson, SC 29634, USA}

\author{Ga\"{e}l Noirot}
\affiliation{Institute for Computational Astrophysics and Department of Astronomy \& Physics, Saint Mary's University, 923 Robie Street, Halifax, NS B3H3C3, Canada}

\author[0000-0003-0607-1136]{Encarni Romero-Colmenero}
\affiliation{South African Astronomical Observatory, PO Box 9, Observatory, South Africa}
\affiliation{Southern African Large Telescope, PO Box 9, Observatory, South Africa}

\author[0000-0001-7568-6412]{Ezequiel Treister}
\affiliation{Instituto de Astrofísica, Facultad de Física, Pontificia Universidad Católica de Chile, Casilla 306, Santiago 22, Chile}

\author[0000-0002-0745-9792]{C. Megan Urry}
\affiliation{Physics Department and Yale Center for Astronomy \& Astrophysics, PO Box 208120, New Haven, CT 06520-8120, USA}

\author[0000-0001-7673-4850]{Petri V$\ddot{\mathrm{A}}$is$\ddot{\mathrm{A}}$nen}
\affiliation{South African Astronomical Observatory, PO Box 9, Observatory, South Africa}
\affiliation{Finnish Centre for Astronomy with ESO (FINCA), University of Turku, FI-20014 Turku, Finland}

\author[0000-0003-1873-7855]{Brian van Soelen}
\affiliation{Department of Physics, University of the Free State, 9300 Bloemfontein, South Africa}




\begin{abstract}
We present a catalog of hard X-ray serendipitous sources detected in the first 80 months of observations by the \textit{Nuclear Spectroscopic Telescope Array (NuSTAR)}.
The \nustar serendipitous survey 80-month (NSS80) catalog has an unprecedented $\sim$\,62\,Ms of effective exposure time over 894 unique fields (a factor of three increase over the 40-month catalog), with an areal coverage of $\sim$\,36\,deg$^2$, larger than all \textit{NuSTAR} extragalactic surveys.
NSS80 provides 1274 hard X-ray sources in the $3-24$\,keV band (822 new detections compared to the previous 40-month catalog). Approximately 76\% of the \textit{NuSTAR} sources have lower-energy ($<10$\,keV) X-ray counterparts from \textit{Chandra}, \textit{XMM-Newton}, and \textit{Swift}-XRT.
We have undertaken an extensive campaign of ground-based spectroscopic follow-up to obtain new source redshifts and classifications for 427 sources.
Combining these with existing archival spectroscopy provides redshifts for 550 NSS80 sources, of which 547 are classified.  
The sample is primarily composed of active galactic nuclei (AGN), detected over a large range in redshift ($z$ = 0.012--3.43), but also includes 58 spectroscopically confirmed Galactic sources. In addition, five AGN/galaxy pairs, one dual AGN system, one BL~Lac candidate, and a hotspot of 4C\,74.26 (radio quasar) have been identified.
The median rest-frame $10-40$\,keV luminosity and redshift of the NSS80 are $\langle{L_\mathrm{10-40\,keV}}\rangle$\,=\,1.2\,$\times$\,10$^{44}$\,erg\,s$^{-1}$ and $\langle z \rangle = 0.56$.
We investigate the optical properties and construct composite optical spectra to search for subtle signatures not present in the individual spectra, finding an excess of redder BL AGN compared to optical quasar surveys predominantly due to the presence of the host-galaxy and, at least in part, due to dust obscuration.
\end{abstract}



\section{Introduction} \label{sec:intro}
A major focus of X-ray surveys over the last few decades has been understanding the origin of the Cosmic X-ray Background (CXB). 
The CXB was first discovered in the early 1960's \citep[see][]{giacconi1962}, several years before the identification of the cosmic microwave background \citep[CMB;][]{penzias1965,dicke1965}.
Unlike the CMB, which is truly diffuse in origin, 
the CXB is found to be dominated by the emission from high-energy distant point
sources \citep[][]{brandt2015,brandt2021}: Active Galactic Nuclei (AGNs), the observed manifestation of the accretion
of gas and dust onto a super-massive black hole \citep[see][]{lyndenbell1969}. 
Therefore the CXB essentially provides a fossil record of mass accretion onto
super-massive black holes throughout cosmic time. Consequently, ever since the discovery of the CXB
over five decades ago, a key objective of high-energy astrophysics has
been to measure the properties and evolution of AGN throughout cosmic time using 
sensitive X-ray observations.

Huge progress in the resolution of the CXB has been made using X-ray telescopes at 
low energies ($\lesssim 10$~keV).
The most sensitive X-ray surveys with \chandra \citep[e.g.,][]{hickox2007,cappelluti2017,luo2017} and \textit{XMM-Newton} \citep[e.g.,][]{moretti2003,deluca2004,worsley2005} have resolved $\approx$~70--90\% of the CXB at low energies into AGNs at $z<$~5--6.
However, the energy flux density of the CXB peaks at 20--30~keV \citep[see e.g. Fig. 2 of][]{accretion_ananna_2020} and, until recently, observatories 
in this energy range (e.g.,\ \textit{Swift}-BAT; \textit{INTEGRAL}) had only resolved $\approx$~1--2\% 
of the CXB at these energies \citep[e.g.,][]{burlon2011}. 
The great breakthrough in resolving the peak 
of the CXB comes from the \textit{Nuclear Spectroscopic Telescope Array} \citep[\textit{NuSTAR};][]{harrison2013}.
\nustar is the first orbiting
observatory with focusing optics and significant collecting area at $>10$~keV, allowing for a $\approx$~2 orders of magnitude improvement in sensitivity 
and an order of magnitude improvement in angular resolution over previous non-focusing hard X-ray missions. Importantly, the high-energy coverage at $3-79$\,keV means that \nustar selects AGNs almost 
irrespective of the absorbing column as it peels back the curtain of gas and dust, 
missing only the most heavily obscured systems (with line-of-sight column densities of $N_\mathrm{H}\geq 10^{23}$~cm$^{-2}$).
This has opened up the possibility to study large, cleanly selected samples of high-energy-emitting AGNs in the 
distant universe.

The \nustar extragalactic survey is the largest scientific project undertaken to date with \textit{NuSTAR} \citep{harrison2013,harrison2016}.
It has resolved $\approx$~35\% of the CXB at $8-24$\,keV \citep{harrison2016}, provided the first measurements of the $>10$~keV AGN luminosity function at $z>0.1$ (Aird et~al. 2015a), and identified heavily obscured AGN \citep[e.g.,][]{civano2015,lansbury2017,masini2018}. 
There are two main components to the \nustar extragalactic survey: (i) dedicated surveys of well-studied blank-fields ($\approx$~3\,deg$^2$) including COSMOS \citep{civano2015}, ECDFS \citep{mullaney2015}, EGS (Aird et al., \textit{in prep}), GOODS-N (Aird et al., \textit{in prep}), and UDS \citep{masini2018}, and (ii) a wide-area ``serendipitous survey''
performed by searching archival \nustar observations for
background X-ray sources \citep[][hereafter \citetalias{lansbury2017_cat}]{alexander2013,lansbury2017_cat}. 
The serendipitous survey is the largest component of the extragalactic survey programme,
providing the majority ($\approx$~75--80\%) of \nustar detected
sources.\footnote{We note that, although Galactic sources are identified in the \nustar serendipitous survey, the majority are extragalactic; consequently, we consider the NSS80 to be predominantly an extragalactic survey. See \citet{tomsick2017,tomsick2018} for results on galactic sources detected in the \nustar serendipitous survey.} It provides a combination of deep and shallow wide-area
coverage, which fills out the $L_\mathrm{X}$--$z$ plane and identifies
rare CXB source populations not detected in the smaller-area dedicated \textit{NuSTAR} surveys. For example, our first full catalog, the 40-month serendipitous survey catalog \citep[][]{lansbury2017_cat} contained 497 sources over 13~deg$^2$, already a factor $>$~4 larger volume than the dedicated surveys.

Here we provide an update to \citetalias{lansbury2017_cat} with the 80-month serendipitous survey catalog, hereafter NSS80. 
Due to an increase in the fraction of general observer (GO) observations compared to the 40-month survey catalog (hereafter NSS40), the areal coverage ($\sim$\,36\,deg$^2$), integrated exposure ($\sim$\,62\,Ms), number of fields (894), and number of sources (1274) in the 80-month catalog are a factor $\approx$~3 larger than those presented in \citetalias{lansbury2017_cat}. The most natural comparison survey to the NSS80 is the {\it Swift}-BAT survey \citep{baumgartner_70_2013,oh2018}, which has identified $\approx$~1600 sources at $>10$~keV in 105-months of observations over the entire sky. Comparable serendipitous X-ray surveys have also been undertaken and regularly updated with the \textit{Chandra} \citep{evans_chandra_2010,evans2019_csc2}, \textit{XMM-Newton} \citep{webb2020}, and {\it Swift}-XRT \citep{evans2014,evans2020_2sxps} but at lower X-ray energies. A substantially greater number of X-ray sources are detected in these surveys ($\approx$~200,000--550,000) due to their larger areal coverage and/or greater relative X-ray sensitivity than NSS80 at $<10$~keV. These catalogs, both individually and in combination, provide a wealth of resource to the X-ray astronomy community, greatly improving the range of possible studies. The \nustar serendipitous survey (higher energy than \textit{Chandra}, \textit{XMM-Newton}, and {\it Swift}-XRT; more sensitive and higher resolution than {\it Swift}-BAT) is an important member of that line-up.

Our aim in this paper is to present an update to the \nustar serendipitous survey catalogue, including salient information on the reduction of the \nustar data and construction of the catalogue, the identification of multi-wavelength counterparts, spectroscopic follow-up observations and identifications, in addition to some brief scientific analyses to motivate further in-depth studies with the NSS80. This approach is consistent with our previous \nustar survey work \citep[e.g.,][]{alexander2013,civano2015,harrison2016,lansbury2017_cat,masini2018}. In Section~\ref{sec:nustar data} we detail the \nustar observations, data reduction and source-detection to construct the 80-month catalog. We search for counterparts at lower X-ray energies from \textit{Chandra}, \textit{XMM-Newton}, and \textit{Swift}-XRT, described in Section~\ref{subsec:soft xray cpart}, and utilize a probabilistic approach with \textsc{Nway} to reliably cross-match to infrared and optical counterparts, described in Section~\ref{subsec:nway infrared and optical cpart}.\footnote{\textsc{Nway} provides an improvement on simple distance-based matching, using a wider array of information to find likely matches; see \cite{salvato2018}.}
To obtain spectroscopic identifications for the NSS80 sources (redshifts and classifications), we undertake an extensive follow-up campaign with ground-based optical telescopes at multiple latitudes (Section~\ref{subsec:optical spectroscopy}). 
To characterise the properties of the NSS80 sources, we use X-ray, multi-wavelength photometry, and optical spectroscopy in Section~\ref{sec:nustar science}. The basic X-ray properties of the extragalactic NSS80 sample are given in Section~\ref{subsec:xray prop}, the MIR properties of the sources are examined in Section~\ref{subsec:wise properties nustar}, and in Section~\ref{subsec:optical prop} we explore the optical properties of the AGN, with a particular focus on red quasars and the utilization of composite spectra to determine the origin of their observed optical colors. Finally, in Section~\ref{sec:summary nustar} we draw conclusions and summarise our results. 
We assume a concordance flat $\Lambda$-cosmology with $H_0$ = 70\,km\,s$^{-1}$\,Mpc$^{-1}$ , $\Omega_{M}$ = 0.3, and $\Omega_\Lambda$ = 0.7.


\section{The \nustar data} \label{sec:nustar data}
The \nustar observatory \citep{harrison2013} was launched in 2012 and consists of two grazing-incidence telescopes that focus X-rays onto two focal plane modules (FPMA and FPMB) which cover a co-aligned field-of-view of $\approx 12'\times12'$.
\nustar is sensitive to photons across the $3-79$\,keV energy range and achieves an angular resolution of 18$''$ FWHM and a half power diameter of 58$''$, which enables 2 orders of magnitude improvement in sensitivity over previous X-ray missions with sensitivity to hard ($\gtrsim10$keV) energies.
In this work we present our analysis and results for the $3-8$\,keV (soft band), $8-24$\,keV (hard band) and $3-24$\,keV (full band) energy bands \citep[following the energy bands adopted in previous \nustar survey work, see][]{alexander2013,luo2014,aird2015,lansbury2015,harrison2016,lansbury2017_cat}, with the $3-24$\,keV band being our main focus since it provides the best sensitivity for the detection of relatively faint sources in the \nustar extragalactic surveys.\footnote{Adopting a broad band pass ensures that faint sources with low photon counts are still likely to be detected, compensating for the drop in sensitivity at higher energies ($\sim8-24$\,keV) due to the decrease in the effective area with increasing energy and the increased relative contribution of instrumental background compared to lower energies.}

In the following subsections we outline the selection of the \nustar observations utilized in NSS80 (Section~\ref{subsec:serendip obs}), describe the data-processing and source-detection approaches (Section~\ref{subsec:data processing}), summarise the properties of the serendipitous survey source catalog (Section~\ref{subsec:source catalog}), and highlight key changes between the 40-month and 80-month catalogs (Section~\ref{subsec: differences in catalog to L17}). 

\subsection{The Serendipitous Survey Observations} \label{subsec:serendip obs}
The NSS80 is comprised of observations taken by \nustar over the period from 2012 July to 2019 March and provides a significant update over the NSS40 reported in \citetalias{lansbury2017_cat} (2012 July to 2015 November).

The \nustar serendipitous survey is constructed by searching the background regions for sources that are not associated with the original science target in almost every \nustar pointing that is not associated with a dedicated survey field. Following \citetalias{lansbury2017_cat} we excluded observations from: 
\begin{itemize}
	\vspace{-0.1cm}
	\setlength\itemsep{0.3em}
	\item dedicated extragalactic survey fields: COSMOS \citep{civano2015}, ECDFS \citep{mullaney2015}, EGS (Aird et al., \textit{in prep}), GOODS-N (Aird et al., \textit{in prep}), and UDS \citep{masini2018};
	\item Galactic surveys \citep[][]{mori2015,hong2016}, i.e., all fields within a 2 degree radius of the Galactic Center;
	\item the Norma Arm survey \citep{fornasini2017};
	\item fields where the total counts exceed $10^6$ within 120$''$ of the on-axis position due to a bright science target.
\end{itemize}
In addition, prior to processing the data, we also excluded solar system fields (i.e., solar, lunar and planetary observations), nebular fields (e.g., supernova remnants), galaxy clusters, and fields of nearby galaxy nuclei (e.g., M31). 
We further excluded fields found to have bad exposure maps (i.e., bad/hot pixels in exposure maps) or if more than two thirds of the field is contaminated by excess background emission (see Section~\ref{subsec:data processing} and Figure~\ref{fig: flowchart data processing}). \\

\begin{figure}
	\centering
	\includegraphics[width=20pc]{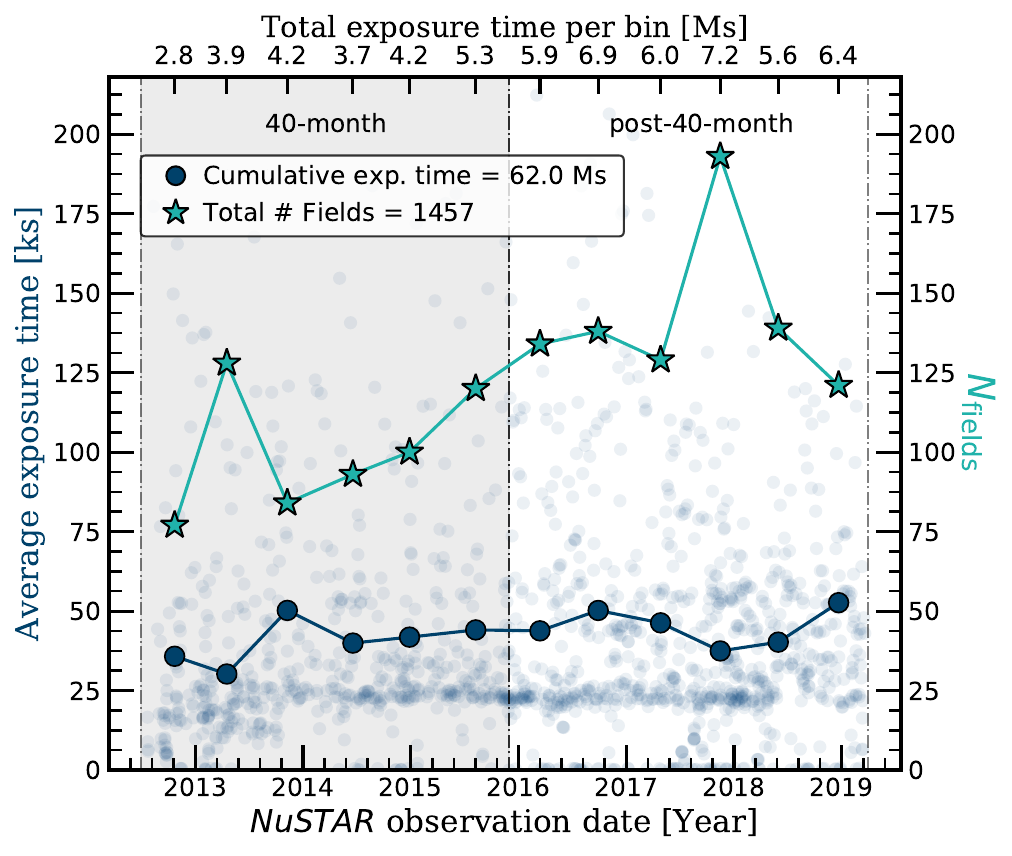}
	\vspace*{-7mm}
	\caption{The average exposure time (blue circles) and the number of \nustar exposures (green stars) roughly per semester over the full 80-month period. 
		In general, the average exposure time shows little variation between the 40-month (shaded grey area) and post-40-month observations, whilst we see an increase in the number of exposures from the 40-month to the post-40-month survey, resulting in an increase in the total exposure time per semester which can be utilized for the \nustar serendipitous survey. 
		In total, the 80-month serendipitous survey includes 1457 individual exposures with a cumulative exposure time of 62~Ms.
	}
	\label{fig:cumulative exptime}	
\end{figure}

Table~\ref{tab:survey stats} provides a summary of the \nustar serendipitous survey information. 
Overall, NSS80 comprises 1457 individual \nustar exposures, performed over 894 unique fields, the majority of which come from post-NSS40 observations. These fields yield an overall sky coverage of 36~deg$^2$ and a cumulative exposure time of 62.0~Ms, both a factor of $\sim 3$ increase over NSS40 as shown in Table~\ref{tab:survey stats}.
Figure~\ref{fig:cumulative exptime} shows how the number of fields included in the NSS samples has gradually increased over time, while the average exposure per pointing has remained roughly constant, driving this overall increase in both the sky coverage and total exposure time.
A key contribution to the increase in the number of fields with time is the larger number of GO versus Legacy observations, since the dedicated Legacy survey fields (described above) are excluded from the serendipitous survey. A further contribution to the increased area of NSS80 can be attributed to the \textit{Swift}-BAT snapshot survey \citep[e.g.,][]{oh2018}, which is a Legacy survey comprising of multiple, short exposures. 
Furthermore, two distinctive spikes in the number of fields are evident: spike 1 (bin 2) includes 128 exposures before GO observations were undertaken with \nustar during the period January -- August 2013, and spike 2 (bin 10) includes 193 Cycle 3 GO observations performed during the period of August 2017 -- February 2018.
These spikes coincide with a decrease in the average exposure time indicating that more shallow observations were scheduled in these periods.

\begin{deluxetable}{lllll}
\tablecolumns{5}
\tablewidth{\textwidth}
\tablecaption{\label{tab:survey stats} Comparison of included observations in the NSS40 and NSS80 surveys.}
\tablehead{& & NSS40 & post-NSS40 & NSS80 \\
			& & (\citetalias{lansbury2017_cat}) & & \\}
			\startdata
			(1)& \thead[l]{Obs. start date} & 2012/08 & 2015/12 & 2012/08 \\ 
			(2)& \thead[l]{Obs. end date} & 2015/11 & 2019/04 & 2019/04 \\
			(3)& \thead[l]{Individual exposures} & 510 & 947 & 1457 \\
			(4)& \thead[l]{Unique fields} & 331 & 563 & 894 \\
			(5)& \thead[l]{Cumulative exp. time} &20.4~Ms & 41.6~Ms & 62.0~Ms \\
			(6)& \thead[l]{Sky coverage } & 13~deg$^2$ & 
            \enddata
	\tablecomments{
		{\sc Rows}:\,{\bf (1) \& (2)} Observation date range for specific survey.
		{\bf(3)} Number of individual exposures. 
		{\bf(4)} Number of unique fields, each with contiguous coverage comprised of one or more \nustar exposures.
		{\bf(5)} Cumulative exposure time in Ms. 
		{\bf(6)} Total sky area coverage in deg$^2$.}
\end{deluxetable}
Table~\ref{tab:fields} lists the individual exposures in alphabetical order for the first 5 unique fields (i.e.~comprising of non-overlapping pointings) in the NSS80 (the full table will be made available online) and provides details including the number of observations and the number of serendipitous sources detected in each unique field.
For 28\% (247/894) of the fields there are multiple \nustar exposures, ranging between two and 15 observations, which are combined together into a single mosaic (see Section~\ref{subsec:data processing} and Figure~\ref{fig: flowchart data processing}). 
The serendipitous survey fields have a median exposure time of 34\,ks but cover a wide range in individual exposure times (from $\sim$\,100\,s to 1\,Ms). 

\begin{deluxetable}{lllllllllll}
\tablecolumns{11}
\tablewidth{\textwidth}
\tablecaption{\label{tab:fields} Details of the individual NSS80 \nustar observations for the first 5 unique fields (the full table will be made available online).}
\tablehead{ID & Science Target & NSS40 & $N_{\rm obs}$ & Obs. ID & Obs. Start Date & R.A. & Decl. & $b$ & $t_{\rm exp}$ & $N_{\rm serendips}$ \\
			& & & & & & (deg) & (deg) & (deg) & (ks) & \\
			(1) & (2) & (3) & (4) & (5) & (6) & (7) & (8) & (9) & (10) & (11) \\}
			\startdata
			1  & 1A\_0535p262\_SADA\_18360  & ... & 2   & ...         & ...                & 82.00  & 26.00  & -4.87 &  0.9   & 0   \\
			1a & ...                        & ... & ... & 90401371001 & 2018 December 26   & ...    & ...    & ...   &  0.2   & ... \\
			1b & ...                        & ... & ... & 90401371002 & 2018 December 27   & ...    & ...    & ...   &  0.7   & ... \\
			2  & 1E1048d1m5937              & ... & 5   & ...         & ...                & 162.53 & -59.89 & -0.52 &  397.9 & 9   \\
			2a & ...                        & ... & ... & 30001024003 & 2013 July 17       & ...    & ...    & ...   &  25.7  & ... \\
			2b & ...                        & ... & ... & 30001024002 & 2013 July 17       & ...    & ...    & ...   &  26.6  & ... \\
			2c & ...                        & ... & ... & 30001024005 & 2013 July 19       & ...    & ...    & ...   &  167.7 & ... \\
			2d & ...                        & ... & ... & 30001024007 & 2013 July 25       & ...    & ...    & ...   &  119.1 & ... \\
			2e & ...                        & ... & ... & 90202032002 & 2016 August 05     & ...    & ...    & ...   &  58.9  & ... \\
			3  & 1E1530m085                 & 303 & 1   & 60061265002 & 2015 August 07     & 233.34 & -8.70  & 36.88 &  23.1  & 0   \\
			4  & 1E161348m5055              & ... & 3   & ...         & ...                & 244.37 & -50.92 & -0.27 &  283.2 & 3   \\
			4a & ...                        & ... & ... & 90201028002 & 2016 June 25       & ...    & ...    & ...   &  70.7  & ... \\
			4b & ...                        & ... & ... & 30301017002 & 2017 June 02       & ...    & ...    & ...   &  70.3  & ... \\
			4c & ...                        & ... & ... & 30301013002 & 2018 April 29      & ...    & ...    & ...   &  142.3 & ... \\
			5  & 1E1841m045                 & ... & 6   & ...         & ...                & 280.33 & -4.94  & -0.01 &  346.3 & 7   \\
			5a & ...                        & 29a & ... & 30001025002 & 2012 November 09   & ...    & ...    & ...   &  52.4  & ... \\
			5b & ...                        & 29b & ... & 30001025004 & 2013 September 05  & ...    & ...    & ...   &  37.8  & ... \\
			5c & ...                        & 29c & ... & 30001025006 & 2013 September 07  & ...    & ...    & ...   &  70.9  & ... \\
			5d & ...                        & 29d & ... & 30001025008 & 2013 September 12  & ...    & ...    & ...   &  41.6  & ... \\
			5e & ...                        & 29e & ... & 30001025010 & 2013 September 14  & ...    & ...    & ...   &  35.4  & ... \\
			5f & ...                        & 29f & ... & 30001025012 & 2013 September 21  & ...    & ...    & ...   &  108.2 & ... \\
		\enddata
	\tablecomments{ {\sc Columns:}\,{\bf (1)} ID assigned to each field. For fields with multiple \textit{NuSTAR} exposures (i.e., $N_\mathrm{obs}$ $>$ 1), each individual component exposure is listed with a letter suffixed to the field ID (e.g., 1a and 1b). 
		{\bf (2)} Object name for the primary science target of the \nustar observation(s). 
		{\bf (3)} ID assigned to each field in the NSS40 (\citetalias{lansbury2017_cat}).
		{\bf (4)} The number of individual \nustar exposures for a given field. 
		{\bf (5)} \nustar observation ID. 
		{\bf (6)} Observation start date. 
		{\bf (7)} and {\bf (8)} Right ascension and declination (J2000) coordinates for the aim-point, in decimal degrees. 
		{\bf (9)} The IAU Galactic latitude for the aim-point, in decimal degrees \citep{blaauw1960}.
		{\bf (10)} Exposure time (“ONTIME” in the \nustar image header; ks), for a single FPM (i.e., averaged over FPMA and FPMB).
		For multiple exposures the total exposure time is recorded. 
		{\bf (11)} The number of serendipitous \nustar sources detected in a given field. }
\end{deluxetable}	

\begin{figure*}
	\centering
	\includegraphics[width=42pc]{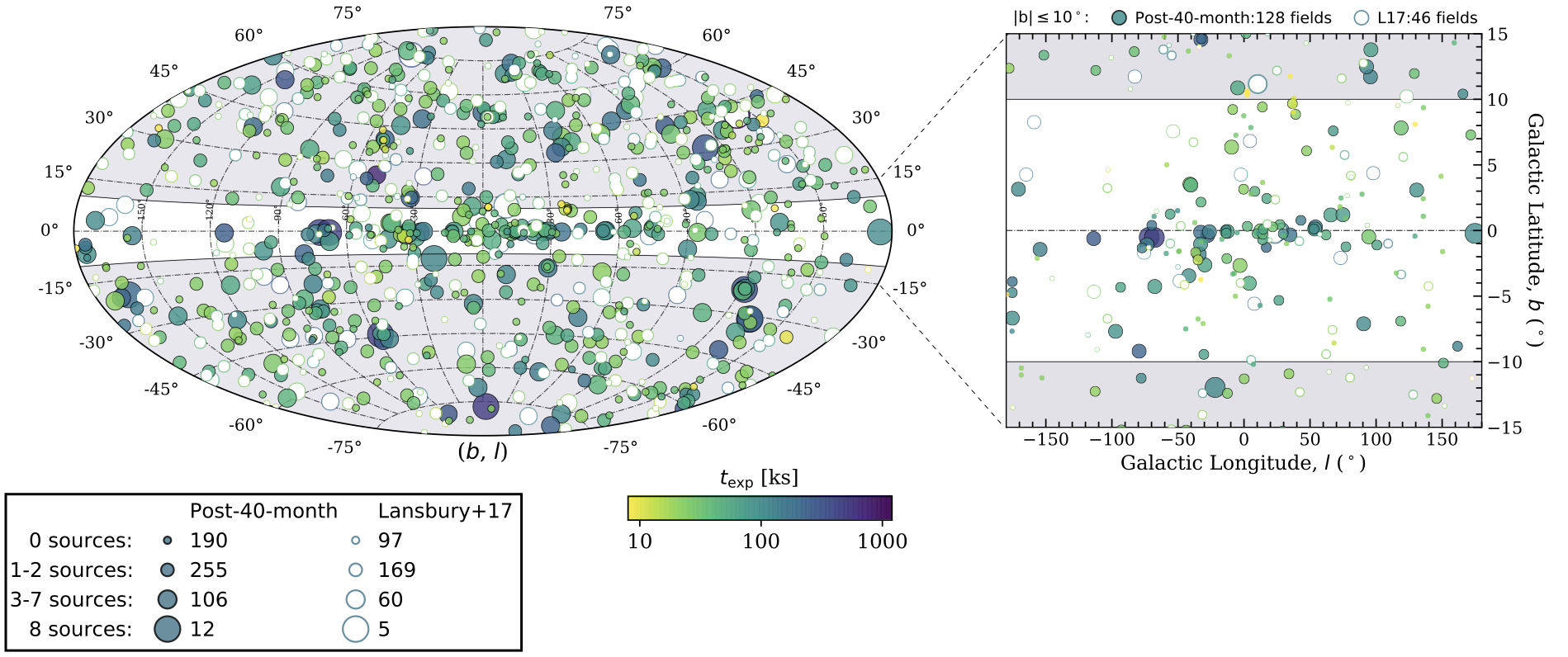}	
	\vspace*{-1mm}
	\caption{Aitoff projection showing the distribution of the \nustar serendipitous survey fields on the sky, in Galactic coordinates.
		The white and color filled circles show the NSS40 and the post-NSS40 data, respectively. 
		The circle sizes correspond to the number of sources detected in a given field, and the colors correspond to the cumulative exposure time (per FPM) for a given field. The white area highlights the region $\pm$10$^\circ$ of the Galactic plane.
	}
	\label{fig:aitoff}	
\end{figure*}

In Figure~\ref{fig:aitoff} we show an all-sky map of the 894 unique \nustar fields, color coded by average exposure time. 
The white filled circles represent the NSS40, whereas the post-NSS40 fields are shown with filled circles.
Both the NSS40 and the post-NSS40 fields comprise pointings across the whole sky, with the latter having a higher density of observations, as also shown in Figure~\ref{fig:cumulative exptime}. 
In comparison to the NSS40, the number of post-NSS40 fields has increased by $\approx$\,50\% for a given amount of serendipitous sources per field; e.g., 169 NSS40 fields have 1-2 detected sources whereas 255 post-NSS40 fields have 1-2 serendipitous detections (see Section~\ref{subsec:data processing} for further information on source detection procedures).  
The zoom-in panel shows the Galactic plane fields with $|b| \leq 10^\circ$: 174 of the 894 NSS80 fields (19\%) lie within the Galactic plane, also a factor of $\sim$\,3 increase over that of NSS40. 

\subsection{Data processing and source-detection} \label{subsec:data processing}
\begin{figure*}
	\centering
	\includegraphics[width=40pc]{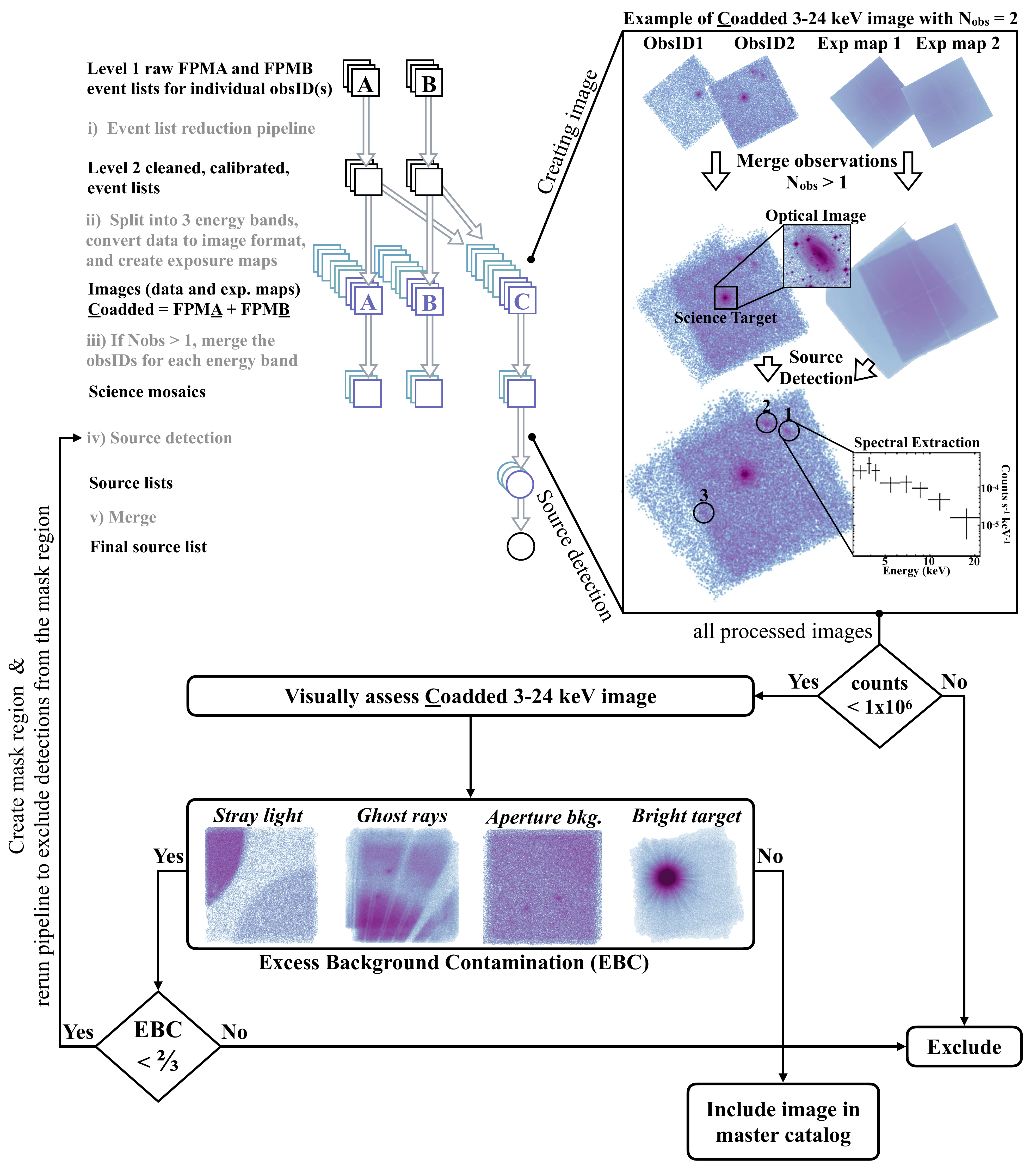}		
	\vspace*{-3mm}
	\caption{ 
		{\bf Upper left panel: }Flowchart schematic illustrating the \nustar data processing steps undertaken to reduce individual fields for the serendipitous survey.
		A field can comprise multiple exposures (obsIDs) mosaicked into a single counts image as illustrated on the right. Data from the two telescopes, FPM\underline{A} and FPM\underline{B}, and the \underline{C}oadded FPMA+B are indicated as ``A'', ``B'' and ``C'', respectively, and the three energy bands are indicated in different colors ($3-8$\,keV, 8$-$24\,keV and $3-24$\,keV).
		{\bf Upper right panel: } Example of real counts images and exposure maps associated with stages (iii) to (v) for one of the included processed fields, i.e., IC\,2560. The data shown are for the coadded images and the $3-24$\,keV energy band only. 
		This illustrates the mosaicking of two exposures with different exposure times and orientations.
		Information from both the image and the exposure map mosaics is used to perform source-detection (step (iv)).  
		In this example, three serendipitous sources (circled) are detected.
		{\bf Lower panel: } Each image mosaic with counts $< 1\times10^{6}$ is visually assessed to identify regions with excess background contamination (EBC) which includes stray light, ghost rays, aperture background and bright science target. If less than two-thirds of the image is contaminated by excess background, a mask region is created and used to mask detections from the final source list. Images with counts $> 1\times10^{6}$ or with EBC covering more than two-thirds of the image are excluded from the catalog (38 exposures were removed in this way). 
	}
	\label{fig: flowchart data processing}	
\end{figure*}

The reduction of the new (i.e.~post-40~month) \nustar serendipitous fields followed the custom pipeline procedure described by  \citetalias{lansbury2017_cat}, which is broadly consistent with the approach adopted in our previous \nustar survey studies \citep{mullaney2015,aird2015,harrison16}. An overview of the \nustar data-reduction steps is shown in Figure~\ref{fig: flowchart data processing} and described briefly here, highlighting updates to the \citetalias{lansbury2017_cat} procedure (see Section~\ref{subsec: differences in catalog to L17} below for further discussion of differences between the final NSS80 and the prior NSS40 catalogs).

Briefly, the raw event files were processed using the {\sc nupipeline} procedure from the \nustar Data Analysis Software (NuSTARDAS) v1.9.2 \footnote{\url{https://heasarc.gsfc.nasa.gov/docs/nustar/analysis/}} (incorporated within the HEASoft v6.24 software suite) and CIAO v4.7.0 to produce calibrated event files.
These event files were used to produce counts images for each individual \nustar exposure (obsID), which comprises FPMA and FPMB data in the full, soft and hard energy bands; we note the pixel size is 2.46$"$. 
We produced exposure maps that account for the vignetting across the field-of-view for each energy band (at fixed representative energies of 9.88~keV, 5.42~keV and 13.02~keV for the full, soft and hard bands, respectively) as well as a single exposure map that does not include vignetting effects (used to estimate the background count rate at each pixel; see below).

To optimise the depth of our datasets, we coadded the images and exposure maps from the FPMA and FPMB detectors.
This resulted in a total of 9 images and 9 exposure maps per field (3 energy bands for FPMA, FPMB and FPMA+B).\footnote{The uncombined FPMA and FPMB exposure maps are useful to access regions of excess background contamination; see lower panel of Figure~\ref{fig: flowchart data processing}.} 
We produced single mosaics for each energy band by combining observations covering the same sky region within 12$'$ of the aim point for each obsID (step (iii) in Figure~\ref{fig: flowchart data processing}).
There are 54 observations previously included in NSS40 which were coadded with more recent observations and re-analysed for NSS80. 
We note that this can lead to small changes in the resulting source lists, including the detection of fainter sources or the loss of sources close to the detection limit (see further details in Section~\ref{subsec:source catalog}).

Source-detection [see step (iv) in Figure~\ref{fig: flowchart data processing}] was performed as described in \citetalias{lansbury2017_cat}. 
To summarize, we first produced a ``false probability" map for each energy band,
\footnote{Regions close to the edge of the FoV, with $<$\,10\% of the maximum exposure in the 3--24~keV band and thus where the background is poorly characterised, are excluded from the source detection process and estimates of survey area coverage.} which gives the probability that the observed counts within a circular aperture of 20\arcsec\ radius \citep[justified by the tight core of the \textit{NuSTAR} PSF; see][]{civano2015,mullaney2015} were produced purely by a fluctuation of the background.
The expected background is estimated by convolving the image counts with an annular aperture of inner radius 45\arcsec\ and outer radius 90\arcsec and re-scaling to the 20\arcsec\ source detection region.
These background estimates incorporate counts from any bright target which will impact the sensitivity to faint serendipitous sources. 
Finally, we created source lists by identifying distinct regions where the false probability is less than $10^{-6}$ \citep[equivalent to $\sim$\,5$\sigma$; see][for full details]{mullaney2015} using the SExtractor software \citep{bertin1996}.
To produce a master source list, we merged the source lists for the three individual energy bands 
and removed any sources within a 90\arcsec\ radius of the science target position. 
See \citetalias{lansbury2017_cat} and references therein for a full description\footnote{As in \citetalias{lansbury2017_cat}, final source positions are (in order of priority) full-band, soft-band, then hard-band.}.

Once the master source list was created, we visually inspected all the post-processed fields (lower panel in Figure~\ref{fig: flowchart data processing}) to identify and mask out extended areas that exhibit a high background rate due to stray light, ghost rays, aperture background\footnote{The aperture background refers to unfocussed X-rays that pass between the optics and the focal plane i.e.~enter via the unbaffled ``sides'' of the telescope \citep[see ][for further detail]{wik2014}.} and/or emission from the science target (illustrated in the lower panel of Figure~\ref{fig: flowchart data processing}). 
These custom-made masks were then applied to both the background estimate and source-detection procedures, to produce the final source list. 
We excluded fields from our analysis when the excess background contamination exceeded two thirds of the field (based on a visual estimation), resulting in the removal of 38 fields.
In addition to masking excess background contamination, \citetalias{lansbury2017_cat} also created custom-made regions to mask out sky areas which are clearly overlapping with extended optical/IR counterparts associated with the \textit{NuSTAR} science target. 
However, for NSS80 we included this as a later post-processing step to obtain an X-ray catalog independent of optical/IR information (see Section~\ref{subsec:source catalog}). 
To be consistent in the construction of NSS80 we therefore removed the masked regions for 31 of the 40-month fields with highly extended optical hosts and reprocessed the data. For further details see Section~\ref{subsec:source catalog}.   

Following \citetalias{lansbury2017_cat}, we measured count rates, fluxes, net source counts and its errors for each detected source, and estimated upper limits for sources undetected in a given band using the Bayesian approach from \citet{kraft1991}.
We also applied the deblending procedure described in Section 2.3.2 of \citet{mullaney2015}, which increases the background estimates for a given source due to the contribution of any other nearby serendipitously detected sources that will not be accounted for in our smoothly varying background maps.
We then re-assessed the false probability of these sources using the updated background estimates and excluded sources that no longer met our false probability detection threshold in at least one of the energy bands.
This process assumes that the sources are all not capable of being resolved by \nustar and are considered effectively point-like.

\begin{figure}
	\centering
	\includegraphics[width=20pc]{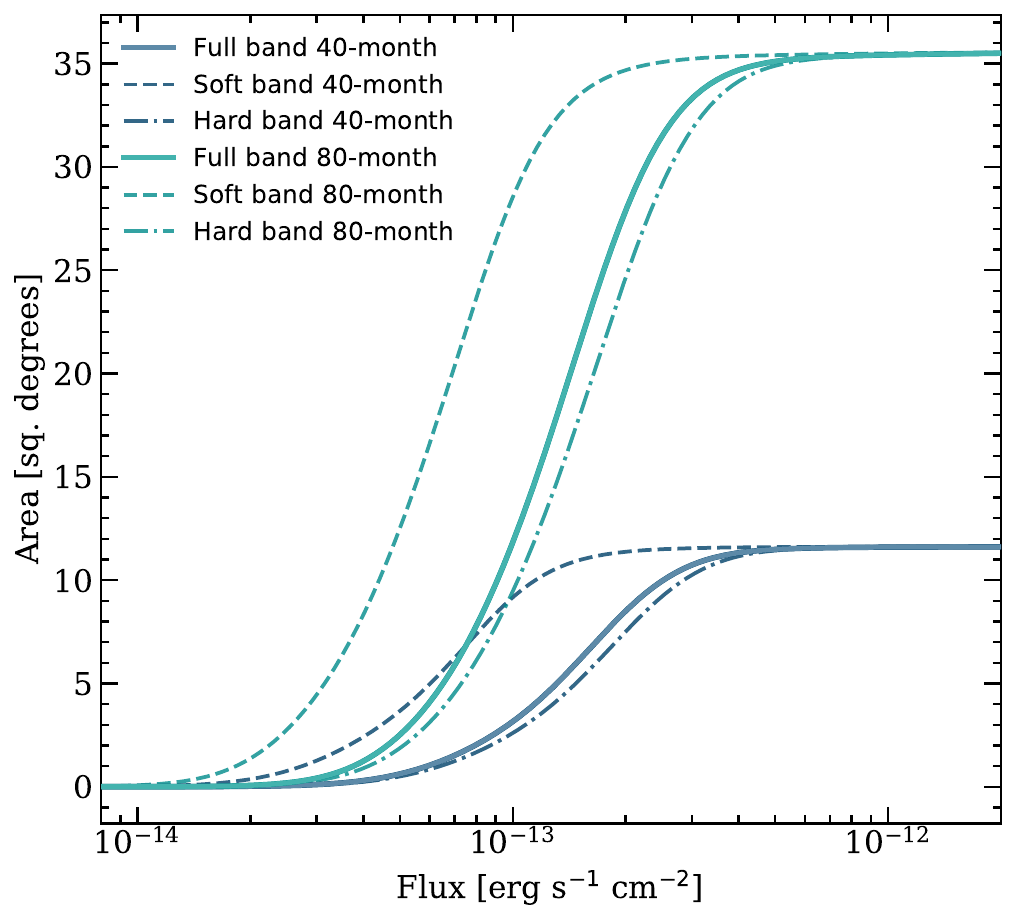}		
	\vspace*{-6mm}
	\caption{Sky coverage (solid angle) of the NSS40 (blue) and NSS80 (green) surveys as a function of aperture-corrected flux sensitivity, for the three main energy bands, i.e., full ($3-24$~keV), soft ($3-8$~keV) and hard ($8-24$~keV). Note the factor of $\sim$3 increase in the sky coverage with \nustar between the 40-month and 80-month catalogs. 
	The sensitivity curves include fields at all Galactic latitudes for both NSS40 and NSS80 (cf. 
	the curves shown in \citetalias{lansbury2017_cat} for both the full survey and the subset of fields that lie outside the Galactic plane, $|b| > 10^\circ$).} 
	\label{fig:sky coverage}	
\end{figure}
\begin{figure}
	\centering
	\includegraphics[width=20pc]{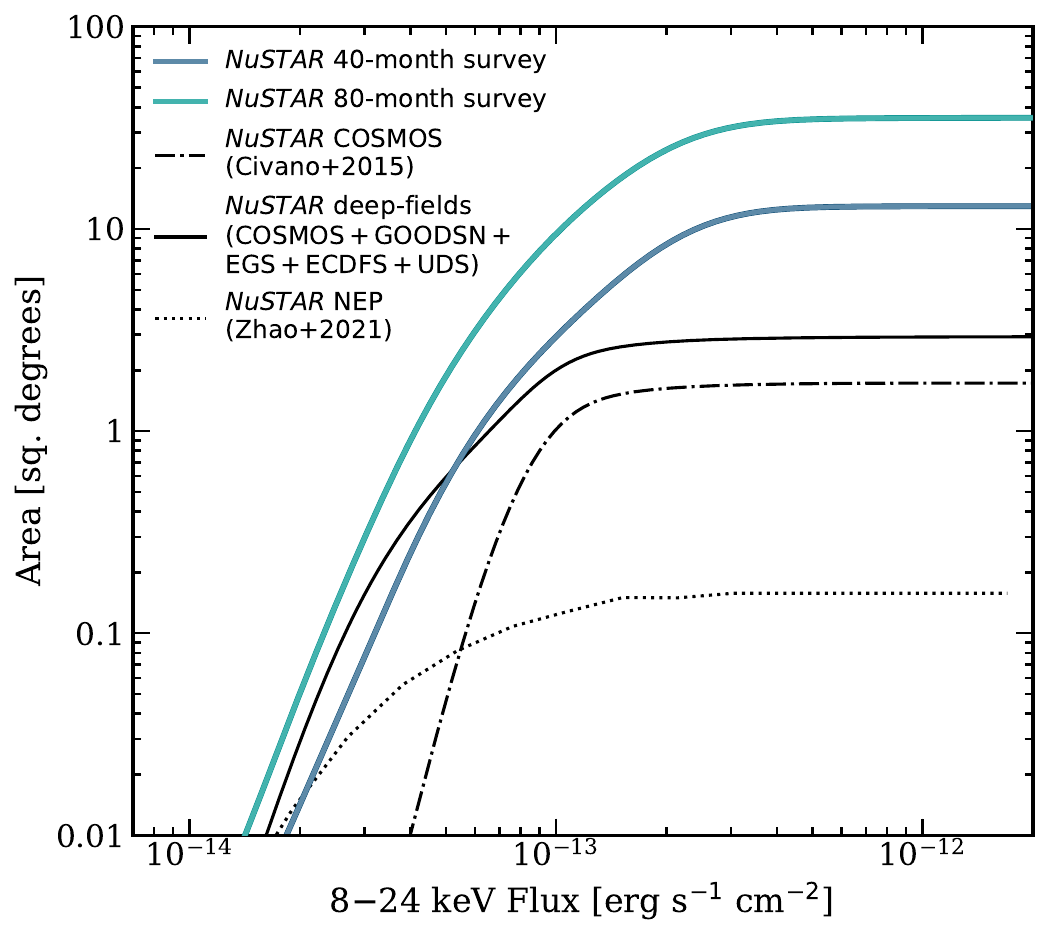}		
	\vspace*{-6mm}
	\caption{Sky coverage of the NSS40 and NSS80 surveys as a function of flux sensitivity, for the hard ($8-24$~keV) energy band. The green and blue solid lines show the area curves for the overall NSS40 and NSS80 surveys, respectively. We compare with the other completed components of the \textit{NuSTAR} extragalactic surveys program, which include the following dedicated deep-field surveys: 
	COSMOS (dash-dotted black line), ECDFS, EGS, GOODSN and UDS}. The total area for these deep-field surveys is shown as a black solid line. We also compare to cycle 5 of the \textit{NuSTAR} extragalactic survey of the James Webb Space Telescope North Ecliptic Pole (NEP) time-domain field shown in a dotted black line \citep{zhao2021}.
	\label{fig:sky coverage hard band}	
\end{figure}

To determine the sensitivity curve for a given background and exposure map, we calculated the flux limit at the detection threshold for every point in the \nustar image, with the exclusion of the peripheral regions and any regions that are masked due to high background as described above or corresponding to extended optical galaxies and nearby galaxy clusters (see Section~\ref{subsec:source catalog} below).
We then summed the sensitivity curves of the 894 unique fields for each of the three energy bands to obtain the total areal coverage of the NSS80, which results in a factor $\sim$3 increase in sky coverage compared to NSS40 (see Figure~\ref{fig:sky coverage}).
In Figure~\ref{fig:sky coverage hard band}, we also compare NSS40 and NSS80 to the dedicated \nustar deep-field surveys collectively and the \nustar survey of the North Ecliptic Pole (NEP) region \citep{zhao2021}. 
The NSS80 has the largest areal coverage at all fluxes but is most comparable to the deep-field surveys near the low flux tail. 
Consequently the combination of NSS80 with the deep-field surveys allows for a factor $\sim$2 improvement in analyses of the faint end of the hard X-ray source population, in addition to an order of magnitude increase at brighter fluxes. 

With the total area coverage, we can estimate the number of spurious X-ray detections in NSS80 due to background fluctuations. Our 36~deg$^2$ survey corresponds to $\approx$370,000 independent 20\arcsec radius regions, which with our strict (i.e.~low) false probability threshold of $10^{-6}$ \citep[cf. the higher thresholds adopted in][]{mullaney2015,civano2015} corresponds to an expectation of 0.37 spurious sources in a given band \citep{nandra2005}. We thus expect 1.11 spurious X-ray sources due to performing source detection independently over three bands, although we note that this number is conservative as the 3--24~keV band overlaps with the other energy bands and is thus not completely independent.

\subsection{The Serendipitous Survey Source catalog} 
\label{subsec:source catalog}
\begin{figure*}
	\centering
	\includegraphics[width=42pc]{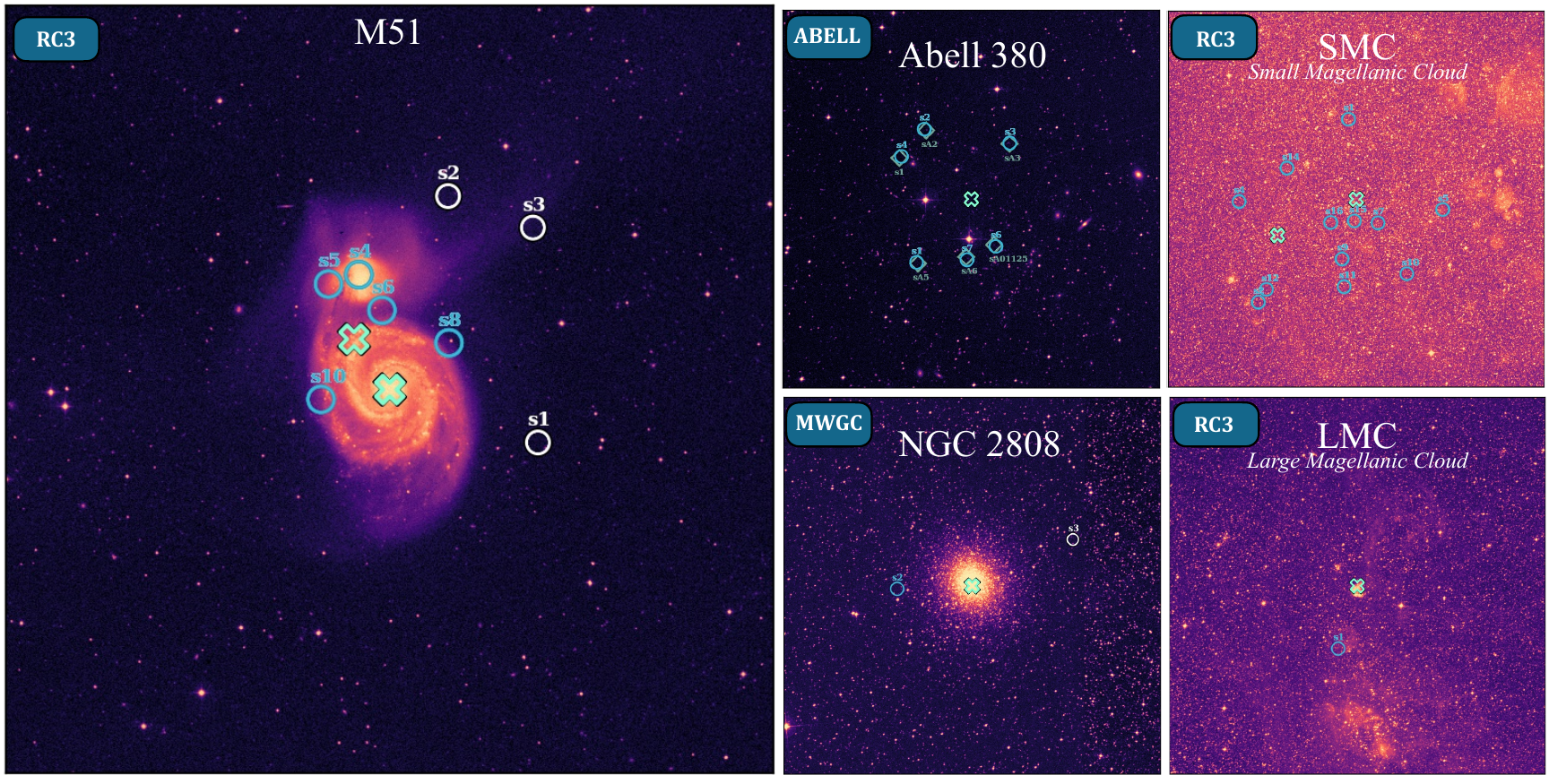}		
	\vspace*{-1mm}
	\caption{Example DSS $R$-band cutouts centred on the \nustar aim point for a given observation (green cross).  Sources within the radius of RC3 galaxies \citep[Third Reference Catalog of Bright Galaxies;][]{deVaucouleurs1991}, Milky Way (Galactic) globular clusters (MWGC) or Abell clusters are flagged as secondary sources and indicated with blue circles. The white circles and the teal-green diamonds mark primary and \citetalias{lansbury2017_cat} serendipitous sources, respectively (see Figure~\ref{fig:cutout example} for further details). 
	}
	\label{fig:cutout optical mask}	
\end{figure*}

\begin{figure*}
	\centering
	\includegraphics[width=42pc]{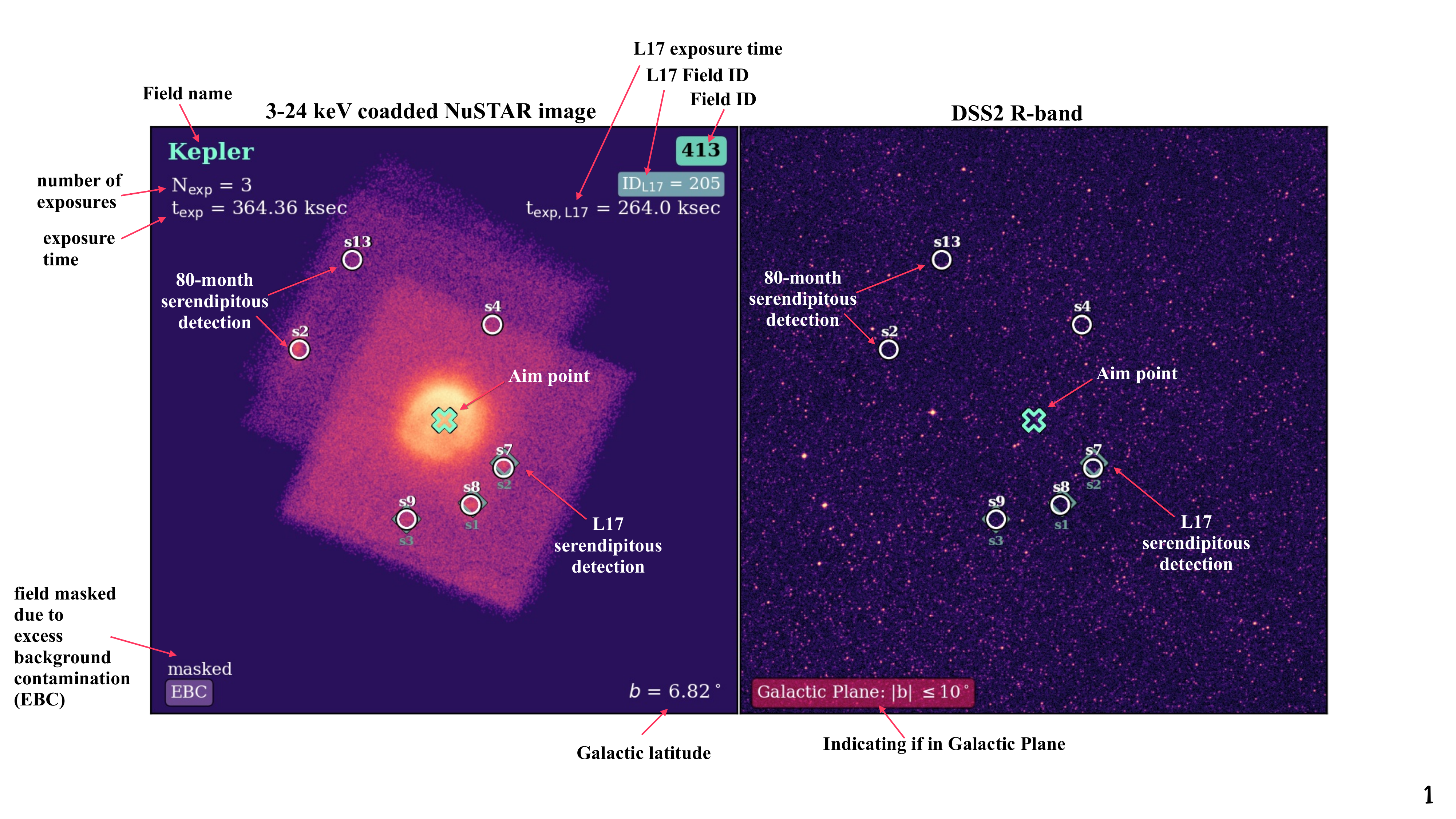}		
	\vspace*{-3mm}
	\caption{Example image of the NSS80 survey library. The online library will be accessible at \url{https://www.nustar.caltech.edu/page/59}.
		\textit{Left cutout:} The $3-24$\,keV coadded \nustar image of one of the 894 NSS80 unique fields, Kepler (field ID 413); $N_\mathrm{exp} = 3$ individual exposures (each 12$'$\,$\times$\,12$'$) are combined with a total exposure time of $t_\mathrm{exp} = 364.36$\,ks. For comparison, the NSS40 field ID and total exposure time is recorded in the top right corner. 
		NSS80 detections are marked with white circles and the corresponding serendip number and NSS40 detections are shown with teal-green diamonds -- 3 additional \nustar serendipitous sources are detected in the new post-NSS40 data. 
		The science target is marked with a green cross at the centre of each field (aim point). Fields which are masked post processing due to excess background contamination (e.g., stray light, ghost rays, bright science target) are flagged in the bottom left corner, and the Galactic latitude of the science target is shown in the bottom right corner.
		\textit{Right cutout:} DSS2 $R$-band image centred on the \nustar science target position. The same labelling as on the left is used to indicate NSS40 and NSS80 serendipitous sources. Galactic plane fields with latitudes $|b| < 10^\circ$ are flagged in the bottom left corner. 
	}
	\label{fig:cutout example}	
\end{figure*}

The master source list comprises 1488 serendipitous \nustar sources that are significantly detected in at least one energy band, independently of any prior multi-wavelength information.
Based on findings in \citetalias{lansbury2017_cat}, the majority of the X-ray detected sources in NSS80 are expected to be AGN which should reside in background field galaxies that are not associated with the science target.
However, due to the high-sensitivity of \nustar and the large areal coverage of NSS80, a small and non-negligible fraction are X-ray emitting sources within nearby highly extended galaxies associated with the science target (e.g.,\ X-ray binaries and ultra-luminous X-ray sources), X-ray AGN residing in nearby galaxy clusters, or X-ray emitting sources within the Galaxy. 
In NSS80 we therefore distinguish between X-ray detected sources lying within highly extended optical galaxies and nearby galaxy clusters from those hosted in fainter field galaxies. To enable easy and efficient use of NSS80, all \nustar sources residing in highly extended optical galaxies and galaxy clusters are placed in a secondary catalog to complement the primary catalog that is dominated by AGN in field galaxies. 
To identify \nustar sources in highly extended optical galaxies or nearby galaxy clusters we selected sources that lay within 
\begin{itemize}
	\vspace{-0.2cm}
	\setlength\itemsep{-0.1em}
	\item the isophotal radius ($D_{25}$) of RC3 galaxies \citep[Third Reference Catalog of Bright Galaxies;][]{deVaucouleurs1991} where the $R$-band surface brightness $\mu_{R}$ = 25 mag arcsec$^{-2}$, including the SMC and LMC; or 
	\item the radii of Abell clusters obtained from \citet{abell1958} or, if unavailable, a median value of 0.5$'$ as a radius; or
	\item the 2$\times$ half-mass radius of Galactic globular clusters \citep{harris1996}. 
\end{itemize}
In addition, source-detections from fields covering the Eta Carinae nebula (e.g., 1E1048d1m5934 and ASASSN\_18fv) are also reported in the secondary catalog. 
Figure~\ref{fig:cutout optical mask} shows example cutouts of identified fields with highly extended optical hosts. 
The flagged secondary serendipitous sources are marked with blue circles (the primary and \citetalias{lansbury2017_cat} sources are indicated with white circles and green diamonds; refer to Figure~\ref{fig:cutout example}) and the respective optical catalog is flagged in the left corner.     
We found 214 sources to be associated with highly extended galaxies, galaxy clusters or globular clusters, and we refer to the overall catalog of these sources as the secondary NSS80 catalog; 22/214 secondary NSS80 sources were included in NSS40. 
By comparison the primary serendipitous survey source catalog contains 1274 sources; hereafter all statistics reported for NSS80 refer to the primary source catalog.   

Table~\ref{tab:source stats} provides the numbers of sources in the primary NSS80 catalog that are detected in different energy bands
as well statistics regarding optical counterparts, spectroscopic follow-up and successful redshift measurements.
The total number detected in the full, soft, and hard bands are 1078 (85\%), 706 (55\%), and 406 (32\%), respectively.
In total, we have obtained redshifts for 550 NSS80 sources, of which 547 can be spectroscopically classified; see Figure~\ref{fig:spec classifications} and Section~\ref{subsec:optical spectroscopy} below.

\begin{deluxetable}{lllllll}
\tablecolumns{7}
\tablewidth{\textwidth}
\tablecaption{\label{tab:source stats} Source statistics for the primary NSS80 catalog.}
\tablehead{Band & $N$ & $N_\mathrm{spec}$& $N_z$ & $N_\mathrm{\textit{z},failed}$ & $N_\mathrm{r,det}$ & $N_\mathrm{r\,<\,22}$  \\
			(1) & (2) & (3) & (4) & (5) & (6) & (7)  \\}
		\startdata
			Any band  & 1274  & 594    & 550 & 44 & 1015 & 765   \\
			F$+$S$+$H & 257  & 174 & 165 & 9 & 221 & 190  \\
			F$+$S only  & 315  & 170 & 157 & 13 & 275  & 212\\
			F$+$H only    & 81   & 35  & 32 & 3 & 68 & 46  \\
			F only         & 422 & 166  & 153  & 13 & 326 & 233  \\
			S only       & 131 & 39 & 35 & 4 & 88 & 59  \\
			H only        & 68  & 10 & 8  & 2 & 37 & 25  \\ 	
		\enddata
	\tablecomments{
		{\sc Columns:}\,\textbf{(1)} F, S, and H refer to sources detected in the full (3–24 keV), soft (3–8 keV), and hard (8–24 keV) energy bands, respectively. “F + H”, for example, refers to sources detected in the full and hard bands only, but not in the soft band, and “S only” refers to sources detected exclusively in the soft band. 
		\textbf{(2)}  The number of sources detected post-deblending for a given band or set of bands. 
		\textbf{(3)} The number of sources for which (ground-based) optical spectroscopic observations were undertaken.
		\textbf{(4)} The number of sources with spectroscopic redshift measurements and the associated percentage (including robust and uncertain counterpart associations based on our \textsc{Nway} analysis; see Sections~\ref{subsec:nway infrared and optical cpart} \& \ref{subsec:optical spectroscopy}). 
		\textbf{(5)} The number of sources for which spectroscopic observations were undertaken, but lack a reliable redshift measurement (the majority of which have faint, red continuum spectra); see Table~\ref{tab:spec results} and Figure~B5. 
		\textbf{(6)} The number of sources with an associated optical counterpart detected in the $r$-band; magnitudes are obtained from SDSS, PanSTARRS, USNOB1 and NSC (NOAO Source Catalog).
		\textbf{(7)} The number of sources with an associated optical counterpart brighter than $r = 22$ (detectable with current groundbased telescopes).}
\end{deluxetable}

Both the primary and secondary source catalogs are provided as electronic tables. 
In Appendix~\ref{appendix:nss80 catalogue} we give a detailed description of the columns that are provided in the catalog. In addition, we also created an online library of the 894 unique fields to allow for quick and easy verification of the X-ray and optical counterpart information for each of the \nustar fields. 
The online library will be accessible at \url{https://www.nustar.caltech.edu/page/59}
and in Figure~\ref{fig:cutout example} we show an example of one of these fields. In Table~\ref{tab:subset_summary} we give a summary of the subsets of this primary catalog as discussed in future sections. 

\begin{deluxetable}{lllll}
\label{tab:subset_summary}
\tablecolumns{5}
\tablewidth{\textwidth}
\tablecaption{Summary of the primary catalog subsets and selection flags.}
\tablehead{Subset & Number & Selection Flag & Section & Description \\
         (1) & (2) & (3) & (4) & (5) \\}
         \startdata
         Primary (all) & 1285 & -- & \ref{subsec:serendip obs} & \makecell[l]{The full catalogue, excluding sources within highly \\ extended optical galaxies and clusters} \\
         Unique & 1274 & \textsc{MainCAT} & \ref{subsec:serendip obs} & \makecell[l]{All unique detections, i.e. excluding objects with \\ multiple optical counterpart candidates} \\
         Reliable & 962 & \textsc{NWAY\_RFlag} & \ref{subsec:nway infrared and optical cpart} & \makecell[l]{Sources with a high probability \textsc{Nway} match in either \\ CatWISE20 or PS1-DR2} \\
         Spectroscopic & 594 & \textsc{SpecCAT} & \ref{subsec:optical spectroscopy} & Sources with spectroscopic observations \\
         BL & 287 & \makecell[l]{\textsc{Classification} is \lq BL\rq\ or \\ \lq BL?\rq\ } & \ref{subsubsec:spectral classification} & \makecell[l]{Sources with broad permitted emission line widths \\ ($FWHM \geq 1000$\,km\,s$^{-1}$)} \\
         NL & 198 & \makecell[l]{\textsc{Classification} is \lq NL\rq\ or \\ \lq NL?\rq\ } & \ref{subsubsec:spectral classification} & Sources with narrow permitted emission line widths \\
         Extra-galactic & 492 & \makecell[l]{\textsc{SpecCAT}, \textsc{zQuality}!=\lq F\rq\ \\ or \lq C\rq\, and \textsc{zspec}$>0$ } & \ref{subsubsec:spectral classification} & \makecell[l]{Sources with spectroscopic redshifts indicating \\ extra-galactic origin} \\
         \enddata
\end{deluxetable}

\newpage

\subsection{Key changes in NSS80 with respect to NSS40} 
\label{subsec: differences in catalog to L17}
As discussed in Section~\ref{subsec:data processing}, when constructing the NSS80 sample, we mainly adopted the same underlying methodology and data processing procedures as in NSS40 to be consistent between the two \nustar serendipitous surveys. 
Nevertheless, there are several significant changes and updates with respect to NSS40. 
The key differences are summarised below. 
\begin{itemize}
	\vspace{-0.1cm}
	\item{NSS40 combined individual exposures of the same science target to increase the sensitivity. In NSS80 we extend this approach to coadd all exposures performed over the 80-month period within a 12$'$ search radius of the aim point (i.e., all overlapping sky regions were automatically identified and coadded), providing improved sensitivity in fields with multiple overlapping observations.} 
	\item{NSS40 excluded obsIDs with exposure times $<$\,1\,ks from the analysis. NSS80 now coadds all of the data (which satisfied our criteria mentioned in Section~\ref{subsec:serendip obs}), including any low-exposure time data.}  
	\item{Preceding source-detection, serendipitous detections which could be associated with highly extended optical/IR hosts were manually masked and removed in NSS40. NSS80 retains these in a secondary catalog, based on a later post-processing step; see Section~\ref{subsec:source catalog}.}
\end{itemize}

To quantify how many serendipitous sources were added/removed due to the aforementioned alterations, we assessed overlapping fields between the NSS40 and NSS80. Of the 331 NSS40 unique fields, we reprocessed 50 fields -- 6 which include only NSS40 observations while 44 fields include post-NSS40 observations as well. 
For these 50 fields, we detected an additional 111 sources: 89 sources arise from the deeper or wider data and 22 by including highly extended optical host mask regions originally excluded. 
36 NSS40 sources were not detected in our coadded fields; we summarise potential reasons to explain these undetected sources in Table~\ref{tab:missed 40-month serendips}.
32 of the 36 undetected NSS40 sources have updated false probabilities, based on the deeper co-added data, that no longer satisfy our detection threshold, indicating that they were likely spurious detections in the NSS40 sample.
This is particularly noticeable for NSS40 sources with detections in a single energy band only.   
Of the remaining 4 undetected NSS40 sources, 1 lies in an excess background region and 3 sources are on the peripheries of coadded fields.

Overall, 444/497 NSS40 sources are included in the primary NSS80 catalog, 17/497 are included in the secondary NSS80 catalog, and 36/497 sources are excluded. Additionally, \citetalias{lansbury2017_cat} constructed a secondary catalog of which 8/64 are included in the primary NSS80 and 5/64 in the secondary NSS80 catalog. Hence, in total, 452 and 22 NSS40 sources are included in the primary and secondary NSS80 catalogs, respectively. 

\section{The Multi-wavelength Data} \label{sec:multiwav data}
The compiled NSS80 presented in this work is independent of prior multi-wavelength information. 
To further explore the source properties, such as luminosities and source classifications, we require multi-wavelength information to draw a more complete picture of the properties and nature of individual sources. 
Since our primary focus for the NSS80 is extragalactic sources, we also require optical counterparts to establish redshift measurements from which a range of other properties can be inferred. 
However, since the positional accuracy of \nustar ranges between $\approx$\,8$''$ to $\approx$\,20$''$ for bright to faint sources \citep[90\% confidence; see e.g.,][]{lansbury2017_cat}, it is desirable to have more accurate X-ray positions to search for reliable optical/IR counterparts. To achieve this we first searched for lower-energy (soft) \xray counterparts with more accurate source positions (see Section~\ref{subsec:soft xray cpart}) and, subsequently, searched for IR/optical counterparts to the \xray sources (see Section~\ref{subsec:nway infrared and optical cpart}), which were then used in our spectroscopic follow-up campaign (see Section~\ref{subsec:optical spectroscopy}). 

\subsection{Lower-energy \xray Counterparts} \label{subsec:soft xray cpart}
\begin{deluxetable}{lllllll}
\tablecolumns{7}
\tablewidth{\textwidth}
\tablecaption{\label{tab:softX stats} The number of \nustar serendipitous sources with lower-energy X-ray counterparts.
		}
\tablehead{& & CSC2.0 & 4XMM-DR10s & 4XMM-DR10 & 2SXPS & Any \\}
\startdata
			(1) & $N_\mathrm{Total}$\,/\,$N_\mathrm{Coverage}$    & 300\,/\,408  & 492\,/\,651  & 394\,/\,553   &  661\,/\,949 & 956\,/\,1249  \\
			(2) & $N_\mathrm{Single}$    &  211 &  382 &  349 &  617 &  907  \\
			(3) & $N_\mathrm{Multiple}$  &   89 &  110 &  45  &  44  &  214   \\
			(4) & $N_\mathrm{Best}$      & 300  & 317  & 168  &  171 & $^\dagger$956    \\
			\\
			(5) & $\langle \Delta$RA$\times \cos$(Decl.)$\rangle$ & $-0.50\pm0.60''$ & $0.55\pm0.52''$ & $0.20\pm0.61''$ & $-0.28\pm0.49''$ & \\
			(6) & $\langle \Delta$Decl.$\rangle$ & $0.36\pm0.49''$ & $-0.13\pm0.40''$ & $0.20\pm0.51''$ & $0.71\pm0.34''$ & \\
			(7) & $\langle  \theta \rangle$ & 10.7$''$ & 11.2$''$ & 12.6$''$ & 10.9$''$ & \\
\enddata
	\tablecomments{ 
		{\sc Rows:}\,\textbf{(1)} The total number of \nustar sources in the primary catalogue with a lower-energy X-ray counterpart within a search radius of 30$''$ ($N_\mathrm{Total}$) compared to the total number of primary \nustar sources with \textit{Chandra}, \textit{XMM-Newton} and/or \textit{Swift}-XRT coverage ($N_\mathrm{Coverage}$); the coverage was determined by matching sources without a soft X-ray counterpart with observations within the default radius on HEASARC, and then checking the exposure maps for non-zero exposure times at the \nustar coordinates. The final column enumerates the number of unique \nustar sources with a match in any catalog.  
		\textbf{(2)} The number of \nustar sources with a single match within 30$''$ to the specific lower-energy X-ray catalog.   
		\textbf{(3)} The number of \nustar sources with multiple matches within 30$''$ to the specific lower-energy X-ray catalog.
		\textbf{(4)} The number of \nustar sources where the position from a given lower-energy X-ray catalog is taken to be the most reliable (i.e., the best), and consequently the adopted, lower-energy X-ray position. The order of preference is: \textit{Chandra}, \xmm and then \textit{Swift}-XRT. 
		$^\dagger$In addition to the 956 unique sources with automatically matched counterparts described in the table, a further 8 sources (1 \textit{Chandra} + 4 \textit{XMM-Newton} + 3 \textit{Swift}-XRT) were manually identified, resulting in a total of 964 NSS80 sources with a lower-energy X-ray counterpart.
		\textbf{(5)}\,\&\,\textbf{(6)} The mean positional offsets in right ascension (5) and declination (6) of the \nustar position relative to the lower-energy X-ray counterpart (see Figure~\ref{fig:softx sep+pfalse} top panel). 
		\textbf{(7)} The mean angular offset between the \textit{NuSTAR} and lower-energy X-ray positions in arcsec.
		These values are computed for all CSC2.0, 4XMM-DR10/s and 2SXPS matches. }
\end{deluxetable}

To search for lower-energy (soft) \xray counterparts, we used \textit{Chandra}, \textit{XMM-Newton}, and \textit{Swift}-XRT observations. We
cross matched the \nustar sources to (1) the \textit{Chandra} Source Catalog Release 2.0 \citep[CSC2.0;][]{evans2019_csc2}, (2) the Fourth \textit{XMM-Newton} Serendipitous Source Catalog, Tenth Data Release \citep[4XMM-DR10;][]{webb2020} and its stacked version \citep[4XMM-DR10s;][]{trauslen2020}, and (3) the \textit{Swift}-XRT Point Source Catalogue \citep[2SXPS;][]{evans2020_2sxps}, using a search radius of 30$''$ for each \nustar source position (consistent with \citetalias{lansbury2017_cat}). 
There is a trade-off between completeness and the number of false associations when cross-matching between different surveys, and thus here and in Section~\ref{subsec:nway infrared and optical cpart} we select cross-matching radii carefully with this balance in mind.
As discussed in \citetalias{lansbury2017_cat}, the uncertainty in the \nustar positions dominates the errors in the source matching.
We would expect to exclude a true match in a very small fraction of cases ($<$0.5\%) and for $\sim$7\% of the associations to be false (\citetalias{lansbury2017_cat}).

We identified lower-energy X-ray counterparts for 956 \nustar sources between the four lower-energy X-ray catalogs.
In addition, we manually identified a potential lower-energy X-ray counterpart for a further 8 sources which have faint lower-energy X-ray emission (yet not statistically significant) in the vicinity of the \textit{NuSTAR} position (1 \textit{Chandra}, 4 \textit{XMM-Newton} and 3 \textit{Swift}-XRT; see Appendix~\ref{appendix:nss80 catalogue}), leaving a total of 964 NSS80 sources (76\%) with an identified lower-energy X-ray counterpart. Accordingly we were unable to identify lower-energy X-ray counterparts for 310 NSS80 sources, of which 94.5\% (293/310) have lower-energy X-ray coverage with either one of the lower-energy X-ray observatories: 34.8\% with \textit{Chandra} (ACIS), 51.3\% with \textit{XMM-Newton}, and 92.9\% with \textit{Swift}-XRT; these sources are flagged in the catalog (see Appendix~\ref{appendix:nss80 catalogue}). The reason for the non-detections could be that the observations were too shallow to detect faint sources, or it could be attributed to variability given that the observations are non-contemporaneous, or it could be the result of absorption of lower-energy X-ray photons along the line-of-sight. 
Only $\sim$1\% (17/1274) lack any form of coverage from all of these three lower-energy X-ray observatories (flagged as \textsc{flag\_softx\_cov}\,=\,\textsc{null} in the catalog).

Of the 964 \nustar sources with lower-energy X-ray counterparts, 142 sources have been detected with more than one of the lower-energy X-ray observatories: 22 from \textit{Chandra}+\textit{XMM-Newton}; 32 from \textit{Chandra}+\textit{Swift}-XRT; 51 from \textit{XMM-Newton}+\textit{Swift}-XRT; and 37 from all three lower-energy X-ray observatories. 
For these sources we adopted the position with the highest accuracy as the `best' lower-energy X-ray counterpart, which are in the following order: CSC2.0, 4XMM-DR10s, 4XMM-DR10 and 2SXPS.
Hence, of the 964 lower-energy X-ray counterparts, we have adopted the positions for 300 from CSC2.0, 317 from 4XMM-DR10s, 168 from 4XMM-DR10, 171 from 2SXPS, and 8 manually measured positions using lower-energy X-ray imaging, i.e., one position from \textit{Chandra}, 4 from \textit{XMM-Newton}, and 3 from \textit{Swift}-XRT; see row 4 in Table~\ref{tab:softX stats}.\footnote{The 8 manually measured lower-energy X-ray positions are used only to identify multi-wavelength counterparts, but are excluded from all further X-ray analysis.} 

Approximately 23\% (288/1274) of the \nustar sources have multiple CSC2.0, 4XMM-DR10/s or 2SXPS matches within our search radius. 
To identify the best counterpart for these cases, we made the assumption that the lower-energy X-ray source with the brightest flux in the highest available energy band (\textit{Chandra}: 2-7\,keV; \textit{XMM-Newton}: 4.5-12\,keV; \textit{Swift}-XRT: 2-10\,keV) is likely to be the correct counterpart.
We note that in some cases we stand the risk of ignoring heavily obscured sources which are faint in lower energies.
Three \nustar sources (one \textit{Chandra} and two \textit{XMM-Newton}) were undetected in solely the highest energy band of the low-energy instrument, for which we used their full band fluxes.

The results from the lower-energy X-ray cross-matching of the primary NSS80 sources are summarized in Table~\ref{tab:softX stats}.
We provide the positions, the angular separation between the lower-energy X-ray counterpart and the \nustar source, and the number of sources with matches in each catalogue.

\begin{figure}
	\centering
	\includegraphics[width=20pc]{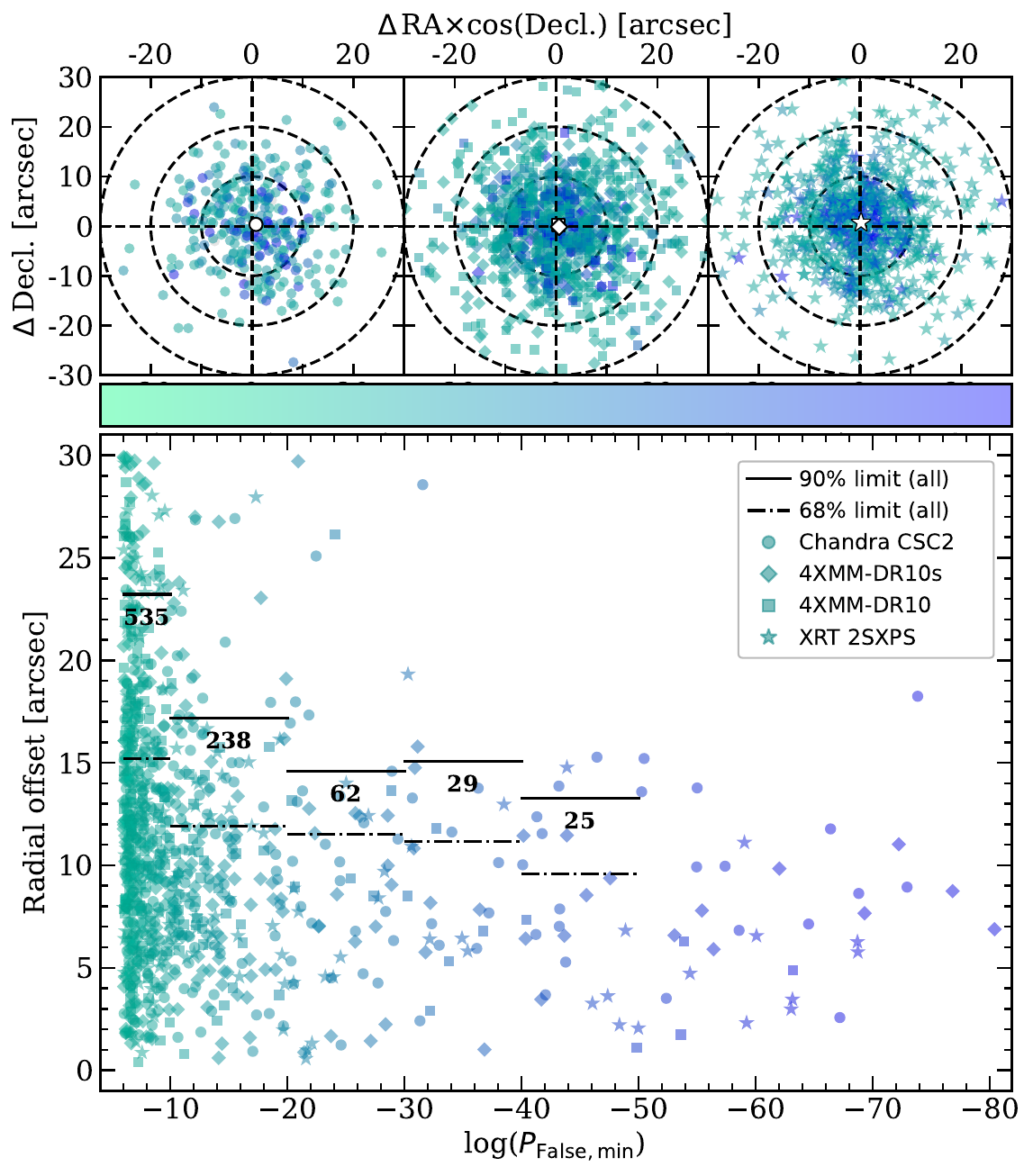}		
	\vspace*{-2mm}
	\caption{The positional accuracy of \nustar as a function of source-detection significance for the 956 NSS80 sources with lower-energy counterparts in pointed archival observations; all matched lower-energy counterparts are plotted.
		\textbf{Top panel:} Astrometric offsets between the \nustar source and its matched lower-energy X-ray counterpart coordinates from (a) \textit{Chandra}/CSC2.0 (circles), (b) \textit{XMM-Newton}/4XMM-DR10/s (squares and diamonds), and (c) \textit{Swift }-XRT/2SXPS (stars), color-coded by source-detection significance. 
		The negligible mean positional offsets are shown with white markers. 
		\textbf{Bottom panel:} The radial offset between the \nustar source and its best lower-energy X-ray counterpart as a function of the minimum source-detection significance ($\Delta p_\mathrm{False,min}$) which increases towards the right (from green to purple color-code). 
		The solid and dash-dot black lines indicate the 90\% and 68\% confidence limits, respectively, on the \nustar positional uncertainty for bin sizes of $\log$($p_\mathrm{False,min}) = -10$.
		The number of sources which each bin contains is given above the solid lines. 
		Since the number of sources becomes small towards high $p_\mathrm{False,min}$ values, we only plot bins for $\log$($p_\mathrm{False,min}) < -50$. }
	\label{fig:softx sep+pfalse}	
\end{figure}

We show the positional offsets between the \nustar sources and their (a) \textit{Chandra}, (b) \xmm and \mbox{(c) \textit{Swift}-XRT} counterparts in the top panel of Figure~\ref{fig:softx sep+pfalse}, and list the mean positional offsets for all CSC2.0, 4XMM-DR10/s and 2SXPS matches in rows 5--6 of Table~\ref{tab:softX stats} (as well as the mean angular offset between the \textit{NuSTAR} and lower-energy X-ray positions; see row 7).
The sources are plotted in color, coded by the source-detection significance (the minimum false probability), and the dashed circles illustrate different search radii: 10$''$ (inner), 20$''$ (middle) and 30$''$ (outer). 
Evidently, the majority of sources for each lower-energy X-ray observatory lie within a 20$''$ separation radius (i.e.,\ 94\%, 84\%, 90\%, 90\% for CSC2.0, 4XMM-DR10s, 4XMM-DR10 and 2SXPS, respectively), particularly those with more significant detections (i.e., lower $\Delta p_\mathrm{False,min}$ values). 
The lack of significant positional offsets (i.e., see the median astrometric offsets for each sample in Figure~\ref{fig:softx sep+pfalse}) are indicative of consistent astrometry between the \xray observatories. 

The bottom panel of Figure~\ref{fig:softx sep+pfalse} shows the positional accuracy of \nustar as a function of the detection significance; i.e.,\ the angular separation between the \nustar position and its best identified lower-energy X-ray counterpart (having a higher likelihood of being correctly matched) versus the minimum false probability of a given source.
By assuming zero uncertainty in the lower-energy X-ray position and that NSS80 sources with lower-energy counterparts are representative of the overall population, we determine the 90\% and 68\% confidence limits on the \nustar positional uncertainty in bin sizes of $\log$($p_\mathrm{False,min}) = -10$, as indicated with solid and dash-dot horizontal black lines, respectively. 
From this analysis we find that the 90\% confidence limit on the \textit{NuSTAR} positional uncertainty varies from 23$''$ to 13$''$ between the least-significant and the most-significant detections. 

\begin{figure*}
	\centering
	\includegraphics[width=42pc]{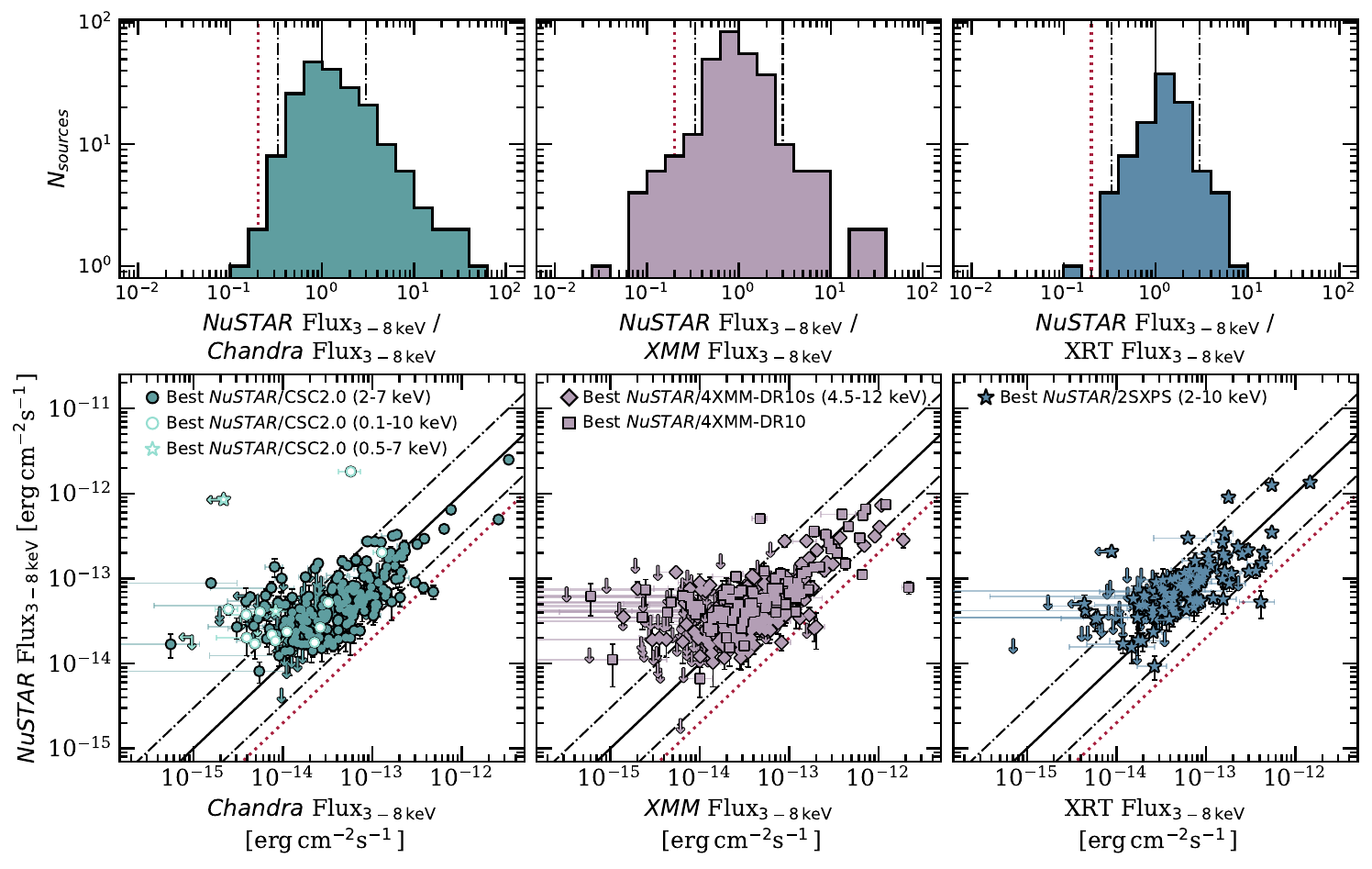}		
	\vspace*{-2mm}
	\caption{Comparison of the $3-8$\,keV \nustar and $<10$\,keV \xray mission fluxes ($F_\mathrm{soft}$) for the best \textit{Chandra}/CSC2.0 (green circles), {XMM-Newton}/4XMM-DR10/s (purple squares and diamonds, respectively), or \textit{Swift}-XRT/2SXPS (blue stars) counterpart matched to the primary NSS80. The different energy bands used to calculate the $3-8$\,keV fluxes for the CSC2.0 are indicated with different symbols, i.e., $2-7$\,keV (solid green circle), $0.1-10$\,keV (white filled, green-edged circle), and $0.5-7$\,keV (white filled, green-edged star).
    \textbf{Top panel:} Distribution of \nustar to $<10$\,keV \xray mission fluxes for each instrument.
		In both panels, the black solid line shows the 1:1 relation, the black dash-dot lines show a factor of three from this relation, and the red dotted line indicates a factor of 5 below the 1:1 relation. 
		Ten sources have \nustar $3-8$\,keV flux measurements below the 1:5 relation.
	}
	\label{fig:softx nustar fluxes}	
\end{figure*}

We estimated the observed-frame $3-8$\,keV flux ($F_\mathrm{soft}$) for the lower-energy X-ray counterparts following the methodology in \citetalias{lansbury2017_cat}. 
For CSC2.0, 4XMM-DR10/s and 2SXPS sources we converted to the $3-8$\,keV flux from the $2-7$\,keV, $4.5-12$\,keV and $2-10$\,keV\footnote{For the 2SXPS sources we calculated the soft-band flux using the available band 3 count rate (2--10\,keV).} flux using a conversion factor of 0.83, 0.92 and 0.62, respectively.\footnote{We estimated the conversion factors in WebPIMMS for a Galactic absorption of $N_\mathrm{H} = 0$ and a photon index of $\Gamma = 1.8$.}   
We show these fluxes relative to the \textit{NuSTAR}-measured fluxes for the best identified lower-energy X-ray counterparts in Figure~\ref{fig:softx nustar fluxes}. 
It should be noted that 21/300 CSC2.0 counterparts are undetected in the $2-7$\,keV (ACIS) energy band: 1/21 source are detected in the $0.5-7.0$\,keV broad band (NuSTARJ184449+7212.1), 2/21 sources (NuSTARJ095712+6904.8 \& NuSTARJ121425+2936.1) only have upper limits in the $0.5-7$\,keV broad band, while the remaining 18/21 sources only have constrained fluxes in the $0.1-10$\,keV HRC wide band\footnote{We note that some of these sources may represent the small fraction of expected false associations between soft X-ray and \nustar sources.}. For these sources we used the respective bands in which they are detected to calculate the $3-8$\,keV fluxes. 
We see a reasonable agreement in the flux measurements between observatories, with the majority of the sources (80\% CSC2.0, 90\% 4XMM-DR10/s and 93\% 2SXPS) lying within a factor of three of the 1:1 relation (see Figure~\ref{fig:softx nustar fluxes}). 
At least a component of the observed scatter is likely to be attributed to intrinsic source variability due to the non-contemporaneous \nustar and lower-energy X-ray observations.
In addition, \citet{civano2015}, \citet{mullaney2015} and \citet{fornasini2017} have shown that Eddington bias affects the lower \nustar fluxes, increasing the spread as the flux limit is approached. Sources at the very lowest fluxes are not commonly detected, contributing to the apparent bias with larger numbers of sources at the lowest X-ray fluxes that are above the 3:1 relation. 

Finally, we assessed the flux contribution from all of the \textit{Chandra}, \xmm and \xrt sources within a radius of 30$''$ from the \nustar source by determining their total combined $3-8$\,keV flux ($F_\mathrm{soft}^{30}$; see Appendix~\ref{appendix:nss80 catalogue}). 
We allows us to compare: (a) $F_\mathrm{soft}$ measured by \textit{NuSTAR} ($F_\mathrm{soft,\nustar}$), in order to assess whether a significant amount of \nustar flux is coming from other sources; and (b) $F_\mathrm{soft}$ from the best lower-energy counterpart ($F_\mathrm{soft,LE}$), to assess the contamination of low energy flux from other sources, less affected by variation and inter-instrument differences.
Overall, 6.9\% of the NSS80 sources with lower-energy X-ray counterparts have $F_\mathrm{soft}^{30} > 1.2\times F_\mathrm{soft,LE}$, and 4.9\% have $F_\mathrm{soft}^{30} > 1.2\times F_\mathrm{soft,\nustar}$. Only 21 sources have combined fluxes that exceed $F_\mathrm{soft,\nustar}$ by a factor of 2.
Therefore, we are confident that the \textit{NuSTAR} source is generally dominated by the emission from the brightest lower-energy X-ray counterpart.


\subsection{Finding IR \& optical counterparts using \textsc{Nway}} \label{subsec:nway infrared and optical cpart}

Obtaining redshifts for the \nustar sources requires the identification of the correct optical counterpart. 
The comparatively large positional uncertainty of \nustar sources often leads to multiple potential optical counterparts. 
Consequently, we require an approach to distinguish between true and unrelated optical counterparts, particularly in the absence of more reliable lower-energy X-ray positions. As discussed further in Section \ref{subsec:wise properties nustar}, MIR emission provides a robust identification of AGN activity, particularly for hidden luminous quasars, since the dusty AGN torus radiates predominantly at these wavelengths, while star formation from the host galaxy peaks at FIR wavelengths and is comparatively weak at MIR wavelengths. Therefore we should not consider star-forming galaxies a major source of contaminants in the MIR distributions found in Figure~\ref{fig:nway distributions}.
The all-sky \textit{WISE} survey therefore provides an excellent
complement to \textit{NuSTAR}: the positional uncertainty of \textit{WISE} sources, particularly in the shorter wavelength $W1$ and $W2$ bands, is sufficient to be able to reliably identify optical counterparts. 
Identifying \textit{WISE} counterparts for hard X-ray selected sources can therefore pave the way to locating the correct optical counterpart even for \nustar sources without a lower-energy X-ray counterpart. This section describes the process of matching MIR and optical counterparts to \nustar sources both with, and without, lower-energy X-ray counterparts.

\citetalias{lansbury2017_cat} adopted a relatively simple closest neighbour approach to identify multi-wavelength counterparts, using the more reliable positions from lower-energy X-ray counterparts, where available, and the distinctive characteristics of AGN with respect to galaxies in the MIR band (as traced using \textit{WISE}). 
Here we adopt a more sophisticated probabilistic approach using \textsc{Nway} \citep[v.4.4.2;][]{salvato2018} to identify IR and optical counterparts for the NSS80 sources.
\textsc{Nway} uses Bayesian methods to probabilistically match multi-wavelength counterparts to X-ray sources by simultaneously matching $N$ catalogs in a multi-dimensional parameter space, e.g., astronomical sky coordinates and positional uncertainties, magnitude and color distributions, source density and morphology, etc. 
Therefore, \textsc{Nway} is a powerful tool for our task of identifying the correct optical and IR counterparts for \nustar detected sources, which can include both Galactic populations, such as stars, and extragalactic objects, such as AGN.
For our \textsc{Nway} matching we use CatWISE20 \citep{marocco2021}, which is a MIR all-sky catalog selected from  \textit{WISE} and NEOWISE at 3.4 and 4.6 $\mu$m (i.e., $W1$ and $W2$).
In addition, we use Pan-STARRS DR2 \citep[PS1-DR2;][]{flewelling2018} which provides coverage at declinations $\gtrsim -30^\circ$ with a single epoch 5$\sigma$ depth of $r < 21.8$\footnote{For sources at lower declinations no optical matching was performed with \textsc{Nway}. Instead we performed positional cross-matching to other optical catalogues;
see case ii) for selection of principal counterparts later in this section for details.}.

\begin{figure*}
	\centering
	\includegraphics[width=43pc]{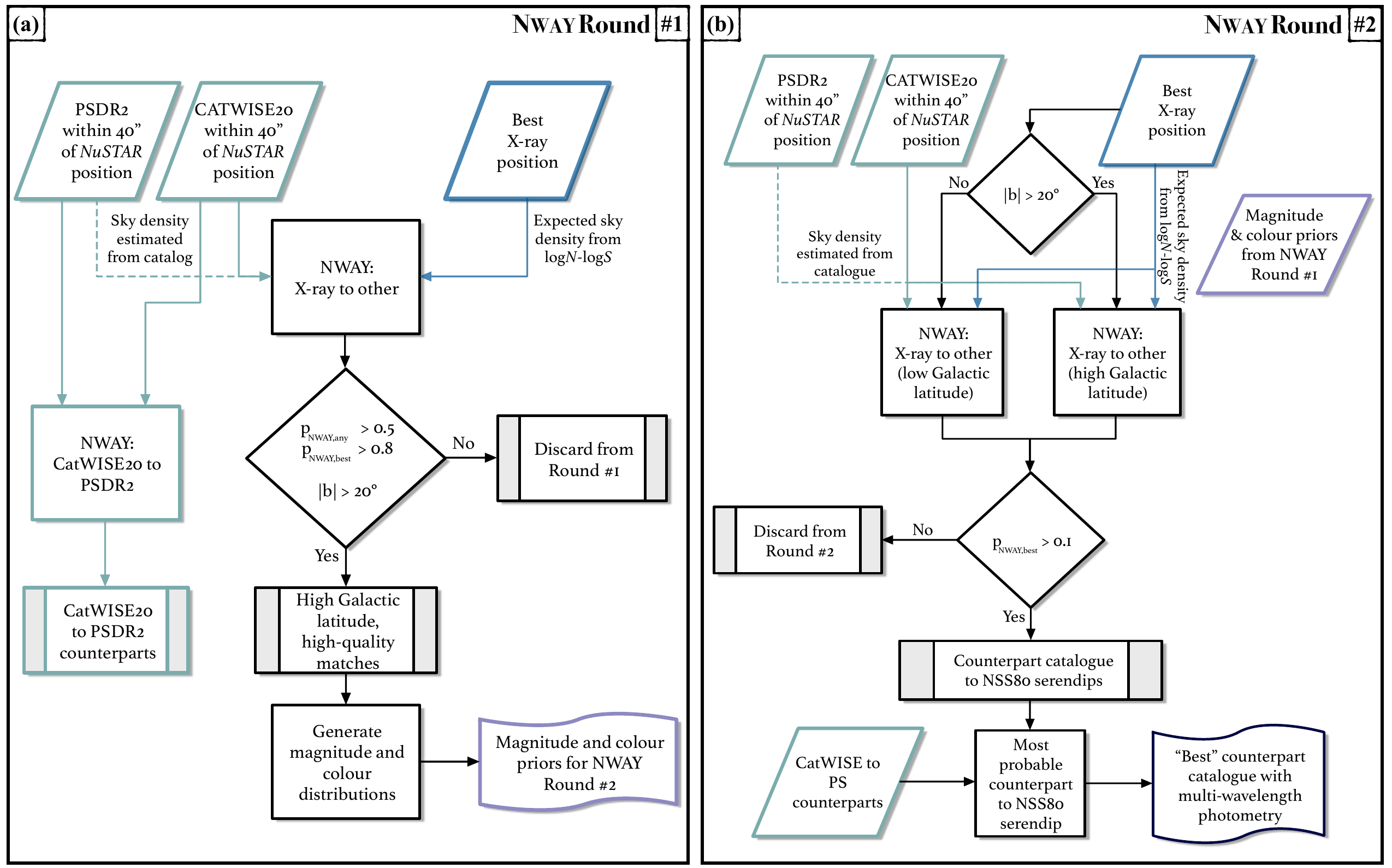}		
	\vspace{-6mm}
	\caption{Flowchart illustrating our \textsc{Nway} matching strategy which we used to identify CatWISE20 and PS1-DR2 counterparts to the NSS80 sources. 
		\textbf{(a)} Round \#1 entails the generation of the magnitude and color priors required for Round \#2 (magenta outlined box; see Figure~\ref{fig:nway flowchart round comb}(b)), by using astrometric information including source positions and their associated uncertainties, and the sky density as a function of magnitude (green outlined boxes) to identify counterparts to the best X-ray position (blue outlined box). Lines are dashed or shown in different colours purely for clarity.
		\textbf{(b)} Round \#2 of \textsc{Nway} which utilizes the priors (magenta) from the matching to lower-energy X-ray counterparts (blue) with  $|b| > 20^\circ$ in Round \#1 and flat priors for sources with $|b| < 20^\circ$).	
		All matches with probabilities $> 10$\% are stored in the final post-processed catalog which include X-ray information, multi-band positions and photometry, and key \textsc{Nway} information such as the Bayesian match probabilities.
	} 
	\label{fig:nway flowchart round comb} 
\end{figure*}

\begin{figure*}
	\centering
	\includegraphics[width=37pc]{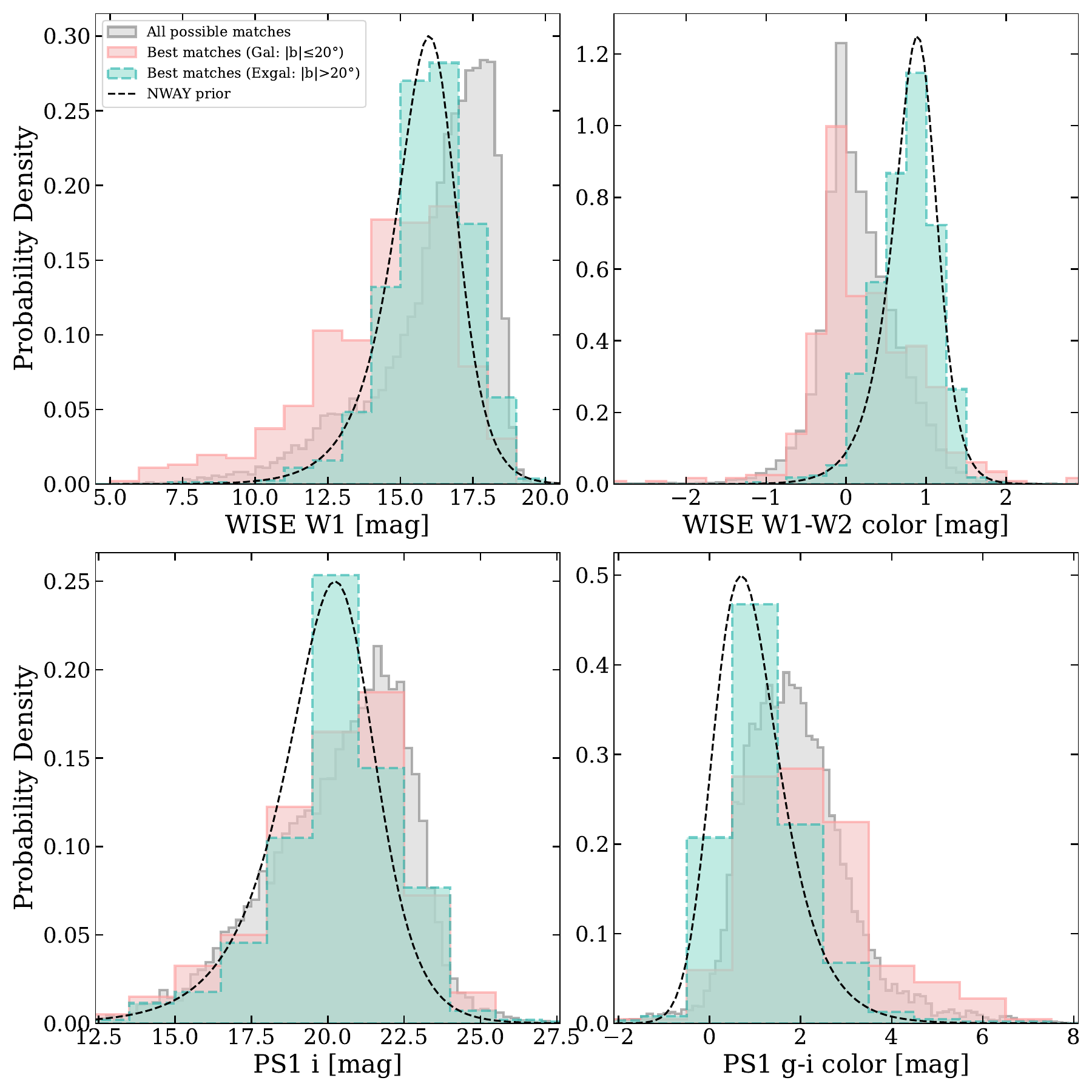}	
	\vspace*{-2mm}
	\caption{ 
		Magnitude and color distributions (integrated to unity) of the CatWISE20 (top) and PanSTARRS (bottom) matched samples from Round~\#1, which are used to evaluate priors in Round~\#2 to identify the ``best'' IR counterpart for each \nustar serendipitous source. \textit{WISE} magnitudes are Vega and optical magnitudes are AB. 
		\textbf{Top left panel:} The $W1$-magnitude distribution for the best matches vs. all sources (grey shaded area), splitting the former by Galactic latitude; high Galactic latitude sources ($|b| > 20^\circ$) in red and low Galactic latitude sources ($|b| < 20^\circ$) in blue. 
		The black line is a Generalised Logistic distribution fit to the data of the high Galactic latitude sources. The $W1$ distribution of the low Galactic latitude sources follows a similar shape to that of the high Galactic latitude sources, with a slightly brighter $W1$ tail.   
		\textbf{Top right panel:} The $W1$-$W2$ color distribution following the same color-code as in the first panel. Here, the \textit{WISE} color of the high Galactic latitude sources, driven by the presence of AGN, is distinct from the low Galactic latitude sources. Moreover, the peak of the distribution is around $W1$-$W2$\,$\approx$\,0.8, which is the \citet{stern2012} criterion for selecting MIR AGN. Therefore, the $W1$-$W2$ color distribution can be used as a prior to identifying AGN candidates for optical/IR follow-up spectroscopy.
        \textbf{Bottom left panel:} The $i$ magnitude distribution following the same color-code as in the previous panels. The distribution of the low Galactic latitude sources follows a similar shape to that of the high Galactic latitude sources, peaking at a slightly less bright magnitude.
        \textbf{Bottom right panel:} The $g$-$i$ color distribution following the same color-code as in the previous panels.
	}
	\label{fig:nway distributions}	
\end{figure*}

In what follows we summarise the main steps of our \textsc{Nway} matching approach to obtain IR/optical counterparts for our spectroscopic follow-up campaign, as outlined in Figures~\ref{fig:nway flowchart round comb}(a) and \ref{fig:nway flowchart round comb}(b).
We begin by constructing colour and magnitude priors that are approximately representative of the population by performing a photometrically unbiased cross-match between the NSS80 sources, MIR, and optical catalogs, and then restricting the results from this to secure counterparts only (i.e. high probability matches). With these expected distributions of magnitude and color in hand we can apply them as priors to a round of cross-matching that includes all possible counterparts (i.e. those that had low probabilities in the first cross-match, as well as the high probability matches) and improve the final cross-match probabilities that inform our selection of principal counterparts.

\begin{figure}
	\centering
	\includegraphics[width=40pc]{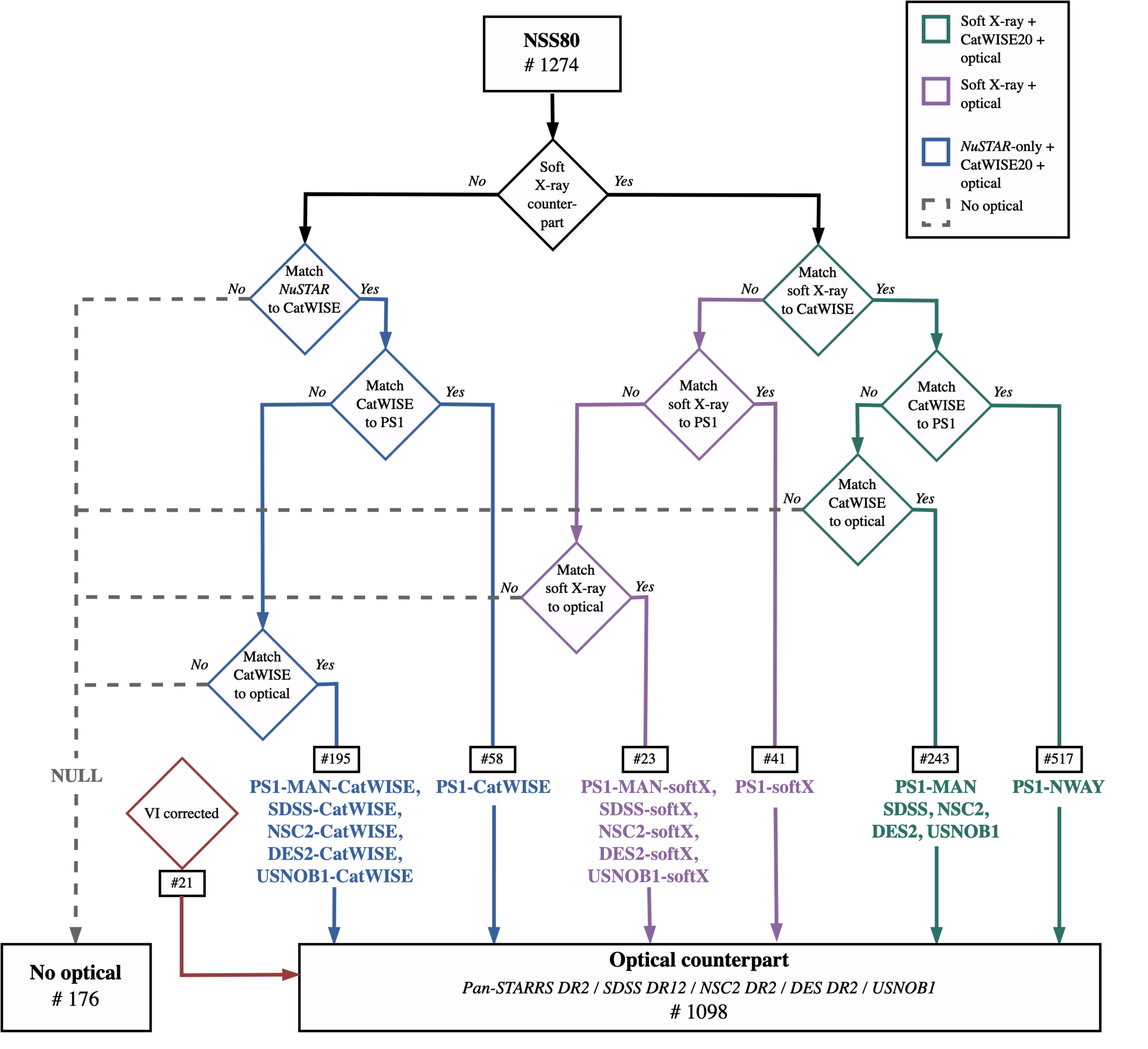}	
	\caption{A flowchart of the principal adopted optical counterparts to the NSS80 sources. The actions in green represent NSS80 sources with lower-energy X-ray\,$+$\,CatWISE20 counterparts with an optical match within 2.7$''$ from the \textit{WISE} position. Optical matches are retrieved from PS1-DR2 (either from our \textsc{Nway} matching or manually), SDSS, NSC2, DES2 or USNOB1.  
		The purple actions indicate lower-energy X-ray\,$+$\,optical matches which lack a CatWISE20 association.
		For these sources we matched the \textit{Chandra}, \textit{XMM-Newton} or \textit{Swift}-XRT positions (softX $\equiv$ lower-energy X-ray) to the aforementioned optical catalogues using characteristic positional uncertainties of 2.5$''$, 5$''$ or 6$''$, respectively. 
		The blue actions represent \textit{NuSTAR}-only X-ray\,$+$\,CatWISE20 positions matched to the aforementioned optical surveys (prioritising \textsc{Nway} matches where available) using a matching radius of 2.7$''$; these optical matches are only used for the purposes of spectroscopic follow-up. 
		NSS80 sources that lack any optical association (some of which may be too faint/distant) are indicated with grey actions. In red is shown the contribution of objects with a visually selected optical match - these include, for example, objects where cross-matching limits for other branches are exceeded slightly but on visual inspection a clear match is found.
		This results in an optical completeness of $\sim$86\% (1098/1274) for the NSS80 catalog: 844/964 lower-energy X-ray and 254/310 \textit{NuSTAR}-only X-ray associations. This information is available in the NSS80 catalog; see Appendix~\ref{appendix:nss80 catalogue}.}
	\label{fig:flowchart best optical} 
\end{figure}

This process and the resulting principal optical counterparts should be understood with some basic information in mind. In general we prioritise counterparts from the X-ray--MIR matching, as AGN are more robustly distinguished in the MIR (see distributions in Figure~\ref{fig:nway distributions}). Selection of optical counterparts follows using a well-defined branching procedure, with the aim of producing the most likely counterpart candidates for the population as a whole. The intention of this is to provide an overview of the properties of counterparts to hard X-ray selected sources (see Sections~\ref{subsec:wise properties nustar} and \ref{subsec:optical prop}). The matches are therefore also manually checked, and in a small minority of cases this overrides the automatic procedure (see red branch \lq VI corrected\rq\ in Figure~\ref{fig:flowchart best optical} and Round \#2 part (v)). Users are advised to consider their specific use cases and the appropriate choice of optical counterpart for their application.

\textbf{$\star$ Preliminary round: constructing base counterpart catalogs} \\
An optical and an IR base catalog are first created (separately) by collecting all IR positions from CatWISE20 and optical positions from PS1-DR2 within 40$''$\footnote{This matching radius is increased to 40$''$ (as compared to matching with soft X-ray sources within 30$''$) to include all potential matches to soft X-ray counterparts, which could be offset by as much as 30$''$.} of the \nustar position (see the green outlined boxes in Figure~\ref{fig:nway flowchart round comb}(a)).

\textbf{$\star$ Round \#1: defining magnitude and color priors | Figure~\ref{fig:nway flowchart round comb}(a)} \\
Round \#1 uses only astrometric and sky-density information (including positional errors) to identify a ``good'' counterpart for the NSS80 sources. The expected local sky densities of optical and \wise counterparts vary greatly with distance from the Galactic plane, and therefore CatWISE20 and PS1 sky densities were calculated from the actual number counts within 40$''$ of each \nustar source. The \nustar source densities were derived using the $\log N$--$\log S$ curves reported in \citet{harrison2016}, adopting the deblended soft-band \nustar flux. 

For X-ray positions, lower-energy coordinates are used where available (see Section~\ref{subsec:soft xray cpart}), otherwise the \nustar position is used. In this work we use 68\% positional uncertainties appropriate for input to \textsc{Nway}, adapted from the various serendipitous catalogue values for sources with soft X-ray matches. For \nustar-only sources we used the 68\% positional uncertainties based on $\log$($p_\mathrm{False,min})$, as shown in Figure~\ref{fig:softx sep+pfalse}. These values can be found in the catalog (column e\_Xdeg).

\textsc{Nway} calculates the probability that each \textit{WISE} or Pan-STARRS source is the correct counterpart to a specific X-ray source  ($p_\mathrm{Cat,best}$ and $p_\mathrm{PS,best}$ for CatWISE20 and PS1-DR2, respectively\footnote{$p_\mathrm{best}$ is the \textit{relative} probability that each potential counterpart is correct; it assumes that one of the matches found is correct. $p_\mathrm{any}$ provides the probability that any one of the potential counterparts is correct. $p_\mathrm{best}$ is therefore equivalent to adopting \textit{p\_i} \citep{salvato2018} as the best potential counterpart to each NSS80 source.}).
For X-ray sources with large positional uncertainties such as pure \nustar sources without lower-energy X-ray information, counterpart identification is often less reliable, which will be captured in the lower probability values returned by \textsc{Nway}.
In addition to the best counterpart match probability, \textsc{Nway} provides the probability that any of the CatWISE/PS1-DR2 sources is the right counterpart ($p_\mathrm{Cat,any}$ and $p_\mathrm{PS,any}$ for CatWISE20 and PS1-DR2, respectively); a higher probability indicates a lower false-association likelihood.
To ensure that we only include sources with a high-probability of a correct match, we applied the following constraints on the matching probabilities and discarded any sources from Round \#1 which do not comply: $p_\mathrm{CAT/PS,any} > 0.5$ and $p_\mathrm{CAT/PS,best} > 0.8$. We treat sources at low Galactic latitudes ($|b| < 20^\circ$) separately since source confusion is high closer to the Galactic plane as a result of the high density of Galactic sources; colour/magnitude priors appropriate to extra-galactic sources do not apply to a population containing a large number of stellar X-ray emitters. A detailed study focusing on Galactic and low latitude sources could improve the counterpart matching for these sources by investigating different potential priors (for example, the top left panel of Figure~\ref{fig:nway distributions} implies that $W1$ may be effective), which may be particularly valuable in these crowded regions. Only two Galactic sources lack a soft X-ray counterpart. No sources outside of the galactic plane that pass these probability cuts lack a soft X-ray counterpart - these matches can therefore be considered reliable and representative of the extra-galactic population.

The high-probability matches of the NSS80 serendipitous sources at high Galactic latitudes are then used to generate \textit{WISE} and optical magnitude distributions in addition to color distributions which all serve as photometric priors for the second round of \textsc{Nway} matching, as described in Figure~\ref{fig:nway flowchart round comb}(b). The $W1$ magnitude and $W1$--$W2$ color distributions for the CatWISE20 base catalog  generated with \textsc{Nway} are shown in the top panels of Figure~\ref{fig:nway distributions}.
$W1$--$W2$ color is known to correlate with the presence of an AGN \citep[e.g.,][]{stern2012, assef_wise_2018} and $W1$ presents the deepest observations, therefore we choose these as our color and magnitude priors. Evidently, the \textit{WISE} color distribution for sources at high Galactic latitudes (blue distribution) provides a valuable way to distinguish between stellar objects and AGN.
A similar result holds for PS1-DR2 $i$-band magnitudes and $g-i$ colors, where reliable counterparts are bluer in the optical compared to the base population from PS1-DR2 (see bottom panels of Figure~\ref{fig:nway distributions}). The $g$ and $i$ bands are both deep in PS1-DR2, and the $g-i$ color can be used to select for reddened AGN \citep[e.g.,][]{klindt2019}.

As an input for prior-based calculations, \textsc{Nway} also requires a distribution of counterparts (WISE or optical) that are not associated with X-ray sources. For this, we use the histograms of all sources in our 40'' search fields around the \nustar positions (grey histograms in the top panels of Figure~\ref{fig:nway distributions}). X-ray source counterparts contribute to $< 2$\% of these histograms, so they are an excellent representation of the background population.

\textit{At this point, we have two ``good'' match catalogues: X-ray--optical, and X-ray--IR, which are used to evaluate priors for the more thorough matching in Round 2.}

\textbf{$\star$ Round \#2: improvement of the \textsc{Nway} matching using priors |  Figure~\ref{fig:nway flowchart round comb}(b)} \\	
Round \#2 now uses the optical and MIR priors evaluated after matching in Round \#1 ($W1$ magnitude, $W1-W2$ color, $i$-band magnitude, and $g-i$ color) to help identify the most probable optical counterparts for the NSS80 sources.
The primary X-ray source catalog is categorized into low and high Galactic latitude sources with the division at $|b| = 20^\circ$. 
$|b| < 20^\circ$ sources are matched geometrically (i.e., including positional errors but not using photometric priors) to the base CatWISE20 and PS1-DR2 catalogs, as in Round 1.
Counterparts for the $|b| > 20^\circ$ X-ray sources are identified using the magnitude and color priors from Round \#1 as inputs to \textsc{Nway}; therefore these matches are weighted towards brighter sources and those with AGN-like colors. 
Figure~\ref{fig:nway distributions} compares the priors used in this round with the distributions used to construct them. It also shows the distributions for $|b| < 20^\circ$ sources and all possible matches; the latter can be taken as a reasonable indication of the \lq background\rq, or unmatched, distributions. We also compare the results of Round~\#2 with a version of the catalogs created using a positional offset to simulate the chances of a random association (as advised in the \textsc{Nway} documentation) and assess the probability of a false association based on these results. This informs our selection of threshold cuts in the remainder of this section. For each NSS80 source, the highest probability CatWISE20/PS1-DR2 match is stored in the catalog, if $p_\mathrm{CAT/PS,best} >$ 10\%.

\textit{We now have two ``improved'' match catalogues: X-ray--optical, and X-ray--IR, which contain a potential match for each NSS80 source.
These catalogues must then be assessed to select a preferred counterpart, which we refer to as the \lq principal counterpart\rq. The goal of this process is to characterise the general optical and MIR properties of the NSS80 sources (see Sections~\ref{subsec:wise properties nustar} and \ref{subsec:optical prop}) and prepare for spectroscopic follow-up (see Section~\ref{subsec:optical spectroscopy}).}

\textbf{Final step: identification of principal counterpart |  Figure~\ref{fig:nway flowchart round comb}(b) and Figure~\ref{fig:flowchart best optical}} \\	
Finally, the two catalogs are combined to form a refined catalog, with a single \textsc{Nway} match in each of CatWISE20 and PS1-DR2 for each NSS80 source.
This refined catalog includes X-ray information, multi-band positions and photometry (which can be expanded according to one's preferences), and key \textsc{Nway} information such as $p_\mathrm{CAT/PS,best}$ and $p_\mathrm{CAT/PS,any}$ -- all of these will be included in the online NSS80 catalog; see Appendix~\ref{appendix:nss80 catalogue}\footnote{Each most probable PS1 \textsc{Nway} match is kept in the catalog (regardless of whether it is selected as the principal counterpart), and are found in the columns with the suffix \textit{-PS1}, columns 119-132.}. The next part of this section describes selecting the best available optical counterpart, which is the \textsc{Nway}-selected source in the majority of cases\footnote{Counterpart information in the catalog has the suffix \textit{-cpart}, columns 134-146.}.

\begin{figure}
	\centering
    \includegraphics[width=45pc]{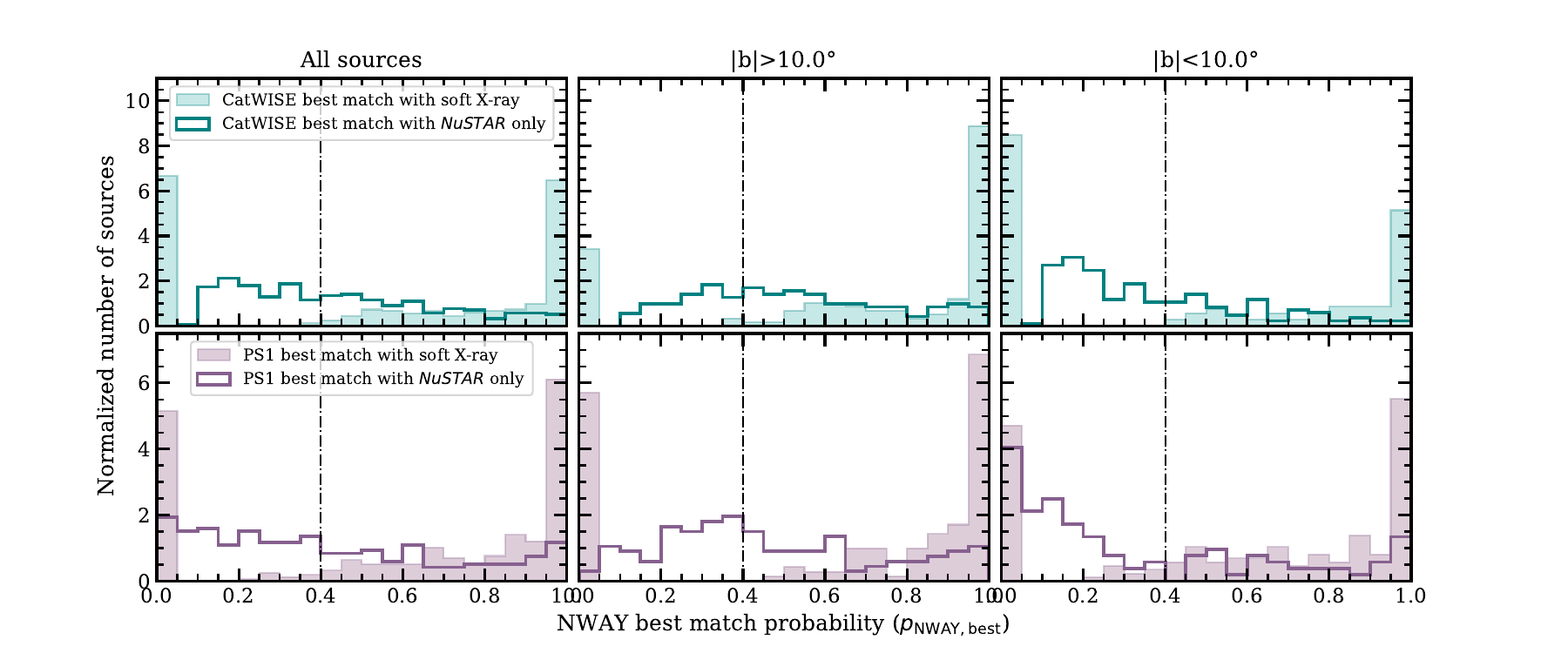}	
	\vspace*{-7mm}
	\caption{A histogram of the distribution of match probabilities for the \textsc{Nway} CatWISE20 (top panel) and PanSTARRS (bottom panel) candidates before applying the $>$10\% $p_\mathrm{Cat/PS1,best}$ cut at the end of Round \#2, both for \textit{NuSTAR}-only (open) and improved lower-energy X-ray positions (filled). 
		The high quality probability cut ($p_\mathrm{Cat/PS1,best} \geq 0.4$) is plotted with a dash-dotted line.
		Evidently, in the absence of more reliable lower-energy X-ray positions, \textsc{Nway} returns low probabilities for the CatWISE20/PS1-DR2 matches. Left panel shows all sources; middle and right panels show high and low Galactic latitude sources respectively.    
	}
	\label{fig:nway prob distributions}	
\end{figure}

When selecting the final principal optical counterparts, subsets of the NSS80 sources are dealt with in different ways, depending on the available optical or lower-energy X-ray data and the results of the \textsc{Nway} matching, as outlined below. Generally we consider the \textit{CatWISE20} match to be the primary counterpart, with optical associations secondary; Figure~\ref{fig:nway distributions} shows that the MIR distribution of best \textsc{Nway} matches is more distinct from that of all possible matches than the equivalent comparison using optical distributions (for extra-galactic sources). We note that in all cases the results were visually inspected (with imaging in all cases, and optical spectra where available; see Section~\ref{subsec:optical spectroscopy}) by several of the authors, to assess the results. If multiple convincing counterparts are seen then the match is not flagged as reliable, although the results are not changed for a significant number of sources. 
We consider the following cases, with each following a branch of Figure~\ref{fig:flowchart best optical}:

\begin{enumerate}[i)]
    \item 
\textbf{\textit{Sources with a lower-energy X-ray match, at least one CatWISE20 match, and at least one PS1 match within 2.7$''$ of the CatWISE20 position:}}
As mentioned in Section~\ref{subsec:soft xray cpart}, 964/1274 NSS80 sources have a lower-energy X-ray counterpart and, therefore, will have more reliable multi-wavelength counterpart associations than those with \textit{NuSTAR}-only positions.
Of the 964 lower-energy X-ray sources, 850 have a CatWISE20 match to the lower-energy X-ray position,\footnote{The majority of these have a greater than 30\% CatWISE20 match probability, $p_\mathrm{CAT,best}$.} of which 573 have at least one PS1 association found by \textsc{Nway} (Pan-STARRS only covers declinations $\gtrsim\,-30^\circ$; 627/850 sources with soft X-ray and CatWISE20 matches fall in this region). 
However, given that the two catalogs were independently matched to the lower-energy X-ray positions, the PS1 position is not necessarily an exact match to the CatWISE20 position. A maximum angular separation of 2.7$''$ between the PS1 and CatWISE20 positions removes counterparts where the probability of a correct match is $\lesssim$0.1\% based on positional uncertainty alone \citep[with probabilities converted from e.g.,][]{lake2012}. 
Hence, for the 573 with lower-energy X-ray\,$+$\,CatWISE20\,$+$a PS1 counterpart from \textsc{Nway} within 2.7$''$ of the CatWISE20 position, we adopt this PS1 source as the principal optical counterpart and it is flagged as \textit{PS1-NWAY}\footnote{A \textit{PS1-NWAY} match within 2.7$''$ of the CatWISE20 position will be selected even if there is a closer potential optical counterpart that is not found by \textsc{Nway}.}. If the position from \textsc{Nway} is too distant but a match can be found in PS1 by matching manually then this is labelled \textit{PS1-MAN} (31 sources\footnote{These numbers can be found for any label by filtering the catalog on the column \textit{OOrig\_cpart}.}).

\item
\textbf{\textit{Sources with a lower-energy X-ray match, at least one CatWISE20 match, but no PS1 match within 2.7$''$:}}
The lower-energy X-ray\,$+$\,CatWISE20\,$+$PS1 sources with angular separations $>$\,2.7$''$ between the CatWISE20 and PS1 positions (with a mean separation of 7.1$''$) are less reliable and possibly different sources altogether.  
To identify potential optical counterparts for the 34 sources with PS1 matches rejected in the previous step, as well as the 277/850 lower-energy X-ray\,$+$\,CatWISE20 counterparts which lack a PS1 association from \textsc{Nway}, we matched, in order of priority, to (1) the SDSS photometric catalog DR12 \citep{alam2015}, (2) the second data release of the NOIRLab Source Catalog \citep[NSC2;][]{nidever2021}, (3) the Dark Energy Survey Data Release 2 \citep[DES-DR2;][]{abbott2016}, and (4) the USNOB1 catalog \citep{monet2003}, using a 2.7$''$ positional uncertainty. These matches were done based on the closest available source and identify 228 optical counterparts. They are labelled according to the catalog used\footnote{The number of sources matched to other catalogs in this step are SDSS: 2; NSC2: 200; DES2: 6; USNOB1: 5.}.

\item
\textbf{Sources with a lower-energy X-ray match, but no CatWISE20 match:}
114/964 lower-energy X-ray counterparts lack any CatWISE20 association (i.e. a CatWISE20 \textsc{Nway} match with $p_\mathrm{Cat,best} > 0.1$); 79\% (90/114) of these sources are at low Galactic latitudes.
If an \textsc{Nway} PS1 match is available for these sources this is used, and if not a PS1-DR2 match based on angular separation alone is attempted using the \textit{Chandra}, \textit{XMM-Newton} and \textit{Swift}-XRT positions, with positional offset limits of 2.5$''$, 5$''$ and 6$''$, respectively. These values are used instead of the individual source probabilities used in \textsc{Nway} because the goal in this case is different; we aim to select a single closest match and exclude distant and therefore unlikely matches, rather than statistically assessing the probabilities. If no PS1-DR2 is found, we move to the aforementioned optical catalogues, identifying counterparts for 86/114 sources in this category.  Matches made in this way have flags with the suffix \textit{-softX}.

In total we have identified optical associations for $\sim$\,85\% (824/964) of the NSS80 sources with lower-energy X-ray counterparts.
The remaining 140/964 lower-energy X-ray sources lack any optical association at this stage: i.e. we find no optical source down to the magnitude limits of the searched optical catalogs within a matching radius of 2.7$''$ (when there is a CatWISE counterpart), or the matching radius of the soft X-ray counterpart (when there is no CatWISE counterpart)\footnote{Sources which lack an optical counterpart may do so because their counterpart is faint/distant.}.

\item
\textbf{\textit{Sources with no lower-energy X-ray match:}}
For the 310/1274 NSS80 sources without lower-energy X-ray counterparts, the X-ray positional error circle from \textit{NuSTAR} is comparatively large, so unique counterparts cannot be identified with high confidence (see Figure~\ref{fig:nway prob distributions}). 
However, from our \textsc{Nway} matching we were able to secure a potential \textit{WISE} association for all 310 sources with a $p_\mathrm{Cat,best} \gtrsim 0.1$; note that NuSTARJ182353+0742.0 is a borderline case with $p_\mathrm{Cat,best} = 0.095$ and $p_\mathrm{Cat,any} = 0.53$.
To search for an optical counterpart for these sources, for the purposes of optical spectroscopic follow-up only (i.e., to retrieve an optical position and $r$-magnitude if detected), we cross-matched the CatWISE20 position to the aforementioned optical surveys using a 2.7$''$ search radius, starting by checking the position of the \textsc{Nway} PS1 match, then moving onto a manual PS1 match, and finally working through the other optical catalogs. In total we obtained a potential optical counterpart for 243/310 \textit{NuSTAR}-only X-ray sources. Matches made in this way have flags with the suffix \textit{-CatWISE}, and should be viewed with more caution than those with soft X-ray counterparts.

\item
\textbf{\textit{Visually inspected sources:}}
Finally, as each source was visually inspected we find a small number (21/1274) where it is appropriate to override the procedural matching explained in this section. The majority of these are sources that fail a matching distance criterion by only a small amount but on visual inspection appear likely to be a correct match. For example, if a PS1 source is further than 2.7$''$ from the CatWISE20 counterpart, but it is within the match radius for the soft X-ray counterpart and in a region with few nearby sources we may choose to manually select the optical counterpart. Matches made in this way have flags with the suffix \textit{-VI}, and should also be treated with appropriate caution.

This step increases the optical completeness of the NSS80 catalog from $\sim$85\% (1077/1274) to $\sim$86\% (1098/1274): 844/964 lower-energy X-ray and 254/310 \textit{NuSTAR}-only X-ray associations; see Appendix~\ref{appendix:nss80 catalogue} for an abbreviated code indicating the origin of the adopted optical counterpart to the \textit{NuSTAR} source. 
\end{enumerate}

Our selection process to identify the principal adopted optical counterpart is summarised in Figure~\ref{fig:flowchart best optical}.
Not all \textsc{Nway} matches will be true counterpart associations since the cross-matching depends on X-ray counterpart information (i.e. positional uncertainty, which is influenced by source and background counts), whether a soft X-ray counterpart can be found (generally these have much smaller positional uncertainties than \nustar and thus improve cross-matching), the Galactic latitude of the source (i.e., source confusion easily occurs for high density fields usually within the Galactic plane) and the depth of the optical/IR imaging surveys (e.g., shallow imaging can miss counterparts that are evident in deeper surveys), to mention but a few. 
Figure~\ref{fig:nway prob distributions} shows the distribution of \textsc{Nway} match probabilities ($p_\mathrm{CAT/PS,best}$) for the ``best'' CatWISE20 (top panel) and PanSTARRS (bottom panel) candidates, both for \textit{NuSTAR}-only (open) and improved lower-energy X-ray positions (filled). Evidently, the CatWISE20/PS1-DR2 probabilities are low in the absence of more reliable lower-energy X-ray positions.

We therefore attempted to identify a subset of our NSS80 that have high-probability matches based on our \textsc{Nway} analysis to minimise false associations---henceforth described as \lq reliable\rq\ counterparts---and provide this information in the catalog (NWAY\_RFlag; Table~\ref{tab:description of NSS80 cat}.
We first identified high-probability CatWISE20 matches with thresholds of $p_\mathrm{Cat,best} > 0.4$ and $p_\mathrm{Cat,any} > 0.5$. 
The expected rate of false associations based on matching with randomised catalogs \citep[as in][]{salvato2018} around each \nustar/soft X-ray position gives a 1.4\% chance of false association if a soft X-ray source is present, and 12.6\% if there is no soft X-ray source, improving to 0.5\% and 9.6\% respectively when limited to $|b| > 10^\circ$.
We supplement the reliable CatWISE20 counterparts with reliable PS1-DR2 counterparts, if present, for sources that fail the CatWISE20 probability criterion, by applying the same higher probability cut on their PS1-DR2 probabilities (i.e., $p_\mathrm{PS,best}>0.4$ and $p_\mathrm{PS,any}>0.5$).
In total, 963/1274 NSS80 sources have a reliable CatWISE20 or PS1-DR2 counterpart, of which 76\% (726) are high Galactic latitude sources. 370 of these only have a reliable CatWISE20 counterpart and 103 only PS1-DR2; many of these do have an \textsc{Nway} match that is below the 0.4/0.5 thresholds.
The remaining 321 NSS80 sources have low-probability \textsc{Nway} counterparts and should, therefore, be used with caution to avoid biasing the results. 

\textit{Finally, we have a single principal match catalog, checked at each stage for the most reasonable match. All selected matches are included, and particularly high probability matches are flagged as reliable.}
This subset can be selected with the flag \textit{NWAY\_RFlag}; see Table~\ref{tab:subset_summary}.

\begin{figure*}
	\centering
	\includegraphics[width=42pc]{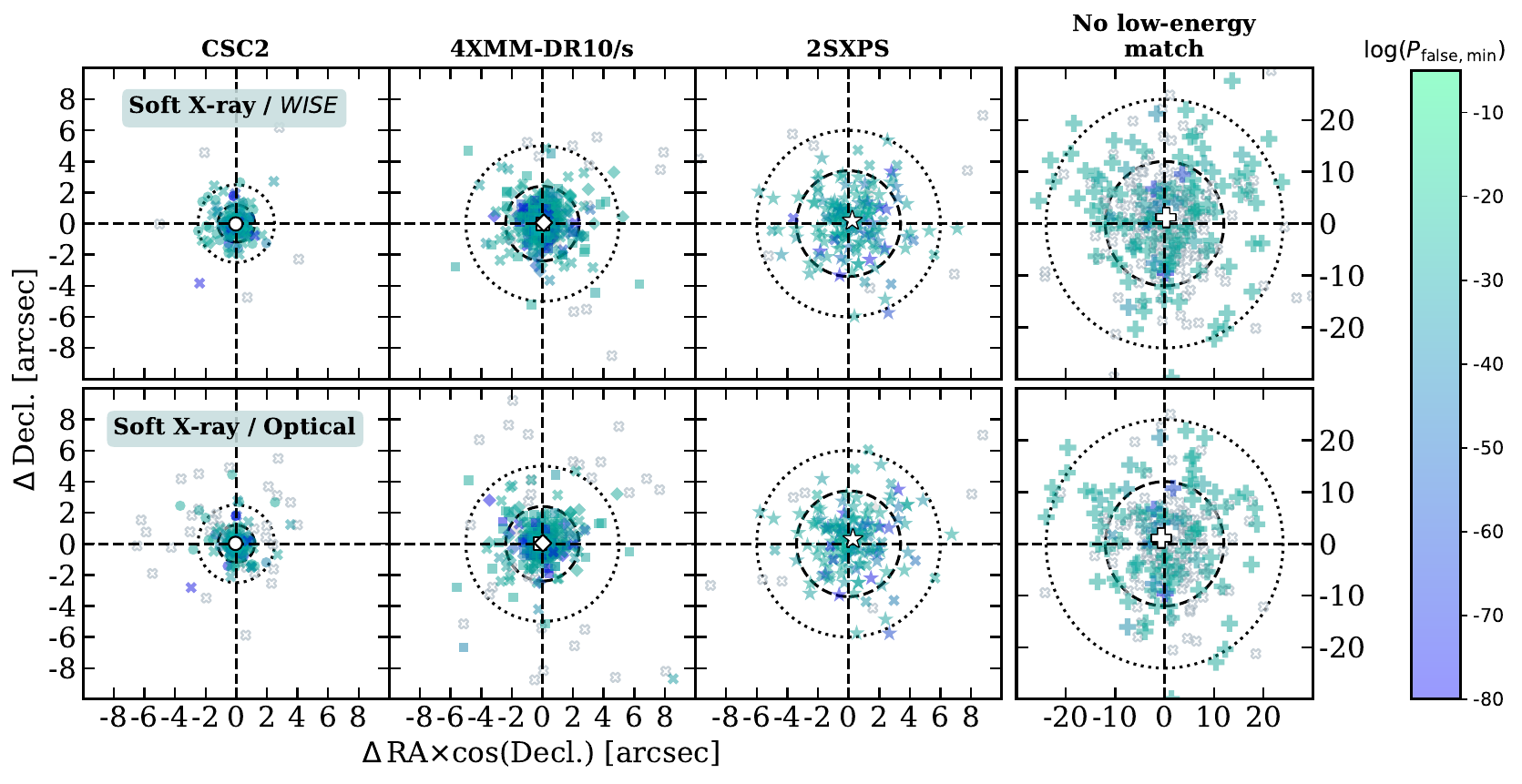}	
	\vspace*{-3mm}
	\caption{Astrometric offsets between the lower-energy X-ray counterpart coordinates and the \textit{WISE} (top row) and the adopted optical (bottom row) coordinates, color-coded using the \nustar detection likelihood ($p_\mathrm{false,min}$), for \textsc{Nway}-identified counterparts flagged as reliable. The high Galactic latitude lower-energy X-ray counterparts are from CSC2 (left column; circles), 4XMM-DR10/s (middle left column; squares and diamonds, respectively), 2SXPS (middle right column; stars), and \nustar only (right column; plus symbols). Low Galactic latitude sources are plotted with color-coded crosses and sources with less reliable counterpart associations are indicated with open grey-edged crosses.
		The dashed (dotted) circles correspond to the 1$\sigma$ (2$\sigma$) search radii for each lower-energy X-ray telescope: 1.2$''$ (2.5$''$), 2.4$''$ (5$''$),  3.4$''$ (6$''$), 12$''$ (24$''$) for \textit{Chandra}, \textit{XMM-Newton}, \textit{Swift}-XRT, and \textit{NuSTAR}, respectively. 
		The white circle, square, diamond, star, and plus indicates the median astrometric offset for each of the extragalactic samples. 
    Note the different scale for the \textit{NuSTAR}-only section.
	}
	\label{fig:nway astrometric offsets}	
\end{figure*}

\begin{figure*}
	\centering
	\includegraphics[width=42pc]{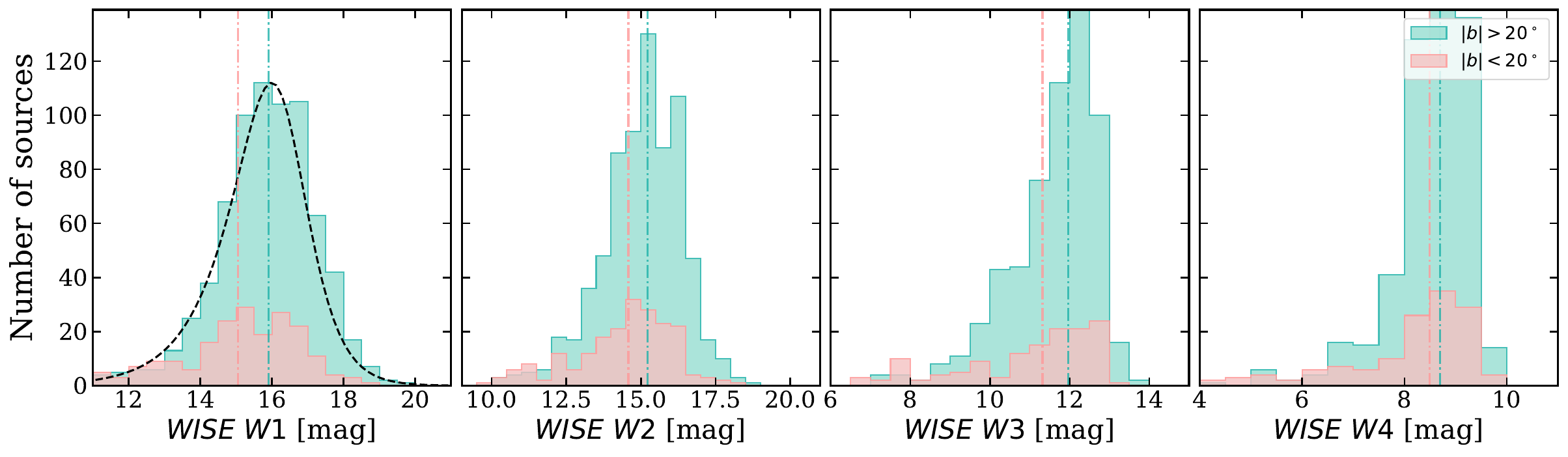}		
	\includegraphics[width=32pc]{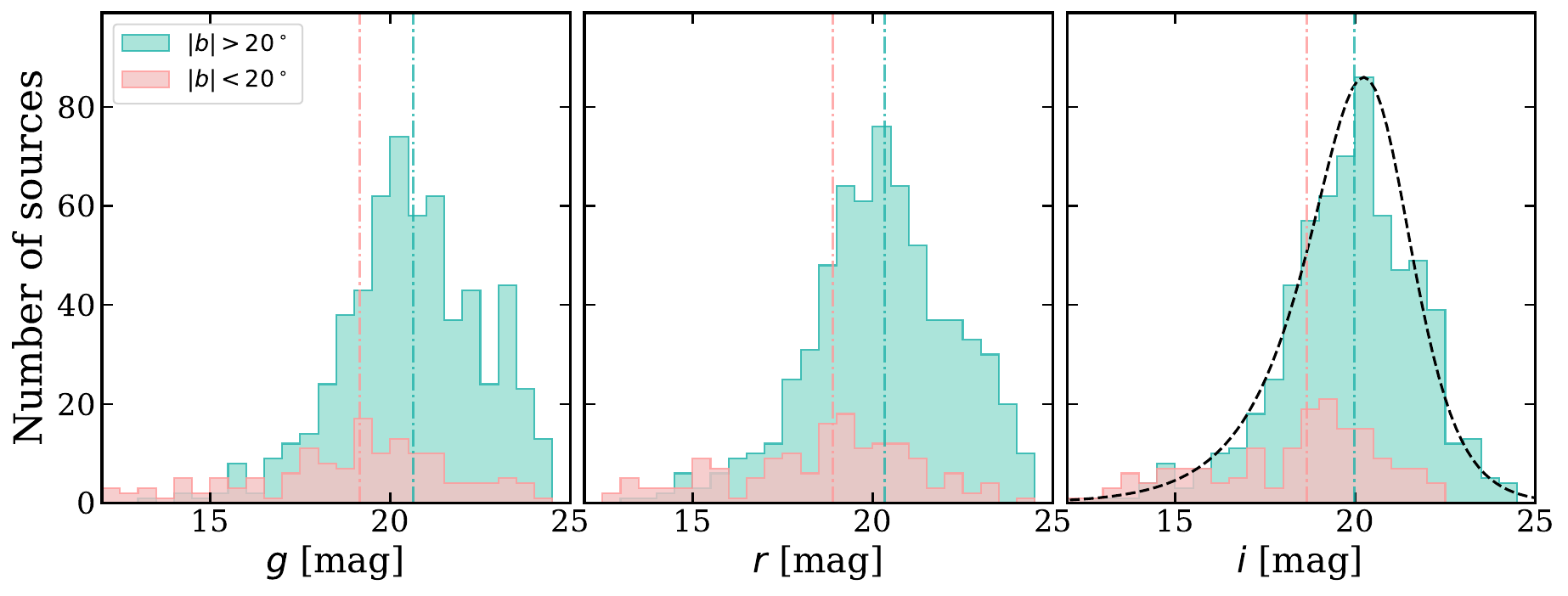}	
	\includegraphics[width=22pc]{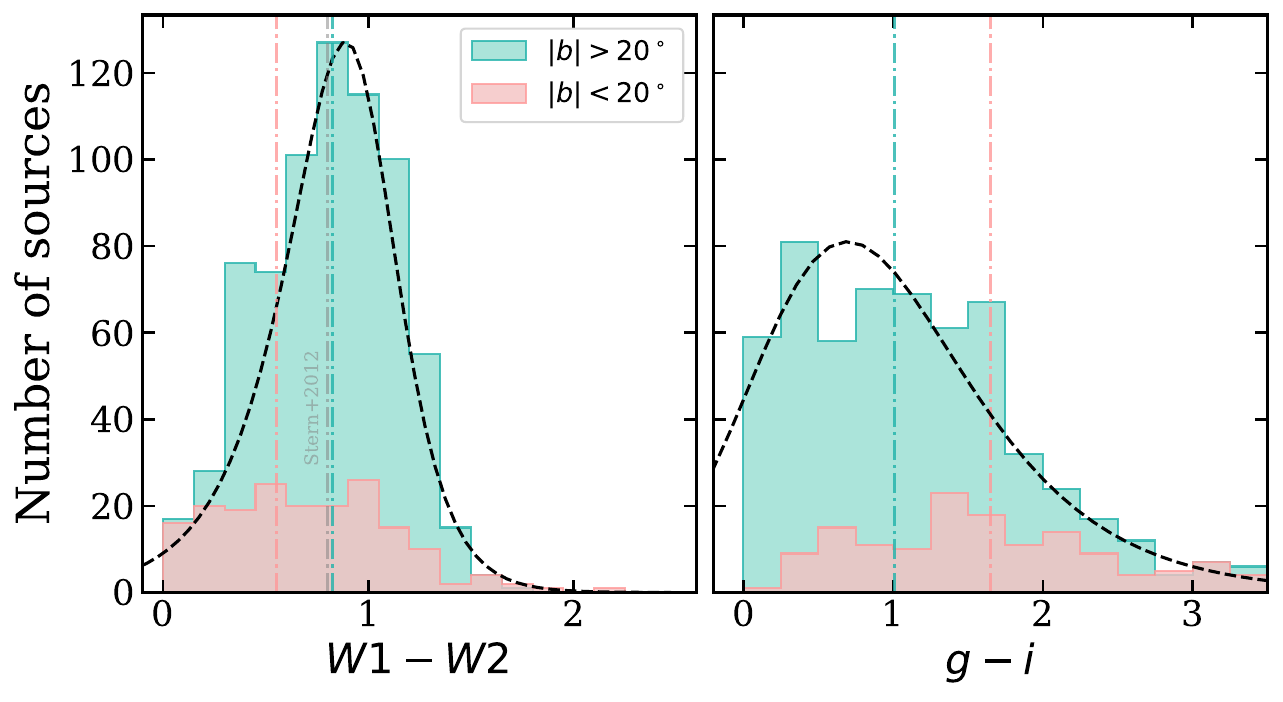}		
	\vspace*{-3mm}
	\caption{Distributions of the MIR and optical magnitudes and colors for the high probability \textsc{Nway}-selected counterpart samples at high Galactic latitudes ($|b| > 20^\circ$; green) and low Galactic latitudes ($|b| < 20^\circ$; peach). The median of each distribution is plotted with a dash-dotted line. The dashed black lines indicate the distributions constructed in \textsc{Nway} round \#1 and used as priors in round \#2.
		\textit{Top four panels:} magnitude distributions for the four photometric \textit{WISE} bands, for the sources with successful CatWISE20 matches satisfying the high probability cut. 
		\textit{Middle three panels:} the $g$-band, $r$-band and $i$-band magnitudes (corrected for Galactic extinction) for all the sources satisfying the \textsc{Nway} high probability cuts.
		\textit{Bottom two panels:} the $W1-W2$ and $g-i$ colors for all the sources satisfying the \textsc{Nway} high probability cuts. These colors are used as as priors for \textsc{Nway}. The dash-dotted, grey line indicates the \citet{stern2012} AGN threshold of $W1-W2 > 0.8$. 
  \textit{WISE} magnitudes are Vega and optical magnitudes are AB.
	}
	\label{fig:nway magnitude hist}	
\end{figure*}
Figure~\ref{fig:nway astrometric offsets} shows the distribution of astrometric offsets between the X-ray source and the  adopted CatWISE20 (top panel) and optical (Pan-STARRS, SDSS, NSC2, DES, and USNOB1; bottom panel) counterpart, for the high probability \textsc{Nway}-selected samples (filled symbols) and for sources with low \textsc{Nway} probabilities (open, grey-edged crosses).
Sources with Galactic latitudes $|b| > 20^\circ$ are plotted with circles, squares/diamonds and stars for NSS80 sources with CSC2, 4XMMDR10/s and 2SXPS counterparts, respectively. The low Galactic latitude sources identified with \textsc{Nway} based solely on geometric information are shown with crosses.

In Figure~\ref{fig:nway magnitude hist} we show histograms of the \textit{WISE} (top panels) and optical ($g$,$r$ and $i$; middle panels) magnitudes\footnote{All PanSTARRS magnitudes quoted in the NSS80 catalog are, in order of preference, Kron, PSF, or the default aperture magnitudes. There are flagged in the columns \textit{mag\_type\_PS1} and \textit{mag\_type\_cpart} for \textsc{Nway} and selected counterpart magnitudes, respectively. The Kron magnitudes are better for extended sources, which would constitute the larger number of counterparts for the \textit{NuSTAR} serendipitous survey given their apparent redshift distributions. Thus if a Kron magnitude is available, it is used, and if not we check for the other types. If users perform a detailed study on individual sources we advise a check that the magnitude is appropriate for their application.} for the high probability \textsc{Nway}-selected samples at high (green) and low (peach) Galactic latitudes, respectively. 
The median magnitudes of the low Galactic latitude sources are on average 1-2 magnitudes brighter than the high Galactic latitude sources.  

In addition the $W1-W2$ and $g-i$ colors for all the sources satisfying the \textsc{Nway} high probability cuts are shown in the bottom panels of Figure~\ref{fig:nway magnitude hist}. The median $W1-W2$ for high Galactic latitude sources is similar to the AGN threshold of $W1-W2 = 0.8$ presented in \citet{stern2012}.

\subsection{Optical spectroscopy} \label{subsec:optical spectroscopy}
To maximise the scientific impact of the \nustar observations and to explore the intrinsic source properties, we carried out a major, coordinated spectroscopic campaign using a broad range of telescopes across the globe to obtain redshifts of the NSS80 sources. 
The analysis of, and classifications obtained from, these new spectroscopic data and those from pre-existing spectroscopy, are described in Section~\ref{subsubsec:spectral classification}.

\subsubsection{Dedicated Follow-up Campaign}  \label{subsubsec:follow-up spectra}
\citetalias{lansbury2017_cat} obtained spectroscopic redshifts for 276 of the 497 \nustar serendipitous sources through multi-year observing programmes in both hemispheres.
For NSS80 we continued the multi-year spectroscopic follow-up campaign with the following telescopes (see Section~4.4.2): 
\begin{itemize}
	\vspace{-0.05cm}
	\setlength\itemsep{-0.1em}
	\item{In the \textbf{Northern hemisphere} we used a combination of the 5.1\,m Hale Telescope at the Palomar Observatory (Decl.\,$\gtrsim$\,-21$^\circ$; P.I. \mbox{F. A. Harrison} and \mbox{D. Stern}) and 10\,m Keck I at the W. M. Keck Observatory (-35$^\circ$\,$\lesssim$\,Decl.\,$\lesssim$\,75$^\circ$; P.I. \mbox{F. A. Harrison}), targeting brighter targets preferentially with the former, and fainter targets with the latter;}
	\item{In the \textbf{Southern hemisphere} we used a combination of the 8.2\,m Very Large Telescope (VLT) at the European Southern Observatory (Unit Telescope 1 (UT1); $-70^\circ$\,$\lesssim$\,Decl.\,$\lesssim$\,$10^\circ$; P.I. \mbox{G. B. Lansbury} and \mbox{L. Klindt}), and the 11\,m Southern African Large Telescope (SALT) which is part of the South African Astronomical Observatory in Sutherland (\mbox{-80$^\circ$}\,$\lesssim$\,Decl.\,$\lesssim$\,20$^\circ$; PI \mbox{L. Klindt}).} 
\end{itemize}

This multi-latitude campaign resulted in spectroscopic redshifts and classifications for 550 NSS80 sources (including those in \citetalias{lansbury2017_cat}, 2 galaxy and 4 AGN pairs), as summarized in Table~\ref{tab:telescope follow-up}.
We provide the date of the observing run, the telescope and instrument used, and the total number of sources observed, for both visitor and service mode observations; this key observing information is listed for each individual source in the spectroscopic catalog (see Appendix~\ref{appendix:optical spectra} and Table~\ref{tab:spec results}).
This subset can be selected from the primary catalog with the flag \textit{SpecCAT}; see Table~\ref{tab:subset_summary}.
The reader can refer to Table~4 in \citetalias{lansbury2017_cat} for a summary of the NSS40 spectroscopic follow-up observations. 

We note that all instruments are long-slit spectrographs. A further 44 NSS80 sources were followed up, but we failed to obtain a reliable redshift measurement: 33/44 sources are optically faint (7 undetected in the $r$-band and 26 with $r > 20$), 9/44 sources have low S/N spectra due to compromised observations (either due to weather conditions or telescope failures), one source is optically-undetected, but given its radio and X-ray emission it is possibly a radio-lobe associated with the science target, and the remaining source is a BL~Lac candidate with a power-law continuum lacking spectral features; see Table~\ref{tab:spec results} and Figure~B5.
Hence, 680/1274 (53\%) NSS80 sources have not yet been targeted for spectroscopic follow-up.

\begin{deluxetable}{lllll}
\tablecolumns{5}
\tablewidth{\textwidth}
\tablecaption{\label{tab:telescope follow-up} Chronological list of the optical spectroscopic follow-up campaign of NSS80 sources post \citetalias{lansbury2017_cat}; see \citetalias{lansbury2017_cat} for the details of the NSS40 spectroscopic follow-up campaign.
		Listed for visitor mode observations are the ID assigned to each observing run, the date of the observing run, the telescope and instrument used and the number of sources observed. 
		For service mode observations we also provide the observing period (semester) and the observing run end date. The different telescope instruments include the Double Spectrograph (DBSP), the Low Resolution Imaging Spectrometer (LRIS), the Focal Reducer and low dispersion Spectrograph (FORS2), and the Robert Stobie Spectrograph (RSS).
	}
 \tablehead{
		Run ID & UT Start Date & Telescope & Instrument & $N_\mathrm{src}$ \\
		}
		\startdata
        \multicolumn{5}{c}{Visitor mode} \\
        \tableline
		1  & 2016 October 02     & Palomar  & DBSP    & 3  \\                                          
		2  & 2016 November 27    & Keck     & LRIS    & 3  \\               
		3  & 2017 April 28       & Keck     & LRIS    & 16 \\                        
		4  & 2017 July 21        & Keck     & LRIS    & 15 \\                    
		5  & 2017 July 27        & Palomar  & DBSP    & 7 \\                   
		6  & 2017 September 14   & Palomar  & DBSP    & 2  \\             
		7  & 2017 September 16   & Keck     & LRIS    & 9 \\             
		8  & 2017 November 22    & Palomar  & DBSP    &2  \\             
		9  & 2018 March 18      & Keck     & LRIS    & 8 \\             
		10 & 2018 June 06        & Palomar  & DBSP    & 9 \\             
		11 & 2018 July 16        & Keck     & LRIS    & 4  \\             
		12 & 2018 September 09   & Palomar  & DBSP    & 2  \\              
		13 & 2018 October 03     & Keck     & LRIS    & 6  \\              
		14 & 2019 March 07       & Keck     & LRIS    & 1  \\             
		15 & 2019 July 22        & Palomar  & DBSP    & 3  \\             
		16 & 2019 July 28        & Palomar  & DBSP    & 1  \\              
		17 & 2019 August 02      & Palomar  & DBSP    & 1 \\             
		18 & 2019 August 23      & Palomar  & DBSP    & 1  \\                       
		19 & 2019 October 26     & Palomar  & DBSP    & 7 \\              
		20 & 2019 December 24    & Keck     & LRIS    & 15 \\             
		21 & 2020 June 16        & Palomar  & DBSP    & 5  \\             
		22 & 2020 September 12   & Palomar  & DBSP    & 5  \\             
		23 & 2020 September 25  & Palomar  & DBSP    & 24 \\             
		24 & 2020 October 23     & Palomar  & DBSP    & 28 \\ 
		\\  
        \tableline
		\multicolumn{5}{c}{Service mode} \\
        \tableline
		25 & P100 &  VLT & FORS2 & 13 \\      
		26 & P101 &  VLT & FORS2 & 12 \\  
		27 & P102 &  VLT & FORS2 & 9 \\ 
		28 & P103 &  VLT & FORS2 & 17 \\
		\\
		29 & 2017-SEM1+SEM2 &   SALT & RSS & 4 \\
		30 & 2018-SEM1+SEM2 &   SALT & RSS & 10 \\  
		31 & 2019-SEM1+SEM2 &  SALT & RSS & 8 \\  
		32 & 2020-SEM1 &   SALT & RSS & 3 \\ 
	\enddata
\end{deluxetable}

\subsubsection{Spectroscopic observations \& data reductions}  
\label{subsubsec:spectroscopic data reductions}
Since the spectral features of the NSS80 sources are not known \textit{a priori}, we adopt a broad wavelength coverage with sufficient spectral resolution to search for, identify and accurately measure any spectral features.
Therefore, two grating angles were selected for each target observation with Keck,\footnote{Keck/LRIS is a double spectrograph, consisting of a blue and red channel. However, only the grating angle on the red side is adjustable, while the blue side grism is fixed.} Palomar and SALT, whereas a broad wavelength coverage was achieved with a single camera station for VLT. 
This allowed us to access the 3200\,\AA{}\,$-$\,9000\,\AA{} visible wavelength range and cover commonly known AGN and quasar emission and absorption lines across a range of redshifts. For example, at lower redshifts (e.g., $z = 0.3$), emission lines such as Mg\,\textsc{ii} $\lambda$2800, [Ne\,\textsc{v}] $\lambda$3346 and $\lambda$3426, [O\,\textsc{ii}] $\lambda$3728, [Ne\,\textsc{iii}] $\lambda$3869, H$\delta$ $\lambda$4102, H$\gamma$ $\lambda$4340, H$\beta$ $\lambda$4861, [O\,\textsc{iii}] $\lambda$4959 and $\lambda$5007, [O\,\textsc{i}] $\lambda$6300 and $\lambda$6364, [N\,\textsc{ii}] $\lambda$6548 and $\lambda$6584, H$\alpha$ $\lambda$6563, and [S\,\textsc{ii}] $\lambda$6716 and $\lambda$6731 are covered, and at higher redshifts (e.g., $z = 2$), lines such as Ly$\alpha$ $\lambda$1216, Si\,\textsc{iv} $\lambda$1398, C\,\textsc{iv} $\lambda$1549, He\,\textsc{ii} $\lambda$1640, C\,\textsc{iii}] $\lambda$1909, C\,\textsc{ii}] $\lambda$2326, and Mg\,\textsc{ii} $\lambda$2800 are covered.    
For all our observations we adopted a slit width of $1-1.5''$, depending on the seeing, and configured the spectrographs to obtain low-resolution spectra (i.e., a resolving power of $R\sim1000$), which is sufficient to achieve our science goals.

For the majority of observations a total of two exposures were obtained (with the exception of very bright targets), and in the case where dithering\footnote{Dithering is important for fringing corrections, especially at longer wavelengths for standard CCD arrays such as SALT/RSS and VLT/FORS2. Keck/LRIS and Palomar/DBSP, on the other hand, have thick, red-sensitive CCDs on their red arms which have negligible fringing.} was requested, 1 horizontal tile and 2 vertical tiles with a maximum offset size of 10$''$ were obtained.
Calibration images (flat fields and arc frames) were also recorded, and spectrophometric standard stars were observed in the different instrument configurations for flux calibration. 
It should be noted that, due to the design of SALT, the spectra cannot be absolute flux calibrated. This is due to the moving pupil during exposures and tracking which consequently changes the effective area of the telescope. 
Therefore, the standards were only used for relative spectral (shape) calibration.

The spectroscopic data reductions included basic CCD pre-reductions, spectral calibration, background subtraction, spectral extraction and flux calibration. 
The tasks available in \textsc{iraf} \citep[see][]{massey1992} were used to perform the reduction and analysis processes for the Keck, Palomar and SALT spectra, and the \textsc{EsoReflex} 2.9.1 pipeline was used to reduce the VLT/FORS2 spectra.
For more details see Section 4.4 of \citet{klindt2022}. The final flux-calibrated optical spectra described herein are available in Appendix~\ref{appendix:optical spectra} and will be made available on the NSS80 webpage; see Section A.2 in \citetalias{lansbury2017_cat} for the NSS40 optical spectra). 


\subsubsection{Spectral Classification and Analysis}  
\label{subsubsec:spectral classification}

To assist in the measurement of spectroscopic redshifts, we used the open-source Manual and Automatic Redshifting Software \citep[\textsc{Marz};][]{hinton2016} by matching the observed flux calibrated input spectra (FITS file format) against a library of stellar, galaxy and AGN templates available in the \textsc{Marz} web application; see fig.~6\ in \citet{hinton2016} for a visual display of the twelve (5 stellar + 6 galaxy + 1 AGN) current templates available in \textsc{Marz}.\footnote{\citet{hinton2016} compiled the library of templates from \textsc{runz} \citep[originally developed for the use of the 2dF galaxy redshift survey;][]{colless2001} and \textsc{autoz} with original templates from 2dF \citep{colless2001}, WiggleZ \citep{drinkwater2010}, the Gemini Deep Survey \citep{abraham2004}, SDSS DR2 \citep{subbarao2002} and galaxy eigenspectra from \citet{bolton2012}.}
Via this manual template comparison approach, the object type can easily be identified and the spectrum can be redshifted to align the observed spectral lines with the template (for moderate to high S/N spectra).
From this, \textsc{Marz} provides a redshift solution, however, it does not assign uncertainties to the redshifts \citep[see][for more details]{yuan2015}.
For low S/N spectra where template comparison and line identification are arduous, we identified potential spectral features and used a look-up table of wavelength ratios based on the emission and absorption lines observed in AGN and galaxy spectra for spectral lines.
The redshift solution is then determined by cross-matching the observed wavelength ratios of the identified lines to the rest wavelength ratios based on the emission and absorption lines observed in AGN. Sources with low S/N spectra or dubious redshift measurements are flagged in the catalogue. 

During the full 80-month period, {including \citetalias{lansbury2017_cat}, we obtained redshift measurements for 550 \nustar sources, of which we spectrally classified 547 (this accounts for 43\% of the NSS80 primary catalog): 427 were obtained via our optical follow-up campaign and 123 from archival data (primarily SDSS DR16; \citealp{ahumada_16th_2020}; see Table~\ref{tab:spec results} for details). 
The source classifications and redshift measurements for all NSS80 sources with optical spectra are provided as supplementary material in Appendix~\ref{appendix:optical spectra}. 
The majority of the sources have robust redshift measurements obtained from two or more spectral lines, whilst 21/550 sources have a single-line measurement; these sources are flagged as ``quality B" redshift measurements.
Sources were selected for follow-up based on target visibility, chances of success given optical magnitude and instrument characteristics, and where possible higher probability counterparts were chosen. However, many targets were included in telescope observing programs as filler targets and thus the selection is not completely uniform.

\begin{figure}
	\centering
	\includegraphics[width=20pc]{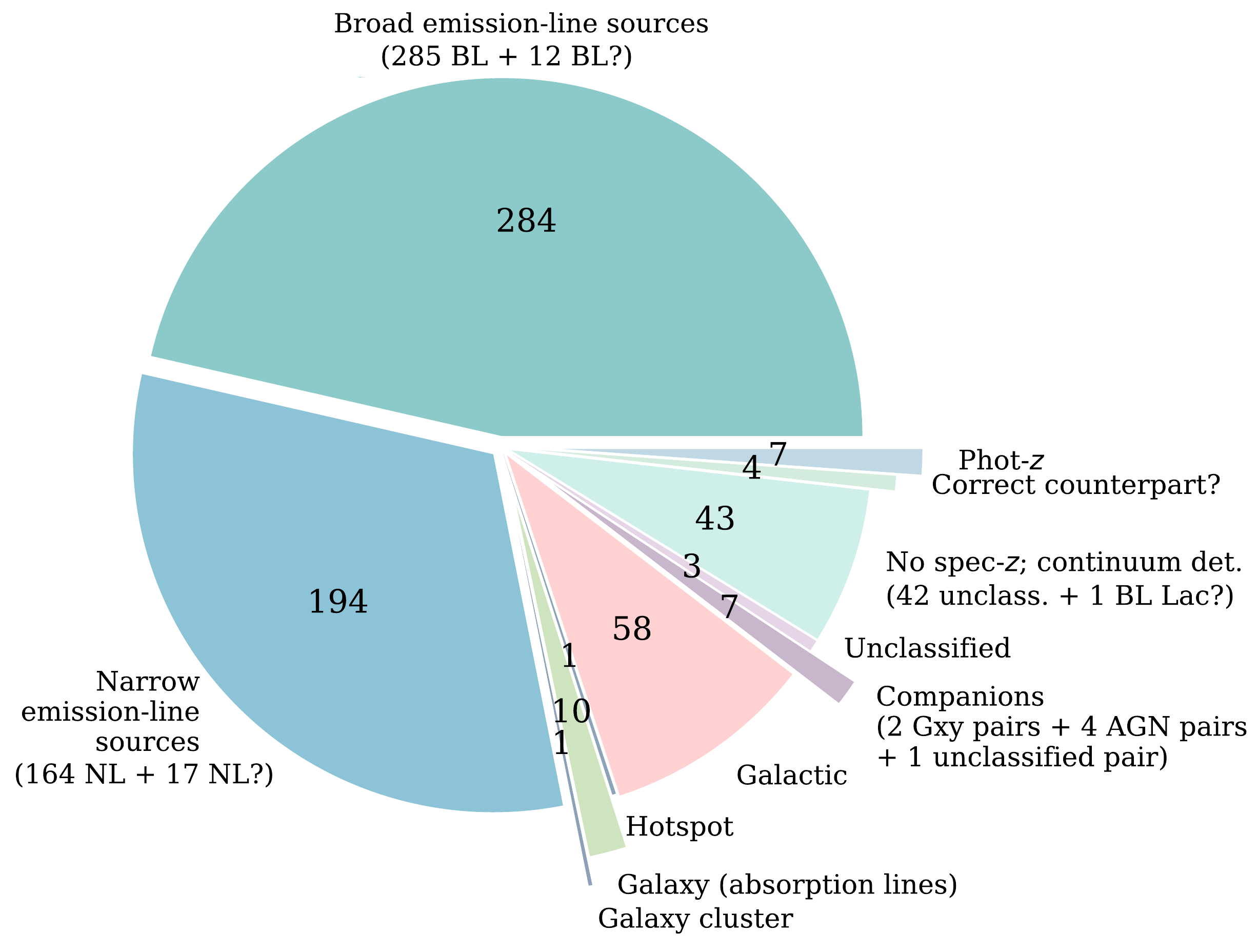}		
	\vspace*{-2mm}
	\caption{ 
		A pie-chart of the NSS80 source classifications obtained via our spectroscopic follow-up campaign. 
		In total, there are 612 source entries in the NSS80 spectroscopic catalog: 594 unique spectroscopic entries and 18 additional entries.
		Of the 594 unique entries, we have classified 492 as extragalactic sources, including 284 BL (or BL?), 194 NL (or NL?), one galaxy cluster, 10 galaxies, and three unclassified sources with redshift measurements from literature. 
		We have also classified 58 Galactic sources at $z = 0$. 
		There are a further 43 sources that lack a reliable redshift measurement, 28 of which are unclassified due to the lack of spectral lines, but a power-law continuum is detected in all of the cases, 14 sources have counterpart uncertainties due to the lack of a lower-energy X-ray counterpart or high optical source density, one source is a \textit{Fermi} BL~Lac candidate. 
		The remaining one source is a hotspot of 4C\,74.26 (radio quasar), which was targeted as the primary \textit{NuSTAR} science target; see Figure~\ref{fig:hotspot}.
		In addition to the unique entries, we obtained spectra for a further 18 sources: three AGN pairs of which one pair is a dual AGN system, two AGN-galaxy pairs, one galaxy pair, and one unclassified pair (see Figure~\ref{fig:spec dual AGN}), 7 sources with photometric redshifts, and four sources which are potentially the correct counterpart oppose to the one selected; 3/4 are fainter lower-energy X-ray sources (2 BLs + 1 phot-$z$ source of unknown type) nominally closer to the \nustar source than the selected (X-ray brighter) BL counterpart, and the remaining source is an optically bright NL galaxy at a different redshift than the \textsc{Nway}-selected NL counterpart (which is associated with the primary \textit{NuSTAR} science target). 
	}
	\label{fig:spec classifications}	
\end{figure}

Based on the spectroscopic redshift measurements, 492/550 NSS80 sources are extragalactic and 58 are Galactic. 
The extragalactic sources are classified through visual inspection of the flux-calibrated spectra into the following general classes, and subsets can be selected from the primary catalog using the \textit{Classification} column (see Table~\ref{tab:subset_summary}): 
\begin{itemize}
	\vspace{-0.2cm}
	\setlength\itemsep{-0.1em}
	\item{\textit{\textbf{Broad line object}} \textbf{(BL)} if any permitted line is significantly broader than the forbidden lines, or if a single line measurement for our quality ``B'' spectra satisfies the standard definition of a broad-line, i.e., $FWHM \geq 1000$\,km\,s$^{-1}$ (measured in \textsc{iraf});}
	\item{\textit{\textbf{Narrow line object}} \textbf{(NL)} if the permitted lines are of similar width to the narrow forbidden lines;}
	\item{\textit{\textbf{Galaxy}} \textbf{(Gxy)} if only absorption lines are detected.}
\end{itemize}
Based on this simple classification scheme, we find that 58\% (284/492) are BL AGN, 39\% (194/492) are NL objects and 2\% (10/492) are galaxies (Gxy); see Figure~\ref{fig:spec classifications}. We also classify 1/492 source as a galaxy cluster (GClstr; NuSTARJ132535-3825.6) and 3/492 sources have redshift measurements obtained from the literature, but lack spectroscopic classification.
We append the optical classification with a ``?" symbol for 29 sources (12 BL? + 17 NL?) where it is ambiguous whether the permitted lines are broad or not.
Regardless of the optical classification, the vast majority of the sources are expected to be AGN due to the detection of X-ray emission at high X-ray energies of $>$\,3\,keV, which is further confirmed for most sources by the identification of strong optical emission lines often superimposed on a power-law spectrum.
The 58 Galactic sources are not classified here but based on the NSS40 results, they are likely to include, for example, cataclysmic variables, X-ray binaries, and active stars; 40\% (23/58) of the Galactic sources are at low Galactic latitudes ($|b|$\,$<$\,$10^\circ$).
The number of Galactic \nustar sources has increased by a factor of $\sim$3 from the NSS40 to the NSS80 catalog. 
These sources will be further investigated through an additional follow-up campaign \citep[see e.g.,][for a study of the \citetalias{lansbury2017_cat} Galactic sources]{tomsick2018}.
There are also 43 NSS80 targets which we followed-up, but failed to obtain a reliable redshift measurement.
For all these cases a faint (often red) continuum is detected, one of which is a \textit{Fermi} BL~Lac candidate (NuSTARJ081003-7527.2) with a featureless power-law spectrum \citep{ackermann_contemporaneous_2016}. 
Possible reasons for the lack of spectroscopic identifications for these sources are given in Section~\ref{subsubsec:spectra vs nway}. 
Finally, the remaining source (NuSTARJ204256+7503.1) is a hotspot associated with the primary \textit{NuSTAR} science target 4C\,74.26, a radio quasar at $z=0.104$ \citep[e.g.,][]{erlund2007,erlund2010}. Figure~\ref{fig:hotspot} shows multi-wavelength imaging of the \textit{NuSTAR}-detected hotspot at a projected distance of $\sim$\,580\,kpc from the quasar. 
Hence, in total, there are 594 unique source entries in the NSS80 spectroscopic catalog given in Table~\ref{tab:spec results}.

\begin{figure*}
	\centering
	\includegraphics[width=34pc]{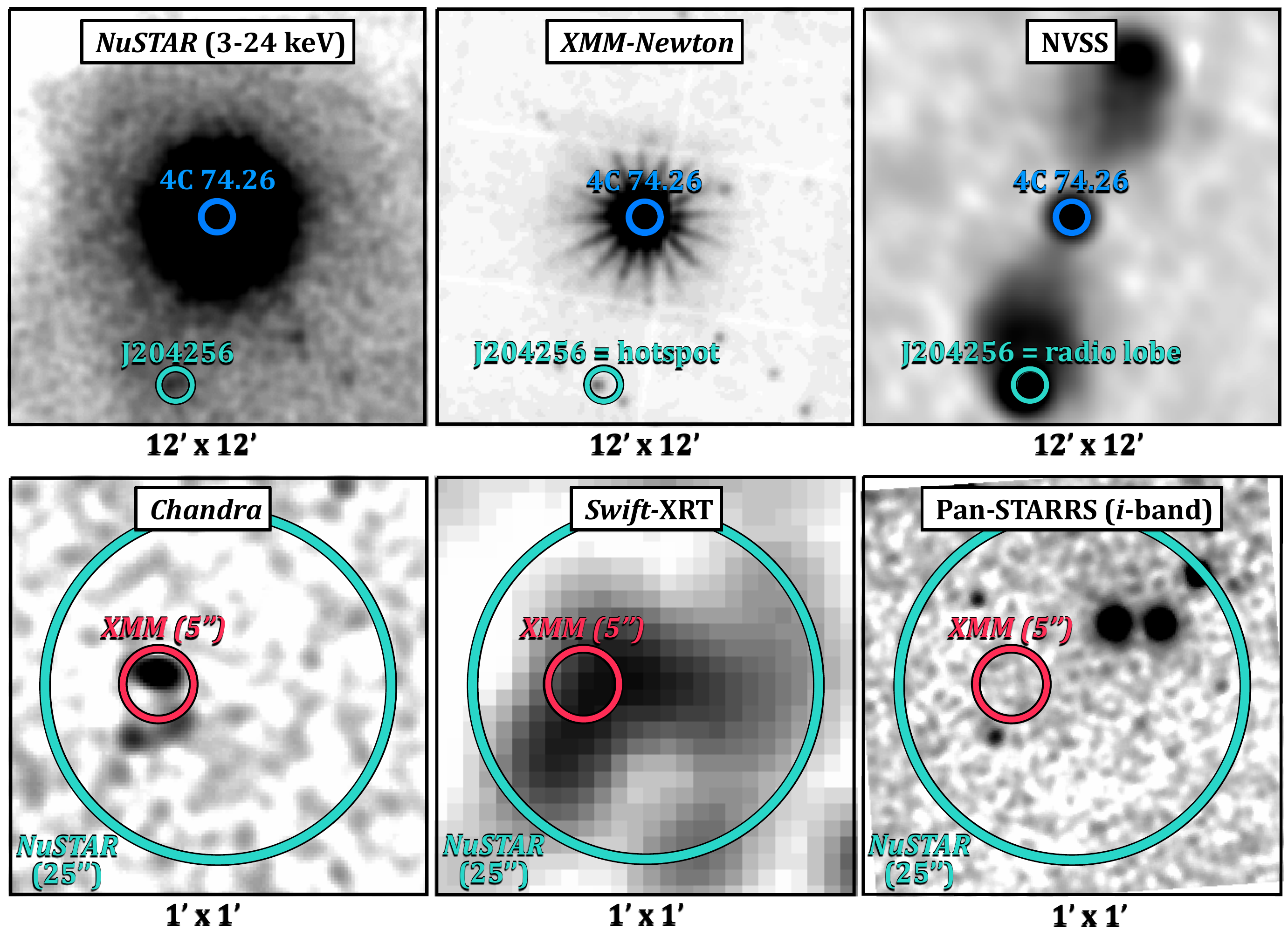}		
	\vspace*{-3mm}
	\caption{Multi-wavelength imaging of the luminous X-ray hotspot in 4C\,74.26 \citep[e.g.,][]{erlund2010}, the primary \textit{NuSTAR} science target. The hotspot, which coincides with a radio-bright lobe of 4C\,74.26, is detected with both lower-energy (i.e., \textit{Chandra}, \textit{XMM-Newton} and \textit{Swift}-XRT) and hard X-ray observatories such as \textit{NuSTAR}.} 
	\label{fig:hotspot}	
\end{figure*}

\begin{deluxetable}{lllllllll}
\tabletypesize{\scriptsize}
\tablecolumns{9}
\tablewidth{\textwidth}
\tablecaption{\label{tab:dual AGN} Source information for the three AGN pairs, two AGN-Gxy pairs, one galaxy pair, and one pair of unknown type.} 
\tablehead{NSS80 name & NSS80 field & \textit{NuSTAR}& $z$ & $L_\mathrm{10-40keV}$ & PFlag & Pair & Angular  & Physical \\
			& & detection  &  &  & & type & distance & distance \\
			& & & & (erg\,s$^{-1}$)  & & & (arcsec) & (kpc) \\
			(1) & (2) & (3) & (4) & (5) & (6) & (7) & (8) & (9) \\}
            \startdata
			NuSTARJ054231+6054.4 & BY\_Cam  & S+H+F & 0.257 & $4.07 \times 10^{44}$ & 0 & NLAGN\,+\,NLAGN & 4.7 & 19.1 \\		
			NuSTARJ091534+4054.6 & NGC\_2785 & F & 1.298 & $6.41 \times 10^{44}$ & 0 & BLAGN\,+\,BLAGN & 16.3 & 138.6\\
			NuSTARJ120530+1649.9 & IRAS12032p1707 & S+H+F & 0.216 & $2.37 \times 10^{43}$  & 1 & BLAGN\,+\,NLAGN & 45.5 & 159.4  \\
			\tableline
			\\
			NuSTARJ022742+3331.5 & 	CXOJ022727d5p333443 & S+H+F & 0.09 & $6.33 \times 10^{42}$ & 0 & NLAGN\,+\,NL & 8.9 & 15.1 \\
			NuSTARJ184552+8428.2 & 1RXSJ184642d2p842506 &  S+H+F  & 0.233 & $4.92 \times 10^{43}$  & 1 & NLAGN\,+\,Gxy & 8.1 & 30.3 \\
			\tableline
			\\
			NuSTARJ021454-6425.9 & RBS0295 & F & 0.068 & $1.21 \times 10^{42}$ & 1 & Gxy\,+\,Gxy & 22.2 & 29.04 \\
			\tableline
			\\
			NuSTARJ020614+6449.3 & 3C\_58 & F & -- & -- & -- & Unknown & 1.99 & -- \\
		\enddata
	\tablecomments{
		{\sc Columns}:\,{\bf (1)} Unique \textit{NuSTAR} source name. 
		{\bf(2)} Object name for the primary science target of the \nustar observation(s), i.e., the field name.
		{\bf(3)} The energy bands for which the source is detected: soft (S; 3-8 keV), hard (H; 8-24 keV), and full (F; 3-24 keV) bands.
		{\bf(4)} The spectroscopic redshift of the \textit{NuSTAR} source.
		{\bf(5)} The rest-frame 10--40\,keV luminosity; see Figure~\ref{fig:Lx vs z}. 
		{\bf(6)} A binary flag indicating sources that show evidence for being associated with the primary science target of their respective \textit{NuSTAR} observations, according to the definition $\Delta(cz) < 0.05\,cz$. 
		{\bf(7)} The type of sources for each pair: BLAGN refers to quasars, NLAGN to narrow-line AGN (from BPT diagnostics), NL to narrow-line objects (e.g., star forming galaxies), and Gxy to galaxies (absorption lines only).
		{\bf(8)} The angular distance  for each pair.
		{\bf(9)} The projected physical distance for each pair.
	}
\end{deluxetable}

In addition to the unique spectroscopic entries, we obtained spectra for a further 18 cases where two potential optical/IR counterparts were identified -- this gives a total of 612 entries in the NSS80 spectroscopic catalog (i.e., 18 \textit{NuSTAR} sources have duplicate entries to capture information for both targeted counterparts). 
Notably, among these there are 6 sources with companions (i.e., in pairs with $\Delta z<$0.1): three AGN pairs, two AGN-galaxy (AGN-Gxy) pairs, and one galaxy pair.
In Figure~\ref{fig:spec dual AGN} optical spectra of the three AGN pairs are shown (the main target spectra are plotted in black and their companions in peach), and Figure~\ref{fig:spec galaxy pair} show the two AGN-galaxy pairs and the galaxy-galaxy pair. We also show the spectra of a candidate source pair (NuSTARJ020614+6449.3) at low Galactic latitude, for which both sources only have a faint, red continuum detected. The source information for these systems is provided in Table~\ref{tab:dual AGN}. For the three AGN and two AGN-galaxy-pairs the projected physical distances are in the range $\sim$\,15--160\,kpc. These include: 
\begin{itemize}
	\vspace{-0.05cm}
	\setlength\itemsep{-0.1em}
	\item{\textbf{NuSTARJ054231+6054.4}, a dual AGN system at $z=0.257$, comprising a pair of likely merging, obscured AGN (in X-rays, there is lower-energy X-ray coverage with \textit{Swift}-XRT but no detection; in optical, their spectra show only narrow lines). The term dual AGN refers to a system where two AGN at the same redshift are identified at a small separation angle.
	\citet{koss2016} reported on the first dual AGN identified with \textit{NuSTAR}, i.e., SWIFTJ2028.5+2543 -- a system where both nuclei are heavily obscured to Compton thick ($N_\mathrm{H} \approx (1-2) \times 10^{24}$\,cm$^{-2}$). }
	\item{\textbf{NuSTARJ091534+4054.6}, a BL AGN pair at $z=1.298$}, comprising two closely associated quasars;
	\item{\textbf{NuSTARJ120530+1649.9}, an AGN pair including a BL AGN \citep[WISEAJ120530.63+164941.4;][]{ahn2012,toba2014} and a NL AGN at $z=0.217$. This pair is at the same redshift as the primary \textit{NuSTAR} science target \citep[WISEAJ120547.71+165107.9;][]{darling2006} and is, therefore, associated with the luminous IR galaxy (see Table~\ref{tab:pflag sources}).} 
	\item{\textbf{NuSTARJ022742+3331.5}, an AGN-Gxy pair at $z=0.09$ which is composed of a ``borderline'' NL AGN and a star-forming galaxy (confirmed by BPT diagnostics). The NL AGN is detected in all three \nustar bands;}
	\item{\textbf{NuSTARJ184552+8428.2}, an AGN-Gxy pair at $z=0.233$ comprising a NL AGN and an early-type galaxy companion, which are in the same halo as the \textit{NuSTAR} science target (see Table~\ref{tab:pflag sources}). The NL AGN is detected in all three \nustar bands with a soft band luminosity of $L_\mathrm{3-8\,keV} = 1.91 \times 10^{43}$\,erg\,s$^{-1}$, indicating that it is indeed an X-ray AGN.}
\end{itemize}

The galaxy pair (NuSTARJ021454-6425.9) is comprised of a galaxy at $z=0.068$ and a galaxy companion at $z=0.075$; the latter is associated with the primary \textit{NuSTAR} science target \citep[RBS0295;][]{schwope2000}; see Table~\ref{tab:pflag sources}. The pair of sources are undetected in the \nustar soft and hard energy bands with a full band luminosity of $L_\mathrm{3-24\,keV} = 1.52 \times 10^{42}$\,erg\,s$^{-1}$ which may be a spurious detection, considering also its low \nustar detection probability [$\log(P_\mathrm{false,min}) \sim -6.2$] and lack of a lower-energy X-ray identification.

\begin{figure*}
	\centering
	\includegraphics[width=40pc]{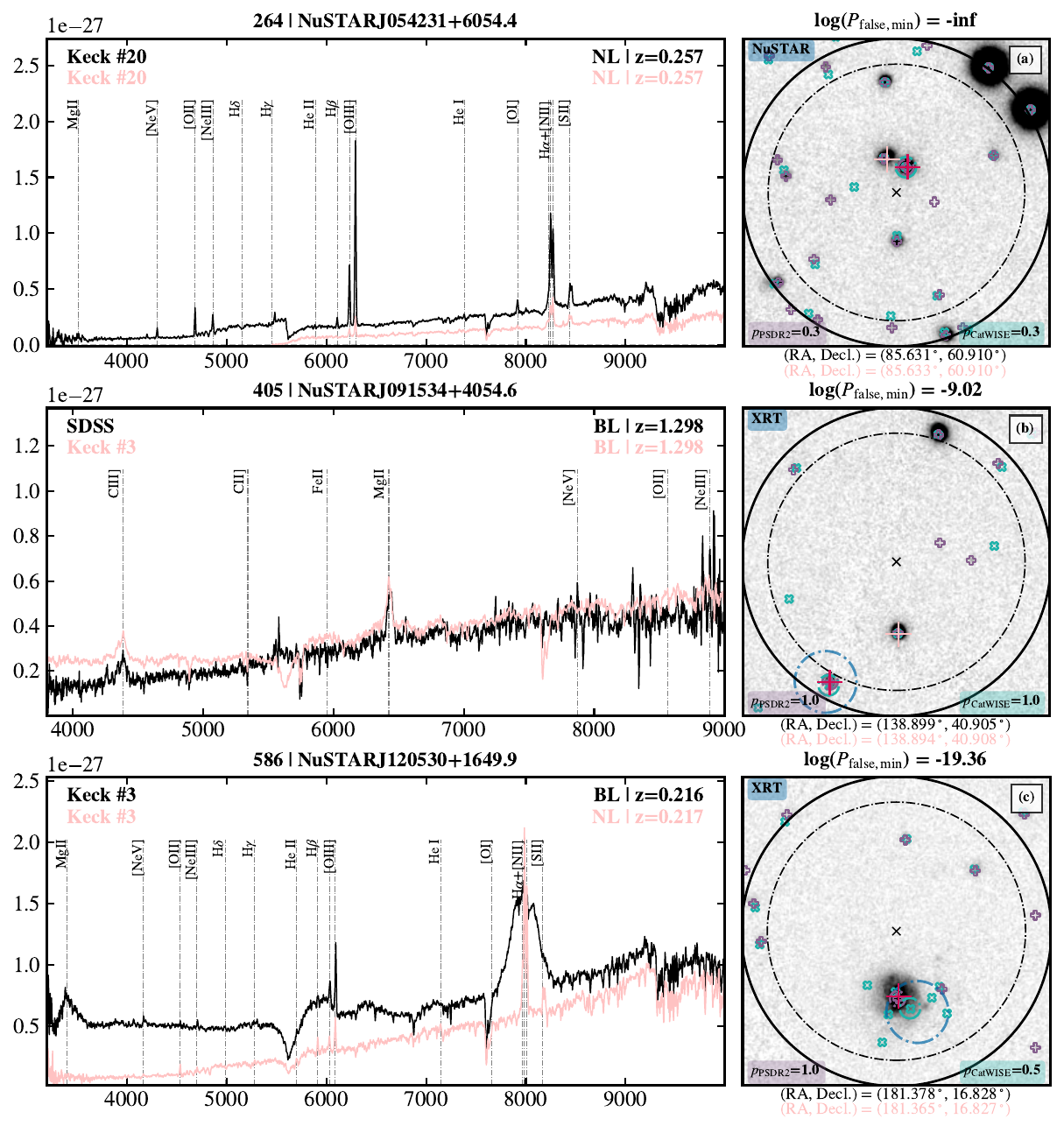}		
	\vspace*{-3mm}
	\caption{Optical spectra for the three NSS80 AGN pairs: \textbf{(a)} a dual AGN system of merging, obscured AGN, \textbf{(b)} a BL AGN (i.e., quasar) pair, and a \textbf{(c)} BL AGN + NL AGN pair.
		For each source pair the two spectra are plotted in the left panel and a $30''$\,$\times$\,$30''$ Pan-STARRS (or DECam for sources with declinations $< -25^\circ$) $i$-band image centred on the \nustar position is shown on the right. 
		\textbf{Spectrum panel:} Shown on the top are the unique \nustar ID and source name, in the upper left corner the observing telescope and run identification number (corresponding to Table~\ref{tab:telescope follow-up}), and in the upper right corner the source classification and redshift. Sky subtraction has been performed, but some features may remain: for example the 7600~\AA\ absorption feature. 
		\textbf{Image panel:} All \textit{WISE} detections are shown with `X' marks, color-coded in $W1-W2$ colors: non-AGN like sources with $W1-W2 < 0.8$ (green), AGN-like sources with $W1-W2 > 0.8$ (red), and non-AGN like sources based on the $W1-W2$ color, but with a bright $W3$ detection (peach). 
		The \nustar 25$''$ and 30$''$ error circles are plotted in a dash-dotted and a solid black line, respectively.
		The lower energy X-ray position is marked with a blue cross and the respective error circle, the CatWISE20 position is indicated with a green crosshair, a \textit{WISE} color-coded star and a green 2.7$''$ error circle, the PS1-DR2 position is shown with a purple 2$''$ error circle, and the spectroscopically observed pair of sources are marked with a red and a peach crosshair corresponding to the spectrum (black and peach).
		\label{fig:spec dual AGN}
	}
\end{figure*}

\begin{figure*}
	\centering
	\includegraphics[width=37pc]{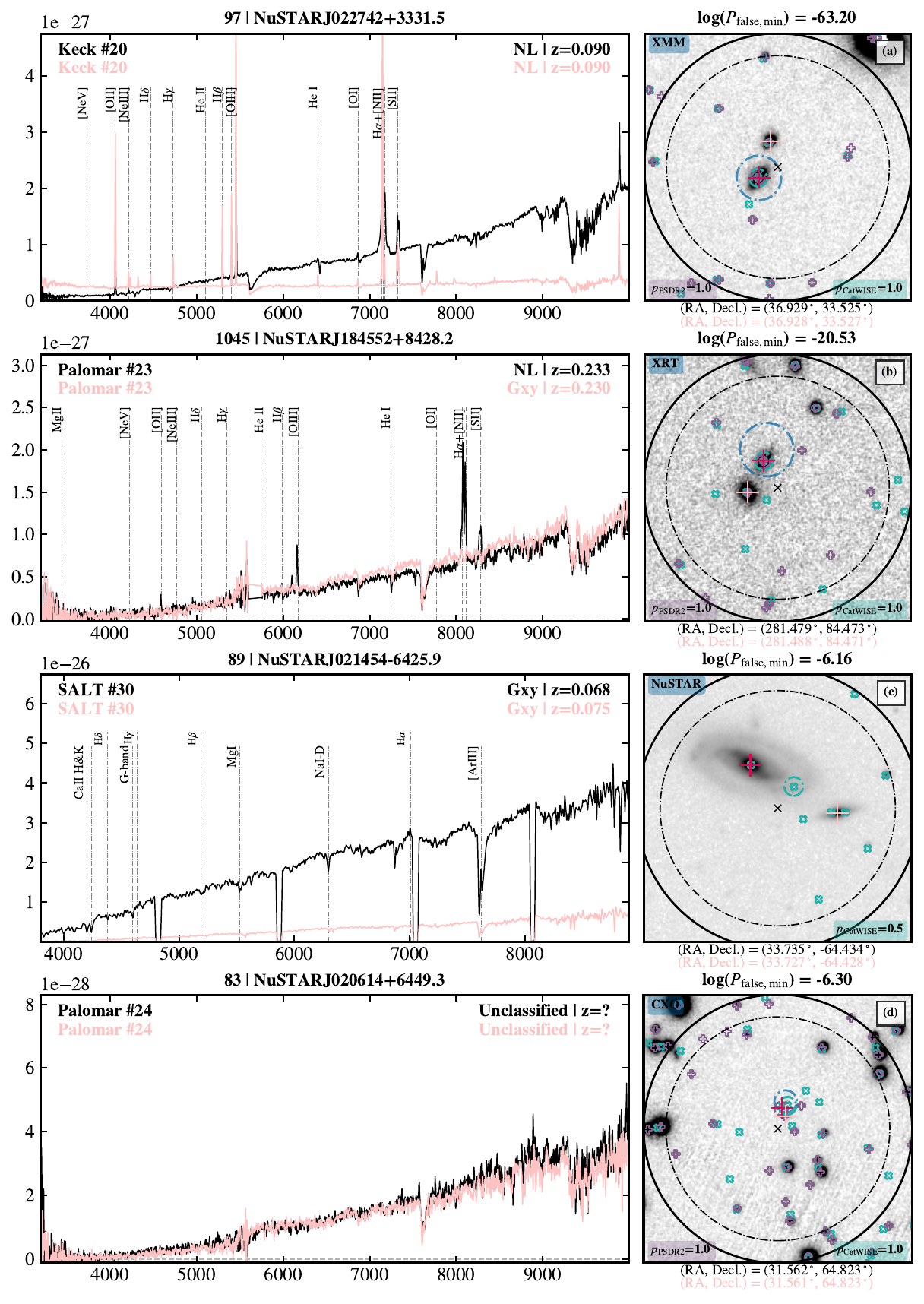}		
	\vspace*{-3mm}
	\caption{Optical spectra for the two AGN-Gxy pairs [\textbf{(a)}+\textbf{(b)}], one galaxy pair \textbf{(c)}, and one pair of sources of unknown type \textbf{(d)}. The spectroscopically observed pair of sources are marked with a red and a peach crosshair corresponding to the spectrum (black and peach), or a red and a green crosshair if one of the sources is the CatWISE20 counterpart. See Figure~\ref{fig:spec dual AGN} for the symbol key. 
	\label{fig:spec galaxy pair}
	}	
\end{figure*}

\begin{figure*}
	\centering
	\includegraphics[width=37pc]{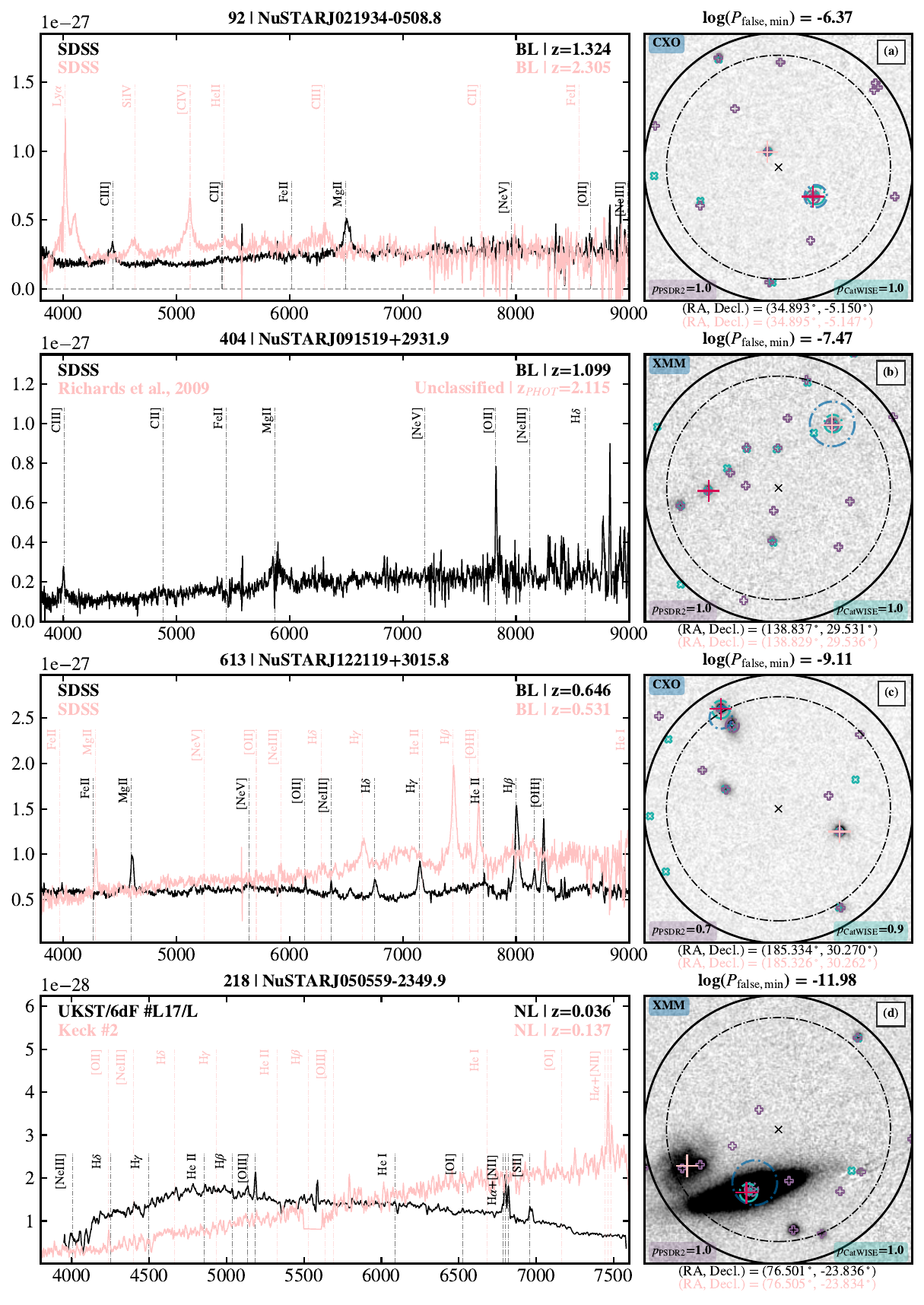}		
	\vspace*{-3mm}
	\caption{Optical spectra for the NSS80 sources with two counterpart candidates. Three of the four candidates [\textbf{(a)}--\textbf{(c)}] are fainter lower-energy X-ray sources nominally closer to the \nustar source than the selected (X-ray brighter) BL counterpart. The correct counterpart for the remaining source \textbf{(d)} is either an optically bright NL AGN ($z=0.036$) associated with the BAT-detected primary \textit{NuSTAR} science target or a NL galaxy ($z=0.137$) which lacks a lower-energy X-ray counterpart. See Figure~\ref{fig:spec dual AGN} for the symbol key. 
	}
	\label{fig:spec extra}
\end{figure*}

The 18 additional spectroscopic entries also include four sources that could potentially be the correct counterparts rather than the one selected in the unique catalog entry; see Figure~\ref{fig:spec extra}. 
Three are fainter lower-energy X-ray sources [\textbf{(a)}-\textbf{(c)}] nominally closer to the \nustar source than the selected (X-ray brighter) BL counterpart. 
The correct counterpart for the remaining source \textbf{(d)} is the choice between an optically bright NL AGN ($z=0.036$) associated with the BAT-detected primary \textit{NuSTAR} science target \citep[LEDA178130;][]{jones2004,jones2009,lansbury2017_cat,koss2022}, or a NL galaxy ($z=0.137$) 12.5$''$ from the \textsc{Nway}-selected NL AGN, which lacks a lower-energy X-ray counterpart; the former takes preference of being the correct counterpart to the NSS80 source since it is a confirmed AGN.
These sources will be further investigated through our multi-wavelength follow-up campaign. 

In addition to the 550 spectroscopically confirmed redshifts, we also obtained photometric redshifts from the literature for an additional 7 NSS80 sources \citep[from][]{salvato2009, richards2009, bilicki2014, masini2020}; flagged as ``quality C" redshift measurements in the catalog and we append the optical classification with a ``C". 
We exclude these 7 sources with phot-$z$ measurements from any redshift-dependent analysis given the larger uncertainty on phot-$z$ measurements.
Overall the current spectroscopic completeness for the primary NSS80 catalog is 43\% (550/1274); this improves to 52\% (508/981) for sources at high Galactic latitudes ($|b| > 10^\circ$).
Further on-going optical spectroscopic campaigns will increase the spectroscopic completeness of the NSS80, in addition to further lower-energy X-ray observations to improve the number of sources with reliable lower-energy X-ray counterparts.

For all \nustar sources with spectroscopic identifications, we assign an ``associated'' flag to those that have a velocity offset from the science target smaller than 5\% of the total science target velocity, i.e., $\Delta(cz) < 0.05\,cz$, following \citetalias{lansbury2017_cat}.
Based on this, 19 spec-$z$ NSS80 sources show evidence for being associated with the primary \textit{NuSTAR} science target and are, therefore, excluded from any subsequent analysis. Table~\ref{tab:pflag sources} provides source information for these serendipitous sources and their associated \textit{NuSTAR} science target \citep[of which 15 are BASS DR2 sources; see][]{koss2022}. 
      
\begin{deluxetable}{lllllll}
\tablecolumns{7}
\tablewidth{\textwidth}
\tablecaption{\label{tab:pflag sources} NSS80 sources which show evidence for being associated with the primary \textit{NuSTAR} science target according to the definition $\Delta(cz) < 0.05\,cz$.} 
\tablehead{Serendip Name   & Serendip Type &  $z_\mathrm{serendip}$  & Primary Target &  Primary Type        &  $z_\mathrm{target}$   & BASS \\
			(1) & (2) & (3) & (4) & (5) & (6) & (7) \\}
\startdata
			NuSTARJ002544+6818.8  & NL    &  0.012  &  LEDA136991               &  Sy2         &  0.012  & 1    \\
			NuSTARJ010736-1732.3  & NL    &  0.021  &  IC1623                   &  GPair       &  0.02   & 0    \\
			NuSTARJ012215+5002.2  & -     &  0.021  &  MCG+8-3-18               &  Sy2         &  0.021  & 1    \\
			NuSTARJ021454-6425.9  & Gxy   &  0.075  &  RBS295                   &  Sy1         &  0.074  & 1    \\
			NuSTARJ024144+0512.3  & NL    &  0.07   &  IRAS02394+0457           &  Sy2         &  0.07   & 1    \\
			NuSTARJ035902-3011.7  & NL    &  0.093  &  SARS059.33488-30.34397   &  Sy1.9       &  0.097  & 1    \\
			NuSTARJ040702+0346.8  & NL    &  0.088  &  3C105                    &  Sy2         &  0.088  & 1    \\
			NuSTARJ050559-2349.9  & NL    &  0.036  &  LEDA178130               &  Sy2         &  0.035  & 1    \\
			NuSTARJ054349-5536.7  & BL    &  0.272  &  WISEJ054357.21-553207.5  &  QSO         &  0.273  & 0    \\
			NuSTARJ065805-5601.2  & NL?   &  0.296  &  Bullet Cluster            &  GxyCluster  &  0.296  & 1    \\
			NuSTARJ065842-5550.2  & NL    &  0.297  &  Bullet Cluster            &  GxyCluster  &  0.296  & 0    \\
			NuSTARJ071422+3523.9  & NL    &  0.015  &  MCG+6-16-28              &  Sy2         &  0.015  & 1    \\
			NuSTARJ120530+1649.9  & BL    &  0.216  &  WISEAJ120547.71+165107.9 &  LIRG       &  0.218  & 0    \\
			NuSTARJ125442-2657.1  & Gxy   &  0.058  &  CTS18                    &  Sy1.2       &  0.059  & 1    \\
			NuSTARJ151253-8124.3  & NL    &  0.069  &  2MASXJ15144217-8123377   &  Sy1.2       &  0.069  & 1    \\
			NuSTARJ165105-0129.4  & NL    &  0.041  &  2MASXJ16510578-0129258   &  Sy2         &  0.04   & 1    \\
			NuSTARJ184552+8428.2  & NL    &  0.233  &  1RXSJ184642.2+842506     &  Sy1         &  0.225  & 1    \\
			NuSTARJ190813-3925.7  & Gxy   &  0.075  &  IGRJ19077-3925           &  Sy1         &  0.075  & 1    \\
			NuSTARJ224536+3947.1  & NL    &  0.081  &  3C452                    &  Sy2         &  0.081  & 1    \\
   \enddata
	\tablecomments{
		{\sc Columns}:\,{\bf (1)} The unique NSS80 name for each serendipitous source. 
		{\bf (2)} Spectroscopic classification of the NSS80 serendipitous source.
		{\bf (3)} Spectroscopic redshift of the NSS80 serendipitous source obtained with our follow-up campaign.
		{\bf (4)} The primary \textit{NuSTAR} science target name.
		{\bf (5)} Source type of the \textit{NuSTAR} science target name.
		{\bf (6)} Redshift of the \textit{NuSTAR} science target.
		{\bf (7)} A binary flag indicating \textit{NuSTAR} science targets which are BASS DR2 sources \citep{koss2022}.
	}
\end{deluxetable}

\subsubsection{Comparison between confirmed spectroscopic and \textsc{Nway} identified AGN}  
\label{subsubsec:spectra vs nway}

A significant fraction of the optical spectroscopic follow-up programme was based on the approach outlined in \citetalias{lansbury2017_cat} using closest counterparts for lower-energy X-ray sources and/or counterparts with AGN-like MIR properties from \textit{WISE}. 
Here we investigate how the optical counterpart identified from that approach compares to that found using the \textsc{Nway} approach described in Section~\ref{subsec:nway infrared and optical cpart}. 
We note that a disagreement between the counterparts does not necessarily mean that the incorrect counterpart was followed-up; due to the probabilistic approach of \textsc{Nway} the \lq best\rq\ selected counterpart will not always be the true one (see discussion of false probabilities in Section~\ref{subsec:nway infrared and optical cpart}).
For example, the clear identification of AGN signatures in the optical spectrum provides compelling evidence that the correct optical source was followed-up, given the low probability of selecting a clear optical AGN by chance.
Therefore, to provide easy access to the IR/optical counterpart information (i.e., magnitudes), we include binary flags in the catalog (NWAY\_CatWISE, NWAY\_PS1) indicating whether the spectroscopic target position matches to the \textsc{Nway}-identified CatWISE20 and PS1-DR2 counterparts.
In total, 515/594 unique spectroscopic targets (Section~\ref{subsubsec:spectral classification}) coincide with the CatWISE20 positions, and 388/594 with the \textsc{Nway}-identified PS1-DR2 positions.

To assist in the interpretation of our results in Section~\ref{sec:nustar science}, we classify the spectroscopically classified sources as either reliable or uncertain (see Cpart\_RFlag in Appendix~\ref{appendix:nss80 catalogue}). 
The majority of the reliable sources will have positional offsets of $<$\,5$''$, mostly assisted by the identification of lower-energy X-ray counterparts, while the majority of the uncertain sources will have larger positional offsets (see Appendix~\ref{appendix:catwise probability}). 
However, as our previous analysis was limited to using catalogued data it will not take into account potential anomalies such as the presence of nearby bright sources, close-separation systems, and image artifacts. 
We therefore supplemented our analyses with a visual inspection of all spectroscopically-targeted NSS80 sources using the detailed optical finding charts; see Appendix~\ref{appendix:optical spectra} for the post-NSS40 finding charts (and optical spectra) and Appendix~\ref{appendix:findercharts} for the NSS40 finding charts.\footnote{The individual spectroscopic data for each followed-up NSS80 source will be made available at \url{https://www.nustar.caltech.edu/page/59}.}

Overall, on the basis of the combination of these analyses, we determined 91\% (449/492) of the spectroscopically classified extragalactic sources to be reliable (274 BLs + 166 NLs); only 13/449 sources fail our high probability cut defined in Section~\ref{subsec:nway infrared and optical cpart}, mostly due to the lack of lower-energy X-ray or MIR information. 
The remaining 43 sources are flagged as uncertain (13 of which do not have any \textsc{Nway} matches satisfying our high probability cut), either due to source confusion as a result of high source density, multiple potential counterpart associations, or shallow IR/optical coverage.
Of the 58 spectroscopically classified Galactic objects, we identify 30 as reliable spectroscopic counterparts (28 of which satisfy the \textsc{Nway} high probability cut) and flagged the other 28 as uncertain.
We note that counterpart identification is more challenging in these cases \citep[see e.g.,][]{tomsick2018}.
We use this reliable spectroscopic sample for our science analysis in Section~\ref{sec:nustar science}, unless otherwise indicated. 

In Figure~\ref{fig:rmag distribution for all + spec} we show $r$-band magnitude distributions for the reliable optical NSS80 sources (i.e. those satisfying our \textsc{Nway} high probability cut; light purple), overlaid with the NSS80 sources with reliable spectroscopic classifications (purple). Of the 958 high probability counterparts, 721 sources have available constrained $r$-band magnitudes with a mean magnitude of $\langle r_\mathrm{Nway} \rangle = 19.73$. 
Of the 492 extragalactic NSS80 sources, 399 are flagged as reliable (based on our visual comparison between the \textsc{Nway} identified counterpart and the spectroscopic target) and have available constrained $r$-magnitudes (excluding 19 sources with evidence for being associated with the \textit{NuSTAR} targets for their respective observations): 261 BLs, 132 NLs and 6 other sources (four galaxies and 2 unclassified), with average magnitudes of $\langle r_\mathrm{BL} \rangle = 19.53$, $\langle r_\mathrm{NL} \rangle = 19.3$ and $\langle r_\mathrm{Other} \rangle = 16.58$, respectively.
Overall, our spectroscopic campaign is targetting the majority of the sources (in terms of $r$-band magnitude) but misses the faintest sources.

\begin{figure}
	\centering
	\includegraphics[width=20pc]{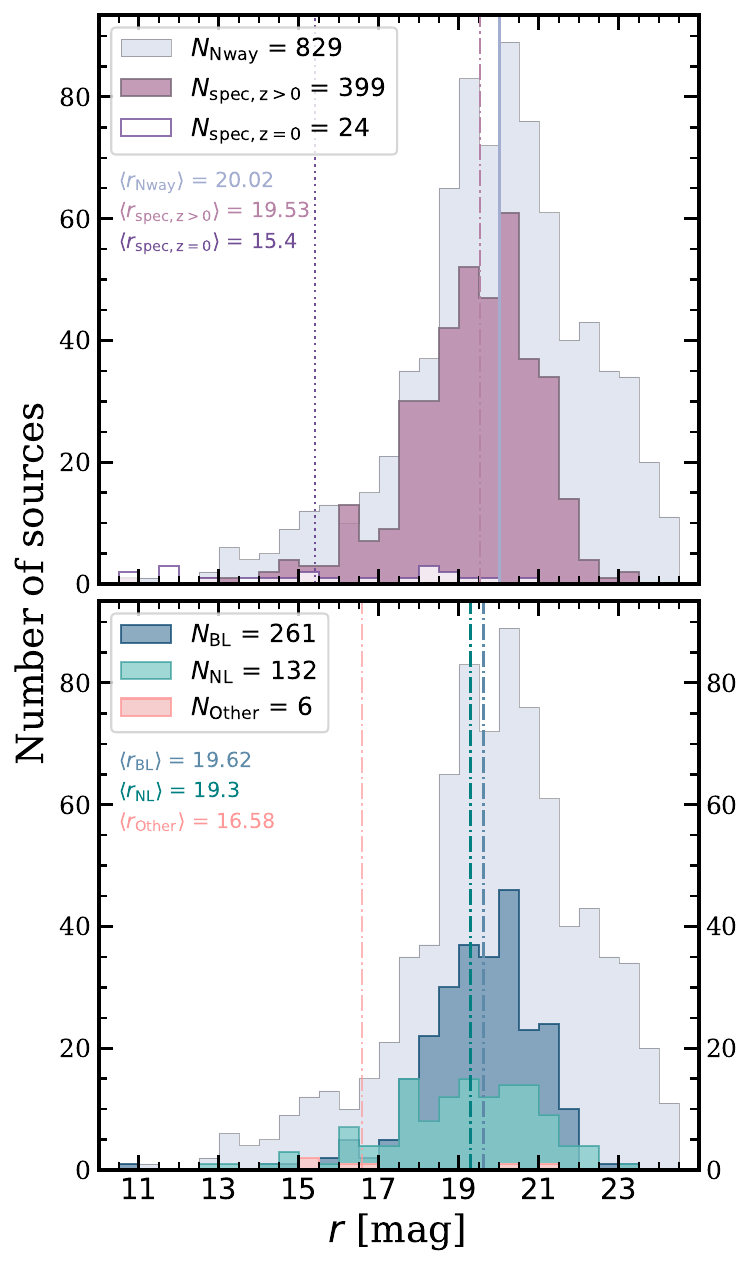}		
	\vspace*{-6mm}
	\caption{\textit{Top panel:} $r$-band magnitude distributions for the high probability NSS80 sources (i.e., adopted optical counterparts satisfying our \textsc{Nway} probability cut) with constrained $r$-magnitudes (721/953; light purple), overlaid with the constrained $r$-band magnitude distribution of the spectroscopically-classified extragalactic (399/492) and Galactic (24/58) NSS80 sources with reliable counterpart associations shown in filled purple and white, purple-edged histograms, respectively. The 19 sources with evidence for being associated with the \textit{NuSTAR} science targets for their respective observations are excluded. 
		\textit{Bottom panel:} The $r$-magnitude distribution separated by spectroscopic classification: BL objects are shown in blue, NL objects are shown in green, and ``Other'' (including 4 galaxies and 2 unclassified sources) are shown in peach. The vertical lines mark the median $r$-magnitude for the respective subsamples. Overall the current spectroscopic completeness for the primary NSS80 catalog is 43\% (550/1274), missing the faintest sources; this improves to 52\% (508/981) for sources at high Galactic latitudes, $|b| > 10^\circ$. } 
	\label{fig:rmag distribution for all + spec}	
\end{figure}

\section{Results and Discussion} \label{sec:nustar science}
In this section we exploit the X-ray data, multi-wavelength data, and optical spectroscopy to characterise the properties of the NSS80 sources.
The basic X-ray properties of the extragalactic NSS80 sources (see selection in Table~\ref{tab:subset_summary}) are given in Section~\ref{subsec:xray prop} while in Section~\ref{subsec:wise properties nustar} we explore the MIR colors of these sources. 
In Section~\ref{subsec:optical prop} we investigate the optical properties of the NSS80 AGN, with particular focus on quasars, utilizing detailed composite spectra to explore the origin of their observed optical colors. 


\subsection{\xray characteristics of the NSS80 sources} \label{subsec:xray prop}

\subsubsection{\nustar source counts, count rates \& fluxes } \label{subsubsec:nustar prop}
Altogether there are 1274 NSS80 sources with significant detections in at least one \nustar energy band -- a factor of three improvement over that of NSS40 reported in \citetalias{lansbury2017_cat}. Similarly to other \nustar surveys \citep[e.g.,][]{civano2015,mullaney2015,lansbury2017_cat,masini2020}, 32\% (412/1274) of the sample is detected in the 8-24\,keV band, which is unique to \nustar amongst focusing X-ray observatories. 

The basic properties of the NSS80 sources are given in Table~\ref{tab:stats xray prop}.
We find a large range ($\sim$\,6--1700\,ks) in the net exposure time per source for the combined telescopes FPMA+B (cleaned and vignetting-corrected; $t_\mathrm{net}$), with a median of $\sim$\,80\,ks.
For the sources with detections in the $3-8$\,keV, $8-24$\,keV, and $3-24$\,keV bands the lowest (deblended) net source counts ($S_\mathrm{net}$) are 12, 11, and 19, respectively. 
The source with the highest $S_\mathrm{net}$ in all three energy bands is still NuSTAR~J043727–4711.5, a BL AGN at z = 0.051, as reported in \citetalias{lansbury2017_cat}, with $S_\mathrm{net}$ values of 11,337, 6,653 and 17,943 counts, respectively.
The median $S_\mathrm{net}$ values in the respective bands are 62, 67, and 82, respectively.
Finally, we find a range in the net count rates of 0.09--37.1, 0.07--44.3, and 0.12--72.2\,ks$^{-1}$, and median values of 0.8, 0.8, and 1.04\,ks$^{-1}$ for the soft, hard and full band energies, respectively. 
\begin{deluxetable}{llll}
\tablecolumns{4}
\tablewidth{\textwidth}
\tablecaption{\label{tab:stats xray prop} X-ray characteristics of the NSS80 sources with significant detections in the soft ($3-8$\,keV), hard ($8-24$\,keV) and full ($3-24$\,keV) bands, respectively. The listed data include the minimum, maximum and median values for the net exposure times ($t_\mathrm{net}$; ks), the net source counts ($S_\mathrm{net}$ counts), the net count rates (CTRT$_\mathrm{net}$; ks$^{-1}$), and the X-ray fluxes.
			The Median Absolute Deviation (MAD) is taken as the uncertainty on the median values. 
		} 
\tablehead{\nustar energy band & $t_\mathrm{net,min}$ (ks) & $t_\mathrm{net,max}$ (ks) & $\langle{t}_\mathrm{net}\rangle$ (ks) \\ }
\startdata
			$3-8$\,keV  &  7.6 & 1657.4 &  79.8 \\
			$8-24$\,keV  & 6.1 & 1617.9 & 76.4 \\
			$3-24$\,keV  & 8.9  & 1656.4 & 78.3 \\
			\\
			\tableline
			& $S_\mathrm{net,min}$ (counts) & $S_\mathrm{net,max}$ (counts) & $\langle{S}_\mathrm{net}\rangle$ (counts) \\
			\tableline
			$3-8$\,keV  & 12\,$\pm$\,7   & 11,337\,$\pm$\,114  & 62\,$\pm$\,45 \\
			$8-24$\,keV & 11\,$\pm$\,5  & 6,653\,$\pm$\,88   & 67\,$\pm$\,48 \\
			$3-24$\,keV &  19\,$\pm$\,29   & 17,943\,$\pm$\,143   & 82\,$\pm$\,55 \\
			\\
			\tableline
			& CTRT$_\mathrm{net,min}$ (ks$^{-1}$) & CTRT$_\mathrm{net,max}$ (ks$^{-1}$)& $\langle{\mathrm{CTRT}}_\mathrm{net}\rangle$ (ks$^{-1}$) \\
			\tableline
			$3-8$\,keV  & 0.09\,$\pm$\,0.04 & 37.1\,$\pm$\,0.4 & 0.8\,$\pm$\,0.5 \\
			$8-24$\,keV & 0.07\,$\pm$\,0.03  & 44.3\,$\pm$\,1.3 & 0.8\,$\pm$\,0.6 \\
			$3-24$\,keV & 0.12\,$\pm$\,0.05 & 72.2\,$\pm$\,1.6 & 1.04\,$\pm$\,0.7 \\
			\\
			\tableline
			& Flux$_\mathrm{X,min}$ (erg\,s$^{-1}$\,cm$^{-2}$) & Flux$_\mathrm{X,max}$ (erg\,s$^{-1}$\,cm$^{-2}$)& $\langle{\mathrm{Flux}}_\mathrm{X}\rangle$ (erg\,s$^{-1}$\,cm$^{-2}$) \\
			\tableline
			$3-8$\,keV  & (6.65\,$\pm$\,2.68)\,$\times$\,10$^{-15}$ & (2.49\,$\pm$\,0.025)\,$\times$\,10$^{-12}$ & (5.18\,$\pm$\,3.56)\,$\times$\,10$^{-14}$ \\
			$8-24$\,keV & (9.93\,$\pm$\,4.07)\,$\times$\,10$^{-15}$ &  (6.15\,$\pm$\,0.17)\,$\times$\,10$^{-12}$ & (11.31\,$\pm$\,7.86)\,$\times$\,10$^{-14}$  \\		
			$3-24$\,keV & (11.55\,$\pm$\,4.43)\,$\times$\,10$^{-15}$ & (6.79\,$\pm$\,0.15)\,$\times$\,10$^{-12}$ & (9.82\,$\pm$\,6.83)\,$\times$\,10$^{-14}$ \\					
		\enddata
\end{deluxetable}

The flux distributions of detected and undetected NSS80 sources for a given band are shown in Figure~\ref{fig:flux distribution} in comparison to the NSS40 flux distributions. 
The faintest fluxes are 6.65 and 9.93\,$\times$\,10$^{-15}$\,erg\,s$^{-1}$\,cm$^{-2}$ for detected sources in the $3-8$\,keV and $8-24$\,keV bands, respectively, and 1.15\,$\times$\,10$^{-14}$\,erg\,s$^{-1}$\,cm$^{-2}$ for full-band source-detections.
The brightest fluxes in the NSS80 catalog correspond to two sources: NuSTARJ043727-4711.5, a BL AGN at $z\,=\,0.051$, with a soft band flux of 2.49\,$\times$\,10$^{-12}$\,erg\,s$^{-1}$\,cm$^{-2}$, and  NuSTARJ153602-5749.0, a Galactic source ($z\,=\,0$), with hard and full band fluxes of 6.15 and 6.79\,$\times$\,10$^{-12}$\,erg\,s$^{-1}$\,cm$^{-2}$, respectively. 
The source with the brightest flux in \citetalias{lansbury2017_cat}, NuSTARJ075800+3920.4 (BL AGN at $z = 0.095$) is recorded in our secondary NSS80 catalog,\footnote{Flagged with our optical masking as within the radius of an Abell cluster.} with soft, hard, and full band fluxes of 3.50, 4.99, and 8.8\,$\times$\,10$^{-12}$ erg\,s$^{-1}$\,cm$^{-2}$; the median fluxes in the NSS80 catalog are (5.18\,$\pm$\,3.56), (11.31\,$\pm$\,7.86), and (9.82\,$\pm$\,6.83)\,$\times$\,10$^{-14}$\,erg\,s$^{-1}$\,cm$^{-2}$, respectively.
The serendipitous survey pushes to fluxes $\sim$\,two orders of magnitude fainter than those achieved by previous-generation hard X-ray observatories such as \textit{INTEGRAL} \citep[e.g.,][]{malizia2012} and \textit{Swift}-BAT \citep{oh2018}; see Section~\ref{subsubsec:redshifts and luminosities}.

\begin{figure}
	\centering
	\includegraphics[width=20pc]{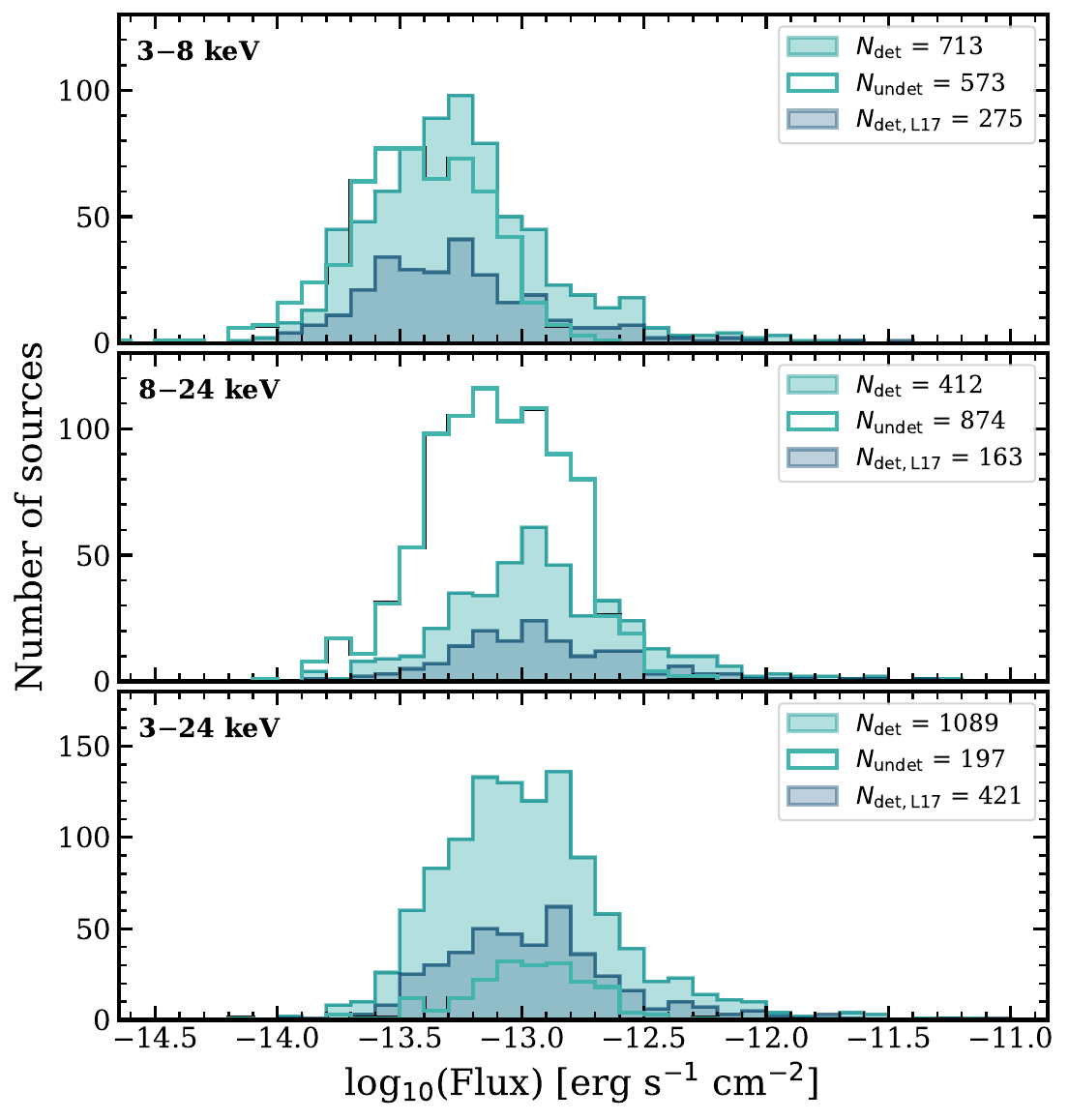}		
	\vspace*{-2mm}
	\caption{ 
		\nustar flux distributions in the soft (top), hard (middle), and full (bottom) energy bands for the NSS40 (blue) and NSS80 (green) samples. 
		For each band, the filled histogram shows the flux distribution for sources independently detected in that band (the number of these sources, $N_\mathrm{det}$, is indicated in the upper right corner), and the open histogram shows the distribution of flux upper limits for sources undetected in that band, but independently detected in at least one other band ($N_\mathrm{undet}$). }
	\label{fig:flux distribution}	
\end{figure}

\subsubsection{Band Ratios} \label{subsubsec:band ratios}

In obscured objects, as optical depth increases with decreasing energy, relatively larger numbers of hard X-ray photons are detected in comparison to soft X-rays due to the preferential obscuration of lower-energy X-ray emission due to photoelectric absorption. 
With its capability of focusing high-energy photons, \textit{NuSTAR} is well-suited
to categorise the obscured population of AGN and to search for heavily obscured sources of up to Compton-thick (CT; $N_\mathrm{H} \gtrsim 1.5 \times 10^{24}$\,cm$^{-2}$) levels of obscuration \citep[e.g.,][]{marchesi2018,marchesi2019,torres_alba2021,zhao2021_obscured}.
The ratio in count rates between the hard and soft X-ray bands, defined as the band ratio, is indicative of the amount of intrinsic obscuration along the line of sight to the nucleus of an X-ray emitting source and can, therefore, be used as a basic estimate of the amount of obscuration to (crudely) identify CT sources from their extreme band ratios. This technique has been successfully demonstrated in several previous \nustar studies \citep[e.g.,][]{gandhi2014,balokovic2014,lansbury2017,torres2021}, and we use this definition in this work for consistency with the literature.

Figure~\ref{fig:band ratios vs count rates} shows the $8-24$\,keV to $3-8$\,keV band ratios (BR$_\mathrm{Nu}$) for the NSS80 sample\footnote{BR$_\mathrm{Nu}$ is calculated as in \citet{lansbury2017_cat}. Where both hard and soft count rates have defined values, uncertainties are propagated as standard. Where one is an upper limit, the resulting BR$_\mathrm{Nu}$ is thus an upper/lower limit and if both are upper limits the value is undefined. These are flagged as described in Appendix~\ref{appendix:nss80 catalogue}.} as a function of the $3-24$\,keV (full band) count rate (CTRT).
In order to examine the results for extragalactic sources only, we remove sources which are spectroscopically confirmed as having $z = 0 $ and exclude sources with Galactic latitudes below $|b| = 10^\circ$, for which there is significant contamination to the non-spectroscopically identified sample from Galactic sources. 
A large variation in BR$_\mathrm{Nu}$  is observed across the sample corresponding to spectral slopes (applying to a single absorbed power law model with fixed Galactic $N_\mathrm{H}$) ranging from $\Gamma_\mathrm{eff} \approx 3$ (at the softest values) to $\Gamma_\mathrm{eff} \approx -0.5$ (at the hardest values). 
To include weak and non-detections in the NSS80 catalog, we also calculate ``stacked'' medians in BR$_\mathrm{Nu}$ per count rate (in bins of $1\times10^{-3}$\,s$^{-1}$) by summing the net count rates of all NSS40 (blue-edged diamonds) and NSS80 (green-edged circles) sources.
The results are consistent with a flat relation in the average band ratio versus count rate, and a constant average effective photon index of $\Gamma_\mathrm{eff} \approx 1.5$, suggesting at least modest amounts of obscuration on average within the sample \citep[compared to $\Gamma_\mathrm{eff} \sim 1.8$ for typical unobscured sources; see e.g.,][]{ricci2017}.
Furthermore, we find no evidence of a relationship between band ratio and count rate in the higher energy 3–24 keV band, as found by previous studies at $< 10$\,keV \citep[e.g.,][]{dellaceca1999,mushotzky2000,alexander2003}.
The absence of such a trend may partly be attributed to the fact that X-ray spectra of AGN are less strongly affected by absorption in the high-energy \nustar band.

\begin{figure}
	\centering
	\includegraphics[width=20pc]{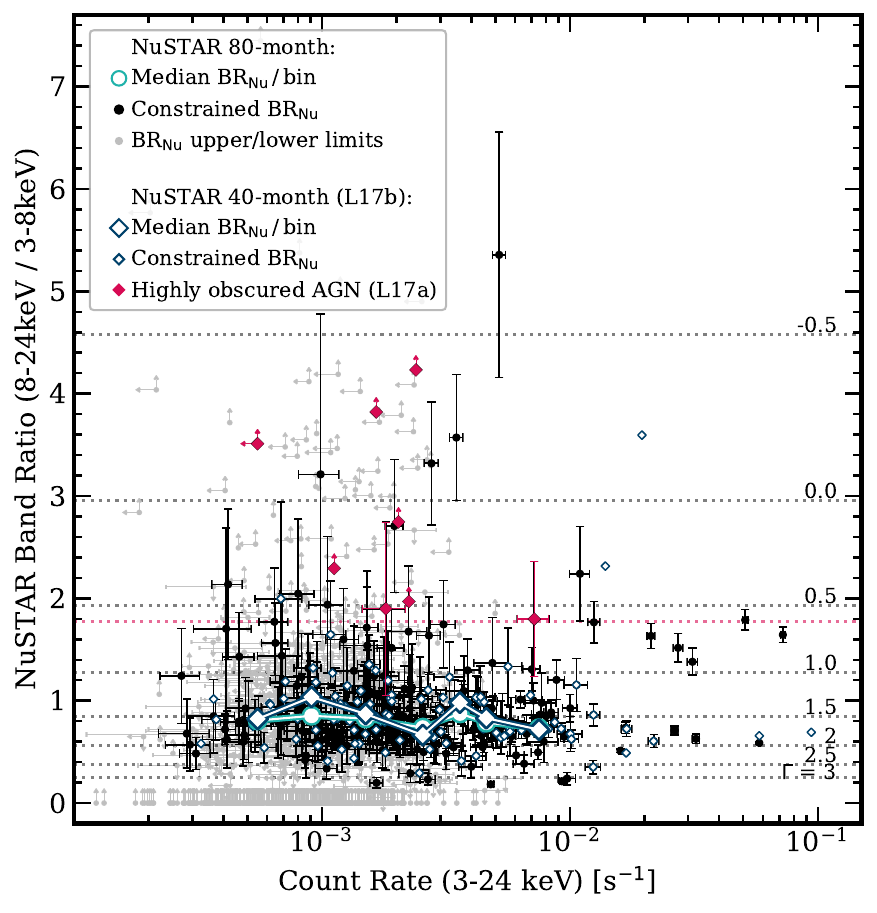}		
	\vspace*{-5mm}
	\caption{The \nustar\ $8-24$ to $3-8$\,keV band ratio (BR$_\mathrm{Nu}$) versus full band ($3-24$\,keV) count rate for the NSS80 sources. Constrained BR$_\mathrm{Nu}$  values are shown in black, and those with upper or lower limits are shown in grey. 
		The dotted horizontal lines indicate the equivalent X-ray spectral slope ($\Gamma_\mathrm{eff}$) for a given band ratio.  
		Highly obscured NSS40 serendipitous sources \citep{lansbury2017}, with BR$_\mathrm{Nu}$  values that correspond to $\Gamma_\mathrm{eff} < 0.7$, are marked with red diamonds. 
		The blue diamonds and the green circles show the median BR$_\mathrm{Nu}$ value per count rate bin of size $1\times10^{-3}$\,s$^{-1}$ for the NSS40 and NSS80 sources, respectively.
	}
	\label{fig:band ratios vs count rates}	
\end{figure}

When using the BR$_\mathrm{Nu}$  alone to identify obscured AGN, additional knowledge of the source is required to estimate the absorbing column density ($N_\mathrm{H}$), as key spectral features (e.g., the photoelectric absorption cut-off) are shifted across the observed energy band for sources at different redshifts. 
Therefore, potentially highly obscured AGN can be identified using the BR$_\mathrm{Nu}$  values complemented by the source's redshift information, as demonstrated in \citet{lansbury2017}.
The BR$_\mathrm{Nu}$  values as a function of redshift are plotted for the spectroscopically-identified NSS80 sample in Figure~\ref{fig:band ratios vs z}.
We plot tracks for a range of column densities ($N_\mathrm{H} = 10^{23-24}$\,cm$^{-2}$) to provide an estimate of the absorbing columns giving rise to the observed band ratios of the NSS80 sources.
We find that the majority of extragalactic NSS80 sources ($z > 0$) have BR$_\mathrm{Nu}$  values in the range of 0.4--1.4, with a median of 0.8\,$\pm$\,0.3 which breaks down into a median of 0.7\,$\pm$\,0.2 for the BLs and a slightly higher value of 0.9\,$\pm$\,0.2 for the NLs (as illustrated with the blue and green histograms, respectively).
In comparison to the column density tracks, the majority of the extragalactic NSS80 serendipitous sources have $N_\mathrm{H}$ values of $< 3$\,$\times$\,10$^{23}$\,cm$^{-2}$, with only a minority (14/82 BL and 13/55 NL constrained sources), predominantly at low redshift, with significantly higher absorbing column densities.

We apply the same basic approach to that used in \citet{lansbury2017} to identify potentially heavily obscured sources, i.e., a BR$_\mathrm{Nu}$ \,$>$\,1.7 cut (red dotted line in Figure~\ref{fig:band ratios vs z}) which corresponds to an effective (i.e., observed) photon index of $\Gamma_\mathrm{eff} \lesssim 0.6$ \citep[motivated by observed CT AGN in other \nustar programs; e.g.,][]{balokovic2014,gandhi2014,civano2015,lansbury2015}.
The sample is limited to NSS80 sources with spectroscopic redshifts and constrained BR$_\mathrm{Nu}$ values (or lower limits).
Based on this analysis, 22 sources stand out as CT-candidates: 10 sources detected in all three \nustar bands, 8 with hard- and full band detections and 4 only detected in the hard band. 
Of these, 7 are reported in \citet{lansbury2017}, increasing the number of NSS-selected candidate CT AGN by a factor of $\sim$\,3; however, the 8$^\mathrm{th}$ source in \citet{lansbury2017}, NuSTAR~J165346+3953.7, is undetected in the NSS80 catalog. 
Note that 3/22 sources show evidence for being associated with the primary \textit{NuSTAR} science target based on $\Delta(cz) < 0.05\,cz$ (see Table~\ref{tab:pflag sources}).
The basic properties of these candidate CT AGN are provided in Table~\ref{tab:hard band sources}. 
The majority (18/22) are spectroscopically classified as NL systems, consistent with expectations for obscured AGN; the other systems are classified as either low-redshift galaxies or BL AGN.
We note that BR$_\mathrm{Nu}$ provides a crude estimate of the absorbing columns, and a more detailed investigation of the \textit{NuSTAR} spectra and multi-wavelength properties of the 15 newly identified CT-candidates is required to strengthen the interpretation of these high-BR$_\mathrm{Nu}$  sources as highly absorbed systems and provide significantly improved constraints on the space density of CT AGN \citep[see e.g.,][]{yan2019}. However, based on the X-ray spectral analysis presented in \citet{lansbury2017}, we expect at least 50\% of these candidates to be CT AGN (i.e.,\ at least 4 of the 8 CT-candidates). Importantly, 3 of these 4 systems would not have been identified as candidate CT AGN without \nustar\ data.

\begin{figure*}
	\centering
	\includegraphics[width=42pc]{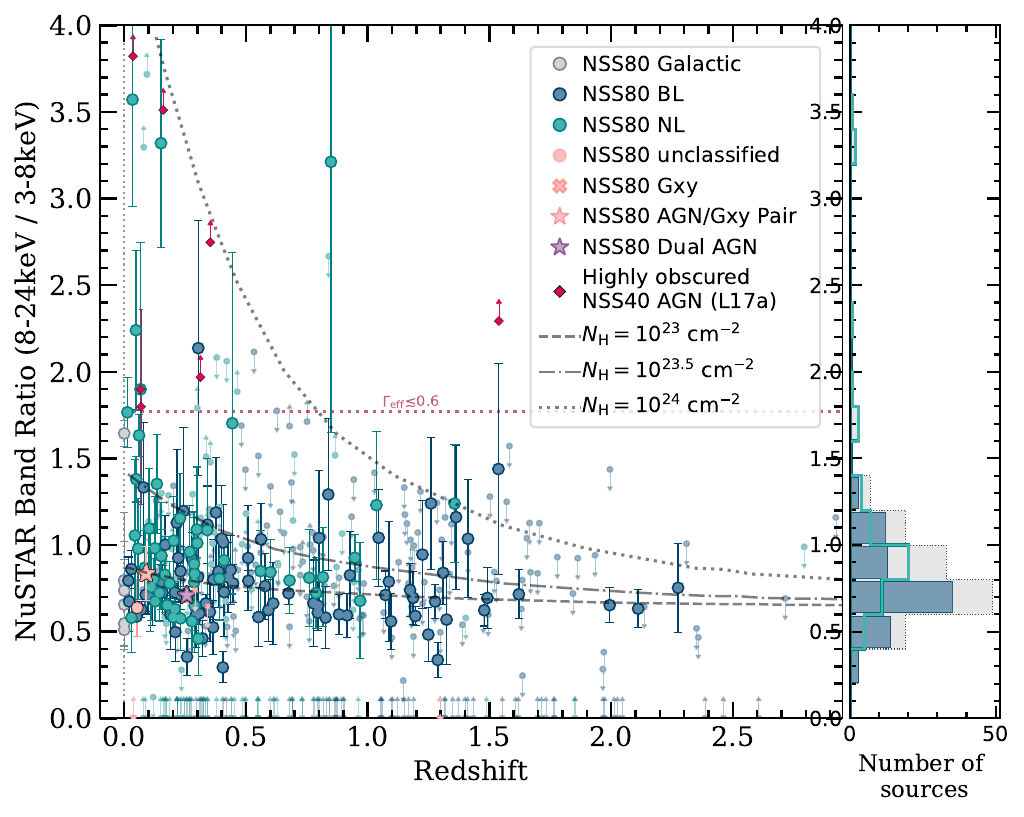}		
	\vspace*{-4mm}
	\caption{\nustar band ratio (BR$_\mathrm{Nu}$) versus redshift for the NSS80 sources, color-coded by source classification: Galactic sources ($z = 0$; grey circle), broad emission-line objects (BL; blue circle), narrow emission-line objects (NL; green circle), unclassified sources (peach circle), galaxies (Gxy; peach `X'), AGN/galaxy pairs (peach star), dual AGN (purple star), and highly obscured AGN observed in NSS40 \citep[red diamond;][]{lansbury2017}. Sources associated with the primary \textit{NuSTAR} science target are excluded. 
	BR$_\mathrm{Nu}$  values with upper or lower limits are faded using the respective classification colors. 
	The dashed lines show tracks for a simple absorbed power law model (assuming $\Gamma = 1.8$) and Galactic absorption of $N_\mathrm{H,Gal}$ = 10$^{20}$\,cm$^{-2}$ for a range of column densities along the line-of-sight to the nucleus.
	The distribution of constrained band ratios for the full NSS80 sample is shown on the right in faded grey, overlaid with the distribution of the BL (filled blue) and NL (open green) objects. The median constrained band ratio of the NSS80 sample is $\sim$\,0.8.
	}
	\label{fig:band ratios vs z}	
\end{figure*}
\begin{deluxetable}{l l l l l l l l l}
\tablecolumns{5}
\tablewidth{\textwidth}
\tablecaption{\label{tab:hard band sources} Candidate obscured NSS80 AGN with BR$_\mathrm{Nu} > 1.7$.
		} 
\tablehead{\nustar object name & Short name & R.A. & Decl. & Det. & BR$_\mathrm{Nu}$  & $z_\mathrm{spec}$ & Type & $L_\mathrm{10-40keV}$ \\
			& & ($^\circ$) & ($^\circ$) & & & & & (erg\,s$^{-1}$) \\
			(1) & (2) & (3) & (4) & (5) & (6) & (7) & (8) & (9) \\}
			\startdata
			NuSTARJ010739-1139.1 & J010739 & 16.914801 & -11.65257  & F\,S\,H& 2.2\,$\pm$\,0.5 &  0.048 & NL   & $7.99 \times 10^{42}$ \\
			NuSTARJ022951-0856.4 & J022951 & 37.46319  & -8.94133   & F\,H & $>$\,1.8        &  0.300 & NL  & $1.63 \times 10^{43}$ \\
			NuSTARJ035951-3009.9 & J035951 & 59.96408  & -30.16580  & H  & $>$\,2.4        &  0.685 & NL & $2.85 \times 10^{44}$ \\
			\rowcolor{extraLightcadetblue} 
			$^\star$NuSTARJ050559-2349.9 & J050559 & 76.49839  & -23.83168  & F\,H & $>$\,3.8        &  0.036 & NL  & $8.97 \times 10^{41}$\\
			\rowcolor{extraLightcadetblue} 
			NuSTARJ082303-0502.7 & J082303 & 125.76385 & -5.04649   & F\,H & $>$\,2.0        &  0.313 & NL  & $9.81 \times 10^{43}$ \\
			NuSTARJ094910+0022.9 & J094910 & 147.29356 & 0.38186    & F\,H & $>$\,3.7        &  0.093 & NL  & $1.63 \times 10^{42}$ \\
			NuSTARJ103456+3939.6 & J103456 & 158.73575 & 39.66031   & F\,S\,H& 3.3\,$\pm$\,0.6 &  0.151 & NL  & $2.41 \times 10^{43}$ \\
			NuSTARJ115658+5508.2 & J115658 & 179.24379 & 55.13830   & F\,H & $>$\,3.3        &  0.080 & NL  & $7.03 \times 10^{42}$ \\
			\rowcolor{extraLightcadetblue} 
			NuSTARJ141056-4230.0 & J141056 & 212.73727 & -42.50139  & F\,S\,H& 1.9\,$\pm$\,0.8 &  0.067 & NL & $2.48 \times 10^{42}$ \\
			\rowcolor{extraLightcadetblue} 
			NuSTARJ144406+2506.3  & J144406 & 221.02819 & 25.10514   & F\,H & $>$\,2.3        &  1.539 & NL? & $1.05 \times 10^{45}$ \\
			NuSTARJ150225-4208.3 & J150225 & 225.60725 & -42.13960  & F\,S\,H& 5.4\,$\pm$\,1.2 &  0.054 & Gxy & $6.18 \times 10^{42}$ \\
			\rowcolor{extraLightcadetblue} 
			NuSTARJ150646+0346.2  & J150646 & 226.69512 & 3.77105    & F\,S\,H& 3.6\,$\pm$\,0.6 &  0.034 & NL  & $1.37 \times 10^{42}$ \\
			\rowcolor{extraLightcadetblue} 
			$^\star$NuSTARJ151253-8124.3  & J151253 & 228.22496 & -81.40501  & F\,S\,H& 1.8\,$\pm$\,0.6 &  0.069 & NL  & $9.26 \times 10^{42}$\\
			\rowcolor{extraLightcadetblue} 
			NuSTARJ153445+2331.5 & J153445 & 233.68763 & 23.52592   & H  & $>$\,3.5        &  0.160 & NL & $4.82 \times 10^{42}$ \\
			NuSTARJ160817+1221.4 & J160817 & 242.07274 & 12.35752   & H  & $>$\,1.9        &  0.181 & NL  & $1.34 \times 10^{43}$ \\
			NuSTARJ163126+2357.0 & J163126 & 247.85845 & 23.95061   & F\,H & $>$\,1.7        &  0.751 & BL  & $3.97 \times 10^{44}$ \\
			$^\star$NuSTARJ190813-3925.7 & J190813 & 287.05529 & -39.42912  & H  & $>$\,3.1        &  0.075 & Gxy & $2.46 \times 10^{42}$ \\
			NuSTARJ194234-1011.9 & J194234 & 295.64177 & -10.19846  & F\,S\,H& 3.2\,$\pm$\,1.6 &  0.849 & NL? & $4.58 \times 10^{44}$ \\
			NuSTARJ214320+4334.8 & J214320 & 325.83368 & 43.58032   & F\,S\,H& 1.8\,$\pm$\,0.2 &  0.013 & NL  & $5.86 \times 10^{41}$ \\
			NuSTARJ224225+2942.0 & J224225 & 340.60580 & 29.70105   & F\,S\,H& 2.1\,$\pm$\,0.7 &  0.304 & BL & $1.57 \times 10^{43}$ \\
			NuSTARJ224925-1917.5 & J224925 & 342.35456 & -19.29294  & F\,S\,H& 1.7\,$\pm$\,0.9 &  0.445 & NL  & $3.32 \times 10^{43}$\\
			NuSTARJ231840-4223.0 & J231840 & 349.66942 & -42.38454  & F\,H & $>$\,1.9        &  0.464 & NL & $9.68 \times 10^{43}$ \\
		\enddata
	\tablecomments{The sources are listed in order of increasing right ascension. 
		The shaded green rows mark the extremely hard NSS40 sources reported in \citet{lansbury2017}. Note that J150646 is now detected in all three bands, whilst in the 40-month it was only detected in the hard band.    
		{\sc Columns:}\,\textbf{(1)} \nustar serendipitous source name. $^\star$Sources which show evidence for being associated with the primary \textit{NuSTAR} science target according to the definition $\Delta(cz) < 0.05\,cz$; see Table~\ref{tab:pflag sources}.
		\textbf{(2)} Abbreviated \nustar source name adopted here.
		\textbf{(3)} and \textbf{(4)} \nustar right ascension and declination J2000 coordinates in decimal degrees.
		\textbf{(5)} The \nustar energy bands for which the source is independently detected. F, S, and H correspond to the full ($3-24$ keV), soft ($3-8$ keV), and hard ($8-24$ keV) bands, respectively. 
		\textbf{(6)} \nustar photometric band ratio.
		\textbf{(7)} Source spectroscopic redshift obtained from emission-line fitting in \textsc{iraf} or by matching the observed flux calibrated input spectra (FITS file format) against a library of stellar, galaxy and AGN templates available in the \textsc{Marz} web application; see Section~\ref{subsubsec:spectral classification}. All redshifts are robust, except for J144406 and J194234 where fewer lines (or low S/N lines) are identified, and J050559 which has two candidate soft X-ray and \textit{WISE} counterparts.
		\textbf{(8)} Source spectroscopic classification. 
		\textbf{(9)} Non-absorption-corrected, rest-frame $10-40$\,keV luminosity. }
\end{deluxetable}

\subsubsection{Redshift-luminosity plane} \label{subsubsec:redshifts and luminosities}

Overall, on the basis of our optical spectroscopic campaign (described in Section~\ref{subsec:optical spectroscopy}) we have classified 492 NSS80 sources as extragalactic: 449 of which have reliable counterpart identifications and 43 sources with uncertain counterpart associations based on our \textsc{Nway} assessment in Section~\ref{subsubsec:spectra vs nway} (which are indicated with white filled symbols in all figures). We exclude from our analysis the 7 additional sources with photometric-redshift measurements from the literature and those with reliable counterparts closely associated with the targets of the \nustar observations, leaving 433 sources.

The redshift distribution for the 433 extragalactic NSS80 sources with reliable counterpart associations is shown in Figure~\ref{fig:z distributions nustar}, excluding sources with evidence for being associated with the \textit{NuSTAR} targets for their respective observations (see Table~\ref{tab:pflag sources}).
The redshifts cover a large range, from $z = 0.012 - 3.43$, with a median of $\langle{z}\rangle$ = 0.56.
For the 166 extragalactic objects with independent detections in the high-energy band ($8-24$\,keV), to which \nustar is uniquely sensitive, the median redshift is $\langle{z}\rangle$  = 0.34.
Roughly comparable numbers of NL and BL objects are identified for $z < 1$ (132 and 147, respectively) but comparison of redshift distributions shows a significant difference, with a larger fraction of BL sources found at higher redshifts. This is supported by a larger median redshift for BL sources of $\langle{z}\rangle$  = 0.80. This result is not unexpected since BL AGN at a given redshift are typically brighter in the optical band than NL AGN of the same intrinsic luminosity (i.e.,\ the BL AGN are less obscured in the optical); see Figures~\ref{fig:rmag distribution for all + spec} \& ~\ref{fig:z distributions nustar}.

\begin{figure*}
	\centering
	\includegraphics[width=42pc]{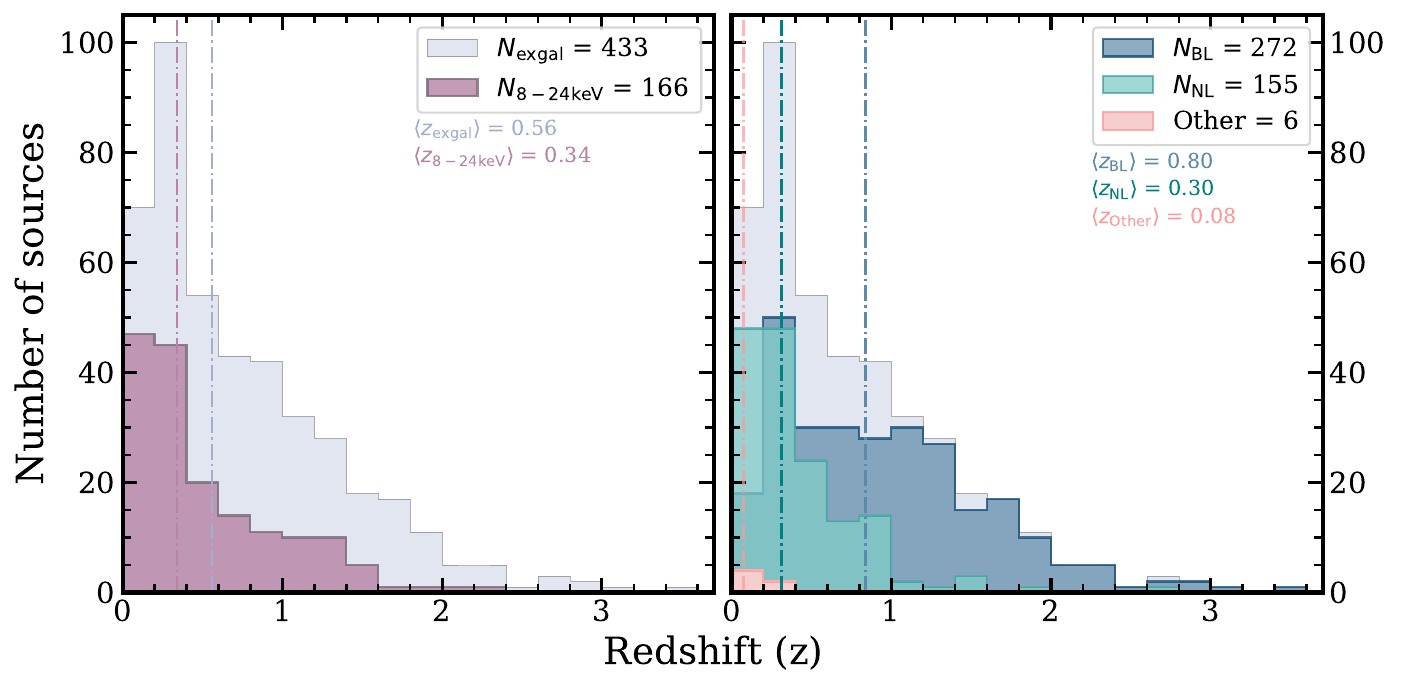}		
	\vspace*{-4.5mm}
	\caption{Redshift distribution for the 433 spectroscopically identified extragalactic NSS80 sources with reliable counterpart identifications (light purple), excluding 19 sources with evidence for being associated with the \textit{NuSTAR} targets for their respective observations. \textit{Left panel:} The distribution for the subset of NSS80 sources which are independently detected in the hard band ($8-24$ keV; dark purple). \textit{Right panel:} The redshift distribution separated by spectroscopic classification: BL objects are shown in blue, NL objects are shown in green, and ``Other'' (including four galaxies and two unclassified sources) are shown in peach. The vertical lines mark the median redshifts for the respective subsamples.
	}
	\label{fig:z distributions nustar}	
\end{figure*}
\begin{figure*}
	\centering
	\includegraphics[width=42pc]{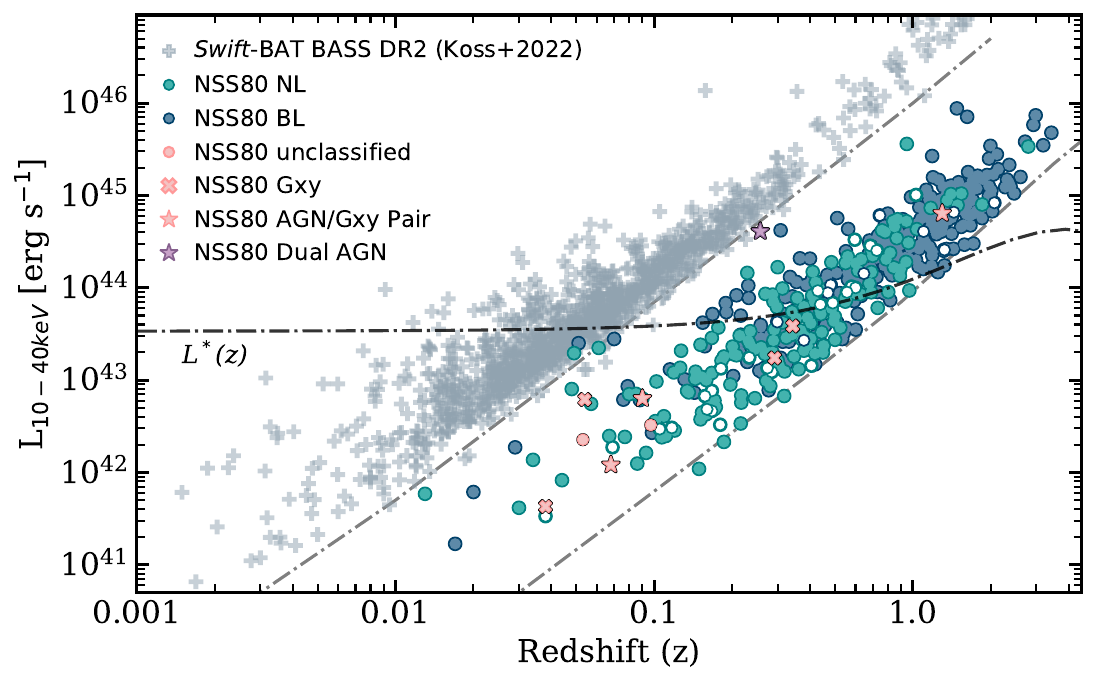}		
	\vspace*{-3mm}
	\caption{Rest-frame 10$-$40\,keV luminosity ($L_\mathrm{10-40\,keV}$) versus redshift for the extragalactic NSS80 sources with reliable counterparts, separated into different spectroscopic classes:
		broad emission-line objects (BL; blue circle), narrow emission-line objects (NL; green circle), unclassified sources (peach circle), galaxies (Gxy; peach `X'), AGN/galaxy pairs (peach star), dual AGN (purple star), and sources with uncertain counterpart identification (white filled circles with green and blue edges).
		For comparison, the \textit{Swift}-BAT AGN Spectropic Survey (BASS) DR2 \citep{koss2022} is plotted with grey crosses. 
		The \textit{Swift}-BAT $14-150$\,keV fluxes were used to calculate $L_\mathrm{10-40\,keV}$ values assuming an effective photon index of $\Gamma_\mathrm{eff} = 2.0$ for the $K$-correction factor.
		The grey dash-dotted lines indicate an observed-frame X-ray flux range of 0.02--2\,$\times$\,10$^{-12}$\,erg\,s$^{-1}$\,cm$^{-2}$, i.e., spanning two orders of magnitude.
		The black dash-dotted line highlights the evolution of the knee of the X-ray luminosity function ($L_*$) with redshift \citep{aird2015}.
	}
	\label{fig:Lx vs z}	
\end{figure*}

In Figure~\ref{fig:Lx vs z} we show the redshift–luminosity plane for the rest-frame $10-40$\,keV band, calculated from the observed frame \nustar fluxes (following the same approach as in \citetalias{lansbury2017_cat}), assuming an effective photon index of $\Gamma_\mathrm{eff} = 1.8$ \citep[typical of AGN detected by \textit{NuSTAR}; see][]{alexander2013}\footnote{If we instead choose a photon index closer to the median of this sample the resulting change in luminosity is not large. For example, a decrease in $\Gamma_\mathrm{eff}$ from $1.8$ to $1.5$ causes a small increase of $\sim$10\% in $10-40$\,keV luminosity, assuming median redshift and $3-24$\,keV flux.}.
For comparison, the \textit{Swift}-BAT AGN Spectroscopic Survey (BASS) DR2 \citep{koss2022} is plotted with grey crosses. 
The \textit{Swift}-BAT $14-150$\,keV fluxes were used to calculate $L_\mathrm{10-40\,keV}$ values assuming an effective photon index of $\Gamma_\mathrm{eff} = 2.0$ for the $K$-correction factor (using the median slope of 2). As can be seen, the NSS80 sources span the knee of the X-ray luminosity function out to $z\approx$~1 \citep{aird2015}, as opposed to $z\approx$~0.1 for the \textit{Swift}-BAT AGN.

Of the 433 extragalactic NSS80 sources with reliable counterpart associations, 99\% (427/433) are within the luminosity range of $L_\mathrm{10-40\,keV} \approx 10^{42} - 10^{46}$\,erg\,s$^{-1}$, with a median luminosity of 1.2\,$\times$\,10$^{44}$\,erg\,s$^{-1}$.
These values are consistent with the NSS40 catalog. 
The faintest source in the \citetalias{lansbury2017_cat} sample, however, with $L_\mathrm{10-40\,keV} = 1.0 \times 10^{39}$\,erg\,s$^{-1}$, is recorded in the secondary NSS80 catalog (see Section~\ref{subsec:source catalog} and Appendix~\ref{appendix:secondary catalogue}), since the NL AGN at $z = 0.002$ is hosted by the galaxy IC750.
Hence, the least luminous source in the primary NSS80 catalog is \mbox{NuSTAR J010736-1732.3} (NL AGN at $z = 0.021$) with $L_\mathrm{10-40\,keV} = 2.6 \times 10^{40}$\,erg\,s$^{-1}$. 
As in \citetalias{lansbury2017_cat}, the source at the other extreme end in luminosity is \mbox{NuSTAR J052531-4557.8}: a radio-bright BL AGN at $z = 1.479$ with $L_\mathrm{10-40\,keV} = 8.8 \times 10^{45}$\,erg\,s$^{-1}$, also classified as a blazar in the literature \citep[e.g.,][]{massaro2009} which means that the X-ray luminosity may be inflated by beaming effects. 
The most distant source detected is still the optically unobscured quasar, NuSTAR J232728+0849.3 at a $z = 3.43$, reported in the NSS40.

We also compare to the luminosity-redshift plane of BASS, as shown in Figure~\ref{fig:Lx vs z}.
The \textit{Swift}-BAT and \nustar serendipitous survey are complementary to one another, with the former providing a statistical sample of AGN in the nearby universe ($z < 0.1$) selected in hard X-rays, and the latter providing its counterpart for the distant universe.
Consequently there is little overlap between the two surveys, which sample different regions of the $L_\mathrm{x}$-$z$ parameter space, with the exception of four \nustar sources outlying in Figure~\ref{fig:Lx vs z} which have very high fluxes at the detection threshold of \textit{Swift}-BAT (all BASS detected):
\begin{itemize}
	\vspace{-0.1cm}
	\setlength\itemsep{-0.3em}
	\item \textbf{NuSTAR J043727-4711.5:} a BL AGN at $z = 0.051$; $L_\mathrm{10-40\,keV} = 2.5 \times 10^{43}$\,erg\,s$^{-1}$.
	\item \textbf{NuSTAR J091912+5527.8:} a NL AGN at $z = 0.049$; $L_\mathrm{10-40\,keV} = 2.0 \times 10^{43}$\,erg\,s$^{-1}$.
	\item \textbf{NuSTAR J103135-4206.0:} a NL AGN at $z = 0.061$; $L_\mathrm{10-40\,keV} = 2.2 \times 10^{43}$\,erg\,s$^{-1}$.
	\item \textbf{NuSTAR J180958-4552.6:} a (beamed) BL AGN at $z = 0.07$; $L_\mathrm{10-40\,keV} = 2.8 \times 10^{43}$\,erg\,s$^{-1}$.
\end{itemize}
Overall, the \nustar serendipitous survey provides the higher-redshift component of \textit{Swift}-BAT and fills out the broadest range of luminosities and redshifts in comparison to other \nustar surveys, e.g., the \textit{NuSTAR}-ECDFS survey \citep{mullaney2015},  \textit{NuSTAR}-COSMOS survey \citep{civano2015} and \textit{NuSTAR}-UDS \citep{masini2018}.

In the following sections we further explore the optical and MIR properties of the extragalactic NSS80 sources. 

\subsection{MIR properties of the NSS80 AGN} \label{subsec:wise properties nustar}

\begin{figure}
	\centering
	\includegraphics[width=25pc]{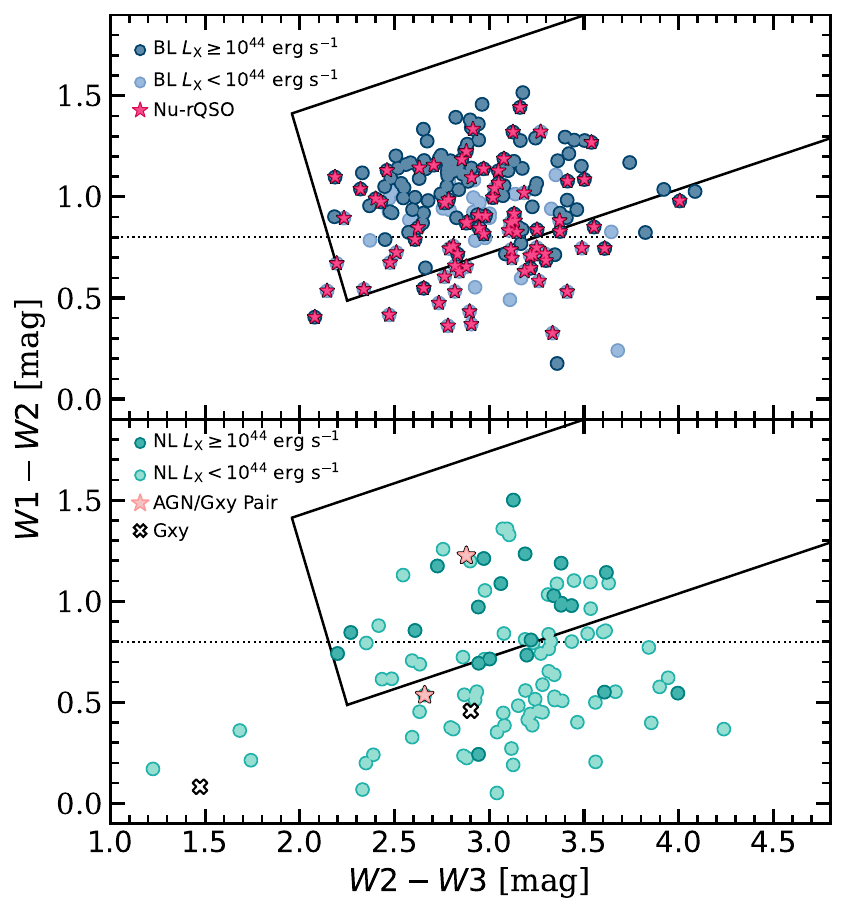}		
	\vspace*{-5mm}
	\caption{\textit{WISE} color-color diagram for the extragalactic NSS80 sources color coded by X-ray luminosity and spectroscopic classification following the color scheme in Figure~\ref{fig:gi vs z nustar}: BL objects (blue circle), NL objects (green circle), and AGN/galaxy pairs (peach star). 
		BL and NL sources with $L_\mathrm{10-40\,keV} \geq 10^{44}$\,erg\,s$^{-1}$ are indicated with dark blue and green circles, whilst the light blue and green circles codify the lower luminosity BLs and NLs with $L_\mathrm{10-40\,keV} < 10^{44}$\,erg\,s$^{-1}$. 
		The \citet{mateos2012} wedge which identifies AGN with red MIR power-law SEDs with a spectral index $\alpha \leq -0.3$ is indicated with a solid black line. We also compare with the AGN color cut of \citet[][$W1-W2 > 0.8$; black dotted line]{stern2012}.
		In the \textit{top panel} we show the \textit{WISE} colors of the BLAGN subdivided into low (light blue) and high (dark blue) $L_\mathrm{10-40\,keV}$. Our identified Nu-rQSOs (i.e., red BL AGN) based on their $g-i$ color are plotted with red stars; see Section~\ref{subsec:optical prop} and Figure~\ref{fig:gi vs z nustar}. 
		The \textit{bottom panel} shows the \textit{WISE} colors for the \nustar NLs separated into $L_\mathrm{10-40\,keV} < 10^{44}$\,erg\,s$^{-1}$ (light green) and $L_\mathrm{10-40\,keV} \geq 10^{44}$\,erg\,s$^{-1}$ (dark green). Two AGN/galaxy pairs are plotted with peach stars and two galaxies with white crosses.
	}

	\label{fig:wise color-color specz}	
\end{figure}

The MIR emission from AGN is typically due to the reprocessing of accretion-disc radiation by circumnuclear dust, and suffers little extinction relative to other wavelengths \citep[e.g.][]{nishiyama_interstellar_2008,netzer2015,hickox2018}; however, a non-negligible fraction can also be produced by star formation in the host galaxy \citep[e.g.][]{stern2005,hickox2018}.
Color selections using the \textit{WISE} telescope bands \citep[3.4\,$\mu$m, 4.6\,$\mu$m, 12\,$\mu$m, and 22\,$\mu$m; e.g.,][]{assef2010,jarrett2011,stern2012,mateos2012,mateos2013,assef2013} can separate bright AGN from host-galaxy light (from stars and the interstellar medium) through the identification of a red MIR spectral slope, and have thus become widely applied.
These selections have the potential to identify large samples of AGN with less bias against heavily obscured systems. 
However, their effectiveness worsens toward lower AGN luminosities, where the AGN component of the MIR spectrum can be swamped by the emission from the host galaxy. 
For example, \citet{cardamone2008} and \citet{lamassa_sdss_2019} found that most X-ray selected AGN in the (deep) GOODS field and Stripe 82-X respectively would not have been found by standard MIR color selection. \citet{lyu_agn_2022} show similar results for SED-selected AGN in the GOODS-S/HUDF region.
Notably, the MIR host emission contribution increases with redshift due to the increase in cosmic SFR density, too, thus it is progressively harder to select AGN at the same luminosity at higher redshifts. 
By comparison, \nustar selects AGN almost irrespective of the relative strength of the AGN to the host galaxy since the X-ray emission from galaxy processes is weak in comparison to the AGN, particularly at the $3-24$\,keV energies probed by \textit{NuSTAR}. 
Here we investigate the MIR properties of our NSS80 sources, and consider the results with respect to the AGN selection criteria.

As CatWISE20 contains only $W1$ and $W2$ photometry, for the purposes of examining the further properties of the sample we match all sources with an \textsc{Nway} CatWISE20 counterpart to their nearest AllWISE \citep[][\footnote{\url{http://wise2.ipac.caltech.edu/docs/release/allwise/}}]{wright2010,mainzer2011} source within a maximum radius of 4\arcsec. This results in 865 matches, including 312/523 upper limits for $W3$/$W4$ respectively. $W1$ and $W2$ magnitudes are roughly consistent between CatWISE20 and AllWISE values except in a small minority of sources, implying that the $W3$ and $W4$ can be used for comparison with the caveat that there may be additional uncertainty due to the different photometric pipelines. In Figure~\ref{fig:wise color-color specz} we plot the \textit{WISE} colors ($W1-W2$ versus $W2-W3$) of the spectroscopically confirmed extragalactic NSS80 sample with reliable counterpart associations and $W3$ detections, i.e., well-defined \textit{WISE} colors (see Section~\ref{subsec:nway infrared and optical cpart}) and compare to the selection ``wedge'' defined by \citet{mateos2012} to identify AGN with red MIR power-law SEDs.
In this comparison we further limit our analysis to the sources with significant detections in all three of the relevant, shorter wavelength \textit{WISE} bands ($W1$, $W2$ and $W3$; centred at 3.4\,$\mu$m, 4.6\,$\mu$m and 12\,$\mu$m, respectively).
Considering sources with optical spectroscopic classifications, the fractions for the overall BL AGN (top panel) and NL AGN (bottom panel) samples are $\mathfrak{f}_\mathrm{wedge,BL}$\,=\,80\% (148/185) and $\mathfrak{f}_\mathrm{wedge,NL}$\,=\,42\% (36/86), respectively.
Therefore, NL AGN are less likely to be identified as AGN based on MIR colors alone.
If we use the ``X-ray quasar'' threshold\footnote{The X-ray quasar threshold is often adopted to define X-ray quasars which roughly agrees with the classical optical quasar definition: $M_\mathrm{B} \leq -23$ \citep[][]{schmidt1983} and the approximate value for $L_\mathrm{X,*}$; see Figure~\ref{fig:Lx vs z}.}  of $10^{44}$\,erg\,s$^{-1}$ in the $10-40$\,keV band to distinguish between faint and luminous X-ray BLs/NLs, we find that this is largely driven by the lower luminosity objects with $L_\mathrm{X}$ below the X-ray quasar threshold: only 35\% (23/66) low-$L_\mathrm{X}$ lie inside the wedge, while 80\% (16/20) of the high-$L_\mathrm{X}$ NL AGN have AGN-like MIR colors.
On the other hand, the bulk of the BL AGN with $L_\mathrm{X} > 10^{44}$\,erg\,s$^{-1}$ lie within the wedge (95\%; 117/123 high-$L_\mathrm{X}$).
Of the $L_\mathrm{X} < 10^{44}$\,erg\,s$^{-1}$ BL AGN, 65\% (40/62) have AGN-like MIR colors.

If we subdivide the BL AGN sample on the basis of their $g-i$ optical color (selecting the reddest 10\%, as in \citealt{klindt2019}; see Section~\ref{subsec:optical prop} and Figure~\ref{fig:gi vs z nustar}), we find a lower fraction of red BL AGN (Nu-rQSOs) lie within the wedge than found for the control BL AGN (Nu-cQSOs): 67\% (49/73) versus 98\% (101/103).
Of the 49 Nu-rQSOs with AGN-like MIR colors, 33 sources have $L_\mathrm{X} > 10^{44}$\,erg\,s$^{-1}$ (85\% of the overall high-$L_\mathrm{X}$ Nu-rQSOs) and the remaining 16 have $L_\mathrm{X}$ values below the X-ray quasar threshold (47\% of the overall low-$L_\mathrm{X}$ Nu-rQSOs). 
Hence, we can deduce from the MIR colors that the majority of Nu-rQSOs (especially at low-$L_\mathrm{X}$) are more host-galaxy dominated than the Nu-cQSOs. This is broadly consistent with our finding on the basis of optical analyses in the following section.

\subsection{Optical photometric and spectroscopic properties} \label{subsec:optical prop}

Optical selection of quasars, using for example SDSS photometry, will miss the most reddened quasars due to their colors overlapping the stellar loci in most SDSS color-color diagrams, and a comparatively shallow optical survey flux limit \citep[e.g.,][]{richards2003}. 
However, since X-rays penetrate circumnuclear obscuration with minimal contribution from the host galaxy, they provide the potential to construct a more complete census of the full quasar population, ranging from heavily obscured sources \citep[e.g., ERQs;][]{goulding2018}, thinly-veiled red quasars \citep[e.g., rQSOs;][hereafter K19]{klindt2019}, and host-galaxy dominated systems.
Here we focus on the optical photometric and spectroscopic properties of the NSS80 sources subdivided on the basis of X-ray luminosity and optical spectroscopic classification.

\begin{figure}
	\centering	
	\includegraphics[width=25pc]{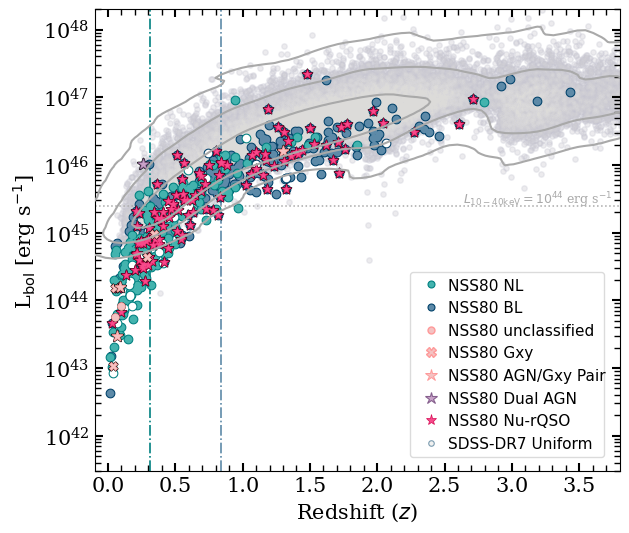}		
	\vspace*{-6mm}
	\caption{The bolometric luminosity-redshift plane for the (spectroscopically reliable) extragalactic NSS80 sources compared to SDSS optical quasars.
		The bolometric luminosity ($L_\mathrm{bol}$) for the NSS80 sources is inferred from the rest-frame 10--40\,keV luminosity using a bolometric correction of BC\,=\,25 (assuming that $L_\mathrm{10-40\,keV}$ makes up $\sim$\,4\% of $L_\mathrm{bol}$).
		Following the color scheme in Figure~\ref{fig:gi vs z nustar} the following subsamples are plotted: BL objects (blue circle), NL objects (green circle), unclassified sources (peach circle), galaxies (Gxy; peach `X'), AGN/galaxy pairs (peach star), dual AGN (purple star), and sources with uncertain counterpart identification (white filled circles with green and blue edges). 
		The $g - i$ selected rQSOs (see Figure~\ref{fig:gi vs z nustar}) are highlighted with red stars and the SDSS DR7 quasars are plotted using shaded grey circles. Compared to the optically-selected SDSS quasars, the X-ray selected \nustar serendipitous sources are typically $\sim$\,0.3\,dex less luminous. The grey dotted line indicates the X-ray quasar threshold of $L_\mathrm{10-40\,keV} = 10^{44}$\,erg\,s$^{-1}$. The median $L_\mathrm{bol}$ for the extragalactic NSS80 sample is 7.19\,$\times$\,10$^{44}$\,erg\,s$^{-1}$ at the median redshift of the NLs ($\langle{z}_\mathrm{NL}\rangle = 0.3$), and 5.94\,$\times$\,10$^{45}$\,erg\,s$^{-1}$ at the BL median redshift ($\langle{z}_\mathrm{BL}\rangle = 0.82$), as indicated with green and blue dash-dotted lines, respectively. Grey lines show KDE contours of the SDSS DR7 quasars at 68\%, 95\%, and 99.7\%, for comparison with the distribution of NSS80 sources.
	}
	\label{fig:Lbol vs z nustar}	
\end{figure}

The NSS80 is the largest-area \nustar survey and picks up the most X-ray luminous AGN over $\sim$\,36\,deg$^2$. 
However, the NSS80 region is a factor $\sim$\,300$\times$ smaller than the SDSS and consequently will miss the most luminous systems.
To place the NSS80 survey into context, and to further motivate our following analyses which make use of SDSS DR7 quasars, in Figure~\ref{fig:Lbol vs z nustar} we compare the bolometric luminosity-redshift plane of both surveys.
For the NSS80 sources we calculate bolometric luminosities from the rest-frame $10-40$\,keV luminosity, assuming that the $10-40$\,keV luminosity makes up 4\% of the total luminosity \citep[see e.g.,][]{lansbury2017}. 
The SDSS bolometric luminosities are available in \citet{shen2011}; however, they are inferred from rest-frame UV-optical continuum measurements and have not been corrected for dust extinction. 
Consequently, the $L_\mathrm{bol}$ values are likely to be significantly underestimated in the SDSS rQSOs \citep[which makes up $~$10\% or more of typical quasar samples; see e.g.,][]{richards2003}.
Overall there is good overlap between NSS80 and SDSS at the lower luminosity end although, as expected, NSS80 misses the most luminous systems.
The median $L_\mathrm{bol}$ for the extragalactic NSS80 sample is 7.62\,$\times$\,10$^{44}$\,erg\,s$^{-1}$ at the median redshift of the NLs ($\langle{z}_\mathrm{NL}\rangle = 0.30$), and 5.93\,$\times$\,10$^{45}$\,erg\,s$^{-1}$ at the BL median redshift ($\langle{z}_\mathrm{BL}\rangle = 0.80$). By comparison, the SDSS sample have median bolometric luminosities of 1.73\,$\times$\,10$^{45}$\,erg\,s$^{-1}$ (0.38 dex higher) and 1.07\,$\times$\,10$^{46}$\,erg\,s$^{-1}$ (0.26 dex higher) at redshifts equal to $\langle{z}_\mathrm{NL}\rangle$ and $\langle{z}_\mathrm{BL}\rangle$, respectively. 

\begin{figure}
	\centering
	\includegraphics[width=25pc]{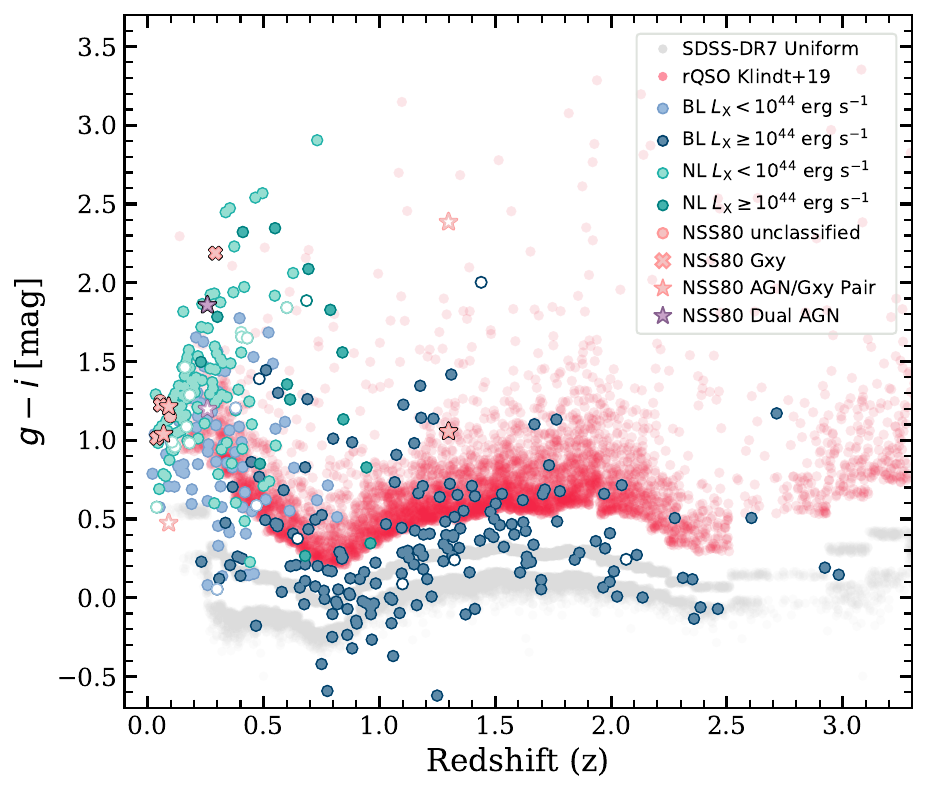}		
	\vspace*{-6mm}
	\caption{The $g-i$ color versus redshift for extragalactic NSS80 sources with reliable counterpart associations (also excluding those that are possible associated with the primary \textit{NuSTAR} science target), plotted in the different spectroscopic classifications: BL objects (blue circle), NL objects (green circle), unclassified sources (peach circle), galaxies (Gxy; peach `X'), AGN/galaxy pairs (peach star), dual AGN (purple star), and sources with uncertain counterpart identification (white filled circles with green and blue edges). The AGN/galaxy and dual AGN companions are plotted with white filled stars with peach and purple edges, respectively.
		BL and NL sources with $L_\mathrm{10-40\,keV} \geq 10^{44}$\,erg\,s$^{-1}$ are indicated with dark blue and green circles, whilst the light blue and green circles codify the lower luminosity BLs and NLs with $L_\mathrm{10-40\,keV} < 10^{44}$\,erg\,s$^{-1}$. 
		SDSS selected rQSOs (reddest 10\% of the $g-i$ distribution; K19) are superimposed on the distribution of the SDSS DR7 uniform sample (shaded grey circles).
		We use the rQSO track to identify 106 red NSS80 BL sources; of these, 67\% (71/106) are at redshifts $z \leq 1$. 
		Using the same track, we identify a control sample comprising 145 BL AGN (52\% of which are at $z \leq 1$). 
		There is a bias against NLs at higher redshifts, given that all NLs (with constrained $g$ and $i$ magnitudes) are at $z \leq 1$.
	}
	\label{fig:gi vs z nustar}	
\end{figure}

To provide a basic characterization of the optical properties of the extragalactic NSS80 sources with reliable optical counterpart associations, in Figure~\ref{fig:gi vs z nustar} we show the $g-i$ color versus redshift of the 371 extragalactic sources with constrained $g-i$ colors and compare to the SDSS DR7 uniformly-selected quasar sample and the rQSOs identified in K19; we note that the optical colors are corrected for Galactic extinction following K19. 
For quasar-dominated systems, $g-i$ color provides a basic measurement of the amount of optical reddening due to dust \citep[e.g.,][]{fawcett2020}, although emission from the host galaxy can have a significant impact on the optical colors for lower luminosity and/or heavily obscured AGN. At a given redshift, a range in $g-i$ color is observed for the BLs and NLs\footnote{As can be seen in Figure~\ref{fig:gi vs z nustar}, the $g-i$ color distribution changes with redshift as different portions of the quasar and host-galaxy emission enter the observed-frame optical bands. Consequently, we only consider the optical color of the sources with a reliable counterpart and redshift. We note that the redshift and luminosity distributions of the subset of BL sources with measured $g-i$ color are similar to the distributions for all BL sources and can therefore be considered representative.}.

We use the $g-i$ color threshold from K19 to identify red quasars (corresponding to the 10\% most reddened in SDSS; threshold is shown as the bottom edge of rQSOs in Figure~\ref{fig:gi vs z nustar}), finding that 42\% (106/251) of the NSS80 BL AGN\footnote{We can use the NSS80 BL AGN as a quasar sample since they have spectral lines with widths $FWHM \gtrsim 1000$\,km\,s$^{-1}$.} with reliable counterpart associations and detections in the $g$- and $i$-band have red $g-i$ colors (we coin these sources as Nu-rQSO); we note that in this basic comparison, we reasonably assume no significant differences in the g-band and i-band pass bands between the various different optical photometric surveys. 
Among these is the higher-redshift AGN pair (NuSTARJ091534+4054.6) at $z = 1.298$, with both BL AGN identified as red quasars based on their $g-i$ color.  
The remaining 145/251 BLs we use as a control sample which represents ``normal" unobscured quasars (Nu-cQSO).
Of the 114 NLs with reliable counterpart associations and constrained $g$ and $i$ magnitudes, 105 are found to lie within the rQSO $g-i$ region, meaning that 92\% of the NLs appear red in their optical colors. 
Since no clear broad emission lines are seen in these sources we expect the optical emission to be dominated by the host galaxy, a result also suggested from the optical spectroscopy (see Section~\ref{subsec:composite spectra}).
As the Nu-rQSOs have identified broad emission lines, to first order we would expect the optical colors to be typically due to dust reddening. Overall, the $g-i$ optical color threshold adopted in Figure~\ref{fig:gi vs z nustar} corresponds to 10\% of the SDSS quasars in K19 while 42\% of the \nustar BL AGN exceed this threshold. 
This larger fraction of rQSOs in NSS80 compared to the SDSS could be due to a greater contribution from dust-reddened quasars; however, since SDSS probes more luminous quasars than NSS80, we should keep in mind that a larger fraction could be optically red due to emission from the host galaxy. A combination of template fitting to the UV--MIR photometry, guided by the optical spectroscopic features (e.g., continuum shape, emission, and absorption lines) would be able to constrain the relative contributions to the reddening from the host galaxy and dust extinction but is beyond the scope of the current study.

\subsection{Composite spectra of the NSS80 AGN}
\label{subsec:composite spectra}

To provide more direct constraints on the origin of the optical colors of the NSS80 sources we utilize the spectroscopic data.
In this analysis we construct composite spectra for subsets of the NSS80 spectroscopic sample, divided on the basis of spectral type, optical color, and X-ray luminosity. 
These composite spectra allow us to search for subtle signatures missed in individual source spectra, such as the absorption features from the host galaxy (e.g., rest-frame Ca\,\textsc{ii} H+K; G-band; Mg\,\textsc{i}; Na\,\textsc{i-d}).
The overall continuum shape of the spectra can provide insights on the relative contributions from dust reddening and host-galaxy emission/absorption: in comparison to typical quasars, a drop in continuum flux at blue wavelengths can indicate reddening due to dust, whereas an increase in flux at longer wavelengths can indicate reddening due to host-galaxy contamination. 

\begin{figure*}
	\centering
	\includegraphics[width=37pc]{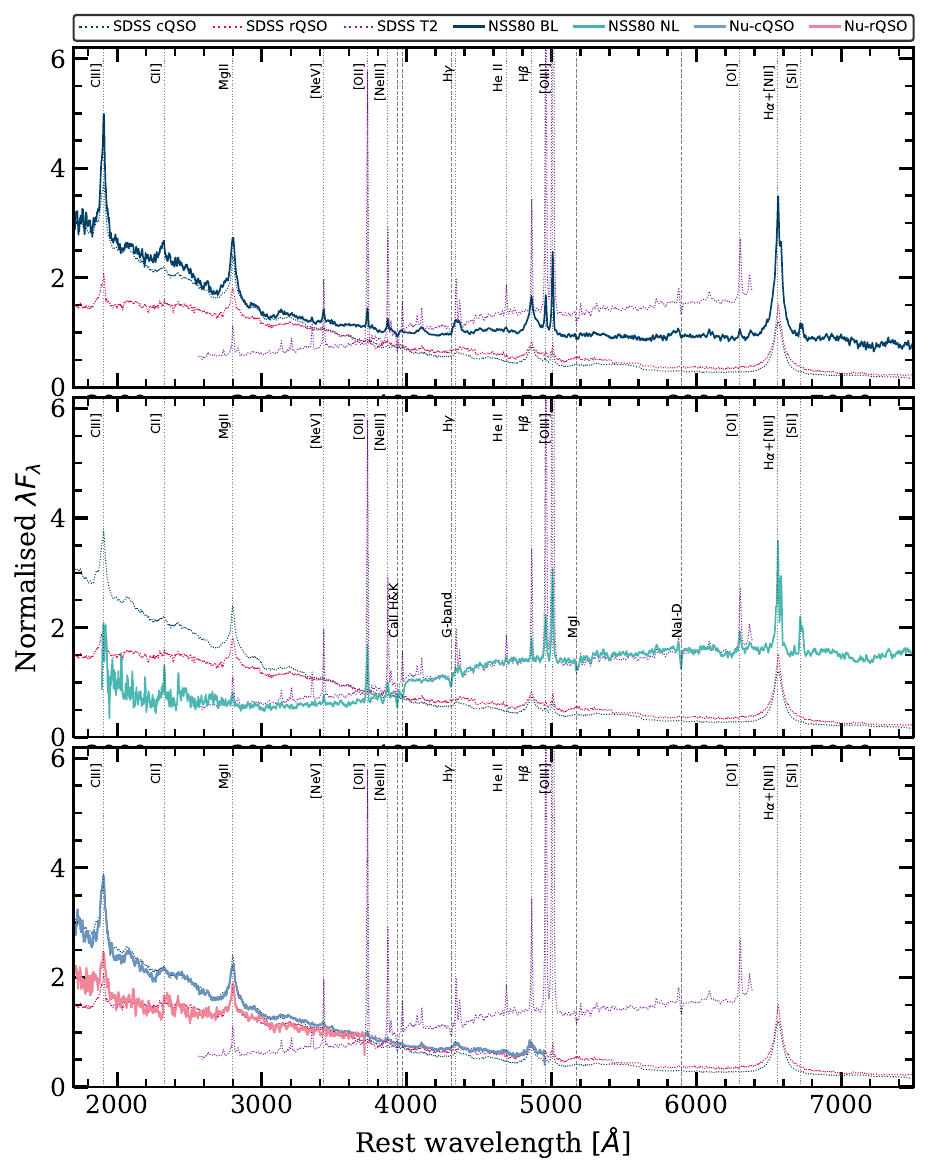}	
	\vspace*{-2mm}
	\caption{Composite spectra of the spectroscopically confirmed NSS80 BL (top panel; $N_\mathrm{stack} = 274$) and NL (middle panel; $N_\mathrm{stack} = 166$). 
		The BL sample is subdivided (bottom panel) into typical quasars (Nu-cQSO; solid blue line) and red quasars (Nu-rQSO; solid peach line). 
		Firstly, we compare to composite X-Shooter spectra of SDSS quasars, plotted with red and blue dotted lines \citep{fawcett2022}. The signature AGN emission lines and galaxy absorption lines are plotted with grey dotted and dashed lines, respectively. We also compare to a composite of SDSS Type II AGN from \citet{yuan_spectroscopic_2016}; purple dotted line.
		The spectra are shifted to rest-frame wavelengths and normalized at 4000\,\AA{} for illustrative purposes. 
		The source threshold when (geometric mean) stacking the data is 15.
		Evidence for a larger host galaxy contribution in the \textit{NuSTAR} objects compared to SDSS are visible, while towards shorter wavelengths strong evidence for reddening in the NL objects and, some in the BL AGN, can be seen.}
	\label{fig:composite nustar all}	
\end{figure*}
In Figure~\ref{fig:composite nustar all} we show rest-frame composite spectra for the extragalactic NSS80 sources subdivided into 274 BLs (dark blue solid line; top panel) and 166 NLs (green solid line; middle panel). 
We compare these composites to the high-quality X-Shooter composite spectra of a subsample of the K19 rQSOs at $1.45 < z < 1.65$, which are luminosity matched in rest-frame 6\,$\mu$m luminosity and redshift to the K19 control quasars (cQSOs).
The X-shooter composite spectra are produced in \citet{fawcett2022} using geometric mean stacks, and have a continuous spectral coverage from $\sim$\,$3000-25,000$\AA{}. 
To stack the NSS80 spectra, we followed the same approach as \citet{fawcett2022}: the spectra were first corrected for Galactic extinction, using the \citet{schlegel1998} map and the \citet{fitzpatrick_correcting_1999} Milky Way extinction law, and shifted to rest frame wavelengths. Each spectrum was then re-binned to a common wavelength grid and normalized at 4000~\AA.
It is worth noting that the spectra contributing to these stacks were obtained via different instruments and therefore will have different spectral resolutions. Therefore, any interpretation should be mainly limited to the spectral shape, with analysis of emission or absorption lines limited to broad comparisons.
To ensure reasonably representative composite spectra we only include data when at least 15 sources contribute at a given rest-frame wavelength. The scatter within each stack subset varies; e.g. the cQSOs have a lower scatter than rQSOs due to the nature of their populations. However, the scatter is not large enough in any subset to make the comparison of their average properties invalid.

The BL composite (dark blue) shows strikingly similar permitted lines to the SDSS quasar X-Shooter composites.
However, the \nustar composite has stronger forbidden lines (e.g., [Ne\,\textsc{v}], [O\,\textsc{ii}] and [Ne\,\textsc{iii}]) and has a different overall continuum shape with a sharper decrease to UV-blue wavelengths and a rise to red wavelengths. 
These differences are consistent with that expected by a modest host-galaxy contribution not present in the more luminous SDSS quasars with a light screen of dust reddening suppressing the UV-blue emission.
The NL composite (green) is distinctly different from the cQSO and rQSO SDSS composites with weak UV-blue emission and strong red emission. It shows a continuum shape more consistent with a composite of Type 2 SDSS quasars \citep[purple;][]{yuan_spectroscopic_2016}.
Furthermore, numerous strong forbidden lines, including [Ne\,\textsc{v}], [O\,\textsc{ii}], [Ne\,\textsc{iii}], [O\,\textsc{iii}], [O\,\textsc{i}] and [S\,\textsc{ii}], and narrow permitted lines are evident. The strong  [Ne\,\textsc{v}] and [O\,\textsc{iii}], in particular, indicate the presence of an optical NLAGN.
The spectral shape of the NL composite is consistent with that expected for a host-galaxy dominated spectrum due to the AGN emission being completely obscured in the optical waveband, as expected given the lack of broad permitted lines. 
Direct evidence for a dominant host-galaxy component is vividly seen from the strong host-galaxy absorption features, i.e., Ca\,\textsc{ii} H\&K, G-band, Mg\,\textsc{i} and Na\,\textsc{i-d}.
Further differences are seen in the BL composite when we split it into Nu-rQSOs (peach) and Nu-cQSOs (light blue); see Figure~\ref{fig:gi vs z nustar}.
Both the Nu-rQSOs and Nu-cQSOs show evidence for modest dust reddening due to suppressed UV-blue emission: the weaker shorter-wavelength broad lines relative to H$\alpha$ (H$\beta$; Mg\,\textsc{ii}; C\,\textsc{iii}]) for the Nu-rQSOs suggest greater dust reddening than that seen in the Nu-cQSOs. These characteristics are qualitatively consistent with those seen in the SDSS X-Shooter composites. However, in stark constrast to the X-Shooter composites, the Nu-rQSOs and Nu-cQSOs show a rise in the continuum emission to red wavelengths, which is most prominent in the Nu-rQSOs, likely due to an increasing contribution from the host galaxy.

Given the evidence for host-galaxy contributions to the NSS80 composites and the lower overall bolometric luminosties of the NSS80 sources in comparison to the SDSS quasars, we subdivided the NSS80 AGN according to their luminosities using the quasar X-ray threshold of $L_\mathrm{10-40\,keV} = 10^{44}$\,erg\,s$^{-1}$.
Figure~\ref{fig:composite nustar Lx split} shows the composites for the NL systems, Nu-cQSOs and Nu-rQSOs split into low-$L_\mathrm{X}$ ($L_\mathrm{10-40\,keV} < 10^{44}$\,erg\,s$^{-1}$) and high-$L_\mathrm{X}$ ($L_\mathrm{10-40\,keV} > 10^{44}$\,erg\,s$^{-1}$) subsamples, with 3300--5300\,\AA{}  zoom-ins plotted in Figure~\ref{fig:composite nustar Lx split zoom} to emphasise the prominent AGN and any host-galaxy absorption features.
The rest-wavelength coverage for each composite is now more limited since the X-ray luminosity selection corresponds broadly to a redshift selection and hence effectively narrower redshift ranges than the overall composites in Figure~\ref{fig:composite nustar all}. 
Despite this limitation, some similarities and differences are apparent across each of the composite pairs. 
Overall, the low-$L_\mathrm{X}$ and high-$L_\mathrm{X}$ systems within each spectral class have similar emission-line features and UV-blue continuum shapes. 
However, the low-$L_\mathrm{X}$ systems all have increased emission at red wavelengths in comparison to the high-$L_\mathrm{X}$ systems, consistent with a relative increase in the host-galaxy contribution for a decrease in the luminosity of the AGN. 
This result is qualitatively similar to that seen in the \textit{WISE} MIR color analyses (see Figure~\ref{fig:wise color-color specz} and Section~\ref{subsec:wise properties nustar}). 
As for the overall composites, more direct evidence for a host-galaxy component is seen from the identification of strong host-galaxy absorption, most strikingly in the low-$L_\mathrm{X}$ NL systems but also evident from the sometimes weak identification of Ca\,\textsc{ii} H+K in all of the composites. Greater insight on the host-galaxy and AGN properties, including constraints on the stellar mass and populations, can be gained from detailed fitting of the composite spectra and SEDs using AGN and stellar population models. However, that goes beyond the scope of this study.

\begin{figure*}
	\centering
	\includegraphics[width=37pc]{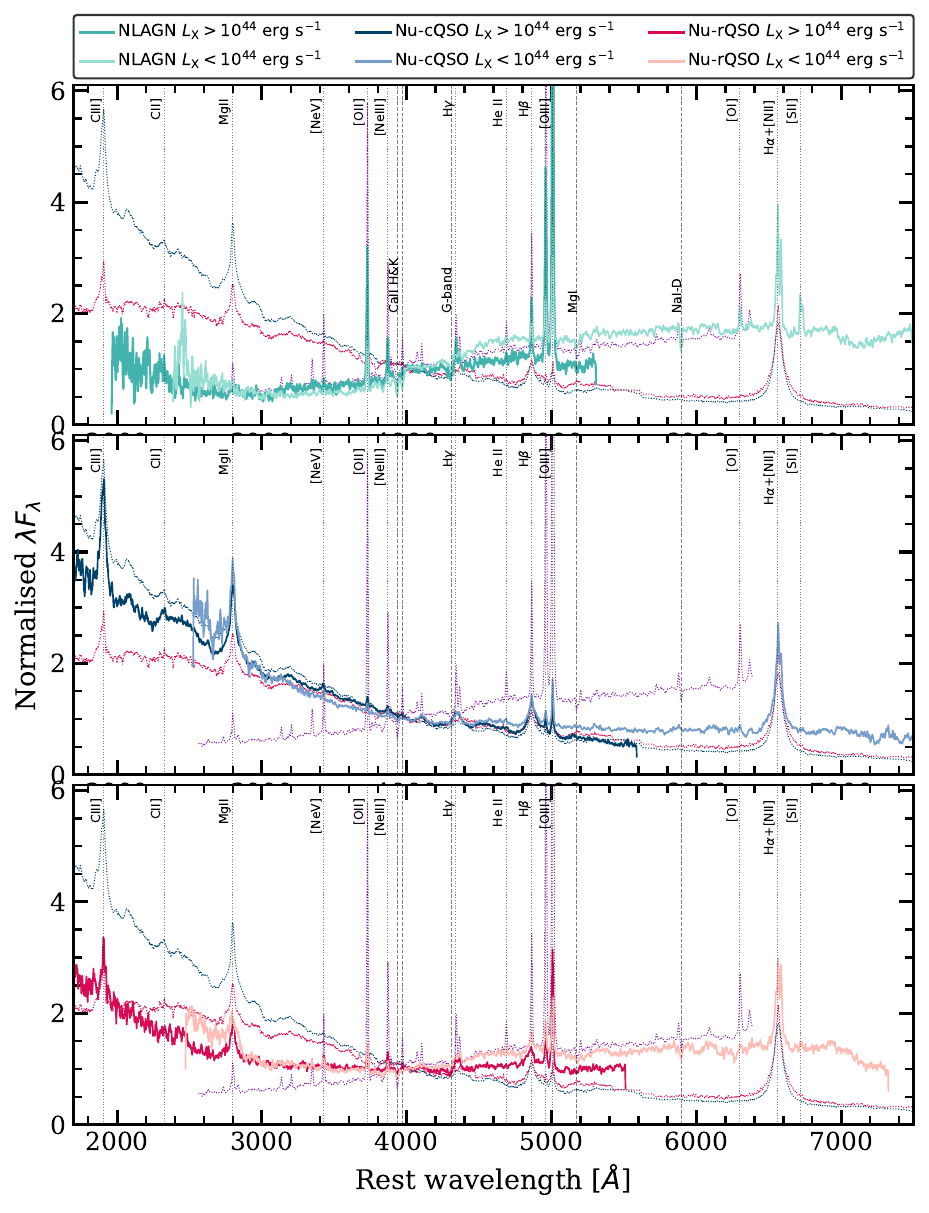}
	\vspace*{-2mm}
	\caption{Composite spectra of the spectroscopically confirmed NLAGN (top), Nu-cQSO (middle) and Nu-rQSO (bottom) subdivided into low (light green, light blue and peach, respectively) and high (dark green, dark blue and red, respectively) luminosity bins.
		As in Figure~\ref{fig:composite nustar all}, we compare to composite X-Shooter spectra of SDSS quasars, plotted with red and blue dotted lines \citep{fawcett2022}, and composite spectra of SDSS Type 2 QSOs \citep{yuan_spectroscopic_2016}, plotted with a purple dotted line. The signature AGN emission lines and galaxy absorption lines are plotted with grey dotted and dashed lines, respectively; see Figure~\ref{fig:composite nustar Lx split zoom} for zoom-ins of the spectral lines.
		The spectra are shifted to rest-frame wavelengths and normalized at 4000\,\AA{} for illustrative purposes.
	}
	\label{fig:composite nustar Lx split}	
\end{figure*}

\begin{figure*}
	\centering
	\includegraphics[width=37pc]{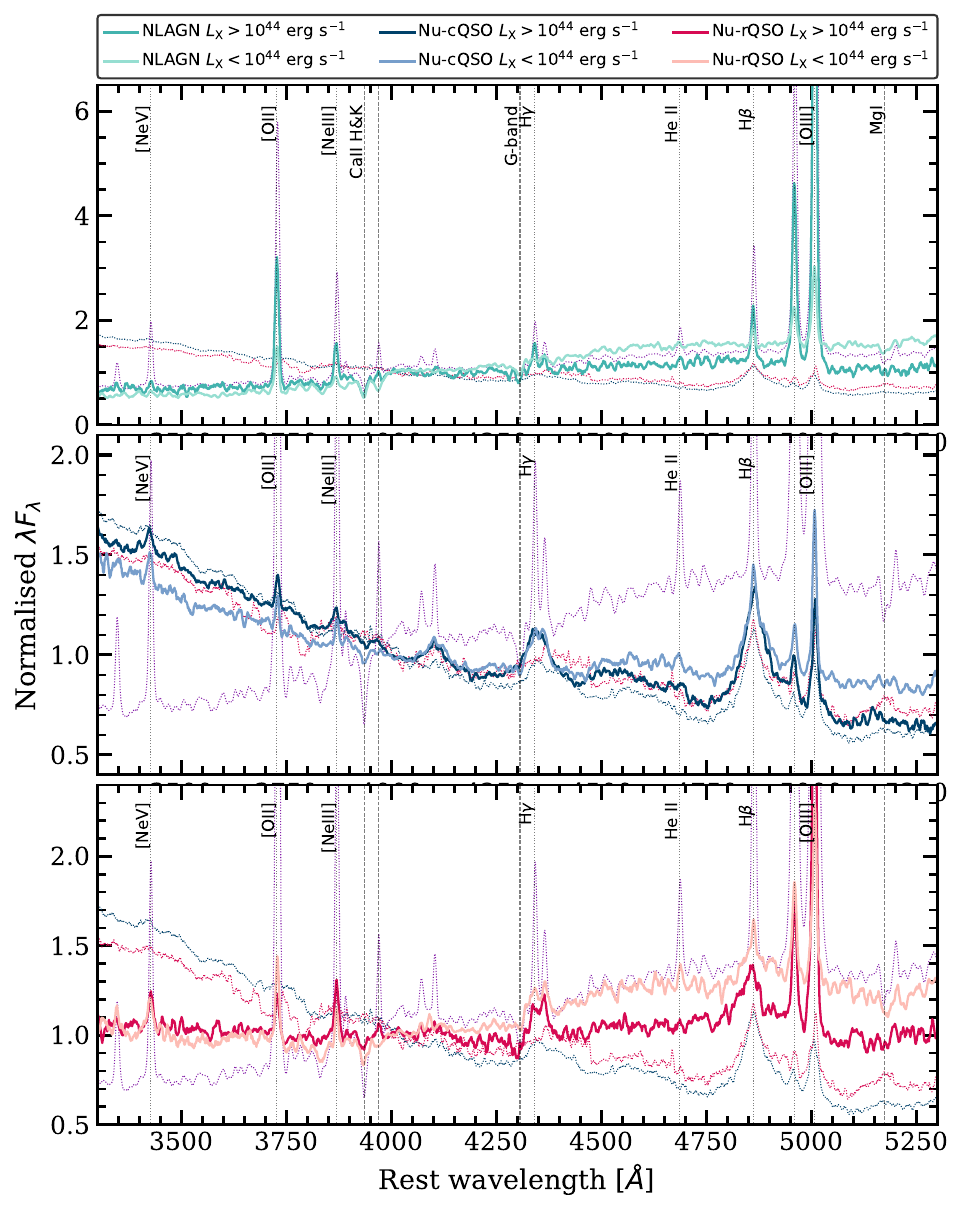}		
	\vspace*{-2mm}
	\caption{Zoom-ins (3300--5300\AA{}) of the prominent AGN spectral lines in Figure~\ref{fig:composite nustar Lx split} of the spectroscopically confirmed NLAGN, Nu-cQSO and Nu-rQSO subdivided into low (light green, light blue and peach, respectively) and high (dark green, dark blue and red, respectively) luminosity bins. The composite X-Shooter spectra of luminosity-redshift matched SDSS rQSOs and cQSOs \citep{fawcett2022} are plotted in red and blue dotted lines, and composite spectra of SDSS Type 2 QSOs \citep{yuan_spectroscopic_2016} as a purple dotted line for comparison. 
	}
	\label{fig:composite nustar Lx split zoom}	
\end{figure*}

In K19 we showed that the red optical colors of SDSS rQSOs are predominantly due to reddening of a normal blue quasar, results which have been subsequently confirmed via detailed broad-band UV-far-IR SED fitting \citep{calistrorivera2021} and broad-band UV-near-IR spectral analysis \citep{fawcett2022}. 
It is worth bearing in mind that these thinly veiled dust-obscured optical quasars may represent the detectable end of a more heavily extinguished luminous AGN population which will be missed by SDSS because of their colors (e.g., \citealp{glikman2004}; \citealp{banerji2012}; \citealp{eisenhardt2012}; \citealp{ross2015}; \citealp{hamann2017}; see Table~2.1 in \citealp{klindt2022} for a summary of dust-reddened quasars).
By comparison, our analyses of the Nu-rQSOs have shown that a substantial contribution of the red optical emission is due to a host-galaxy component rather than dust-reddening of the quasar in the lower luminosity objects. On the other hand, dust obscuration (which dominates the shape at the blue end of the spectrum) plays a key role in the excess red colors for the more luminous BL objects.  
Hence, it is important to keep in mind the differences in the typical luminosities between the SDSS quasars and the NSS80 sources (see Figure~\ref{fig:Lbol vs z nustar}). 
Indeed, in their broad-band SED fitting of SDSS quasars, \citet{calistrorivera2021} found that the host galaxy is likely to dominate in the lower-luminosity rQSOs (and to make a significant contribution in cQSOs), corresponding broadly to the luminosities of the NSS80 sources \citep[see Fig.~5 of][]{calistrorivera2021}. 
Consequently, caution must be applied when adopting a simple $g-i$ optical color cut for lower-luminosity quasars and AGN.

\section{Summary}
\label{sec:summary nustar}

In this work, we present the \nustar serendipitous survey 80-month catalogue (NSS80) -- the most recent and largest survey undertaken with \textit{NuSTAR}. 
The NSS80 succeeds the 40-month catalogue (NSS40) reported in \citet{lansbury2017_cat} and incorporates data from the full 80 month period (2012 July to 2019 March) of telescope operation. 
The data include 894 unique fields (563 newly published data from the post-NSS40 period), with a total areal coverage of 36\,deg$^2$, and a cumulative exposure time of $\approx$\,62 Ms.
Due to an increase in the fraction of general observer (GO) observations over the NSS40 catalogue, the areal coverage, integrated exposure, number of fields, and number of sources in the NSS80 are a factor $\approx$~3 larger than those presented in \citet{lansbury2017_cat}.
Furthermore, we have characterized the \nustar detected AGN in terms of their X-ray, optical, and IR properties.
Below we summarize the main results:
\begin{itemize}

	\item 
	Overall, we detect 1,488 sources which are significant post-deblending (i.e., after accounting for contamination of the photon counts from nearby sources). 
	To enable easy and efficient use of the NSS80, 214 \nustar sources residing in highly extended optical galaxies and galaxy clusters are placed in a secondary catalogue available in Appendix~\ref{appendix:secondary catalogue} to complement the primary catalogue which constitutes 1274 X-ray sources (Appendix~\ref{appendix:nss80 catalogue}), the majority of which are dominated by AGN in fainter field galaxies. 
	Of these, 412 are independently detected in the hard (8--24\,keV) energy band; see Section~\ref{subsec:source catalog}.

	\item 
	Key improvements made in the construction of the NSS80 catalogue over those adopted for the NSS40 catalogue include (1) mosaicing of overlapping fields from different observational programs to increase the sensitivity and (2) masking of fields with large optical/IR hosts post-processing to construct a secondary catalogue of sources which previously were excluded altogether. 
	Overall, we find 91\% of NSS40 sources are retained in NSS80, while 36 NSS40 sources are undetected in NSS80. 
	For the majority of these undetected sources, the deeper data have improved the false probability estimates and, consequently, eliminate low-significance detections.

	\item 
	The full band (3--24\,keV) fluxes cover a range of \textit{f}$_\mathrm{3-24\,keV} \approx 10^{-14}$ to $10^{-11}$\,erg\,s$^{-1}$\,cm$^{-2}$, with a median value of 	$\langle{\textit{f}}_\mathrm{3-24\,keV}\rangle$\,$=$\,9.82$\times$\,10$^{-14}$\,erg\,s$^{-1}$\,cm$^{-2}$; 
	see Section~\ref{subsubsec:nustar prop}.
	The \nustar serendipitous survey has the largest areal coverage at all fluxes compared to the \nustar deep-field surveys in well-studied fields (e.g., COSMOS, ECDFS, EGS, GOODS-N, and UDS), reaching similar flux depths. Consequently, the combination of NSS80 with the deep-field surveys allows for a factor $\approx$\,two improvement in analyses of the faint end of the hard X-ray source population compared to that from just the deep-field surveys; see Section~\ref{subsec:data processing}. Furthermore, in Section~\ref{subsubsec:redshifts and luminosities} we show that NSS80 reaches $\approx$\,two orders of magnitude fainter than the \textit{Swift}-BAT all sky survey.
	
	\item  
	A large range in the observed band ratios of the NSS80 spectroscopically confirmed AGN is seen at 3--24\,keV (Section~\ref{subsubsec:band ratios}).
	This implies a range of observed photon indices going from very soft $\Gamma_\mathrm{eff} \approx 3$ to very hard $\Gamma_\mathrm{eff} \approx 0$.
	Contrary to that previously found for lower energy X-ray bands, our results show no evidence for an anti-correlation between band ratio and X-ray count rate.
	
	\item  
	To study the multi-wavelength properties of the \nustar serendipitous sources, we required lower-energy X-ray ($< 10$\,keV) counterparts with higher positional accuracies to reliably match to optical and IR counterparts.
	To search for lower-energy X-ray counterparts, we utilised \textit{Chandra} (CSC2), \textit{XMM-Newton} (4XMM-DR10/s), and \textit{Swift}-XRT (2SXPS).
	In total, we identified a lower-energy X-ray counterpart for 76\% (964/1274) of the primary NSS80 catalogue detected in surveys or archival data from the four lower-energy X-ray catalogues. 
	The remaining 310 NSS80 sources lack lower-energy X-ray counterparts, which can be attributed to either insufficient or zero lower-energy X-ray coverage or, in the minority of cases, a false \nustar detection (e.g., sources with detection probabilities close to the threshold; $\log(P_\mathrm{false,min}) \approx -6$).
	We find that the lower-energy X-ray counterpart fluxes are generally in agreement with the \nustar fluxes for the 3-8\,keV (soft) energy band. 
	A maximum variation of a factor of $\approx$\,5 between the lower-energy X-ray and \nustar flux observations is identified, which can be attributed to source variability detected in the non-contemporaneous X-ray observations and Eddington bias; see Section~\ref{subsec:xray prop}.
	This variability can have two origins: either a change in intrinsic AGN luminosity, or a change in the line-of-sight column density, due to the non-uniform distribution of the obscuring material surrounding the accreting SMBH.
	
	\item  
	In NSS40 a relatively simple closest neighbour approach was used to identify multi-wavelength counterparts for follow-up spectroscopy. 
	In the NSS80 catalogue, however, we adopted a more sophisticated probabilistic approach with \textsc{Nway} to identify IR (CatWISE20) and optical (PS1-DR2) counterparts; see Section~\ref{subsec:nway infrared and optical cpart}.
	The bulk of the NSS80 sources (95\%) have at least one \textsc{Nway} match with a match probability $>$\,10\%, 74\% of which coincide with a lower-energy X-ray counterpart, and the remaining 26\% have \textit{NuSTAR}-only X-ray positions (due to insufficient or no lower-energy X-ray coverage). 
	Overall we find 953 high probability CatWISE20/PS1 counterpart matches to the NSS80 sources, of which $\sim$\,76\% are at extragalactic latitudes $|b| > 20^\circ$.  
	
	\item  
	Optical spectroscopic identifications (i.e., redshift measurements and source classifications) have successfully been obtained for 547 sources; see Section~\ref{subsubsec:spectral classification}. An additional three sources have redshift estimates, but lack source classifications. 
	We obtained spectroscopic identifications for the majority of sources (427) via our extensive campaign of ground-based spectroscopic follow-up, using a range of observatories at multiple geographic latitudes; see Section~\ref{subsubsec:spectroscopic data reductions}.
	We spectroscopically confirm 58 sources as Galactic objects.
	Of the 492 extragalactic sources (AGN), 284 (57.7\%) are classified as broad-line AGN (BL), 194 (39.4\%) are narrow-line sources, 10 are galaxies (Gxy; absorption line spectra), one is a galaxy cluster, and three are unclassified, but have redshift measurements from literature.
	Among these there are three AGN pairs of which one pair is a dual AGN system, two AGN-galaxy pairs, and one galaxy pair.
	For a further 43 sources a faint continuum (often red) is detected, lacking spectral features and, consequently, spectroscopic identifications. 
	One of these sources is a BL~Lac candidate, and another is a pair of (extremely faint) sources of unknown type.
	The remaining one source is a hotspot of 4C\,74.26 (radio quasar), which was targeted as the primary \textit{NuSTAR} science target.
	While similar numbers of NLs and BLs are identified at lower redshifts ($z \lesssim 1$), there is a bias against detections of high-redshift NLs that are optically fainter, and against the detection of highly obscured AGN; see Section~\ref{subsubsec:redshifts and luminosities}.
	
	\item 
	The NSS80 AGN have redshifts covering a wide range, from 0.012 to 3.43, with a median redshift of $\langle{z}\rangle$\,=\,0.56; see Section~\ref{subsubsec:redshifts and luminosities}.
	The rest-frame 10--40 keV luminosities also span a wide range of $L_\mathrm{10-40\,keV} \approx 10^{42-46}$\,erg\,s$^{-1}$, with a median value of $\langle{L_\mathrm{10-40\,keV}}\rangle$\,=\,1.2\,$\times$\,10$^{44}$\,erg\,s$^{-1}$.
	Previous X-ray missions with sensitivity at $>$\,10\,keV were able to sample the AGN population below the knee of the X-ray luminosity function ($L_*$) for redshifts up to $z \approx 0.05$, and \textit{NuSTAR} extends this to $z \approx 1$.
	Using the \nustar band ratio, we identify 22 Compton-thick candidate sources: 7 are identified in the NSS40 catalogue and 15 are newly identified sources with the post-NSS40 observations; see Section~\ref{subsubsec:band ratios}.
	
	\item 
	We use the distribution of \textit{WISE} $W1-W2$ and $W2-W3$ colors for the extragalactic NSS80 sources, since the reprocessed emission from the AGN's circumnuclear dust is distinguishable from star light (which peaks in the FIR regime) at MIR wavelengths (predominantly for high luminosity AGN); see Section~\ref{subsec:wise properties nustar}.
	We find that 95\% of high-$L_\mathrm{X}$ BL AGN will successfully be selected using the AGN wedge, whilst NL AGN are have a significantly lower chance of being identified as AGN based on their MIR colors alone. This is largely driven by the lower luminosity objects with $10-40$\,keV luminosities below the X-ray quasar threshold.
	It is notable that a number of luminous \textit{NuSTAR}-selected BL AGN are not selected in the MIR -- this appears to be driven by the intrinsic AGN properties. 
	Furthermore, from the MIR colors, we can deduce that $>$\,90\% of the MIR light in two-thirds of the objects come from an AGN, while approximately a third are host dominated.  
	
	\item 
	Given that hard X-rays are largely unbiased against dust obscuration up to CT levels, \nustar facilitates the discovery of rare and extreme sources such as red quasars.
	We therefore explore the optical and IR properties of the NSS80 AGN sample, with particular focus on red quasars and the narrative that X-rays may be telling about this peculiar subpopulation of AGN; see Section~\ref{subsec:optical prop}. 
	Forty-two percent (106/251) of the BL AGN with robust counterpart associations have red $g-i$ colors; half of which have luminosities exceeding the X-ray quasar threshold and, therefore, are potentially dust reddened quasars.
	The remaining 145 BLs have $g-i$ colors consistent with the typical quasar population. 
	Furthermore, 92\% of the NLs appear red in their optical colors; however, the majority are low luminosity objects with $L_\mathrm{10-40\,keV} < 10^{44}$\,erg\,s$^{-1}$ and, therefore, will have host-galaxy dominated emission. 
	Altogether, the NSS80 AGN are typically 0.3\,dex less luminous than SDSS quasars, with a median bolometric luminosity of $\langle{L_\mathrm{bol,X}}\rangle = 5.93 \times 10^{4}$\,erg\,s$^{-1}$ at the median BL AGN redshift of $\langle{z}\rangle \approx 0.8$. 

	\item 
	Finally, we present optical spectra composites to study the spectral properties of the NSS80 AGN to investigate the driving force behind the colors of the NSS80 red quasars; see Section~\ref{subsec:composite spectra}.
	Overall, for all spectral types, the host-galaxy features are more prominent in the low-$L_\mathrm{X}$ when compared to the high-$L_\mathrm{X}$ systems: these signatures are enhanced emission at red wavelengths and host-galaxy absorption features, although these are weak in some of the composites. This is consistent with basic expectations for a decreasing AGN contribution relative to the host galaxy emission. 
	Hence, our spectroscopic analysis indicates that reddening from the presence of a host galaxy can have a large contribution to the optical colors in the hard X-ray selected population of lower-luminosity quasars. 
	
	Our future work will focus on further follow-up studies of the current NSS80 sources, aiming to complete the spectroscopic follow-up of the catalog and to obtain detailed analysis of sources with $>8$\,keV detections, which are unique to \textit{NuSTAR}.
	\\
	\\

\end{itemize}

The NSS80 is a valuable legacy of the \emph{NuSTAR} observatory and provides a powerful sample for future studies of the hard X-ray emitting population. Our total sample size (1,488 sources) is relatively modest compared to the samples of hundreds of thousands of sources compiled with \emph{XMM-Newton}, \emph{Chandra} or \emph{Swift} \citep[e.g.][]{webb2020,evans2019_csc2,evans2020_2sxps}---achieved thanks to the larger fields-of-view, higher sensitivities and longer lifetimes of these observatories compared to \emph{NuSTAR}---or the much larger samples (at soft X-ray energies) that are now being provided by dedicated surveys with \emph{eROSITA} \citep{merloni_erosita_2012,predehl_erosita_2021,brunner_erosita_2022}.
In contrast to these prior studies, we have been careful to ensure our sample consists only of truly serendipitous detections and excludes any source associated with the targets of the observations, which is vital for carrying out scientific studies that require an unbiased sampling of the population \citep[see also][]{delaney2023}.
Our sample remains unique in accessing faint sources at hard ($\gtrsim8$~keV) X-ray energies with a high degree of spectroscopic completeness, providing an important constraints on the obscured AGN population outside the local Universe.

\vspace{5mm}


{\large \sc  ~~~~~~~~~~~acknowledgements} \\
This paper is jointly led by L.K. and C.L.G.: the original draft was primarily produced by L.K., and formed a major component of her PhD thesis, and the modifications made to the draft following an extensive referee report were carried out by C.L.G..
We acknowledge financial support from: the Faculty of Science Durham Doctoral Scholarship (L.K.), European Southern Observatory (ESO) postdoctoral research fellowship (G.B.L. and M.H.), the Science and Technology Facilities Council (STFC; D.M.A. and D.J.R. through grant code ST/P000541/1; D.M.A. and C.L.G. through grant codes ST/T000244/1 and ST/X001075/1), and STFC Ernest Rutherford Fellowship (J.A. through grant code ST/P004172/1). The work of D.S. and T.C. was carried out at the Jet Propulsion Laboratory, California Institute of Technology, under a contract with the National Aeronautics and Space Administration (80NM0018D0004).
F.E.B. acknowledges support from ANID-Chile BASAL AFB-170002, ACE210002, and FB210003, FONDECYT Regular 1200495 and 1190818, and Millennium Science Initiative Program  – ICN12\_009.  
P.G.B. acknowledges financial support from the Czech Science Foundation project No. 22-22643S.
W.N.B.  acknowledges financial support from the V.M. Willaman Endowment.
V.A.F. acknowledges funding from a United Kingdom Research and Innovation grant (code: MR/V022830/1).
E.S.K. acknowledges financial support from the Centre National d’Etudes Spatiales (CNES). 
S.M. acknowledges funding from the the INAF ``Progetti di Ricerca di Rilevante Interesse Nazionale'' (PRIN), Bando 2019 (project: ``Piercing through the clouds: a multiwavelength study of obscured accretion in nearby supermassive black holes'').
G.N. acknowledges funding support from the Natural Sciences and Engineering Research Council (NSERC) of Canada through a Discovery Grant and Discovery Accelerator Supplement, and from the Canadian Space Agency through grant 18JWST-GTO1.
C.R. acknowledges support from Fondecyt Regular grant 1230345 and ANID BASAL project FB210003. 
E.R.C. and P.V. acknowledges the support of the South African National Research Foundation. 
E.T. acknowledges support from FONDECYT Regular 1190818 and 1200495, ANID grants CATA-Basal AFB-170002, ACE210002, and FB210003, and Millennium Nucleus NCN19\_058.
C.M.U. would like to acknowledge support from the National Aeronautics and Space Administration via ADAP Grant 80NSSC18K0418.

We extend gratitude to Sam Tweety Anthony, Subhash Bose, Tea Freedman-Susskind, Tyler George, Elena González Egea, Feiyang Liu, Dan MacMillan, Jeff Maggio, Adric Riedel, Michael Rivkin, Manika Sidhu, Jamie Soon, Thomas Venville, Yerong Xu, and Emily Zhang for their support during the ground-based follow-up observations. 
We thank Victoria Fawcett for her valuable contribution by producing the composite spectra presented in Section~\ref{subsec:composite spectra}.

This work was supported under NASA Contract No. NNG08FD60C, and made use of data from the \textit{NuSTAR} mission, a project led by the California Institute of Technology, managed by the Jet Propulsion Laboratory, and funded by the National Aeronautics and Space Administration.
We thank the \textit{NuSTAR} Operations, Software and Calibration teams for support with the execution and analysis of these observations. This research has made use of the \textit{NuSTAR} Data Analysis Software (NuSTARDAS) jointly developed by the ASI Science Data Center (ASDC, Italy) and the California Institute of Technology (USA).

Facilities: \textit{Chandra}, ESO La Silla, Keck, \textit{NuSTAR}, Palomar, SALT, SDSS, \textit{Swift}, \textit{WISE}, \textit{XMM}-Newton. \\
Some of the data presented herein were obtained at the W. M. Keck Observatory, which is operated as a scientific partnership among the California Institute of Technology, the University of California and the National Aeronautics and Space Administration. The Observatory was made possible by the generous financial support of the W. M. Keck Foundation. 
Some of the observations reported in this paper were obtained with the Southern African Large Telescope (SALT) under programs 2017-1-SCI-053,  2017-2-SCI-037, and 2018-2-MLT-006  (PI: Lizelke Klindt).
The Wide-field Infrared Survey Explorer (\textit{WISE}) is a joint project of the University of California, Los Angeles, and the Jet Propulsion Laboratory/California Institute of Technology, funded by the National Aeronautics and Space Administration.
The National Radio Astronomy Observatory is a facility of the National Science Foundation operated under cooperative agreement by Associated Universities, Inc.
Funding for the SDSS and SDSS-II has been provided by the Alfred P. Sloan Foundation, the Participating Institutions, the National Science Foundation, the U.S. Department of Energy, the National Aeronautics and Space Administration, the Japanese Monbukagakusho, the Max Planck Society, and the Higher Education Funding Council for England. 
The SDSS is managed by the Astrophysical Research Consortium for the Participating Institutions. The Participating Institutions are the American Museum of Natural History, Astrophysical Institute Potsdam, University of Basel, University of Cambridge, Case Western Reserve University, University of Chicago, Drexel University, Fermilab, the Institute for Advanced Study, the Japan Participation Group, Johns Hopkins University, the Joint Institute for Nuclear Astrophysics, the Kavli Institute for Particle Astrophysics and Cosmology, the Korean Scientist Group, the Chinese Academy of Sciences (LAMOST), Los Alamos National Laboratory, the Max-Planck-Institute for Astronomy (MPIA), the Max-Planck-Institute for Astrophysics (MPA), New Mexico State University, Ohio State University, University of Pittsburgh, University of Portsmouth, Princeton University, the United States Naval Observatory, and the University of Washington.
This research uses services or data provided by the Astro Data Lab, which is part of the Community Science and Data Center (CSDC) program at NSF's National Optical-Infrared Astronomy Research Laboratory. NOIRLab is operated by the Association of Universities for Research in Astronomy (AURA), Inc. under a cooperative agreement with the National Science Foundation.

\software{Astropy \citep{astropy:2013, astropy:2018, astropy:2022}, Matplotlib \citep{hunter_computing_2007}, NumPy \citep{harris_array_2020} and SciPy \citep{virtanen_scipy_2020}.}

\clearpage

\appendix
\section{Description of the primary NSS80 source catalogue} 
\label{appendix:nss80 catalogue}
The primary \textit{NuSTAR} serendipitous survey source catalogue, containing 1274 sources in total; available as an electronic table on the NSS80 webpage. 
Here we describe the columns of the catalogue, which are summarized in Table~\ref{tab:description of NSS80 cat}. \\
\textbf{Column 1:} The unique source identification number, in order of increasing right ascension (R.A.). \\
\textbf{Column 2:} The unique \textit{NuSTAR} source name, following the IAU-approved format: \textit{NuSTAR} JHHMMSS±DDMM.m, where m is the truncated fraction of one arcminute for the arcseconds component of the declination (decl.). \\
\textbf{Column 3:} The unique \textit{NuSTAR} field and source index. \\
\textbf{Column 4:} The unique \citetalias{lansbury2017_cat} source identification number. \\
\textbf{Column 5:} The unique \citetalias{lansbury2017_cat} \textit{NuSTAR} field and source index. \\
\textbf{Columns 6,7: } The \textit{NuSTAR} R.A. and decl. coordinates (J2000), as described in Section~\ref{subsec:source catalog}. \textit{Units:} degr. \\
\textbf{Column 8,9:} The IAU 1958 Galactic latitude and longitude. \textit{Units:} degr. \\ 
\textbf{Columns 10–12:} A binary flag indicating whether the source is detected with a false probability lower than the threshold of $\log(P_\mathrm{False})$ = -6, for the soft ($3-8$ keV), hard ($8-24$ keV), and full ($3-24$ keV) bands. These three bands are abbreviated as SB, HB, and FB, respectively, throughout the source catalogue. \\
\textbf{Columns 13–15:} The same as columns 10–12, after deblending has been performed to account for contamination of the source counts from very nearby sources \citep[Section 2.3.2 of][]{mullaney2015}. Deblending only affects a very small fraction of the overall sample (e.g., see Section 2.4 in \citetalias{lansbury2017_cat}). \\
\textbf{Columns 16–18:} The logarithm of the false probabilities ($P_\mathrm{false}$) of the \textit{NuSTAR} detected source, for the three standard energy bands. \\
\textbf{Columns 19:} The \textit{NuSTAR} detection likelihood ($P_\mathrm{false,min}$). \\
\textbf{Columns 20–22:} The same as columns 16–18, after deblending has been performed. \\
\textbf{Columns 23:} The same as column 19, after deblending has been performed. \\
\textbf{Column 24:} A binary flag indicating whether the \textit{NuSTAR} detected source remains significant after deblending, in at least one of the three standard energy bands. \\
\textbf{Columns 25–39:} Photometric quantities, calculated at the \textit{NuSTAR} source coordinates, and using a source aperture of 30$''$ radius (see Section~\ref{subsec:data processing}). The values are non-aperture-corrected; i.e., they correspond to the 30$''$ values, and have not been corrected to the full PSF values. We provide the total counts (i.e., all counts within the source aperture) and associated errors (84\% CL), the background counts scaled to the source aperture, and the net source counts (i.e., total minus background) and associated errors. For the net source counts, we give 90\% CL upper limits for sources not detected in a given band. Throughout the table, upper limits are flagged with a -99 value in the error column. \\ 
\textbf{Columns 40–51:} The same as columns 25–39, after deblending has been performed. \\
\textbf{Columns 52–54:} The average net, vignetting-corrected exposure time at the source coordinates (columns 3 and 4), for each energy band. These correspond to the A+B data, so should be divided by two to obtain the average exposure per FPM. \textit{Units:} s. \\
\textbf{Columns 55–69:} The non-aperture-corrected total, background, and net count rates (and associated errors; 84\% CL) determined from the photometric values in columns 25–39, and the exposure times in columns 52–54. \textit{Units:} s$^{-1}$. \\
\textbf{Columns 70–75:} The deblended net count rates, and associated errors, determined from the photometric values in columns 40–51, and the exposure times in columns 52–54. \textit{Units:} s$^{-1}$. \\
\textbf{Columns 76–78:} The \textit{NuSTAR} band ratio (BRNu) and associated errors. Upper limits, lower limits, and sources with no constraints are flagged with -99, -88, and -77 values, respectively, in the error columns. \\
\textbf{Columns 79–81:} The effective photon index ($\Gamma_\mathrm{eff}$ ), and associated errors, estimated from the band ratio values in columns 76–78. \\
\textbf{Columns 82–87:} The observed-frame fluxes and associated errors (84\% CL) for the three standard energy bands, after deblending has been performed. These are aperture corrected values (i.e., they correspond to the full \textit{NuSTAR} PSF), and are calculated from the count rates in columns 70–75 using the conversion factors listed in \citetalias{lansbury2017_cat}. \textit{Units:} erg\,s$^{-1}$\,cm$^{-2}$. \\
\textbf{Column 88:} An abbreviated code indicating the lower-energy ($<$\,10\,keV) X-ray coverage. ``C'', ``X'', and ``S'' indicate sources which have \textit{Chandra}, \textit{XMM-Newton} and \textit{Swift}-XRT) coverage, respectively. \\
\textbf{Column 89:} An abbreviated code indicating the origin of the adopted lower-energy X-ray counterpart: CSC2 indicates counterparts from the \textit{Chandra} Source Catalog Release 2.0 \citep[][]{evans2019_csc2}, 4XMMDR10 and 4XMMDR10s indicate the Fourth \textit{XMM-Newton} Serendipitous Source Catalog, Tenth Data Release \citep[][]{webb2020} and its stacked version \citep[][]{trauslen2020}, and 2SXPS indicates the \textit{Swift} XRT Point Source Catalogue \citep[][]{evans2020_2sxps}, respectively. CXO\_MAN, XMM\_MAN, and XRT\_MAN indicate sources manually identified using archival \textit{Chandra}, \textit{XMM-Newton}, and \textit{Swift}-XRT data, respectively. Sources that lack a soft X-ray counterpart are indicated with the code NULL; Section~\ref{subsec:soft xray cpart} details the counterpart matching. \\
\textbf{Columns 90,91:} The R.A. and decl. coordinates (J2000) of the adopted lower-energy X-ray counterpart. Sources that lack a lower-energy X-ray counterpart have R.A. and decl. values of (-999,-999). \textit{Units:} degr. \\
\textbf{Column 92:} Positional uncertainty for lower-energy (where available) or \nustar coordinates used for \textsc{Nway} matching. \textit{Units:} degree. \\
\textbf{Column 93:} The angular offset between the \textit{NuSTAR} position (columns 6 and 7) and the lower-energy X-ray counterpart position (columns 90 and 91). Sources lacking a soft X-ray counterpart are flagged with a value of -77. \textit{Units:} arcsec. \\
\textbf{Column 94:} Reference for the lower-energy X-ray flux used to compute the $3-8$\,keV fluxes in columns 95-97: $2-7$\,keV, $0.5-7$\,keV or $0.1-10$\,keV for CSC2 fluxes, $4.5-12$\,keV for 4XMMDR10/s fluxes, and $2-10$\,keV for 2SXPS fluxes. \\ 
\textbf{Columns 95-97:} The observed-frame $3-8$\,keV flux of the lower-energy X-ray counterpart (with assocated lower and upper errors) for sources with counterparts in the CSC2, 4XMMDR10/s and 2SXPS catalogues. For CSC2 sources we convert to the $3-8$\,keV flux from the $2-7$\,keV flux using a conversion factor of 0.83, for the 4XMMDR10/s sources we convert from the $4.5-12$\,keV flux using a conversion factor of 0.92, and for the 2SXPS sources we convert from the $2-10$\,keV flux using a conversion factor of 0.62. Sources lacking a soft X-ray counterpart have values of 0. \textit{Units:} erg\,s$^{-1}$\,cm$^{-2}$. \\
\textbf{Columns 98-100:} The total combined $3-8$\,keV flux of all (CSC2, 4XMMDR10/s, or 2SXPS) sources within 30$''$ of the \textit{NuSTAR} position, with associated lower and upper errors. Sources lacking soft X-ray counterparts have values of 0. \textit{Units:} erg\,s$^{-1}$\,cm$^{-2}$. \\
\textbf{Column 101:} Galactic extinction, $A_\mathrm{v}$; i.e., the total absorption in magnitudes at $V$. \\
\textbf{Columns 102-104:} Galactic line-of-sight absorption in $g$, $r$, and $i$ magnitudes. \\
\textbf{Columns 105,106:} The \textsc{Nway} calculated probability that each CaTWISE20 \citep{marocco2021} source is the correct counterpart to the (adopted lower-energy or \textit{NuSTAR}) X-ray source ($\rho_\mathrm{Cat,best}$ for CatWISE20, and the probability that any of the CatWISE20 sources is the right counterpart ($\rho_\mathrm{Cat,any}$ for CatWISE20; a higher probability indicates a lower false association likelihood. Sources lacking a CatWISE20 counterpart are flagged with a value of -77. \\
\textbf{Columns 107,108:} The CatWISE20 R.A. and decl. coordinates (J2000) for the most probable \textsc{Nway} match \textit{WISE} counterpart. Sources that lack a CatWISE20 counterpart have R.A. and decl. values of (-999,-999). \textit{Units:} degr.\\
\textbf{Column 109:} The angular offset between the best \textsc{Nway}-matched CatWISE20 position and the \textit{NuSTAR} position. Sources lacking a CatWISE20 match are flagged with a -77 value. \textit{Units:} arcsec. \\
\textbf{Column 110:} The angular offset between the best \textsc{Nway}-matched CatWISE20 position and the adopted lower-energy X-ray counterpart position. Sources that lack a CatWISE20 counterpart are flagged with values of -77. \textit{Units:} arcsec. \\
\textbf{Columns 111-114:} The CatWISE20 \textit{WISE} profile-fit magnitudes and associated errors, for the two standard \textit{WISE} bands available in CatWISE20: $W1$ ($\lambda \approx 3.4 \mu$m) and $W2$ ($\lambda \approx 4.6 \mu$m). Sources with no constraints are flagged with a -77 value, and upper limits are flagged with a -99 value in the error column. \textit{Units:} Vega mag.Columns 115,116: The AllWISE R.A. and decl. coordinates (J2000) for crossmatches with CatWISE20 coordinates. Sources that lack an AllWISE counterpart have R.A. and decl. values of (-999,-999). \textit{Units:} degr.\\
\textbf{Columns 117-120:} The AllWISE \textit{WISE} profile-fit magnitudes and associated errors, for the two remaining standard \textit{WISE} bands: $W3$ ($\lambda \approx 12 \mu$m) and $W4$ ($\lambda \approx 22 \mu$m). Sources with no constraints are flagged with a -77 value, and upper limits are flagged with a -99 value in the error column. \textit{Units:} Vega mag. \\ 
\textbf{Columns 121,122:} The \textsc{Nway} calculated probability that each PS1-DR2 \citep{flewelling2018} source is the correct counterpart to the (adopted lower-energy or \textit{NuSTAR}) X-ray source ($\rho_\mathrm{PS1,best}$, and the probability that any of the PS1-DR2 sources is the right counterpart ($\rho_\mathrm{PS1,any}$. A higher probability indicates a lower false association likelihood. Sources lacking a PS1-DR2 counterpart are flagged with a value of -77. \\
\textbf{Columns 123,124: } The PS1-DR2 R.A. and decl. coordinates (J2000) for the best \textsc{Nway}-matched Pan-STARRS counterpart. Sources that lack a PS1-DR2 \textsc{Nway} match have R.A. and decl. values of (-999,-999). \textit{Units:} degr. \\
\textbf{Column 125:} The angular offset between the best \textsc{Nway}-matched PS1-DR2 position and the \textit{NuSTAR} position. Sources lacking a PS1-DR2 match are flagged with a -77 value. \textit{Units:} arcsec. \\
\textbf{Column 126:} The angular offset between the best \textsc{Nway}-matched PS1-DR2 position and the adopted lower-energy X-ray position. Sources lacking a PS1-DR2 match are flagged with a -77 value. \textit{Units:} arcsec. \\
\textbf{Column 127:} The angular offset between the best \textsc{Nway}-matched PS1-DR2 and CatWISE20 positions. Sources lacking a PS1-DR2/CatWISE20 match are flagged with a -77 value. \textit{Units:} arcsec. \\
\textbf{Column 128-134:} The $g$-, $r$- and $i$-band magnitudes (with associated errors) of the best \textsc{Nway}-matched PS1-DR2 counterpart. Magnitudes are not corrected for Galactic extinction. All PanSTARRS magnitudes quoted in the NSS80 catalog are, in order of preference, Kron, PSF, or the default aperture magnitudes, with the selected type for each band indicated in column 132. Sources with unconstrained magnitudes are denoted with -77, and limits with -99 values in the error columns. \textit{Units:} AB mag. \\
\textbf{Columns 135:} A binary flag indicating the subsample of higher probability CatWISE20 or PS1-DR2 matches (i.e., \textsc{Nway} reliable sample) with thresholds of $\rho_\mathrm{Cat/PS1,best} > 0.4$ and $\rho_\mathrm{Cat/PS1,any} > 0.5$. Section~\ref{subsec:nway infrared and optical cpart} details the IR/optical counterpart matching using \textsc{Nway}. \\
\textbf{Column 136:} Reference indicating the origin of the adopted principal optical counterpart to the \textit{NuSTAR} source. The code PS1-NWAY indicates sources with lower-energy X-ray counterparts and successful \textsc{Nway} matches in the PS1-DR2 catalog \citep{flewelling2018}. The codes PS1-MAN, SDSS, NSC2, DES2 and USNOB1 indicate sources with lower-energy X-ray counterparts and successful matches in the PS1-DR2 catalog (where OOff\_PS1\_CatWISE\,$>$\,$2.7''$ and therefore rematched to search for nominally closer PS1 matches), the SDSS photometric catalogue DR12 \citep{alam2015}, the second data release of the NOIRLab Source Catalog \citep[][]{nidever2021}, the Dark Energy Survey DR2 \citep[][]{abbott2016}, and the USNOB1 catalogue \citep{monet2003}, respectively. PS1-CatWISE, SDSS-CatWISE, NSC2-CatWISE, DES2-CatWISE, and USNOB1-CatWISE20 indicate the cases where there is no lower-energy X-ray counterpart to the \textit{NuSTAR} position, but a CatWISE20 AGN candidate is identified within the \textit{NuSTAR} error circle and successfully matched to the PS1-DR2, SDSS-DR12, NSC2, DES-DR2 or USNOB1 catalogue. PS1-softX, SDSS-softX, NSC2-softX, DES2-softX, and USNO1-softX indicate the cases where there is a lower-energy X-ray counterpart to the \textit{NuSTAR} position, but no CatWISE20 candidate is identified within the \textit{NuSTAR} error circle. NULL indicates sources that lack an optical identification. We give a detailed description of the procedure used to identify optical counterparts in Section~\ref{subsec:nway infrared and optical cpart}. \\
\textbf{Column 137,138:} The R.A. and decl. coordinates (J2000) of the adopted optical counterpart, for sources with PS1-DR2, SDSS-DR12, NSC2, DES-DR2 and USNOB1 matches. Sources that lack an identified optical counterpart have R.A. and decl. values of (-999,-999). \textit{Units:} degr. \\
\textbf{Column 139-145:} The $g$-, $r$- and $i$-band magnitudes (and associated errors) of the adopted optical counterpart. If the counterpart is sourced from PS1-DR2 the selected magnitude type for each band is indicated in column 143. Sources with no constraints are flagged with a -77 value, and limits are flagged with -99, respectively, in the error column. \textit{Units:} AB mag.  \\
\textbf{Column 146:} The angular offset between the adopted optical position and the \textit{NuSTAR} position. Sources lacking an optical identification are flagged with a -77 value. \textit{Units:} arcsec. \\
\textbf{Column 147:} The angular offset between the adopted optical position and the adopted lower-energy X-ray position. Sources lacking a PS1-DR2 match are flagged with a -77 value. \textit{Units:} arcsec. \\
\textbf{Column 148:} The angular offset between the adopted optical position and the best \textsc{Nway}-identified CatWISE20 position. Sources lacking a PS1-DR2/CatWISE20 match are flagged with a -77 value. \textit{Units:} arcsec. \\
\textbf{Column 149:} A flag indicating whether the source is in the main spectroscopic catalog presented in Table~\ref{tab:spec results}: a value of 1 indicates the unique source entries for the spectroscopic catalog; a value of -1 flags sources with photometric redshifts; and sources which have not been targeted during our spectroscopic campaign are flagged with a value of 0. \\
\textbf{Column 150,151:} The R.A. and decl. coordinates (J2000) of the spectroscopic target observed during the ground-based follow-up program; see Section~\ref{subsec:optical spectroscopy}. Untargeted sources have R.A. and decl. values of (-999,-999).\textit{Units:} degr. \\
\textbf{Column 152:} The angular offset between the spectroscopic target position and the adopted optical position. Sources lacking spectroscopic follow-up or an optical identification are flagged with a -77 value. \textit{Units:} arcsec. \\
\textbf{Column 153:} The spectroscopic redshift of the \textit{NuSTAR} source. The large majority of the redshifts were obtained through our own campaign of ground-based spectroscopic follow-up of \textit{NuSTAR} serendipitous survey sources (see Section~\ref{subsec:optical spectroscopy}). Sources which have been followed-up, but lack redshift measurements due to low signal-to-noise spectra are flagged with -99 values, while unobserved sources have entries of -77. \\
\textbf{Column 154:} Quality flag of the spectroscopic redshift measurement. Single-line measurements are flagged as ``quality B'' redshift measurements, photometric redshifts obtained from literature are flagged as ``quality C'', and sources with a faint continuum detected (but lack a redshift measurement) are flagged as ``quality F''. \\
\textbf{Column 155:} Spectroscopic classification of the \textit{NuSTAR} source: BL $\equiv$ Broad Line object (i.e., quasar); NL $\equiv$ Narrow Line object (AGN or galaxy); Gxy $\equiv$ Galaxy (absorption lines); Galactic $\equiv$ Galactic sources at $z = 0$. See Section~\ref{subsubsec:spectral classification}.  \\
\textbf{Column 156:} Additional classification information for the duplicate source entries, e.g., AGN pairs (BL\_pair \& NL\_pair), dual AGN (Dual\_AGN), galaxy pairs (Gxy\_pair), and more than one counterpart candidate (2cpart\_candidates). \\
\textbf{Columns 157,158:} Binary flags indicating whether the spectroscopic target position matches to the best \textsc{Nway}-identified counterpart for CatWISE20 and PS1-DR2. \\
\textbf{Column 159:} A flag indicating the reliability of the spectroscopic target as the correct counterpart: 1\,$\equiv$\,correct counterpart; 0\,$\equiv$\,counterpart uncertainty; -1\,$\equiv$\,unobserved.\\
\textbf{Column 160:} Reference indicating the catalog origin of the spectroscopic counterpart.\\
\textbf{Columns 161-166:} The $g$-, $r$- and $i$-band magnitudes (with associated errors) of the spectroscopic target. Magnitudes are not corrected for Galactic extinction. Sources with unconstrained magnitudes are denoted with -77, and limits with -99 values in the error column. \textit{Units:} AB mag. \\ 
\textbf{Column 167:} The rest-frame $10-40$\,keV luminosity, estimated from the fluxes in columns 82-87. The luminosities are observed values, uncorrected for any absorption along the line of sight. The intrinsic luminosities may therefore be higher, for highly absorbed AGN. Sources with no constraints are flagged with a -77 value. \textit{Units:} erg\,s$^{-1}$. \\
\textbf{Column 168:} A binary flag indicating the few sources that show evidence for being associated with the primary science target of their respective \textit{NuSTAR} observations, according to the definition in Section~\ref{subsubsec:spectral classification} [$\Delta(cz) < 0.05\,cz$]. 1\,$\equiv$\,associated with primary target; 0\,$\equiv$\,not associated; -1\,$\equiv$\ no spectroscopy.\\
\textbf{Column 169:} A binary flag indicating the \textit{NuSTAR} sources which are Compton-thick candidates (CT; $N_\mathrm{H} \gtrsim 1.5 \times 10^{24}$\,cm$^{-2}$). These sources can be (crudely) identified from their extreme band ratios as demonstrated in Section~\ref{subsubsec:band ratios}. 

\startlongtable
\begin{deluxetable}{lll}
\tablecolumns{3}
\tablewidth{\textwidth}
\tablecaption{\label{tab:description of NSS80 cat} Column descriptions for the primary NSS80 catalogue. }
\tablehead{Column number & Column name & \colhead{Description} \\}
\startdata
			1             &     ID                                    &   \makecell[{{p{0.5\textwidth}}}]{The unique source identification number, in order of increasing right ascension (R.A.).} \\                                 
			2             &     NSS80\_Name                           &   \makecell[{{p{0.5\textwidth}}}]{The unique \textit{NuSTAR} source name, following the IAU-approved format: \textit{NuSTAR} JHHMMSS±DDMM.m, where m is the truncated fraction of one arcminute for the arcseconds component of the declination (decl.).} \\                        
			3             &     Serendip                              &   \makecell[{{p{0.5\textwidth}}}]{The unique \textit{NuSTAR} field and source index.} \\                        
			4             &     IDL17                                 &   \makecell[{{p{0.5\textwidth}}}]{The unique \citetalias{lansbury2017_cat} source identification number.} \\                        
			5             &     Serendip\_L17                         &   \makecell[{{p{0.5\textwidth}}}]{The unique \citetalias{lansbury2017_cat} \textit{NuSTAR} field and source index.} \\                        
			6,7           &     RAJ2000,DECJ2000                      &   \makecell[{{p{0.5\textwidth}}}]{The \textit{NuSTAR} R.A. and decl. coordinates (J2000), as described in Section~\ref{subsec:source catalog}.} \\                        
			8,9           &     Gal\_lat,Gal\_Long                    &   \makecell[{{p{0.5\textwidth}}}]{IAU 1958 Galactic latitude and longitude.} \\                 
			10,11,12      &     SFlag,HFlag,HFlag                     &   \makecell[{{p{0.5\textwidth}}}]{A binary flag indicating the energy bands for which the source is detected for the three standard \textit{NuSTAR} energy bands.} \\                 
			13,14,15      &     SdbFlag,HdbFlag,FdbFlag               &   \makecell[{{p{0.5\textwidth}}}]{Same as columns 10–12, post-deblending. }    \\            
			16,17,18      &     logSP,logHP,logFP                     &   \makecell[{{p{0.5\textwidth}}}]{The logarithm of the false probabilities for the three standard \textit{NuSTAR} energy bands.}   \\              
			19            &     logMinP                               &   \makecell[{{p{0.5\textwidth}}}]{The \textit{NuSTAR} detection likelihood.}\\                 
			20,21,22      &     logSdbP,logHdbP,logFdbP               &   \makecell[{{p{0.5\textwidth}}}]{The same as columns 16–18, post-deblending.} \\                 
			23            &     logMindbP                             &   \makecell[{{p{0.5\textwidth}}}]{The same as column 19, post-deblending.} \\                  
			24            &     dbFlag                                &   \makecell[{{p{0.5\textwidth}}}]{A binary flag indicating whether the source is significant post-deblending, for at least one energy band.} \\                    
			25,26         &     Scts,e\_Scts                          &  \makecell[{{p{0.5\textwidth}}}]{ Total source count for the soft energy band, and associated errors.} \\                
			27            &     SBck                                  &  \makecell[{{p{0.5\textwidth}}}]{ Background count for the soft energy band.} \\                   
			28,29         &     SNet,e\_SNet                          &   \makecell[{{p{0.5\textwidth}}}]{Net source count for the soft energy band, and associated errors.} \\                   
			30,31         &     Hcts, e\_Hcts                         &   \makecell[{{p{0.5\textwidth}}}]{Total source count for the hard energy band, and associated errors. }\\                 
			32            &     HBck                                  &   \makecell[{{p{0.5\textwidth}}}]{Background count for the hard energy band.} \\                   
			33,34         &     HNet,e\_HNet                          &   \makecell[{{p{0.5\textwidth}}}]{Net source count for the hard energy band, and associated errors.} \\                   
			35,36         &     Fcts, e\_Fcts                         &   \makecell[{{p{0.5\textwidth}}}]{Total source count for the full energy band, and associated errors.} \\                   
			37            &     FBck                                  &   \makecell[{{p{0.5\textwidth}}}]{Background count for the full energy band.} \\                                    
			38,39         &     FNet,e\_FNet                          &   \makecell[{{p{0.5\textwidth}}}]{Net source count for the full energy band, and associated errors. }\\                     
			40            &     Sdbcts                                &   \makecell[{{p{0.5\textwidth}}}]{Total source count for the soft energy band, post-deblending.} \\                 
			41            &     SdbBck                                &   \makecell[{{p{0.5\textwidth}}}]{Background count for the soft energy band, post-deblending.} \\                      
			42,43         &     SdbNet,e\_SdbNet                      &   \makecell[{{p{0.5\textwidth}}}]{Net source count for the soft energy band, post-deblending, and associated errors.} \\                  
			44            &     Hdbcts                                &   \makecell[{{p{0.5\textwidth}}}]{Total source count for the hard energy band, post-deblending.} \\                                  
			45            &     HdbBck                                &   \makecell[{{p{0.5\textwidth}}}]{Background count for the hard energy band, post-deblending.} \\                                       
			46.47         &     HdbNet,e\_HdbNet                      &   \makecell[{{p{0.5\textwidth}}}]{Net source count for the hard energy band, post-deblending, and associated errors.} \\                 
			48            &     Fdbcts                                &   \makecell[{{p{0.5\textwidth}}}]{Total source count for the hafullrd energy band, post-deblending.} \\                                        
			49            &     FdbBck                                &   \makecell[{{p{0.5\textwidth}}}]{Background count for the full energy band, post-deblending.} \\                                          
			50,51         &     FdbNet,e\_FdbNet                      &   \makecell[{{p{0.5\textwidth}}}]{Net source count for the full energy band, post-deblending, and associated errors.} \\                   
			52,53,54      &     Sexp,Hexp,Fexp                        &   \makecell[{{p{0.5\textwidth}}}]{Net vignetting-corrected exposure times at the source position, for the combined A+B data.}                 \\
			55,56         &     SCR,e\_SCR                            &   \makecell[{{p{0.5\textwidth}}}]{Total source count rate for the soft energy band, and associated errors.} \\                 
			57            &     SCRBck                                &   \makecell[{{p{0.5\textwidth}}}]{Background count rate for the soft energy band.} \\                  
			58,59         &     SNetCR,e\_SNetCR                      &   \makecell[{{p{0.5\textwidth}}}]{Net source count rate for the soft energy band, and associated errors.} \\                  
			60,61         &     HCR,e\_HCR                            &   \makecell[{{p{0.5\textwidth}}}]{Total source count rate for the hard energy band, and associated errors.} \\                     
			62            &     HCRBck                                &   \makecell[{{p{0.5\textwidth}}}]{Background count rate for the hard energy band.} \\                            
			63,64         &     HNetCR,e\_HNetCR                      &   \makecell[{{p{0.5\textwidth}}}]{Net source count rate for the hard energy band, and associated errors.} \\                       
			65,66         &     FCR,e\_FCR                            &  \makecell[{{p{0.5\textwidth}}}]{ Total source count rate for the full energy band, and associated errors.} \\                    
			67            &     FCRBck                                &   \makecell[{{p{0.5\textwidth}}}]{Background count rate for the full energy band.} \\                                             
			68,69         &     FNetCR,e\_FNetCR                      &   \makecell[{{p{0.5\textwidth}}}]{Net source count rate for the full energy band, and associated errors.} \\                      
			70,71         &     SdbNetCR,e\_SdbNetCR                  &   \makecell[{{p{0.5\textwidth}}}]{Deblended net source count rate for the soft energy band, and associated errors.} \\                  
			72,73         &     HdbNetCR,e\_HdbNetCR                  &   \makecell[{{p{0.5\textwidth}}}]{Deblended net source count rate for the hard energy band, and associated errors.} \\                                   
			74,75         &     FdbNetCR,e\_FdbNetCR                  &   \makecell[{{p{0.5\textwidth}}}]{Deblended net source count rate for the full energy band, and associated errors.} \\                                   
			76            &     H/S                                   &   \makecell[{{p{0.5\textwidth}}}]{Band ratio computed from using the \textit{NuSTAR} hard and soft energy bands.}       \\          
			77            &     upH/S                                 &   \makecell[{{p{0.5\textwidth}}}]{Band ratio upper error.} \\                 
			78            &     lowH/S                                &   \makecell[{{p{0.5\textwidth}}}]{Band ratio lower error. }\\                  
			79            &     gamma                                 &   \makecell[{{p{0.5\textwidth}}}]{Effective photon index.} \\                 
			80            &     upgamma                               &   \makecell[{{p{0.5\textwidth}}}]{Effective photon index upper error.} \\                 
			81            &     lowgamma                              &   \makecell[{{p{0.5\textwidth}}}]{Effective photon index lower error.} \\                 
			82,83         &     SdbFlux,e\_SdbFlux                    &   \makecell[{{p{0.5\textwidth}}}]{Deblended fluxes in the soft energy band, and associated errors.} \\                
			84,85         &     HdbFlux,e\_HdbFlux                    &   \makecell[{{p{0.5\textwidth}}}]{Deblended fluxes in the hard energy band, and associated errors.} \\                
			86,87         &     FdbFlux,e\_FdbFlux                    &   \makecell[{{p{0.5\textwidth}}}]{Deblended fluxes in the full energy band, and associated errors.} \\     
			88            &     FLAG\_softX\_cov                      &   \makecell[{{p{0.5\textwidth}}}]{Abbreviated code indicating the lower-energy X-ray coverage with \textit{Chandra}, \textit{XMM-Newton} and/or \textit{Swift}-XRT).}\\                 
			89            &     XOrig                                 &   \makecell[{{p{0.5\textwidth}}}]{Reference for the adopted lower-energy X-ray (\textit{Chandra}, \textit{XMM-Newton} or \textit{Swift}-XRT) counterpart.} \\                 
			90,91         &     XRAdeg,XDEdeg                         &   \makecell[{{p{0.5\textwidth}}}]{R.A. and decl. of the lower-energy X-ray counterpart.} \\                 
			92   &     e\_Xdeg                       &   \makecell[{{p{0.5\textwidth}}}]{Positional uncertainty the lower-energy X-ray counterpart (where available, otherwise of the \nustar source) used for \sc{Nway}.} \\                 
			93            &     XOff\_NuSTAR                          &   \makecell[{{p{0.5\textwidth}}}]{Angular separation between the \textit{NuSTAR} and lower-energy X-ray counterpart positions.} \\                 
			94            &     E\_SoftX\_keV                         &   \makecell[{{p{0.5\textwidth}}}]{Reference for the lower-energy X-ray flux used to compute the $3-8$\,keV fluxes (i.e., XFlux).} \\  
			95,96,97      &     XFlux,XFlux\_Lerr,XFlux\_Uerr         &   \makecell[{{p{0.5\textwidth}}}]{$3-8$\,keV (CSC2, 4XMM-DR10/s, or 2SXPS) flux of the lower-energy X-ray counterpart, with associated lower and upper errors.} \\
			98,99,100      &     TXFlux,TXFlux\_Lerr,TXFlux\_Uerr      &   \makecell[{{p{0.5\textwidth}}}]{Total $3-8$\,keV flux of all (CSC2, 4XMM-DR10/s, or 2SXPS) sources within 30$''$ of the \textit{NuSTAR} position, with associated lower and upper errors.} \\         
			101           &     gal\_ext\_Av                          &   \makecell[{{p{0.5\textwidth}}}]{Galactic extinction, $A_\mathrm{v}$.} \\
			102,103,104   &     gal\_ext\_g,gal\_ext\_r,gal\_ext\_i   &   \makecell[{{p{0.5\textwidth}}}]{Galactic line-of-sight absorption in the $g$-, $r$-, and $i$-bands.} \\
			105,106       &     Pbest\_CatWISE,Pany\_CatWISE          &   \makecell[{{p{0.5\textwidth}}}]{The \textsc{Nway} best match probability and probability that any of the CatWISE20 sources is the right counterpart.} \\
			107,108       &     RA\_CatWISE,DEC\_CatWISE              &   \makecell[{{p{0.5\textwidth}}}]{R.A. and decl. of the ``best'' \textsc{Nway}-identified CatWISE20 counterpart.} \\
			109           &     IROff\_NuSTAR                         &   \makecell[{{p{0.5\textwidth}}}]{Angular separation between the \textsc{Nway}-matched \textit{WISE} counterpart position and the \textit{NuSTAR} position.} \\
			110           &     IROff\_SoftX                          &  \makecell[{{p{0.5\textwidth}}}]{ Angular separation between the \textsc{Nway}-matched \textit{WISE} counterpart position and the adopted lower-energy X-ray counterpart. }\\
			111,112       &     W1\_CatWISE,e\_W1\_CatWISE            &   \makecell[{{p{0.5\textwidth}}}]{\textit{WISE} $W1$ magnitude (3.4\,$\mu$m), with associated errors.} \\	
			113,114       &     W2\_CatWISE,e\_W2\_CatWISE            &   \makecell[{{p{0.5\textwidth}}}]{\textit{WISE} $W2$ magnitude (4.6\,$\mu$m), with associated errors. }\\
			115,116       &     RA\_AllWISE,DEC\_AllWISE              &   \makecell[{{p{0.5\textwidth}}}]{R.A. and decl. of the closest AllWISE counterpart. }\\
			117,118       &     W3\_AllWISE,e\_W3\_AllWISE            &   \makecell[{{p{0.5\textwidth}}}]{\textit{WISE} $W3$ magnitude (12\,$\mu$m), with associated errors. }\\
			119,120       &     W4\_AllWISE,e\_W4\_AllWISE            &   \makecell[{{p{0.5\textwidth}}}]{\textit{WISE} $W4$ magnitude (22\,$\mu$m), with associated errors.} \\
			121,122       &     Pbest\_PS1,Pany\_PS1                  &   \makecell[{{p{0.5\textwidth}}}]{The \textsc{Nway} best match probability and probability that any of the PS1-DR2 sources is the right counterpart.} \\
			123,124       &     RA\_PS1,DEC\_PS1                      &  \makecell[{{p{0.5\textwidth}}}]{ R.A. and decl. of the ``best'' \textsc{Nway}-identified PS1-DR2 counterpart.} \\
			125           &     OOff\_PS1\_NuSTAR                     &   \makecell[{{p{0.5\textwidth}}}]{Angular separation between the \textsc{Nway}-matched PS1-DR2 counterpart position and the \textit{NuSTAR} position.} \\
			126           &     OOff\_PS1\_SoftX                      &  \makecell[{{p{0.5\textwidth}}}]{ Angular separation between the \textsc{Nway}-matched PS1-DR2 counterpart position and the adopted lower-energy X-ray position.} \\
			127           &     OOff\_PS1\_CatWISE                    &   \makecell[{{p{0.5\textwidth}}}]{Angular separation between the \textsc{Nway}-matched PS1-DR2 and CatWISE20 counterpart positions.} \\
			128,129       &     gmag\_PS1,e\_gmag\_PS1                &   \makecell[{{p{0.5\textwidth}}}]{$g$-band magnitude for the \textsc{Nway}-identified PS1-DR2 counterpart, with associated errors.} \\  
			130,131       &     rmag\_PS1,e\_rmag\_PS1                &   \makecell[{{p{0.5\textwidth}}}]{$r$-band magnitude for the \textsc{Nway}-identified PS1-DR2 counterpart, with associated errors.} \\  
			132,133       &     imag\_PS1,e\_imag\_PS1                &   \makecell[{{p{0.5\textwidth}}}]{$i$-band magnitude for the \textsc{Nway}-identified PS1-DR2 counterpart, with associated errors.} \\  
			134           &     mag\_type\_PS1                        &   \makecell[{{p{0.5\textwidth}}}]{Three letter code indicating the type of PS1 magnitude in each band: \lq K\rq\ (Kron), \lq P\rq\ (PSF), \lq A\rq\ (aperture), \lq -\rq\ (unavailable).} \\
			135           &     NWAY\_RFlag                           &   \makecell[{{p{0.5\textwidth}}}]{A binary flag indicating the \textsc{Nway} reliable sample following $p_\mathrm{Cat/PS,best}>0.4$ and $p_\mathrm{Cat/PS,any}>0.5$.} \\
			136           &     OOrig\_cpart                          &   \makecell[{{p{0.5\textwidth}}}]{Reference for the adopted optical counterpart, in order of preference: PS1-DR2, SDSS, NSC2, DES-DR2 or USNOB1.} \\               
			137,138       &     ORAdeg\_cpart,ODEdeg\_cpart           &   \makecell[{{p{0.5\textwidth}}}]{R.A. and decl. of the adopted optical counterpart.} \\                     
			139,140       &     gmag\_cpart,e\_gmag\_cpart            &  \makecell[{{p{0.5\textwidth}}}]{ $g$-band magnitude for the adopted optical counterpart, with associated errors.} \\                      
			141,142       &     rmag\_cpart,e\_rmag\_cpart            &   \makecell[{{p{0.5\textwidth}}}]{$r$-band magnitude for the adopted optical counterpart, with associated errors.} \\                      
			143,144       &     imag\_cpart,e\_imag\_cpart            &   \makecell[{{p{0.5\textwidth}}}]{$i$-band magnitude for the adopted optical counterpart, with associated errors. }\\ 
			145           &     mag\_type\_cpart                      &  \makecell[{{p{0.5\textwidth}}}]{ Three letter code indicating the type of PS1 magnitude in each band (if used for counterpart): \lq K\rq\ (Kron), \lq P\rq\ (PSF), \lq A\rq\ (aperture), \lq -\rq\ (unavailable).} \\      
			146           &     OOff\_NuSTAR                          &   \makecell[{{p{0.5\textwidth}}}]{Angular separation between the adopted optical counterpart position and the \textit{NuSTAR} position.} \\
			147           &     OOff\_SoftX                           &   \makecell[{{p{0.5\textwidth}}}]{Angular separation between the adopted optical counterpart position and the best soft X-ray position.} \\
			148           &     OOff\_CatWISE                         &  \makecell[{{p{0.5\textwidth}}}]{ Angular separation between the adopted optical counterpart position and the best CatWISE20 position.} \\
			149           &     SpecCAT                               &   \makecell[{{p{0.5\textwidth}}}]{Flag indicating sources which are included in the spectroscopic catalog.} \\
			150,151       &     RAdeg\_spec,DEdeg\_spec               &   \makecell[{{p{0.5\textwidth}}}]{R.A. and decl. of the spectroscopic target.} \\ 
			152           &     OOff\_Spec\_Cpart                     &   \makecell[{{p{0.5\textwidth}}}]{Angular separation between the spectroscopic target and the adopted optical counterpart.} \\
			153           &     zspec                                 &   \makecell[{{p{0.5\textwidth}}}]{Spectroscopic redshift.} \\
			154           &     zQuality                              &   \makecell[{{p{0.5\textwidth}}}]{Quality flag of the spectroscopic redshift measurement.} \\
			155,156       &     Classification,Class\_extra           &   \makecell[{{p{0.5\textwidth}}}]{Spectroscopic classification and additional source class information.} \\
			157,158       &     NWAY\_CatWISE,NWAY\_PS1               &   \makecell[{{p{0.5\textwidth}}}]{A binary flag indicating a \textsc{Nway} CatWISE20/PS1-DR2 match to the spectroscopic target.} \\
			159           &     Cpart\_RFlag                          &   \makecell[{{p{0.5\textwidth}}}]{A flag indicating the reliability of the spectroscopic target as the correct counterpart.} \\
			160           &     SpecRef              &   \makecell[{{p{0.5\textwidth}}}]{Reference for the information used for the spectroscopic target.} \\
			161,162       &     gmag\_spec,e\_gmag\_spec              &   \makecell[{{p{0.5\textwidth}}}]{$r$-band magnitude for the spectroscopic target, with associated errors.} \\
			163,164       &     rmag\_spec,e\_rmag\_spec              &   \makecell[{{p{0.5\textwidth}}}]{$r$-band magnitude for the spectroscopic target, with associated errors.} \\
			165,166       &     imag\_spec,e\_imag\_spec              &   \makecell[{{p{0.5\textwidth}}}]{$r$-band magnitude for the spectroscopic target, with associated errors.} \\
			167           &     L10\_40keV                            &   \makecell[{{p{0.5\textwidth}}}]{Non-absorption-corrected, rest-frame 10-40\,keV luminosity.} \\
			168           &     PFlag                                 &   \makecell[{{p{0.5\textwidth}}}]{A binary flag indicating sources that show evidence for being associated with the primary science targets of their respective \textit{NuSTAR} observations.} \\
			169           &     CTFlag                                &   \makecell[{{p{0.5\textwidth}}}]{A binary flag indicating Compton-thick candidates.} \\
		\enddata
\end{deluxetable}
\begin{deluxetable}{lllllllll}
\tablecolumns{9}
\tablewidth{\textwidth}
\tablecaption{\label{tab:NSS80 example} Individual sources in order of R.A. for the first 25 NSS80 catalogue entries. 
		The column numbers corresponds to the descriptions given in Table~\ref{tab:description of NSS80 cat}. 
		The full NSS80 catalogue is available on the NSS80 webpage. }
\tablehead{ID & \textit{NuSTAR} name & R.A. & decl. & XOrig & \textsc{Nway} & \textsc{Nway$_\mathrm{clean}$} & $z$ & $L_\mathrm{10-40\,keV}$ \\
		& & ($^\circ$) & ($^\circ$) & & & & & (erg\,s$^{-1}$) \\
		(1) & (2) & (6) & (7) & (85) & (91) & (112) & (113) & (114) \\}
		\startdata
		1  & NuSTARJ000011-7652.5 & 0.04947511 & -76.8755213  & 2SXPS     &  ...       &  0 &  0.05   & $2.26 \times 10^{42}$ \\
		2  & NuSTARJ000303+7017.7 & 0.76529404 & 70.29623148  & ...       &  ...       &  0 &  ...     & ...        \\
		3  & NuSTARJ000333+7013.7 & 0.89107631 & 70.22940726  & 4XMMDR10  &  CatWISE20 &  1 &  0.39  & $4.75 \times 10^{43}$  \\
		4  & NuSTARJ000423+7017.2 & 1.09736601 & 70.28740531  & 4XMMDR10  &  CatWISE20 &  1 &  0.77  & $1.73 \times 10^{44}$ \\
		5  & NuSTARJ000549+2013.0 & 1.45692910 & 20.21813114  & 4XMMDR10s &  CatWISE20 &  1 &  1.97  & $7.79 \times 10^{44}$  \\
		6  & NuSTARJ000613+2018.3 & 1.55586492 & 20.30626396  & 4XMMDR10s &  CatWISE20 &  1 &  ...     & ...        \\
		7  & NuSTARJ000634+2013.0 & 1.64247764 & 20.21818081  & 4XMMDR10s &  CatWISE20 &  1 &  ...     & ...        \\
		8  & NuSTARJ001003+1057.5 & 2.51268295 & 10.95874867  & 4XMMDR10s &  CatWISE20 &  1 &  ...     & ...        \\
		9  & NuSTARJ001026+1054.5 & 2.60935872 & 10.90892409  & ...       &  ...       &  0 &  ...     & ...        \\
		10 & NuSTARJ001443+8131.9 & 3.68230459 & 81.53320406  & 4XMMDR10  &  CatWISE20 &  1 &  0.37  & $6.63 \times 10^{43}$  \\
		11 & NuSTARJ001542+8134.4 & 3.92760279 & 81.57499432  & 4XMMDR10  &  CatWISE20 &  1 &  ...     & ...        \\
		12 & NuSTARJ001852-1026.1 & 4.71867576 & -10.4352099  & CSC2      &  CatWISE20 &  1 &  0.33   & $2.61 \times 10^{43}$ \\
		13 & NuSTARJ001858-1022.7 & 4.74574713 & -10.3799008  & CSC2      &  CatWISE20 &  1 &  1.17   & $7.32 \times 10^{44} $  \\
		14 & NuSTARJ001908-1020.7 & 4.78738895 & -10.3457535  & ...       &  CatWISE20 &  1 &  ...     & ...        \\
		15 & NuSTARJ001952-7057.0 & 4.96672553 & -70.9511159  & CSC2      &  CatWISE20 &  1 &  0.81  & $1.48 \times 10^{44} $ \\
		16 & NuSTARJ002028+5910.3 & 5.12058379 & 59.17219231  & ...       &  ...       &  0 &  ...     & ...        \\
		17 & NuSTARJ002046-7057.6 & 5.19269674 & -70.9614382  & CSC2      &  CatWISE20 &  1 &  0.27  & $9.81 \times 10^{42} $ \\
		18 & NuSTARJ002102+5546.2 & 5.26237238 & 55.77061297  & 2SXPS     &  CatWISE20 &  1 &  ...     & ...        \\
		19 & NuSTARJ002112+5542.5 & 5.30106223 & 55.70913570  & 2SXPS     &  CatWISE20 &  1 &  0.41  & $4.38 \times 10^{43} $  \\
		20 & NuSTARJ002113-7053.5 & 5.30725801 & -70.8924101  & 4XMMDR10s &  ...       &  0 &  ...     & ...        \\
		21 & NuSTARJ002151-1859.7 & 5.46388437 & -18.9957774  & ...       &  PSDR2     &  1 &  ...     & ...        \\
		22 & NuSTARJ002214-1851.0 & 5.56207390 & -18.8503530  & 4XMMDR10  &  CatWISE20 &  1 &  ...     & ...        \\
		23 & NuSTARJ002227-1854.7 & 5.61405050 & -18.9131576  & 4XMMDR10  &  CatWISE20 &  1 &  0.29   & $1.99 \times 10^{43} $ \\
		24 & NuSTARJ002544+6818.8 & 6.43592415 & 68.31345208  & CSC2      &  CatWISE20 &  1 &  ...     & ...        \\
		25 & NuSTARJ002624+6819.9 & 6.60239773 & 68.33175607  & 2SXPS     &  ...       &  0 &  ...     & ...        \\
	\enddata
\end{deluxetable}

\section{Description of the secondary NSS80 source catalogue} 
\label{appendix:secondary catalogue}
The secondary \textit{NuSTAR} serendipitous survey source catalogue, contains 214 sources in total, and is available as an electronic table on the NSS80 webpage. 
Here we describe the columns of the catalogue, which are summarized in Table~\ref{tab:description of NSS80 cat sec}.

\begin{deluxetable}{ll}
\tablecolumns{2}
\tablewidth{\textwidth}
\tablecaption{\label{tab:description of NSS80 cat sec} Column descriptions for the secondary NSS80 catalogue. }
\tablehead{Column number & Description \\}
			\startdata
			1 & Unique source identification number (ID). \\
			2 & Unique \textit{NuSTAR} source name. \\
			3 & Unique \textit{NuSTAR} field and source ID. \\
			4,5 & Unique \citetalias{lansbury2017_cat} \textit{NuSTAR} source ID and field; the secondary \citetalias{lansbury2017_cat} catalog is appended with a ``S''.  \\
			6,7 & Right ascension (R.A.) and declination (decl.). \\
			8,9 & Flags indicating optical mask and optical source name. \\
			10-12 & Flags indicating the energy bands for which the source is detected. \\
			13-15 & Same as columns 10–12, post-deblending. \\
			16-18 & The logarithm of the false probabilities for the three standard \textit{NuSTAR} energy bands.\\
			19 & The \textit{NuSTAR} detection likelihood. \\
			20-22 & Same as columns 16–18, post-deblending. \\
			23 &  The same as column 19, post-deblending. \\
			24 & Flag indicating whether the source is significant post-deblending, for at least one energy band. \\
			25-39 & Total, background, and net source counts for the three standard energy bands, and associated errors. \\
			40-51 & Same as columns 25–39, post-deblending. \\
			52-54 & Net vignetting-corrected exposure times at the source position, for the combined A+B data. \\
			55-69 & Total, background, and net source count rates for the three standard energy bands, and associated errors. \\
			70-75 & Deblended net source count rates for the three standard energy bands, and associated errors. \\
			76-78 & Band ratio and upper and lower errors. \\
			79-81 & Effective photon index and upper and lower errors. \\
			82-87 & Deblended fluxes in the three standard bands and associated errors. \\
		\enddata
\end{deluxetable}

\section{Comparison to the NSS40}
\label{appendix: comparison to 40-month}
Overall, we find a 91\% match between the primary NSS80 catalog and NSS40 for a search radius of 25$''$ (which is the positional accuracy of \textit{NuSTAR} for faint sources), with 36 primary NSS40 sources reported in \citetalias{lansbury2017_cat} undetected in the NSS80. 
For the majority of these sources the deeper data has improved the false probability estimates and, consequently, eliminates false detections. This is particularly noticeable for \citetalias{lansbury2017_cat} sources with detections in a single energy band only.    
We summarise potential reasons to explain the undetected \citetalias{lansbury2017_cat} sources in Table~\ref{tab:missed 40-month serendips}: 32 sources have improved false probabilities, 1 source lies in an excess background region, and 3 sources are on the peripheries of coadded fields.

\begin{deluxetable}{llllllll}
\tablecolumns{8}
\tablewidth{\textwidth}
\tabletypesize{\scriptsize}
\tablecaption{\label{tab:missed 40-month serendips} A list of the NSS40 serendipitous sources reported in \citetalias{lansbury2017_cat} which are undetected in the NSS80 catalogue. }
\tablehead{ID & \textit{NuSTAR} name & Science target name & R.A. & decl. & Type & $z_\mathrm{spec}$ & Explanation \\
			& & & ($^\circ$) & ($^\circ$) & & & \\
			(1) & (2) & (3) & (4) & (5) & (6) & (7) & (8) \\  }
   \startdata
			$2  $ & NuSTARJ$001130+0057.8$  &  SDSSJ001111d97p005626d3      &  $2.88   $ &  $ 0.96  $  & BL     &  1.492 &  P$_\textrm{false}$ condition not met in NSS80 reductions  \\  
			$5  $ & NuSTARJ$001717+8127.3$  &  S50014p81                    &  $4.32   $ &  $ 81.46 $  & -      &  -     &  Undetected in HB+FB; logSdbP $\approx$ -6                           \\  
			$9  $ & NuSTARJ$001953+5911.8$  &  IC10\_X1                     &  $4.97   $ &  $ 59.2  $  & -      &  -     &  P$_\textrm{false}$ condition not met in NSS80 reductions  \\                                          
			$78 $ & NuSTARJ$040730+0344.2$  &  3C105                        &  $61.88  $ &  $ 3.74  $  & -      &  -     &  Undetected in SB+FB; logHdbP $\approx$ -6                           \\                                          
			$79 $ & NuSTARJ$042349+0410.9$  &  2MASXJ04234080p0408017       &  $65.96  $ &  $ 4.18  $  & -      &  -     &  P$_\textrm{false}$ condition not met in NSS80 reductions  \\                                          
			$81 $ & NuSTARJ$042509-5709.5$  &  1H\_0419m577                 &  $66.29  $ &  $ -57.16$  & -      &  -     &  Undetected in SB+HB; logFdbP $\approx$ -6                           \\                                          
			$89 $ & NuSTARJ$043724-4722.2$  &  PSRJ0437m4715                &  $69.35  $ &  $ -47.37$  & -      &  -     &  Undetected in SB+HB; logFdbP $\approx$ -6                           \\                                          
			$99 $ & NuSTARJ$052100-2528.8$  &  IRAS05189m2524               &  $80.25  $ &  $ -25.48$  & BL     &  1.666 &  Possible false detection on the edge of an exposure                 \\  
			$121$ & NuSTARJ$065922-5558.6$  &  Bullet\_Bullet\_shock        &  $104.84 $ &  $ -55.98$  & -      &  -     &  Undetected in SB+HB; logFdbP $\approx$ -6                           \\                                          
			$179$ & NuSTARJ$095440+6942.6$  &  SN2014J                      &  $148.67 $ &  $ 69.71 $  & -      &  -     &  Undetected in SB+HB; logFdbP $\approx$ -6                           \\                                          
			$192$ & NuSTARJ$095801+6859.4$  &  M81\_X9                      &  $149.51 $ &  $ 68.99 $  & -      &  -     &  Undetected in SB+HB; logFdbP $\approx$ -6                           \\                                          
			$193$ & NuSTARJ$095806+6910.3$  &  M81\_X9                      &  $149.53 $ &  $ 69.17 $  & -      &  -     &  Undetected in HB+FB; logSdbP $\approx$ -6                           \\  
			$196$ & NuSTARJ$095853+6901.3$  &  M81\_X9                      &  $149.72 $ &  $ 69.02 $  & -      &  -     &  Undetected in SB+HB; logFdbP $\approx$ -6                           \\                                          
			$211$ & NuSTARJ$102318+0036.5$  &  PSRJ1023p0038                &  $155.83 $ &  $ 0.61  $  & Gal    &  0.0   &  Undetected in SB+HB; logFdbP $\approx$ -6                           \\                                          
			$212$ & NuSTARJ$102328+0043.9$  &  PSRJ1023p0038                &  $155.87 $ &  $ 0.73  $  & -      &  -     &  Undetected in SB+HB; logFdbP $\approx$ -6                           \\                                          
			$219$ & NuSTARJ$102802-4351.0$  &  NGC3256                      &  $157.01 $ &  $ -43.85$  & BL     &  1.784 &  P$_\textrm{false}$ condition not met in NSS80 reductions  \\                                          
			$227$ & NuSTARJ$105931+2429.8$  &  IRAS\_10565p2448             &  $164.88 $ &  $ 24.5  $  & BL     &  0.908 &  P$_\textrm{false}$ condition not met in NSS80 reductions  \\                                          
			$233$ & NuSTARJ$110632+7225.9$  &  NGC3516                      &  $166.64 $ &  $ 72.43 $  & -      &  -     &  Undetected in SB+FB; logHdbP $\approx$ -6                           \\                                          
			$236$ & NuSTARJ$110752+7230.7$  &  NGC3516                      &  $166.97 $ &  $ 72.51 $  & BL     &  0.901 &  Undetected in SB+HB; logFdbP $\approx$ -6                           \\                                          
			$256$ & NuSTARJ$120242+4437.2$  &  NGC4051                      &  $180.68 $ &  $ 44.62 $  & NL?    &  0.296 &  Possible false detection on the edge of an exposure                 \\                                          
			$261$ & NuSTARJ$120613+4957.2$  &  2MASXJ12055599p4959561       &  $181.56 $ &  $ 49.95 $  & BL     &  0.784 &  Undetected in SB; logFdbP $\approx$ -6  \\
			$302$ & NuSTARJ$125657+5644.6$  &  Mrk231                       &  $194.24 $ &  $ 56.74 $  & NL     &  2.073 &  Undetected in SB+HB; logFdbP $\approx$ -6                           \\                                          
			$303$ & NuSTARJ$130108-6356.2$  &  2RXPJ130159d6m635806         &  $195.28 $ &  $ -63.94$  & -      &  -     &  Undetected in SB+FB; logHdbP $\approx$ -6                           \\  
			$313$ & NuSTARJ$133311-3406.8$  &  ESO383\_18                   &  $203.3  $ &  $ -34.11$  & NL?    &  0.091 &  Possible false detection on the edge of an exposure                 \\                                          
			$330$ & NuSTARJ$141215-6519.6$  &  Circinus\_XMM2               &  $213.07 $ &  $ -65.33$  & -      &  -     &  Undetected in SB+HB; logFdbP $\approx$ -6                           \\                                          
			$375$ & NuSTARJ$165346+3953.7$  &  Mkn501                       &  $253.44 $ &  $ 39.9  $  & NL     &  0.354 &  Undetected in SB+FB; logHdbP $\approx$ -6                           \\                                          
			$376$ & NuSTARJ$165351+3938.5$  &  Mkn501                       &  $253.46 $ &  $ 39.64 $  & Gal    &  0.0   &  Undetected in SB+HB; logFdbP $\approx$ -6                           \\                                          
			$384$ & NuSTARJ$171718-6239.1$  &  NGC6300                      &  $259.33 $ &  $ -62.65$  & -      &  -     &  Undetected in SB+HB; logFdbP $\approx$ -6                           \\  
			$389$ & NuSTARJ$172755-1417.4$  &  PDS\_456                     &  $261.98 $ &  $ -14.29$  & -      &  -     &  Undetected in HB+FB; logSdbP $\approx$ -6                           \\  
			$390$ & NuSTARJ$172803-1423.0$  &  PDS\_456                     &  $262.02 $ &  $ -14.38$  & BL     &  1.555 &  Contaminated by stray light                                         \\                                          
			$395$ & NuSTARJ$172843-1419.0$  &  PDS\_456                     &  $262.18 $ &  $ -14.32$  & -      &  -     &  Undetected in HB+FB; logSdbP $\approx$ -6                           \\                                          
			$407$ & NuSTARJ$182447-2444.5$  &  PSR\_B1821m24                &  $276.2  $ &  $ -24.74$  & -      &  -     &  P$_\textrm{false}$ condition not met in NSS80 reductions  \\                                          
			$414$ & NuSTARJ$183422-0840.6$  &  HESSJ1834m087\_TeV           &  $278.59 $ &  $ -8.68 $  & -      &  -     &  P$_\textrm{false}$ condition not met in NSS80 reductions  \\                                          
			$413$ & NuSTARJ$183422-0841.4$  &  HESSJ1834m087\_TeV           &  $278.59 $ &  $ -8.69 $  & -      &  -     &  P$_\textrm{false}$ condition not met in NSS80 reductions  \\  
			$485$ & NuSTARJ$224302+2942.1$  &  Ark\_564                     &  $340.76 $ &  $ 29.7  $  & -      &  -     &  Undetected in SB+FB; logHdbP $\approx$ -6                           \\                                          
			$491$ & NuSTARJ$231811-4228.8$  &  NGC7582                      &  $349.55 $ &  $ -42.48$  & -      &  -     &  Undetected in SB+HB; logFdbP $\approx$ -6                         \\  
		\enddata
	\tablecomments{
		{\sc Columns:}\,\textbf{(1)} Unique source identification number in \citetalias{lansbury2017_cat}.
		\textbf{(2)} Unique \textit{NuSTAR} source name in \citetalias{lansbury2017_cat}. 
		\textbf{(3)} Object name for the primary science target of the \textit{NuSTAR} observation in \citetalias{lansbury2017_cat}.
		\textbf{(4) } Right ascension as recorded in \citetalias{lansbury2017_cat}.  
		\textbf{(5) } Declination as recorded in \citetalias{lansbury2017_cat}.
		\textbf{(6)} Spectroscopic classification of the optical counterpart.
		\textbf{(7)} Spectroscopic redshift.
		\textbf{(8) } Explanation for non-detection of the NSS40 serendipitous source in the NSS80 catalogue. }
\end{deluxetable}
\section{Comparison of CatWISE20 counterparts to spectroscopic targets}\label{appendix:catwise probability}

\begin{figure}
	\centering
	\includegraphics[width=20pc]{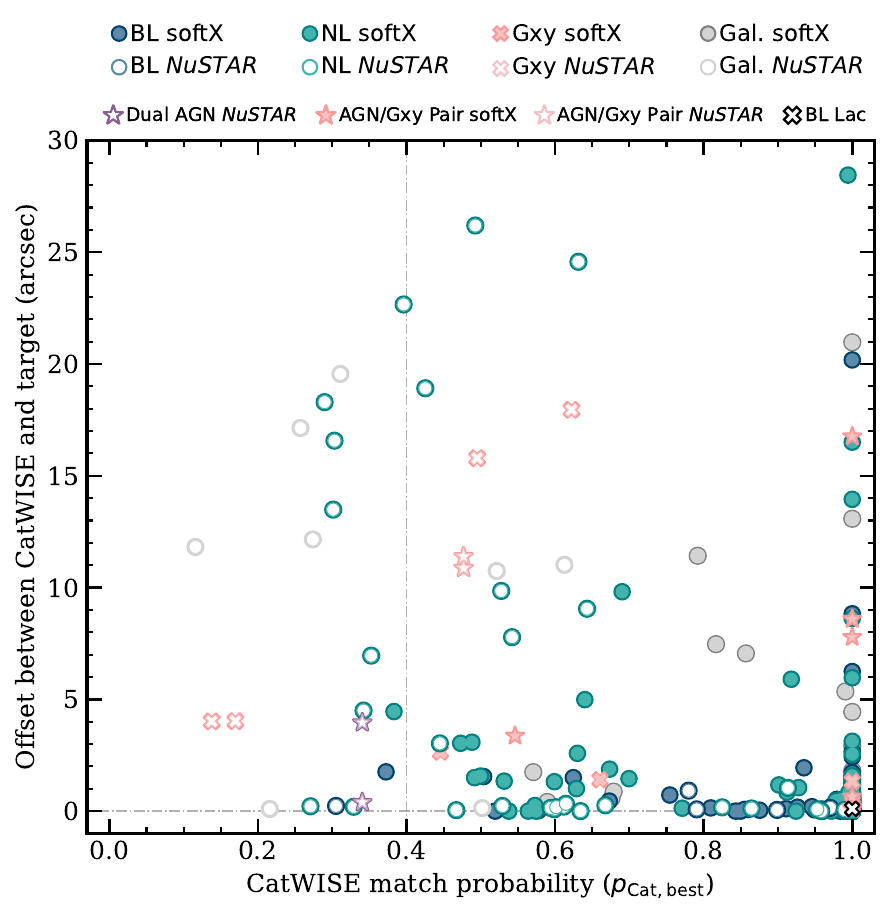}		
    \\
	{\textbf{Figure~D1.} Separation in arcsec between the \textsc{Nway}-identified CatWISE20 counterpart and the spectroscopic target versus the best CatWISE20 match probability, color-coded by spectral classifications: BL (blue circle), NL (green circle), galaxy (Gxy; peach `X'), Galactic sources (grey circle), AGN/Gxy pairs (peach star), dual AGN (purple star), and a BL~Lac candidate (white, black-edged `X').
		Solid symbols indicate NSS80 sources with lower-energy (i.e., soft) X-ray counterparts and white filled symbols mark NSS80 sources which lack lower-energy X-ray counterparts and, therefore, have a lower match probability given the large uncertainty in the \nustar position (i.e., $\sim$\,25$''$).
		The vertical black dash-dotted line indicates the probability cut we applied in Section~\ref{subsec:nway infrared and optical cpart} to obtain reliable counterpart associations to the X-ray source. 
    }
	\label{fig:nway spec comparison}	
\end{figure}
Our assessment of the reliability of the followed-up spectroscopic sources takes into account the positional offset between the best CatWISE20 counterpart and the spectroscopic counterpart in combination with the CatWISE20 probability. 
In Figure~\ref{fig:nway spec comparison} we plot the offset between the best CatWISE20 counterpart and the followed-up spectroscopic target as a function of the \textsc{Nway} best match probability, color-coded by the different spectral classifications. 
Each spectroscopic class is further divided into NSS80 sources with lower-energy X-ray (solid symbols) and without lower-energy X-ray (white filled, color-edged symbols) counterparts.
Overall, there is a good match to \textsc{Nway} for the majority of the spectroscopically classified sources, indicating that the correct spectroscopic counterpart has been identified: the positional offset for the majority of the sources is $<$\,5$''$; we consider 5$''$ as a reasonable threshold since it broadly corresponds to the CatWISE20 positional uncertainty. 
The reliability of the targets with larger positional offsets is less certain. 
The \textsc{Nway} probability for the best CatWISE20 counterpart is strongly dependent on the availability of a lower-energy X-ray counterpart. For example, the dual AGN system in the NSS80 catalog (NuSTARJ054231+6054.4; open, purple-edged star) is the correct counterpart based on its AGN-like \textit{WISE} colours, but it has a low \textsc{Nway} probability due to the lack of lower-energy X-ray information (\textit{Swift}-XRT coverage, but undetected).
Most of the large-offset NSS80 sources without lower-energy X-ray counterparts have low-to-intermediate \textsc{Nway} probabilities and sensitive lower-energy X-ray observations are required to improve the counterpart identification. 
By comparison, most of the large-offset NSS80 sources with lower-energy X-ray counterparts have high \textsc{Nway} probabilities, but \textsc{Nway} has selected a different source as the best match, typically a brighter lower-energy X-ray counterpart to that spectroscopically followed-up (see e.g., Figure~\ref{fig:spec extra}).
To better determine the reliability between the best CatWISE20 counterpart and the followed-up counterpart of these more uncertain sources requires further investigation including, for example, spectroscopic follow-up of the best CatWISE20 counterpart.

\section{Optical spectroscopic properties of individual post-NSS40 objects} 

\subsection{Optical spectroscopic properties of individual post-NSS40 objects} 
\label{appendix:optical spectra}

Here we provide details of the optical spectroscopic properties of individual sources from the \textit{NuSTAR} serendipitous survey. 
As described in Section~\ref{subsubsec:follow-up spectra}, these largely result from our dedicated follow-up campaign using the Keck, Palomar, VLT, and SALT facilities, and also from existing publicly available spectroscopy (primarily SDSS spectroscopy). 

Details for individual sources are tabulated in Table~\ref{tab:spec results}, the columns of which are as follows: 
\begin{enumerate}[(1)]
	\vspace{0.1cm}
	\setlength\itemsep{0.1em}
	\item Unique \textit{NuSTAR} source identification number, in order of increasing right ascension;
	\item Unique \textit{NuSTAR} source name as listed in source catalogue;
	\item The unique observing run identification number, as defined in Table~\ref{tab:telescope follow-up} (``S'' and ``Lit'' mark spectra obtained from the SDSS and from elsewhere in the literature, respectively);
	\item Source redshift (mainly from spectroscopic data);
	\item Source classification as described in Section~\ref{subsubsec:spectral classification} (sources with photometric redshift measurements are appended with a ``C'' symbol and sources with a level of ambiguity in the optical classification are appended with a ``?'' symbol);
	\item Quality flag of spectroscopic redshift. Single line measurements are annotated with ``B'' and photometric redshift measurements with ``C''; 
	``F'' annotates sources which lack a reliable redshift measurement (usually faint, red continuum spectra);
	\item \textsc{Nway} association flag: 0 = mismatch between spectroscopic target and \textsc{Nway} source; 1 = match between spectroscopic target and \textsc{Nway} source; -99 = Secondary post-NSS40 objects.
	\item A binary flag indicating sources that show evidence for being associated with the primary science target of their respective \textit{NuSTAR} observations, according to the definition $\Delta(cz) < 0.05\,cz$. 
	\item An abbreviated flag indicating a \textsc{Nway} CatWISE20 (C)  and/or PS1-DR2 (P) counterpart match to the spectroscopic target.
	\item Notes based on our visual assessment of the spectra.
\end{enumerate}

Individual source spectra are available online and example spectra of the first four sources for the different spectral classes are shown in Figures~E1--E5: BL AGN, NL AGN, galaxies, Galactic sources, and unclassified objects with a faint continuum detected, respectively. The spectra for each class are given in order of increasing right ascension.
Shown on the top are the unique \nustar ID and source name, in the upper left corner the observing telescope and run identification number (corresponding to Table~\ref{tab:telescope follow-up}), and in the upper right corner the source classification and redshift. 
All \textit{CatWISE} detections are shown with `X' marks, colored green. 
The \nustar 25$''$ and 30$''$ error circles are plotted in a dash-dotted and solid black line, respectively. All PS1-DR2 detections are shown with `plus' marks, colored pink.
The X-ray position is marked with the respective error circle, the CatWISE20 position is indicated with a green 2.7$''$ error circle, the PS1-DR2 position is shown with a purple 2$''$ error circle, and the spectroscopic target is marked with a red crosshair corresponding to the black spectrum in the left panel. Where a second spectrum is available, the spectrum, observation details, and target position are shown in peach.

\subsection{Optical finding charts for the spectroscopic NSS40 sources} 
\label{appendix:findercharts}
Figure~E6 shows optical finding charts for the first 12 spectroscopically followed-up NSS40 sources reported in \citetalias{lansbury2017_cat}; the finding charts for the full sample are available online. 
Following the colour scheme of Section~\ref{appendix:optical spectra}, all \textit{WISE} detections are shown with `X' marks, and PS1-DR2 detections with `plus' marks. 
The \nustar 25$''$ and 30$''$ error circles are plotted in a dash-dotted and solid black line, respectively.
The X-ray position is marked with the respective error circle, the CatWISE20 position is indicated with a green 2.7$''$ error circle, the PS1-DR2 position is shown with a purple 2$''$ error circle, and the spectroscopic target is marked with a red crosshair.


\startlongtable
\begin{deluxetable}{llllllllll}
\tablecolumns{10}
\tablewidth{\textwidth}
\tabletypesize{\scriptsize}
\tablecaption{\label{tab:spec results} Summary of the optical spectroscopy for the post-NSS40 serendipitous survey sources. 
			The columns are described in Section~\ref{appendix:optical spectra}.
			Details for NSS40 sources are tabulated in Table~6 in \citetalias{lansbury2017_cat}.}
\tablehead{ID & \textit{NuSTAR} name & Run & $z$ & Type & Quality & PFlag & \textsc{Nway}   & F$_\mathrm{Cpart}$ & \colhead{Notes} \\
			(1) & (2) & (3) & (4) & (5) & (6) & (7) & (8) & (9) &  \\ }
   \startdata
			2      &  NuSTARJ000303+7017.7  & 24       &    0       &   Galactic     &        ...    &     0   &     C+P  &            2   &           \makecell[{{p{0.22\textwidth}}}]{Possibly incorrect counterpart due to high source density  }                                                                                               \\ 
			3      &  NuSTARJ000333+7013.7  & 20       &    0.391   &   NL           &        ...    &     0   &     C+P  &            1   &           \makecell[{{p{0.22\textwidth}}}]{...}                                                                                                                                                       \\ 
			4      &  NuSTARJ000423+7017.2  & 20       &    0.775   &   BL           &        ...    &     0   &     C+P  &            1   &           \makecell[{{p{0.22\textwidth}}}]{...}                                                                                                                                                       \\ 
			5      &  NuSTARJ000549+2013.0  & S        &    1.974   &   BL           &        ...    &     0   &     C+P  &            1   &           \makecell[{{p{0.22\textwidth}}}]{Mg\,\textsc{ii} BALQSO}                                                                                                                                    \\
			6      &  NuSTARJ000613+2018.3  & 23       &    ...     &   ...          &        F      &     0   &     C+P  &            1   &           \makecell[{{p{0.22\textwidth}}}]{Continuum detected}                                                                                                                                        \\
\enddata
	\tablecomments{
		{\sc Columns}:\,{\bf (1)} Unique \textit{NuSTAR} source ID. 
		{\bf(2)} Unique \textit{NuSTAR} source name.
		{\bf(3)} Spectroscopic follow-up observing code; see Table~\ref{tab:telescope follow-up}.
		{\bf(4)} Spectroscopic redshift.
		{\bf(5)} Spectroscopic classification: BL $\equiv$ Broad Line object (i.e., quasar); NL $\equiv$ Narrow Line object (AGN or galaxy); Gxy $\equiv$ Galaxy (absorption lines); Galactic $\equiv$ Galactic sources at $z = 0$. See Section~\ref{subsubsec:spectral classification}.
		{\bf(6)} Spectroscopic redshift quality: F $\equiv$ Faint continuum detected; B $\equiv$ single-line redshift measurement; C $\equiv$ photometric redshift.
		{\bf(7)} Binary flag to indicate association to the \textit{NuSTAR} science target, i.e., sources with a velocity offset from the science target smaller than 5\% of the total science target velocity.
		{\bf(8)} Flag indicating whether the spectroscopic target matches to the CatWISE20 and/or PS1 \textsc{Nway}-identified counterparts: C+P $\equiv$ CatWISE20\,+\,PS1; C $\equiv$ CatWISE20 only; P $\equiv$ PS1 only.
		{\bf(9)} Flag indicating reliability of identified spectroscopic counterpart: 0 $\equiv$ unreliable; 1 $\equiv$ reliable; 2 $\equiv$ counterpart uncertainty.
	}
 \end{deluxetable}

\newpage
\begin{figure*}
	\centering
	\includegraphics[width=40pc]{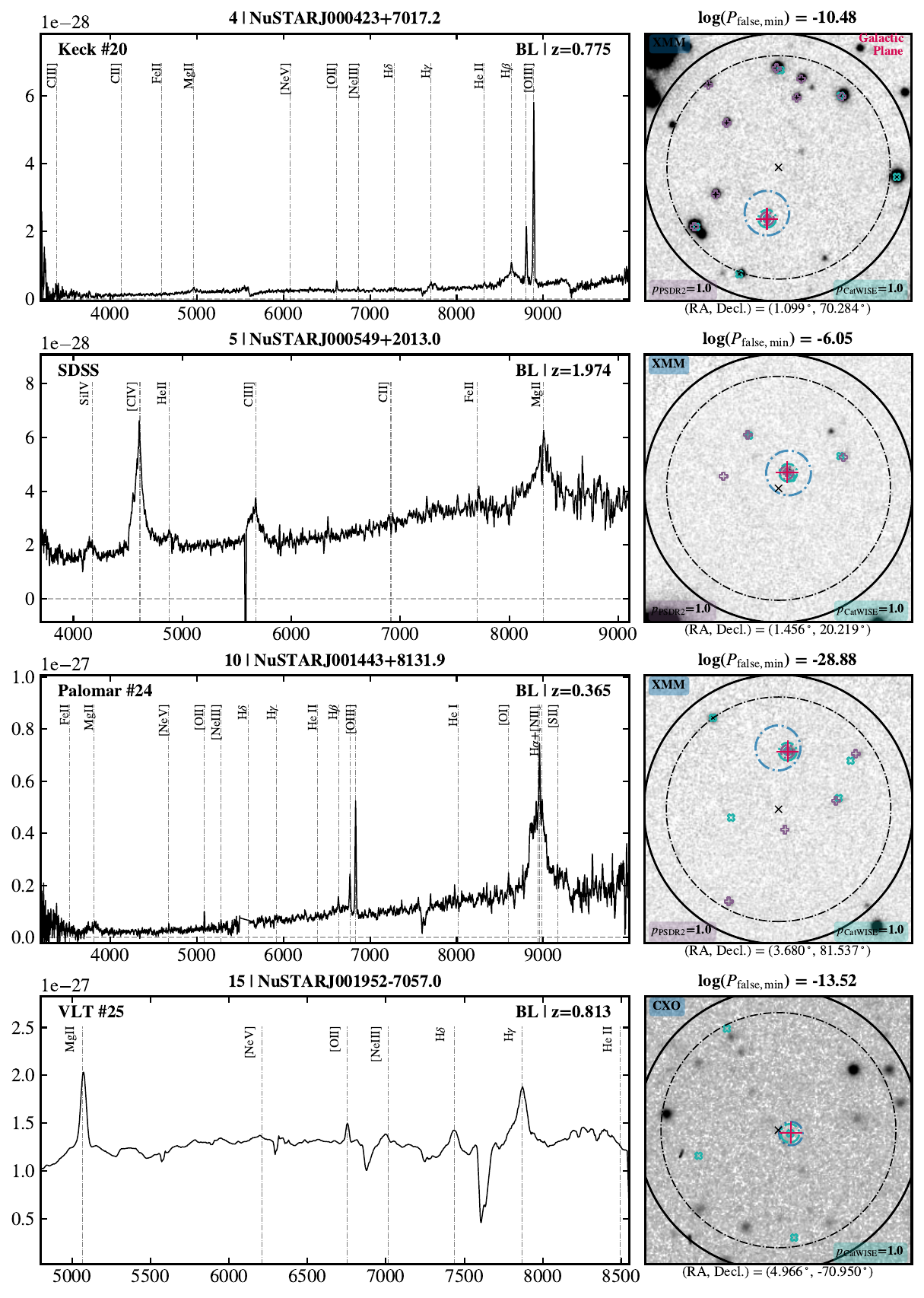}		
	{\textbf{Figure~E1.} Optical spectra for the first four extragalactic post-NSS40 sources classified as BL AGN in order of \nustar ID.
		The symbol key is described in Appendix~\ref{appendix:optical spectra}. 	}
	\label{fig:spec BLAGN}
\end{figure*}

\begin{figure*}
	\centering
	\includegraphics[width=40pc]{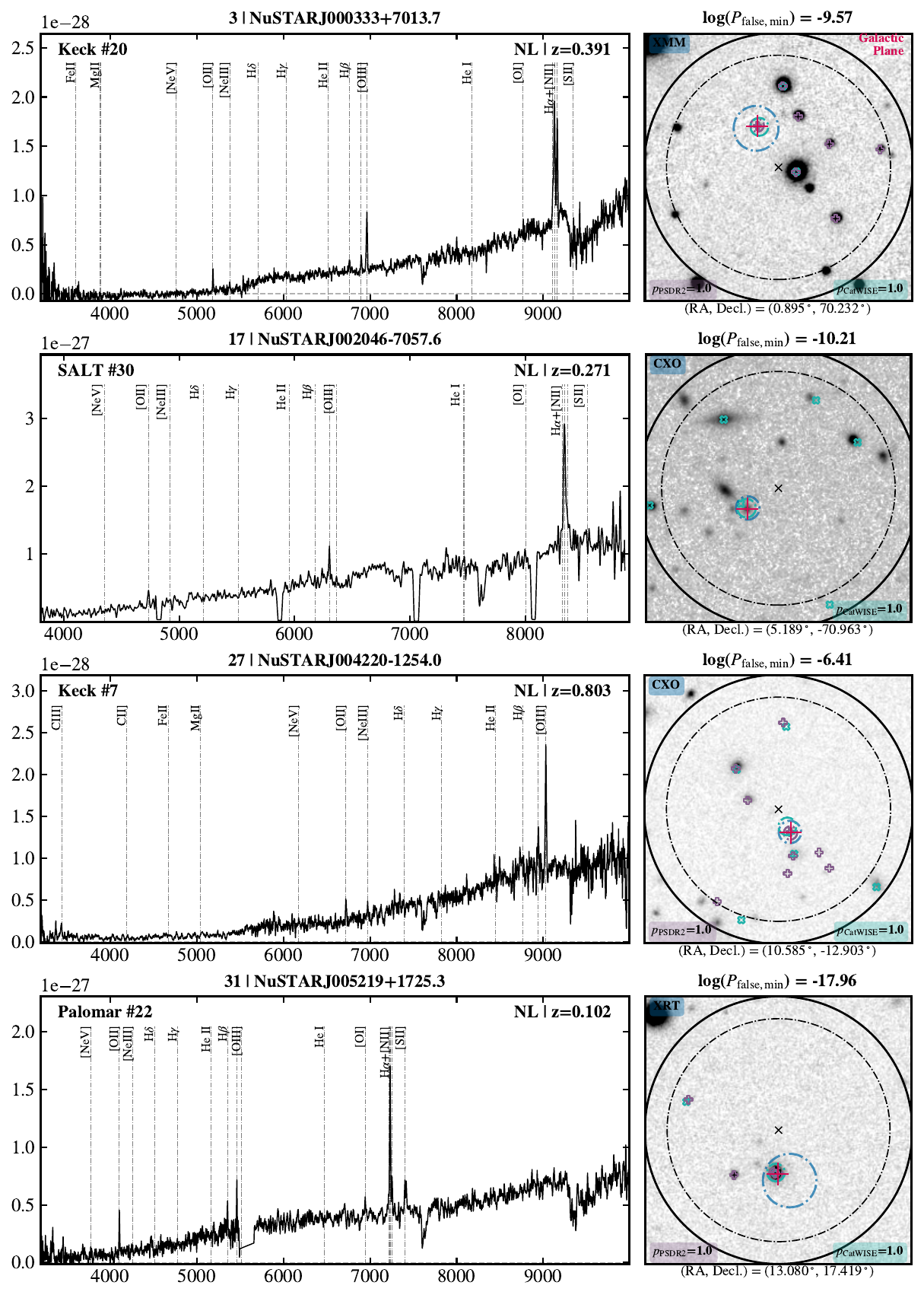}		
	{\textbf{Figure~E2.} Optical spectra for the first four extragalactic post-NSS40 sources classified as NL AGN.
		The symbol key is described in Appendix~\ref{appendix:optical spectra}. }
	\label{fig:spec NLAGN}
\end{figure*}

\begin{figure*}
	\centering
	\includegraphics[width=40pc]{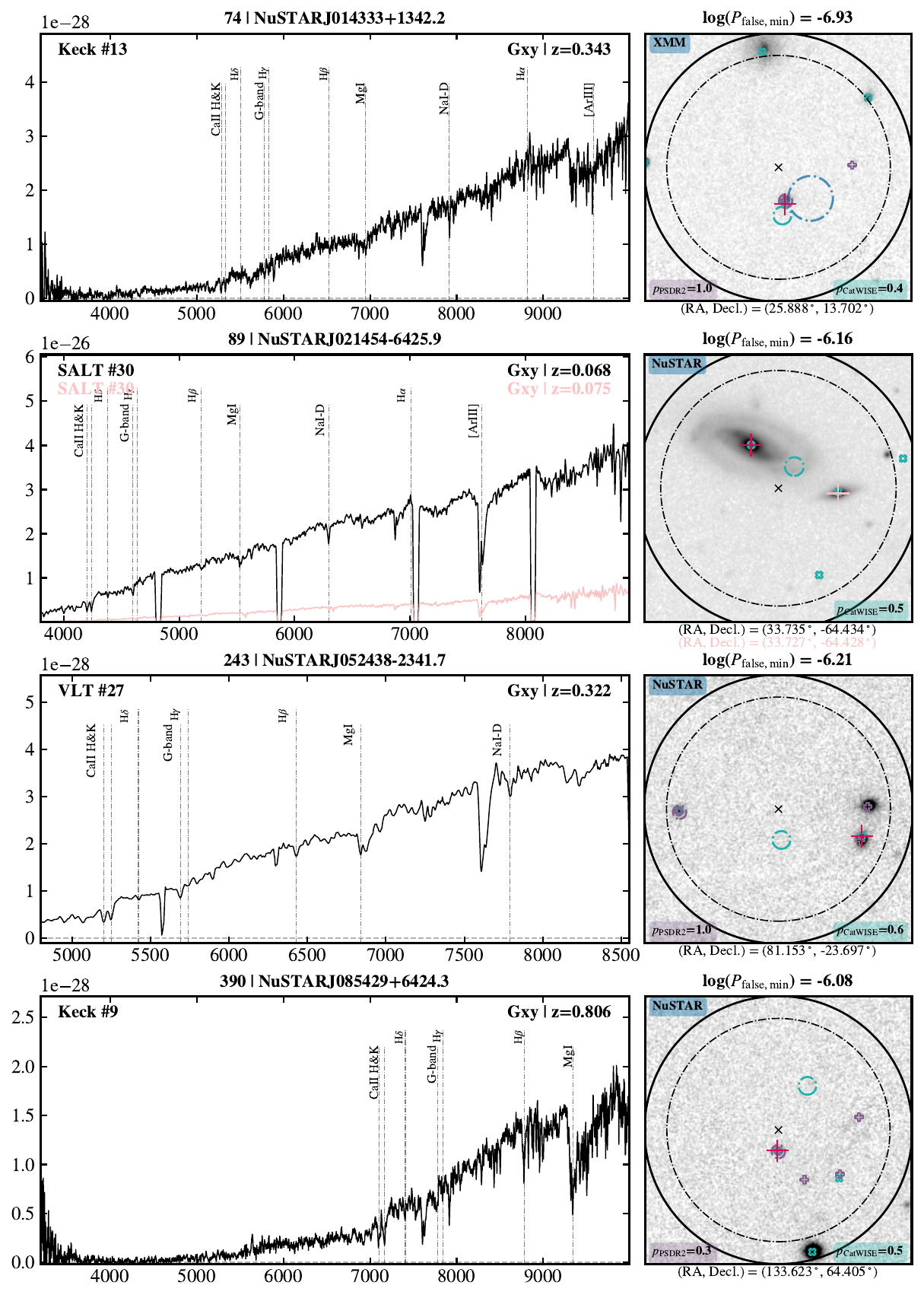}		
	{\textbf{Figure~E3.} Optical spectra for the first four extragalactic post-NSS40 sources classified as galaxies (i.e., absorption spectral features).
		The symbol key is described in Appendix~\ref{appendix:optical spectra}. }
	\label{fig:spec GXY}
\end{figure*}

\begin{figure*}
	\centering
	\includegraphics[width=40pc]{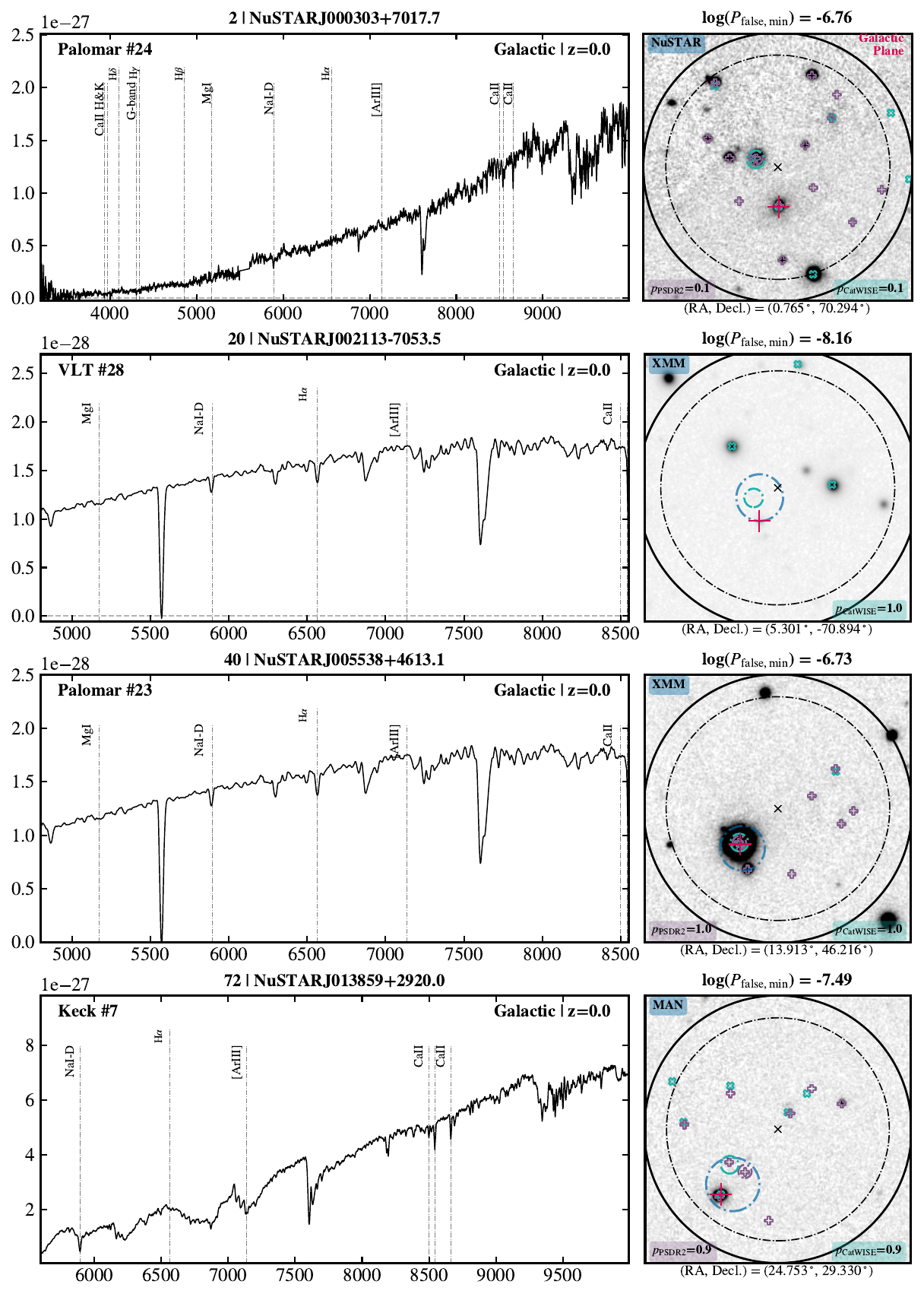}		
	{\textbf{Figure~E4.} Optical spectra for the first four Galactic post-NSS40 sources ($z = 0$). The symbol key is described in Appendix~\ref{appendix:optical spectra}. }
	\label{fig:spec Galactic}
\end{figure*}

\begin{figure*}
	\centering
	\includegraphics[width=40pc]{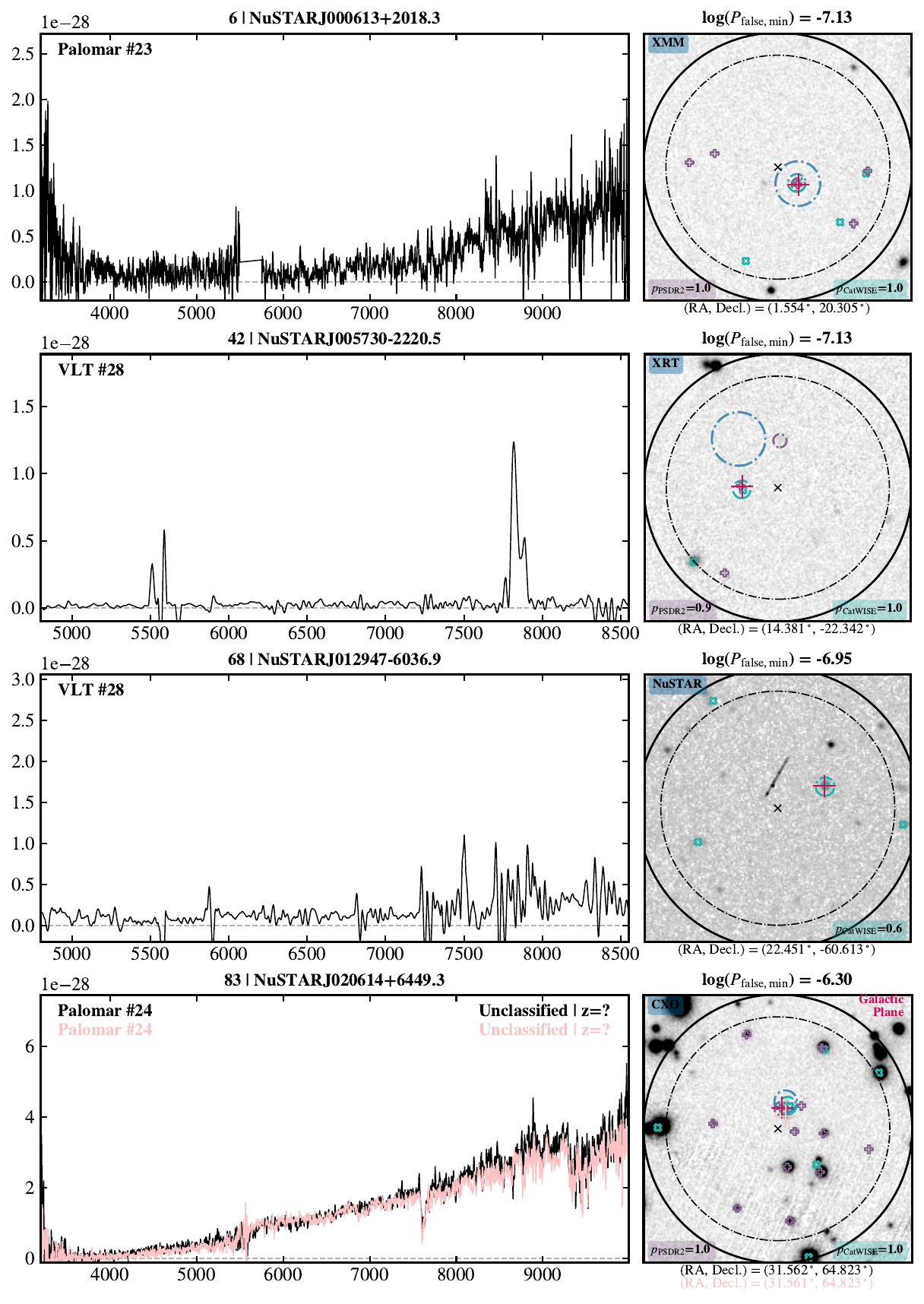}		
	{\textbf{Figure~E5.} Optical spectra for the first four unclassified post-NSS40 sources which lacks a redshift measurement; in some cases a red faint continuum is detected. The symbol key is described in Appendix~\ref{appendix:optical spectra}. } 
	\label{fig:spec Fcontinuum}
\end{figure*}

\begin{figure*}
	\centering
	\includegraphics[width=0.9\textwidth]{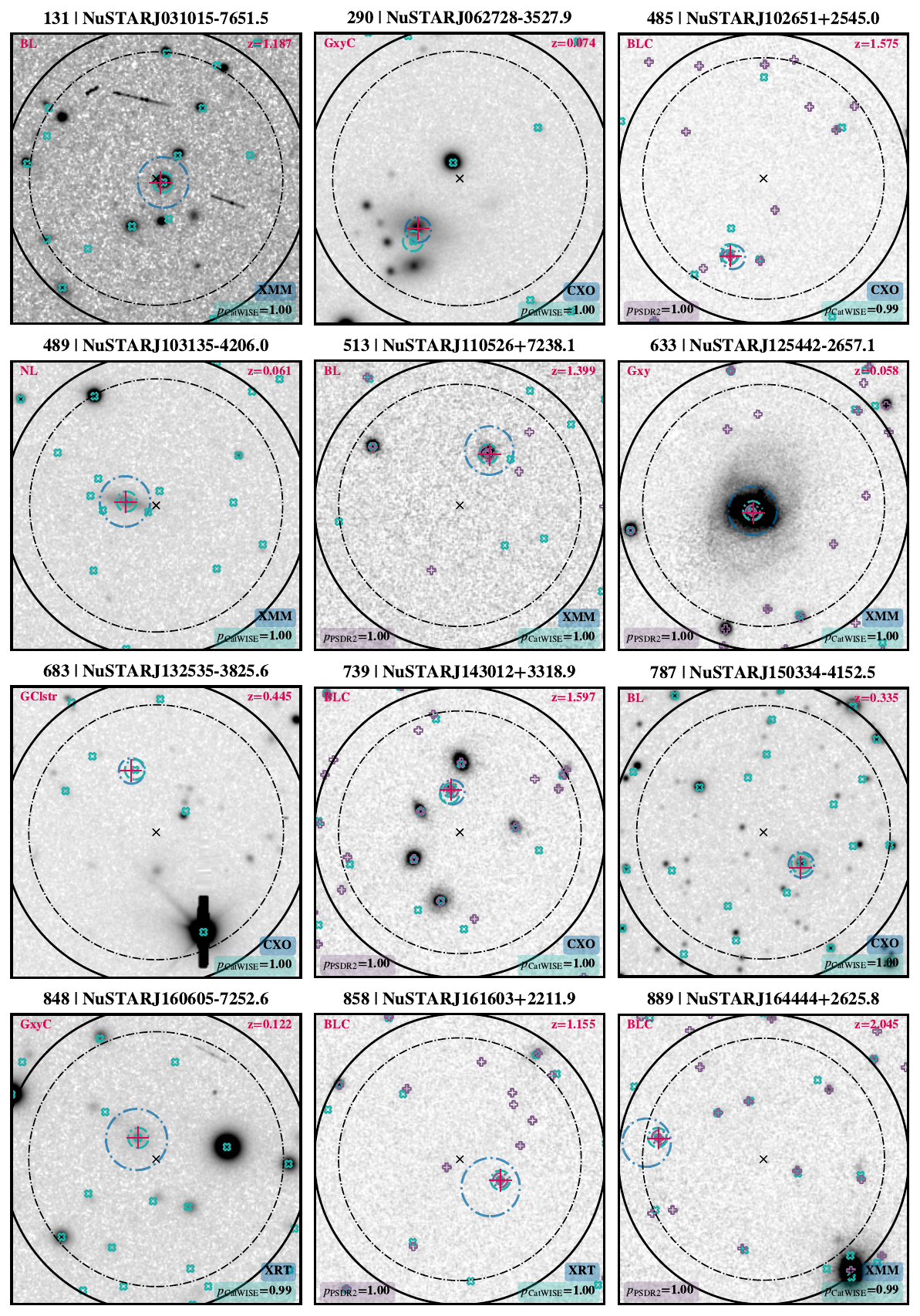}		
	{\textbf{Figure~E6.} 30$''$\,$\times$\,30$''$ Finding charts for the first 12 NSS80 sources with redshift measurements from literature.
		For sources with decl. $>$ -30$^\circ$ Pan-STARRS $i$-band imaging is used, and for sources with decl. $<$ -30$^\circ$ Astro Data Lab \citep{fitzpatrick_noao_2014, nikutta_data_2020} $i$-band imaging is retrieved. The symbol key is described in Appendix~\ref{appendix:optical spectra}. Optical classifications for sources with photometric redshifts are appended with a ``C" (i.e., Gxy\underline{C} = Galaxy \underline{C}andidate; BL\underline{C} = BL \underline{C}andidate). Also identified is a galaxy cluster (GClstr) at $z = 0.445$. } 
	\label{fig:nss80 lit findercharts}
\end{figure*}

\begin{figure*}
	\centering
	\includegraphics[width=0.9\textwidth]{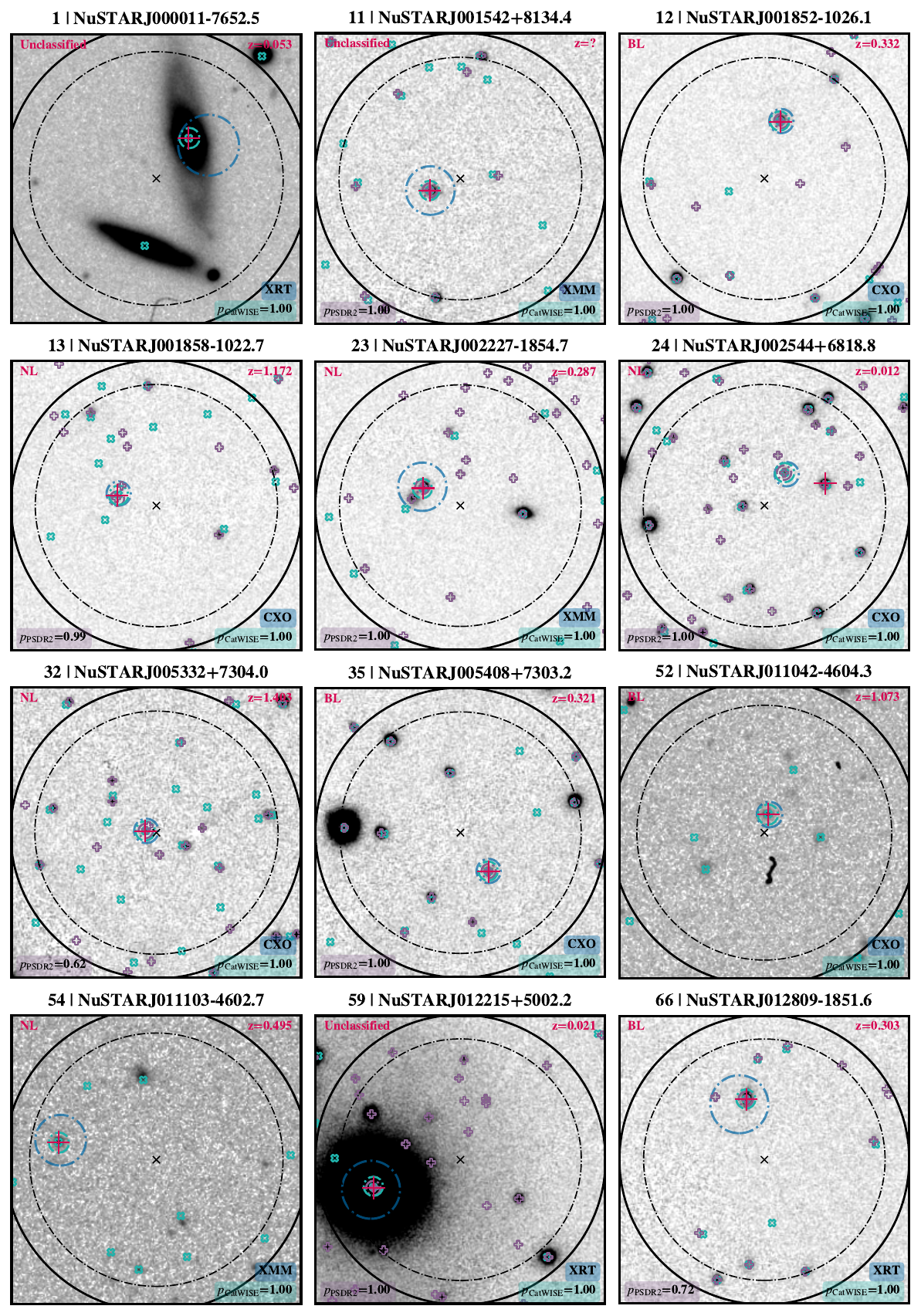}		
	{\textbf{Figure~E7.} 30$''$\,$\times$\,30$''$ Finding charts for the first 12 optically followed-up NSS40 sources reported in \citetalias{lansbury2017_cat}. 
		For sources with decl. $>$-30$^\circ$ Pan-STARRS $i$-band imaging is used, and for sources with decl. $<$ -30$^\circ$ Astro Data Lab $i$-band imaging is retrieved. The symbol key is described in Appendix~\ref{appendix:optical spectra}. } 
	\label{fig:nss40 findercharts}
\end{figure*}

\clearpage

\bibliography{ref} 

\begin{thebibliography}{}
\expandafter\ifx\csname natexlab\endcsname\relax\def\natexlab#1{#1}\fi
\providecommand{\url}[1]{\href{#1}{#1}}
\providecommand{\dodoi}[1]{doi:~\href{http://doi.org/#1}{\nolinkurl{#1}}}
\providecommand{\doeprint}[1]{\href{http://ascl.net/#1}{\nolinkurl{http://ascl.net/#1}}}
\providecommand{\doarXiv}[1]{\href{https://arxiv.org/abs/#1}{\nolinkurl{https://arxiv.org/abs/#1}}}

\bibitem[{{Abazajian} {et~al.}(2009){Abazajian}, {Adelman-McCarthy}, {Ag{\"u}eros}, {Allam}, \& et~al.}]{abazajian2009}
{Abazajian}, K.~N., {Adelman-McCarthy}, J.~K., {Ag{\"u}eros}, M.~A., {Allam}, S.~S., \& et~al. 2009, ApJS, 182, 543, \dodoi{10.1088/0067-0049/182/2/543}

\bibitem[{{Abbott} {et~al.}(2016){Abbott}, {Abbott}, {Abbott}, \& et~al.}]{abbott2016}
{Abbott}, B.~P., {Abbott}, R., {Abbott}, T.~D., \& et~al. 2016, Physical Review Letters, 116, 061102, \dodoi{10.1103/PhysRevLett.116.061102}

\bibitem[{{Abell}(1958)}]{abell1958}
{Abell}, G.~O. 1958, ApJS, 3, 211, \dodoi{10.1086/190036}

\bibitem[{{Abraham} {et~al.}(2004){Abraham}, {Glazebrook}, {McCarthy}, {Crampton}, {Murowinski}, {J{\o}rgensen}, {Roth}, {Hook}, {Savaglio}, {Chen}, {Marzke}, \& {Carlberg}}]{abraham2004}
{Abraham}, R.~G., {Glazebrook}, K., {McCarthy}, P.~J., {et~al.} 2004, AJ, 127, 2455, \dodoi{10.1086/383557}

\bibitem[{{Ackermann} {et~al.}(2016){Ackermann}, {Ajello}, {An}, {Baldini}, {Barbiellini}, {Bastieri}, {Bellazzini}, {Bissaldi}, {Blandford}, {Bonino}, {Bregeon}, {Britto}, {Bruel}, {Buehler}, {Caliandro}, {Cameron}, {Caragiulo}, {Caraveo}, {Cavazzuti}, {Cecchi}, {Charles}, {Chekhtman}, {Chiaro}, {Ciprini}, {Cohen-Tanugi}, {Costanza}, {Cutini}, {D'Ammando}, {de Angelis}, {de Palma}, {Desiante}, {Di Mauro}, {Di Venere}, {Dom{\'\i}nguez}, {Drell}, {Favuzzi}, {Fegan}, {Ferrara}, {Finke}, {Fusco}, {Gargano}, {Gasparrini}, {Giglietto}, {Giordano}, {Giroletti}, {Green}, {Grenier}, {Guiriec}, {Horan}, {J{\'o}hannesson}, {Katsuragawa}, {Kuss}, {Larsson}, {Latronico}, {Li}, {Li}, {Longo}, {Loparco}, {Lovellette}, {Lubrano}, {Magill}, {Maldera}, {Manfreda}, {Mayer}, {Mazziotta}, {Michelson}, {Mirabal}, {Mitthumsiri}, {Mizuno}, {Monzani}, {Morselli}, {Moskalenko}, {Negro}, {Nuss}, {Ohsugi}, {Okada}, {Orlando}, {Paneque}, {Pesce-Rollins}, {Piron}, {Pivato}, {Porter}, {Rain{\`o}}, {Rando}, {Razzano}, {Reimer}, {Rau},
  {Romani}, {Schady}, {Sgr{\`o}}, {Simone}, {Siskind}, {Spada}, {Spandre}, {Spinelli}, {Stern}, {Takahashi}, {Thayer}, {Torres}, {Tosti}, {Troja}, {Vianello}, {Wood}, \& {Wood}}]{ackermann_contemporaneous_2016}
{Ackermann}, M., {Ajello}, M., {An}, H., {et~al.} 2016, \apj, 820, 72, \dodoi{10.3847/0004-637X/820/1/72}

\bibitem[{{Adelman-McCarthy} {et~al.}(2008){Adelman-McCarthy}, {Ag{\"u}eros}, {Allam}, \& et~al.}]{adelmanmccarthy2008}
{Adelman-McCarthy}, J.~K., {Ag{\"u}eros}, M.~A., {Allam}, S.~S., \& et~al. 2008, ApJS, 175, 297, \dodoi{10.1086/524984}

\bibitem[{{Ahn} {et~al.}(2012){Ahn}, {Alexandroff}, {Allende Prieto}, {Anderson}, \& et~al.}]{ahn2012}
{Ahn}, C.~P., {Alexandroff}, R., {Allende Prieto}, C., {Anderson}, S.~F., \& et~al. 2012, ApJS, 203, 21, \dodoi{10.1088/0067-0049/203/2/21}

\bibitem[{{Ahumada} {et~al.}(2020){Ahumada}, {Allende Prieto}, {Almeida}, {Anders}, {Anderson}, {Andrews}, {Anguiano}, {Arcodia}, {Armengaud}, {Aubert}, {Avila}, {Avila-Reese}, {Badenes}, {Balland}, {Barger}, {Barrera-Ballesteros}, {Basu}, {Bautista}, {Beaton}, {Beers}, {Benavides}, {Bender}, {Bernardi}, {Bershady}, {Beutler}, {Bidin}, {Bird}, {Bizyaev}, {Blanc}, {Blanton}, {Boquien}, {Borissova}, {Bovy}, {Brandt}, {Brinkmann}, {Brownstein}, {Bundy}, {Bureau}, {Burgasser}, {Burtin}, {Cano-D{\'\i}az}, {Capasso}, {Cappellari}, {Carrera}, {Chabanier}, {Chaplin}, {Chapman}, {Cherinka}, {Chiappini}, {Doohyun Choi}, {Chojnowski}, {Chung}, {Clerc}, {Coffey}, {Comerford}, {Comparat}, {da Costa}, {Cousinou}, {Covey}, {Crane}, {Cunha}, {Ilha}, {Dai}, {Damsted}, {Darling}, {Davidson}, {Davies}, {Dawson}, {De}, {de la Macorra}, {De Lee}, {Queiroz}, {Deconto Machado}, {de la Torre}, {Dell'Agli}, {du Mas des Bourboux}, {Diamond-Stanic}, {Dillon}, {Donor}, {Drory}, {Duckworth}, {Dwelly}, {Ebelke}, {Eftekharzadeh}, {Davis
  Eigenbrot}, {Elsworth}, {Eracleous}, {Erfanianfar}, {Escoffier}, {Fan}, {Farr}, {Fern{\'a}ndez-Trincado}, {Feuillet}, {Finoguenov}, {Fofie}, {Fraser-McKelvie}, {Frinchaboy}, {Fromenteau}, {Fu}, {Galbany}, {Garcia}, {Garc{\'\i}a-Hern{\'a}ndez}, {Garma Oehmichen}, {Ge}, {Geimba Maia}, {Geisler}, {Gelfand}, {Goddy}, {Gonzalez-Perez}, {Grabowski}, {Green}, {Grier}, {Guo}, {Guy}, {Harding}, {Hasselquist}, {Hawken}, {Hayes}, {Hearty}, {Hekker}, {Hogg}, {Holtzman}, {Horta}, {Hou}, {Hsieh}, {Huber}, {Hunt}, {Ider Chitham}, {Imig}, {Jaber}, {Jimenez Angel}, {Johnson}, {Jones}, {J{\"o}nsson}, {Jullo}, {Kim}, {Kinemuchi}, {Kirkpatrick}, {Kite}, {Klaene}, {Kneib}, {Kollmeier}, {Kong}, {Kounkel}, {Krishnarao}, {Lacerna}, {Lan}, {Lane}, {Law}, {Le Goff}, {Leung}, {Lewis}, {Li}, {Lian}, {Lin}, {Long}, {Longa-Pe{\~n}a}, {Lundgren}, {Lyke}, {Mackereth}, {MacLeod}, {Majewski}, {Manchado}, {Maraston}, {Martini}, {Masseron}, {Masters}, {Mathur}, {McDermid}, {Merloni}, {Merrifield}, {M{\'e}sz{\'a}ros}, {Miglio}, {Minniti},
  {Minsley}, {Miyaji}, {Mohammad}, {Mosser}, {Mueller}, {Muna}, {Mu{\~n}oz-Guti{\'e}rrez}, {Myers}, {Nadathur}, {Nair}, {Nandra}, {Correa do Nascimento}, {Nevin}, {Newman}, {Nidever}, {Nitschelm}, {Noterdaeme}, {O'Connell}, {Olmstead}, {Oravetz}, {Oravetz}, {Osorio}, {Pace}, {Padilla}, {Palanque-Delabrouille}, {Palicio}, {Pan}, {Pan}, {Parker}, {Paviot}, {Peirani}, {Ram{\'r}ez}, {Penny}, {Percival}, {Perez-Fournon}, {P{\'e}rez-R{\`a}fols}, {Petitjean}, {Pieri}, {Pinsonneault}, {Poovelil}, {Povick}, {Prakash}, {Price-Whelan}, {Raddick}, {Raichoor}, {Ray}, {Rembold}, {Rezaie}, {Riffel}, {Riffel}, {Rix}, {Robin}, {Roman-Lopes}, {Rom{\'a}n-Z{\'u}{\~n}iga}, {Rose}, {Ross}, {Rossi}, {Rowlands}, {Rubin}, {Salvato}, {S{\'a}nchez}, {S{\'a}nchez-Menguiano}, {S{\'a}nchez-Gallego}, {Sayres}, {Schaefer}, {Schiavon}, {Schimoia}, {Schlafly}, {Schlegel}, {Schneider}, {Schultheis}, {Schwope}, {Seo}, {Serenelli}, {Shafieloo}, {Shamsi}, {Shao}, {Shen}, {Shetrone}, {Shirley}, {Silva Aguirre}, {Simon}, {Skrutskie}, {Slosar},
  {Smethurst}, {Sobeck}, {Sodi}, {Souto}, {Stark}, {Stassun}, {Steinmetz}, {Stello}, {Stermer}, {Storchi-Bergmann}, {Streblyanska}, {Stringfellow}, {Stutz}, {Su{\'a}rez}, {Sun}, {Taghizadeh-Popp}, {Talbot}, {Tayar}, {Thakar}, {Theriault}, {Thomas}, {Thomas}, {Tinker}, {Tojeiro}, {Toledo}, {Tremonti}, {Troup}, {Tuttle}, {Unda-Sanzana}, {Valentini}, {Vargas-Gonz{\'a}lez}, {Vargas-Maga{\~n}a}, {V{\'a}zquez-Mata}, {Vivek}, {Wake}, {Wang}, {Weaver}, {Weijmans}, {Wild}, {Wilson}, {Wilson}, {Wolthuis}, {Wood-Vasey}, {Yan}, {Yang}, {Y{\`e}che}, {Zamora}, {Zarrouk}, {Zasowski}, {Zhang}, {Zhao}, {Zhao}, {Zheng}, {Zheng}, {Zhu}, \& {Zou}}]{ahumada_16th_2020}
{Ahumada}, R., {Allende Prieto}, C., {Almeida}, A., {et~al.} 2020, \apjs, 249, 3, \dodoi{10.3847/1538-4365/ab929e}

\bibitem[{{Aird} {et~al.}(2015){Aird}, {Alexander}, {Ballantyne}, {Civano}, {Del-Moro}, {Hickox}, {Lansbury}, {Mullaney}, {Bauer}, {Brandt}, {Comastri}, {Fabian}, {Gandhi}, {Harrison}, {Luo}, {Stern}, {Treister}, {Zappacosta}, {Ajello}, {Assef}, {Balokovi{\'c}}, {Boggs}, {Brightman}, {Christensen}, {Craig}, {Elvis}, {Forster}, {Grefenstette}, {Hailey}, {Koss}, {LaMassa}, {Madsen}, {Puccetti}, {Saez}, {Urry}, {Wik}, \& {Zhang}}]{aird2015}
{Aird}, J., {Alexander}, D.~M., {Ballantyne}, D.~R., {et~al.} 2015, ApJ, 815, 66, \dodoi{10.1088/0004-637X/815/1/66}

\bibitem[{{Akiyama} {et~al.}(2003){Akiyama}, {Ueda}, {Ohta}, {Takahashi}, \& {Yamada}}]{akiyama2003}
{Akiyama}, M., {Ueda}, Y., {Ohta}, K., {Takahashi}, T., \& {Yamada}, T. 2003, ApJS, 148, 275, \dodoi{10.1086/376441}

\bibitem[{{Alam} {et~al.}(2015){Alam}, {Albareti}, {Allende Prieto}, {Anders}, {Anderson}, \& et~al.}]{alam2015}
{Alam}, S., {Albareti}, F.~D., {Allende Prieto}, C., {et~al.} 2015, ApJS, 219, 12, \dodoi{10.1088/0067-0049/219/1/12}

\bibitem[{{Albareti} {et~al.}(2017){Albareti}, {Allende Prieto}, {Almeida}, {Anders}, \& et~al.}]{albareti2017}
{Albareti}, F.~D., {Allende Prieto}, C., {Almeida}, A., {Anders}, F., \& et~al. 2017, ApJS, 233, 25, \dodoi{10.3847/1538-4365/aa8992}

\bibitem[{{Alexander} {et~al.}(2003){Alexander}, {Bauer}, {Brandt}, {Hornschemeier}, {Vignali}, {Garmire}, {Schneider}, {Chartas}, \& {Gallagher}}]{alexander2003}
{Alexander}, D.~M., {Bauer}, F.~E., {Brandt}, W.~N., {et~al.} 2003, \aj, 125, 383, \dodoi{10.1086/346088}

\bibitem[{{Alexander} {et~al.}(2013){Alexander}, {Stern}, {Del Moro}, {Lansbury}, {Assef}, {Aird}, {Ajello}, {Ballantyne}, {Bauer}, {Boggs}, {Brandt}, {Christensen}, {Civano}, {Comastri}, {Craig}, {Elvis}, {Grefenstette}, {Hailey}, {Harrison}, {Hickox}, {Luo}, {Madsen}, {Mullaney}, {Perri}, {Puccetti}, {Saez}, {Treister}, {Urry}, {Zhang}, {Bridge}, {Eisenhardt}, {Gonzalez}, {Miller}, \& {Tsai}}]{alexander2013}
{Alexander}, D.~M., {Stern}, D., {Del Moro}, A., {et~al.} 2013, ApJ, 773, 125, \dodoi{10.1088/0004-637X/773/2/125}

\bibitem[{{Ananna} {et~al.}(2020){Ananna}, {Treister}, {Urry}, {Ricci}, {Hickox}, {Padmanabhan}, {Marchesi}, \& {Kirkpatrick}}]{accretion_ananna_2020}
{Ananna}, T.~T., {Treister}, E., {Urry}, C.~M., {et~al.} 2020, \apj, 889, 17, \dodoi{10.3847/1538-4357/ab5aef}

\bibitem[{{Arp}(1966)}]{Halton1966}
{Arp}, H. 1966, ApJS, 14, 1, \dodoi{10.1086/190147}

\bibitem[{{Assef} {et~al.}(2018){Assef}, {Stern}, {Noirot}, {Jun}, {Cutri}, \& {Eisenhardt}}]{assef_wise_2018}
{Assef}, R.~J., {Stern}, D., {Noirot}, G., {et~al.} 2018, \apjs, 234, 23, \dodoi{10.3847/1538-4365/aaa00a}

\bibitem[{{Assef} {et~al.}(2010){Assef}, {Kochanek}, {Brodwin}, {Cool}, {Forman}, {Gonzalez}, {Hickox}, {Jones}, {Le Floc'h}, {Moustakas}, {Murray}, \& {Stern}}]{assef2010}
{Assef}, R.~J., {Kochanek}, C.~S., {Brodwin}, M., {et~al.} 2010, ApJ, 713, 970, \dodoi{10.1088/0004-637X/713/2/970}

\bibitem[{{Assef} {et~al.}(2013){Assef}, {Stern}, {Kochanek}, {Blain}, {Brodwin}, {Brown}, {Donoso}, {Eisenhardt}, {Jannuzi}, {Jarrett}, {Stanford}, {Tsai}, {Wu}, \& {Yan}}]{assef2013}
{Assef}, R.~J., {Stern}, D., {Kochanek}, C.~S., {et~al.} 2013, ApJ, 772, 26.
\newblock \doarXiv{1209.6055}

\bibitem[{{Astropy Collaboration} {et~al.}(2013){Astropy Collaboration}, {Robitaille}, {Tollerud}, {Greenfield}, {Droettboom}, {Bray}, {Aldcroft}, {Davis}, {Ginsburg}, {Price-Whelan}, {Kerzendorf}, {Conley}, {Crighton}, {Barbary}, {Muna}, {Ferguson}, {Grollier}, {Parikh}, {Nair}, {Unther}, {Deil}, {Woillez}, {Conseil}, {Kramer}, {Turner}, {Singer}, {Fox}, {Weaver}, {Zabalza}, {Edwards}, {Azalee Bostroem}, {Burke}, {Casey}, {Crawford}, {Dencheva}, {Ely}, {Jenness}, {Labrie}, {Lim}, {Pierfederici}, {Pontzen}, {Ptak}, {Refsdal}, {Servillat}, \& {Streicher}}]{astropy:2013}
{Astropy Collaboration}, {Robitaille}, T.~P., {Tollerud}, E.~J., {et~al.} 2013, \aap, 558, A33, \dodoi{10.1051/0004-6361/201322068}

\bibitem[{{Astropy Collaboration} {et~al.}(2018){Astropy Collaboration}, {Price-Whelan}, {Sip{\H{o}}cz}, {G{\"u}nther}, {Lim}, {Crawford}, {Conseil}, {Shupe}, {Craig}, {Dencheva}, {Ginsburg}, {Vand erPlas}, {Bradley}, {P{\'e}rez-Su{\'a}rez}, {de Val-Borro}, {Aldcroft}, {Cruz}, {Robitaille}, {Tollerud}, {Ardelean}, {Babej}, {Bach}, {Bachetti}, {Bakanov}, {Bamford}, {Barentsen}, {Barmby}, {Baumbach}, {Berry}, {Biscani}, {Boquien}, {Bostroem}, {Bouma}, {Brammer}, {Bray}, {Breytenbach}, {Buddelmeijer}, {Burke}, {Calderone}, {Cano Rodr{\'\i}guez}, {Cara}, {Cardoso}, {Cheedella}, {Copin}, {Corrales}, {Crichton}, {D'Avella}, {Deil}, {Depagne}, {Dietrich}, {Donath}, {Droettboom}, {Earl}, {Erben}, {Fabbro}, {Ferreira}, {Finethy}, {Fox}, {Garrison}, {Gibbons}, {Goldstein}, {Gommers}, {Greco}, {Greenfield}, {Groener}, {Grollier}, {Hagen}, {Hirst}, {Homeier}, {Horton}, {Hosseinzadeh}, {Hu}, {Hunkeler}, {Ivezi{\'c}}, {Jain}, {Jenness}, {Kanarek}, {Kendrew}, {Kern}, {Kerzendorf}, {Khvalko}, {King}, {Kirkby}, {Kulkarni},
  {Kumar}, {Lee}, {Lenz}, {Littlefair}, {Ma}, {Macleod}, {Mastropietro}, {McCully}, {Montagnac}, {Morris}, {Mueller}, {Mumford}, {Muna}, {Murphy}, {Nelson}, {Nguyen}, {Ninan}, {N{\"o}the}, {Ogaz}, {Oh}, {Parejko}, {Parley}, {Pascual}, {Patil}, {Patil}, {Plunkett}, {Prochaska}, {Rastogi}, {Reddy Janga}, {Sabater}, {Sakurikar}, {Seifert}, {Sherbert}, {Sherwood-Taylor}, {Shih}, {Sick}, {Silbiger}, {Singanamalla}, {Singer}, {Sladen}, {Sooley}, {Sornarajah}, {Streicher}, {Teuben}, {Thomas}, {Tremblay}, {Turner}, {Terr{\'o}n}, {van Kerkwijk}, {de la Vega}, {Watkins}, {Weaver}, {Whitmore}, {Woillez}, {Zabalza}, \& {Astropy Contributors}}]{astropy:2018}
{Astropy Collaboration}, {Price-Whelan}, A.~M., {Sip{\H{o}}cz}, B.~M., {et~al.} 2018, \aj, 156, 123, \dodoi{10.3847/1538-3881/aabc4f}

\bibitem[{{Astropy Collaboration} {et~al.}(2022){Astropy Collaboration}, {Price-Whelan}, {Lim}, {Earl}, {Starkman}, {Bradley}, {Shupe}, {Patil}, {Corrales}, {Brasseur}, {N{"o}the}, {Donath}, {Tollerud}, {Morris}, {Ginsburg}, {Vaher}, {Weaver}, {Tocknell}, {Jamieson}, {van Kerkwijk}, {Robitaille}, {Merry}, {Bachetti}, {G{"u}nther}, {Aldcroft}, {Alvarado-Montes}, {Archibald}, {B{'o}di}, {Bapat}, {Barentsen}, {Baz{'a}n}, {Biswas}, {Boquien}, {Burke}, {Cara}, {Cara}, {Conroy}, {Conseil}, {Craig}, {Cross}, {Cruz}, {D'Eugenio}, {Dencheva}, {Devillepoix}, {Dietrich}, {Eigenbrot}, {Erben}, {Ferreira}, {Foreman-Mackey}, {Fox}, {Freij}, {Garg}, {Geda}, {Glattly}, {Gondhalekar}, {Gordon}, {Grant}, {Greenfield}, {Groener}, {Guest}, {Gurovich}, {Handberg}, {Hart}, {Hatfield-Dodds}, {Homeier}, {Hosseinzadeh}, {Jenness}, {Jones}, {Joseph}, {Kalmbach}, {Karamehmetoglu}, {Ka{l}uszy{'n}ski}, {Kelley}, {Kern}, {Kerzendorf}, {Koch}, {Kulumani}, {Lee}, {Ly}, {Ma}, {MacBride}, {Maljaars}, {Muna}, {Murphy}, {Norman}, {O'Steen},
  {Oman}, {Pacifici}, {Pascual}, {Pascual-Granado}, {Patil}, {Perren}, {Pickering}, {Rastogi}, {Roulston}, {Ryan}, {Rykoff}, {Sabater}, {Sakurikar}, {Salgado}, {Sanghi}, {Saunders}, {Savchenko}, {Schwardt}, {Seifert-Eckert}, {Shih}, {Jain}, {Shukla}, {Sick}, {Simpson}, {Singanamalla}, {Singer}, {Singhal}, {Sinha}, {Sip{H{o}}cz}, {Spitler}, {Stansby}, {Streicher}, {{{S}}umak}, {Swinbank}, {Taranu}, {Tewary}, {Tremblay}, {Val-Borro}, {Van Kooten}, {Vasovi{'c}}, {Verma}, {de Miranda Cardoso}, {Williams}, {Wilson}, {Winkel}, {Wood-Vasey}, {Xue}, {Yoachim}, {Zhang}, {Zonca}, \& {Astropy Project Contributors}}]{astropy:2022}
{Astropy Collaboration}, {Price-Whelan}, A.~M., {Lim}, P.~L., {et~al.} 2022, \apj, 935, 167, \dodoi{10.3847/1538-4357/ac7c74}

\bibitem[{{Balokovi{\'c}} {et~al.}(2014){Balokovi{\'c}}, {Comastri}, {Harrison}, {Alexander}, {Ballantyne}, {Bauer}, {Boggs}, {Brandt}, {Brightman}, {Christensen}, {Craig}, {Del Moro}, {Gandhi}, {Hailey}, {Koss}, {Lansbury}, {Luo}, {Madejski}, {Marinucci}, {Matt}, {Markwardt}, {Puccetti}, {Reynolds}, {Risaliti}, {Rivers}, {Stern}, {Walton}, \& {Zhang}}]{balokovic2014}
{Balokovi{\'c}}, M., {Comastri}, A., {Harrison}, F.~A., {et~al.} 2014, ApJ, 794, 111, \dodoi{10.1088/0004-637X/794/2/111}

\bibitem[{{Banerji} {et~al.}(2012){Banerji}, {McMahon}, { }, {Alaghband-Zadeh}, {Gonzalez-Solares}, {Venemans}, \& {Hawthorn}}]{banerji2012}
{Banerji}, M., {McMahon}, R.~G., { }, P.~C., {et~al.} 2012, MNRAS, 427, 2275.
\newblock \doarXiv{1203.5530}

\bibitem[{{Bauer} {et~al.}(2000){Bauer}, {Condon}, {Thuan}, \& {Broderick}}]{bauer2000}
{Bauer}, F.~E., {Condon}, J.~J., {Thuan}, T.~X., \& {Broderick}, J.~J. 2000, ApJS, 129, 547, \dodoi{10.1086/313425}

\bibitem[{{Baumgartner} {et~al.}(2013){Baumgartner}, {Tueller}, {Markwardt}, {Skinner}, {Barthelmy}, {Mushotzky}, {Evans}, \& {Gehrels}}]{baumgartner_70_2013}
{Baumgartner}, W.~H., {Tueller}, J., {Markwardt}, C.~B., {et~al.} 2013, \apjs, 207, 19, \dodoi{10.1088/0067-0049/207/2/19}

\bibitem[{{Bertin} \& {Arnouts}(1996)}]{bertin1996}
{Bertin}, E., \& {Arnouts}, S. 1996, A\&AS, 117, 393, \dodoi{10.1051/aas:1996164}

\bibitem[{{Bilicki} {et~al.}(2014){Bilicki}, {Jarrett}, {Peacock}, {Cluver}, \& {Steward}}]{bilicki2014}
{Bilicki}, M., {Jarrett}, T.~H., {Peacock}, J.~A., {Cluver}, M.~E., \& {Steward}, L. 2014, ApJS, 210, 9, \dodoi{10.1088/0067-0049/210/1/9}

\bibitem[{{Blaauw} {et~al.}(1960){Blaauw}, {Gum}, {Pawsey}, \& {Westerhout}}]{blaauw1960}
{Blaauw}, A., {Gum}, C.~S., {Pawsey}, J.~L., \& {Westerhout}, G. 1960, MNRAS, 121, 123, \dodoi{10.1093/mnras/121.2.123}

\bibitem[{{Bolton} {et~al.}(2012){Bolton}, {Schlegel}, {Aubourg}, {Bailey}, {Bhardwaj}, {Brownstein}, {Burles}, {Chen}, {Dawson}, {Eisenstein}, {Gunn}, {Knapp}, {Loomis}, {Lupton}, {Maraston}, {Muna}, {Myers}, {Olmstead}, {Padmanabhan}, {P{\^a}ris}, {Percival}, {Petitjean}, {Rockosi}, {Ross}, {Schneider}, {Shu}, {Strauss}, {Thomas}, {Tremonti}, {Wake}, {Weaver}, \& {Wood-Vasey}}]{bolton2012}
{Bolton}, A.~S., {Schlegel}, D.~J., {Aubourg}, {\'E}., {et~al.} 2012, AJ, 144, 144, \dodoi{10.1088/0004-6256/144/5/144}

\bibitem[{{Brandt} \& {Alexander}(2015)}]{brandt2015}
{Brandt}, W.~N., \& {Alexander}, D.~M. 2015, A\&ARv, 23, 1, \dodoi{10.1007/s00159-014-0081-z}

\bibitem[{{Brandt} \& {Yang}(2021)}]{brandt2021}
{Brandt}, W.~N., \& {Yang}, G. 2021, arXiv e-prints, arXiv:2111.01156.
\newblock \doarXiv{2111.01156}

\bibitem[{{Brunner} {et~al.}(2022){Brunner}, {Liu}, {Lamer}, {Georgakakis}, {Merloni}, {Brusa}, {Bulbul}, {Dennerl}, {Friedrich}, {Liu}, {Maitra}, {Nandra}, {Ramos-Ceja}, {Sanders}, {Stewart}, {Boller}, {Buchner}, {Clerc}, {Comparat}, {Dwelly}, {Eckert}, {Finoguenov}, {Freyberg}, {Ghirardini}, {Gueguen}, {Haberl}, {Kreykenbohm}, {Krumpe}, {Osterhage}, {Pacaud}, {Predehl}, {Reiprich}, {Robrade}, {Salvato}, {Santangelo}, {Schrabback}, {Schwope}, \& {Wilms}}]{brunner_erosita_2022}
{Brunner}, H., {Liu}, T., {Lamer}, G., {et~al.} 2022, \aap, 661, A1, \dodoi{10.1051/0004-6361/202141266}

\bibitem[{{Burke} {et~al.}(2003){Burke}, {Collins}, {Sharples}, {Romer}, \& {Nichol}}]{burke2003}
{Burke}, D.~J., {Collins}, C.~A., {Sharples}, R.~M., {Romer}, A.~K., \& {Nichol}, R.~C. 2003, MNRAS, 341, 1093, \dodoi{10.1046/j.1365-8711.2003.06378.x}

\bibitem[{{Burlon} {et~al.}(2011){Burlon}, {Ajello}, {Greiner}, {Comastri}, {Merloni}, \& {Gehrels}}]{burlon2011}
{Burlon}, D., {Ajello}, M., {Greiner}, J., {et~al.} 2011, ApJ, 728, 58, \dodoi{10.1088/0004-637X/728/1/58}

\bibitem[{{Caccianiga} {et~al.}(2008){Caccianiga}, {Severgnini}, {Della Ceca}, {Maccacaro}, {Cocchia}, {Barcons}, {Carrera}, {Matute}, {McMahon}, {Page}, {Pietsch}, {Sbarufatti}, {Schwope}, {Tedds}, \& {Watson}}]{caccianiga2008}
{Caccianiga}, A., {Severgnini}, P., {Della Ceca}, R., {et~al.} 2008, A\&A, 477, 735, \dodoi{10.1051/0004-6361:20078568}

\bibitem[{{Calistro Rivera} {et~al.}(2021){Calistro Rivera}, {Alexander}, {Rosario}, {Harrison}, {Stalevski}, {Rakshit}, {Fawcett}, {Morabito}, {Klindt}, {Best}, {Bonato}, {Bowler}, {Costa}, \& {Kondapally}}]{calistrorivera2021}
{Calistro Rivera}, G., {Alexander}, D.~M., {Rosario}, D.~J., {et~al.} 2021, A\&A, 649, A102, \dodoi{10.1051/0004-6361/202040214}

\bibitem[{{Cappelluti} {et~al.}(2017){Cappelluti}, {Li}, {Ricarte}, {Agarwal}, {Allevato}, {Tasnim Ananna}, {Ajello}, {Civano}, {Comastri}, {Elvis}, {Finoguenov}, {Gilli}, {Hasinger}, {Marchesi}, {Natarajan}, {Pacucci}, {Treister}, \& {Urry}}]{cappelluti2017}
{Cappelluti}, N., {Li}, Y., {Ricarte}, A., {et~al.} 2017, ApJ, 837, 19, \dodoi{10.3847/1538-4357/aa5ea4}

\bibitem[{{Cardamone} {et~al.}(2008){Cardamone}, {Urry}, {Damen}, {van Dokkum}, {Treister}, {Labb{\'e}}, {Virani}, {Lira}, \& {Gawiser}}]{cardamone2008}
{Cardamone}, C.~N., {Urry}, C.~M., {Damen}, M., {et~al.} 2008, ApJ, 680, 130, \dodoi{10.1086/587800}

\bibitem[{{Chang} {et~al.}(2019){Chang}, {Arsioli}, {Giommi}, {Padovani}, \& {Brandt}}]{chang2019}
{Chang}, Y.~L., {Arsioli}, B., {Giommi}, P., {Padovani}, P., \& {Brandt}, C.~H. 2019, A\&A, 632, A77, \dodoi{10.1051/0004-6361/201834526}

\bibitem[{{Chu} {et~al.}(1998){Chu}, {Wei}, {Hu}, {Zhu}, \& {Arp}}]{chu1998}
{Chu}, Y., {Wei}, J., {Hu}, J., {Zhu}, X., \& {Arp}, H. 1998, ApJ, 500, 596, \dodoi{10.1086/305779}

\bibitem[{{Civano} {et~al.}(2015){Civano}, {Hickox}, {Puccetti}, {Comastri}, {Mullaney}, {Zappacosta}, {LaMassa}, {Aird}, {Alexander}, {Ballantyne}, {Bauer}, {Brandt}, {Boggs}, {Christensen}, {Craig}, {Del-Moro}, {Elvis}, {Forster}, {Gandhi}, {Grefenstette}, {Hailey}, {Harrison}, {Lansbury}, {Luo}, {Madsen}, {Saez}, {Stern}, {Treister}, {Urry}, {Wik}, \& {Zhang}}]{civano2015}
{Civano}, F., {Hickox}, R.~C., {Puccetti}, S., {et~al.} 2015, ApJ, 808, 185, \dodoi{10.1088/0004-637X/808/2/185}

\bibitem[{{Colless} {et~al.}(2001){Colless}, {Dalton}, {Maddox}, {Sutherland}, {Norberg}, {Cole}, {Bland-Hawthorn}, {Bridges}, {Cannon}, {Collins}, {Couch}, {Cross}, {Deeley}, {De Propris}, {Driver}, {Efstathiou}, {Ellis}, {Frenk}, {Glazebrook}, {Jackson}, {Lahav}, {Lewis}, {Lumsden}, {Madgwick}, {Peacock}, {Peterson}, {Price}, {Seaborne}, \& {Taylor}}]{colless2001}
{Colless}, M., {Dalton}, G., {Maddox}, S., {et~al.} 2001, MNRAS, 328, 1039, \dodoi{10.1046/j.1365-8711.2001.04902.x}

\bibitem[{{da Costa} {et~al.}(1998){da Costa}, {Willmer}, {Pellegrini}, {Chaves}, {Rit{\'e}}, {Maia}, {Geller}, {Latham}, {Kurtz}, {Huchra}, {Ramella}, {Fairall}, {Smith}, \& {L{\'\i}pari}}]{dacosta1998}
{da Costa}, L.~N., {Willmer}, C.~N.~A., {Pellegrini}, P.~S., {et~al.} 1998, AJ, 116, 1, \dodoi{10.1086/300410}

\bibitem[{{Darling} \& {Giovanelli}(2006)}]{darling2006}
{Darling}, J., \& {Giovanelli}, R. 2006, AJ, 132, 2596, \dodoi{10.1086/508513}

\bibitem[{{De Luca} \& {Molendi}(2004)}]{deluca2004}
{De Luca}, A., \& {Molendi}, S. 2004, A\&A, 419, 837, \dodoi{10.1051/0004-6361:20034421}

\bibitem[{{de Vaucouleurs} {et~al.}(1995){de Vaucouleurs}, {de Vaucouleurs}, {Corwin}, {Buta}, {Paturel}, \& {Fouque}}]{deVaucouleurs1991}
{de Vaucouleurs}, G., {de Vaucouleurs}, A., {Corwin}, H.~G., {et~al.} 1995, VizieR Online Data Catalog, VII/155

\bibitem[{{Delaney} {et~al.}(2023){Delaney}, {Aird}, {Evans}, {Barlow-Hall}, {Osborne}, \& {Watson}}]{delaney2023}
{Delaney}, J.~N., {Aird}, J., {Evans}, P.~A., {et~al.} 2023, \mnras, \dodoi{10.1093/mnras/stac3703}

\bibitem[{{Della Ceca} {et~al.}(1999){Della Ceca}, {Castelli}, {Braito}, {Cagnoni}, \& {Maccacaro}}]{dellaceca1999}
{Della Ceca}, R., {Castelli}, G., {Braito}, V., {Cagnoni}, I., \& {Maccacaro}, T. 1999, ApJ, 524, 674, \dodoi{10.1086/307836}

\bibitem[{{Dicke} {et~al.}(1965){Dicke}, {Peebles}, {Roll}, \& {Wilkinson}}]{dicke1965}
{Dicke}, R.~H., {Peebles}, P.~J.~E., {Roll}, P.~G., \& {Wilkinson}, D.~T. 1965, ApJ, 142, 414, \dodoi{10.1086/148306}

\bibitem[{{Drinkwater} {et~al.}(2010){Drinkwater}, {Jurek}, {Blake}, {Woods}, {Pimbblet}, {Glazebrook}, {Sharp}, {Pracy}, {Brough}, {Colless}, {Couch}, {Croom}, {Davis}, {Forbes}, {Forster}, {Gilbank}, {Gladders}, {Jelliffe}, {Jones}, {Li}, {Madore}, {Martin}, {Poole}, {Small}, {Wisnioski}, {Wyder}, \& {Yee}}]{drinkwater2010}
{Drinkwater}, M.~J., {Jurek}, R.~J., {Blake}, C., {et~al.} 2010, MNRAS, 401, 1429, \dodoi{10.1111/j.1365-2966.2009.15754.x}

\bibitem[{{Driver} {et~al.}(2011){Driver}, {Hill}, {Kelvin}, {Robotham}, {Liske}, {Norberg}, {Baldry}, {Bamford}, {Hopkins}, {Loveday}, {Peacock}, {Andrae}, {Bland-Hawthorn}, {Brough}, {Brown}, {Cameron}, {Ching}, {Colless}, {Conselice}, {Croom}, {Cross}, {de Propris}, {Dye}, {Drinkwater}, {Ellis}, {Graham}, {Grootes}, {Gunawardhana}, {Jones}, {van Kampen}, {Maraston}, {Nichol}, {Parkinson}, {Phillipps}, {Pimbblet}, {Popescu}, {Prescott}, {Roseboom}, {Sadler}, {Sansom}, {Sharp}, {Smith}, {Taylor}, {Thomas}, {Tuffs}, {Wijesinghe}, {Dunne}, {Frenk}, {Jarvis}, {Madore}, {Meyer}, {Seibert}, {Staveley-Smith}, {Sutherland}, \& {Warren}}]{driver2011}
{Driver}, S.~P., {Hill}, D.~T., {Kelvin}, L.~S., {et~al.} 2011, MNRAS, 413, 971, \dodoi{10.1111/j.1365-2966.2010.18188.x}

\bibitem[{{Eckart} {et~al.}(2006){Eckart}, {Stern}, {Helfand}, {Harrison}, {Mao}, \& {Yost}}]{eckart2006}
{Eckart}, M.~E., {Stern}, D., {Helfand}, D.~J., {et~al.} 2006, ApJS, 165, 19, \dodoi{10.1086/504524}

\bibitem[{{Eisenhardt} {et~al.}(2012){Eisenhardt}, {Wu}, {Tsai}, {Assef}, {Benford}, {Blain}, {Bridge}, {Condon}, {Cushing}, {Cutri}, {Evans}, {Gelino}, {Griffith}, {Grillmair}, {Jarrett}, {Lonsdale}, {Masci}, {Mason}, {Petty}, {Sayers}, {Stanford}, {Stern}, {Wright}, \& {Yan}}]{eisenhardt2012}
{Eisenhardt}, P. R.~M., {Wu}, J., {Tsai}, C.-W., {et~al.} 2012, ApJ, 755, 173, \dodoi{10.1088/0004-637X/755/2/173}

\bibitem[{{Erlund} {et~al.}(2010){Erlund}, {Fabian}, {Blundell}, {Crawford}, \& {Hirst}}]{erlund2010}
{Erlund}, M.~C., {Fabian}, A.~C., {Blundell}, K.~M., {Crawford}, C.~S., \& {Hirst}, P. 2010, MNRAS, 404, 629, \dodoi{10.1111/j.1365-2966.2010.16304.x}

\bibitem[{{Erlund} {et~al.}(2007){Erlund}, {Fabian}, {Blundell}, {Moss}, \& {Ballantyne}}]{erlund2007}
{Erlund}, M.~C., {Fabian}, A.~C., {Blundell}, K.~M., {Moss}, C., \& {Ballantyne}, D.~R. 2007, MNRAS, 379, 498, \dodoi{10.1111/j.1365-2966.2007.11962.x}

\bibitem[{{Evans} {et~al.}(2010){Evans}, {Primini}, {Glotfelty}, {Anderson}, {Bonaventura}, {Chen}, {Davis}, {Doe}, {Evans}, {Fabbiano}, {Galle}, {Gibbs}, {Grier}, {Hain}, {Hall}, {Harbo}, {He}, {Houck}, {Karovska}, {Kashyap}, {Lauer}, {McCollough}, {McDowell}, {Miller}, {Mitschang}, {Morgan}, {Mossman}, {Nichols}, {Nowak}, {Plummer}, {Refsdal}, {Rots}, {Siemiginowska}, {Sundheim}, {Tibbetts}, {Van Stone}, {Winkelman}, \& {Zografou}}]{evans_chandra_2010}
{Evans}, I.~N., {Primini}, F.~A., {Glotfelty}, K.~J., {et~al.} 2010, \apjs, 189, 37, \dodoi{10.1088/0067-0049/189/1/37}

\bibitem[{{Evans} {et~al.}(2019){Evans}, {Allen}, {Anderson}, {Budynkiewicz}, {Burke}, {Chen}, {Civano}, {D'Abrusco}, {Doe}, {Evans}, {Fabbiano}, {Gibbs}, {Glotfelty}, {Graessle}, {Grier}, {Hain}, {Hall}, {Harbo}, {Houck}, {Lauer}, {Laurino}, {Lee}, {Martinez-Galarza}, {McCollough}, {McDowell}, {Miller}, {McLaughlin}, {Morgan}, {Mossman}, {Nguyen}, {Nichols}, {Nowak}, {Paxson}, {Plummer}, {Primini}, {Rots}, {Siemiginowska}, {Sundheim}, {Tibbetts}, {Van Stone}, \& {Zografou}}]{evans2019_csc2}
{Evans}, I.~N., {Allen}, C., {Anderson}, C.~S., {et~al.} 2019, in AAS/High Energy Astrophysics Division, Vol.~17, AAS/High Energy Astrophysics Division, 114.01

\bibitem[{{Evans} {et~al.}(2014){Evans}, {Osborne}, {Beardmore}, {Page}, {Willingale}, {Mountford}, {Pagani}, {Burrows}, {Kennea}, {Perri}, {Tagliaferri}, \& {Gehrels}}]{evans2014}
{Evans}, P.~A., {Osborne}, J.~P., {Beardmore}, A.~P., {et~al.} 2014, ApJS, 210, 8, \dodoi{10.1088/0067-0049/210/1/8}

\bibitem[{{Evans} {et~al.}(2020){Evans}, {Page}, {Osborne}, {Beardmore}, {Willingale}, {Burrows}, {Kennea}, {Perri}, {Capalbi}, {Tagliaferri}, \& {Cenko}}]{evans2020_2sxps}
{Evans}, P.~A., {Page}, K.~L., {Osborne}, J.~P., {et~al.} 2020, ApJS, 247, 54, \dodoi{10.3847/1538-4365/ab7db9}

\bibitem[{{Fawcett} {et~al.}(2020){Fawcett}, {Alexander}, {Rosario}, {Klindt}, {Fotopoulou}, {Lusso}, {Morabito}, \& {Calistro Rivera}}]{fawcett2020}
{Fawcett}, V.~A., {Alexander}, D.~M., {Rosario}, D.~J., {et~al.} 2020, MNRAS, 494, 4802, \dodoi{10.1093/mnras/staa954}

\bibitem[{{Fawcett} {et~al.}(2022){Fawcett}, {Alexander}, {Rosario}, {Klindt}, {Lusso}, {Morabito}, \& {Calistro Rivera}}]{fawcett2022}
---. 2022, arXiv e-prints, arXiv:2201.04139.
\newblock \doarXiv{2201.04139}

\bibitem[{{Fiore} {et~al.}(2000){Fiore}, {La Franca}, {Vignali}, {Comastri}, {Matt}, {Perola}, {Cappi}, {Elvis}, \& {Nicastro}}]{fiore2000}
{Fiore}, F., {La Franca}, F., {Vignali}, C., {et~al.} 2000, New Astronomy, 5, 143, \dodoi{10.1016/S1384-1076(00)00017-8}

\bibitem[{{Fitzpatrick}(1999)}]{fitzpatrick_correcting_1999}
{Fitzpatrick}, E.~L. 1999, \pasp, 111, 63, \dodoi{10.1086/316293}

\bibitem[{Fitzpatrick {et~al.}(2014)Fitzpatrick, Olsen, Economou, Stobie, Beers, Dickinson, Norris, Saha, Seaman, Silva, Swaters, Thomas, \& Valdes}]{fitzpatrick_noao_2014}
Fitzpatrick, M.~J., Olsen, K., Economou, F., {et~al.} 2014, in Observatory Operations: Strategies, Processes, and Systems V, ed. A.~B. Peck, C.~R. Benn, \& R.~L. Seaman, Vol. 9149, International Society for Optics and Photonics (SPIE), 91491T, \dodoi{10.1117/12.2057445}

\bibitem[{{Flewelling}(2018)}]{flewelling2018}
{Flewelling}, H. 2018, in American Astronomical Society Meeting Abstracts, Vol. 231, American Astronomical Society Meeting Abstracts \#231, 436.01

\bibitem[{{Fornasini} {et~al.}(2017){Fornasini}, {Tomsick}, {Hong}, {Gotthelf}, {Bauer}, {Rahoui}, {Stern}, {Bodaghee}, {Chiu}, {Clavel}, {Corral-Santana}, {Hailey}, {Krivonos}, {Mori}, {Alexander}, {Barret}, {Boggs}, {Christensen}, {Craig}, {Forster}, {Giommi}, {Grefenstette}, {Harrison}, {Hornstrup}, {Kitaguchi}, {Koglin}, {Madsen}, {Mao}, {Miyasaka}, {Perri}, {Pivovaroff}, {Puccetti}, {Rana}, {Westergaard}, \& {Zhang}}]{fornasini2017}
{Fornasini}, F.~M., {Tomsick}, J.~A., {Hong}, J., {et~al.} 2017, ApJS, 229, 33, \dodoi{10.3847/1538-4365/aa61fc}

\bibitem[{{Gandhi} {et~al.}(2014){Gandhi}, {Lansbury}, {Alexander}, {Stern}, {Ar{\'e}valo}, {Ballantyne}, {Balokovi{\'c}}, {Bauer}, {Boggs}, {Brandt}, {Brightman}, {Christensen}, {Comastri}, {Craig}, {Del Moro}, {Elvis}, {Fabian}, {Hailey}, {Harrison}, {Hickox}, {Koss}, {LaMassa}, {Luo}, {Madejski}, {Ptak}, {Puccetti}, {Teng}, {Urry}, {Walton}, \& {Zhang}}]{gandhi2014}
{Gandhi}, P., {Lansbury}, G.~B., {Alexander}, D.~M., {et~al.} 2014, ApJ, 792, 117, \dodoi{10.1088/0004-637X/792/2/117}

\bibitem[{{Giacconi} {et~al.}(1962){Giacconi}, {Gursky}, {Paolini}, \& {Rossi}}]{giacconi1962}
{Giacconi}, R., {Gursky}, H., {Paolini}, F.~R., \& {Rossi}, B.~B. 1962, \prl, 9, 439, \dodoi{10.1103/PhysRevLett.9.439}

\bibitem[{{Glikman} {et~al.}(2004){Glikman}, {Gregg}, {Lacy}, {Helfand}, {Becker}, \& {White}}]{glikman2004}
{Glikman}, E., {Gregg}, M.~D., {Lacy}, M., {et~al.} 2004, \apj, 607, 60, \dodoi{10.1086/383305}

\bibitem[{{Goulding} {et~al.}(2018){Goulding}, {Zakamska}, {Alexandroff}, {Assef}, {Banerji}, {Hamann}, {Wylezalek}, {Brandt}, {Greene}, {Lansbury}, {P{\^a}ris}, {Richards}, {Stern}, \& {Strauss}}]{goulding2018}
{Goulding}, A.~D., {Zakamska}, N.~L., {Alexandroff}, R.~M., {et~al.} 2018, ApJ, 856, 4, \dodoi{10.3847/1538-4357/aab040}

\bibitem[{{Gu} {et~al.}(1997){Gu}, {Huang}, {Su}, \& {Shang}}]{gu1997}
{Gu}, Q.~S., {Huang}, J.~H., {Su}, H.~J., \& {Shang}, Z.~H. 1997, A\&A, 319, 92.
\newblock \doarXiv{astro-ph/9810098}

\bibitem[{{Hamann} {et~al.}(2017){Hamann}, {Zakamska}, {Ross}, {Paris}, {Alexandroff}, {Villforth}, {Richards}, {Herbst}, {Brandt}, {Cook}, {Denney}, {Greene}, {Schneider}, \& {Strauss}}]{hamann2017}
{Hamann}, F., {Zakamska}, N.~L., {Ross}, N., {et~al.} 2017, MNRAS, 464, 3431.
\newblock \doarXiv{1609.07241}

\bibitem[{Harris {et~al.}(2020)Harris, Millman, van~der Walt, Gommers, Virtanen, Cournapeau, Wieser, Taylor, Berg, Smith, Kern, Picus, Hoyer, van Kerkwijk, Brett, Haldane, del R{\'{i}}o, Wiebe, Peterson, G{\'{e}}rard-Marchant, Sheppard, Reddy, Weckesser, Abbasi, Gohlke, \& Oliphant}]{harris_array_2020}
Harris, C.~R., Millman, K.~J., van~der Walt, S.~J., {et~al.} 2020, Nature, 585, 357, \dodoi{10.1038/s41586-020-2649-2}

\bibitem[{{Harris}(1996)}]{harris1996}
{Harris}, W.~E. 1996, AJ, 112, 1487, \dodoi{10.1086/118116}

\bibitem[{{Harrison} {et~al.}(2013){Harrison}, {Craig}, {Christensen}, \& et~al.}]{harrison2013}
{Harrison}, F.~A., {Craig}, W.~W., {Christensen}, F.~E., \& et~al. 2013, ApJ, 770, 103, \dodoi{10.1088/0004-637X/770/2/103}

\bibitem[{{Harrison} {et~al.}(2016{\natexlab{a}}){Harrison}, {Aird}, {Civano}, {Lansbury}, {Mullaney}, {Ballantyne}, {Alexander}, {Stern}, {Ajello}, {Barret}, {Bauer}, {Balokovi{\'c}}, {Brandt}, {Brightman}, {Boggs}, {Christensen}, {Comastri}, {Craig}, {Del Moro}, {Forster}, {Gandhi}, {Giommi}, {Grefenstette}, {Hailey}, {Hickox}, {Hornstrup}, {Kitaguchi}, {Koglin}, {Luo}, {Madsen}, {Mao}, {Miyasaka}, {Mori}, {Perri}, {Pivovaroff}, {Puccetti}, {Rana}, {Treister}, {Walton}, {Westergaard}, {Wik}, {Zappacosta}, {Zhang}, \& {Zoglauer}}]{harrison2016}
{Harrison}, F.~A., {Aird}, J., {Civano}, F., {et~al.} 2016{\natexlab{a}}, ApJ, 831, 185, \dodoi{10.3847/0004-637X/831/2/185}

\bibitem[{{Harrison} {et~al.}(2016{\natexlab{b}}){Harrison}, {Aird}, {Civano}, {Lansbury}, {Mullaney}, {Ballantyne}, {Alexander}, {Stern}, {Ajello}, {Barret}, {Bauer}, {Balokovi{\'c}}, {Brandt}, {Brightman}, {Boggs}, {Christensen}, {Comastri}, {Craig}, {Del Moro}, {Forster}, {Gandhi}, {Giommi}, {Grefenstette}, {Hailey}, {Hickox}, {Hornstrup}, {Kitaguchi}, {Koglin}, {Luo}, {Madsen}, {Mao}, {Miyasaka}, {Mori}, {Perri}, {Pivovaroff}, {Puccetti}, {Rana}, {Treister}, {Walton}, {Westergaard}, {Wik}, {Zappacosta}, {Zhang}, \& {Zoglauer}}]{harrison16}
---. 2016{\natexlab{b}}, ApJ, 831, 185, \dodoi{10.3847/0004-637X/831/2/185}

\bibitem[{{Healey} {et~al.}(2007){Healey}, {Romani}, {Taylor}, {Sadler}, {Ricci}, {Murphy}, {Ulvestad}, \& {Winn}}]{healey2007}
{Healey}, S.~E., {Romani}, R.~W., {Taylor}, G.~B., {et~al.} 2007, ApJS, 171, 61, \dodoi{10.1086/513742}

\bibitem[{{Hickox} \& {Alexander}(2018)}]{hickox2018}
{Hickox}, R.~C., \& {Alexander}, D.~M. 2018, ARA\&A, 56, 625, \dodoi{10.1146/annurev-astro-081817-051803}

\bibitem[{{Hickox} \& {Markevitch}(2007)}]{hickox2007}
{Hickox}, R.~C., \& {Markevitch}, M. 2007, ApJ, 671, 1523, \dodoi{10.1086/522918}

\bibitem[{{Hinton} {et~al.}(2016){Hinton}, {Davis}, {Lidman}, {Glazebrook}, \& {Lewis}}]{hinton2016}
{Hinton}, S.~R., {Davis}, T.~M., {Lidman}, C., {Glazebrook}, K., \& {Lewis}, G.~F. 2016, Astronomy and Computing, 15, 61, \dodoi{10.1016/j.ascom.2016.03.001}

\bibitem[{{Hong} {et~al.}(2016){Hong}, {Mori}, {Hailey}, {Nynka}, {Zhang}, {Gotthelf}, {Fornasini}, {Krivonos}, {Bauer}, {Perez}, {Tomsick}, {Bodaghee}, {Chiu}, {Clavel}, {Stern}, {Grindlay}, {Alexander}, {Aramaki}, {Baganoff}, {Barret}, {Barri{\`e}re}, {Boggs}, {Canipe}, {Christensen}, {Craig}, {Desai}, {Forster}, {Giommi}, {Grefenstette}, {Harrison}, {Hong}, {Hornstrup}, {Kitaguchi}, {Koglin}, {Madsen}, {Mao}, {Miyasaka}, {Perri}, {Pivovaroff}, {Puccetti}, {Rana}, {Westergaard}, {Zhang}, \& {Zoglauer}}]{hong2016}
{Hong}, J., {Mori}, K., {Hailey}, C.~J., {et~al.} 2016, ApJ, 825, 132, \dodoi{10.3847/0004-637X/825/2/132}

\bibitem[{{Hunter}(2007)}]{hunter_computing_2007}
{Hunter}, J.~D. 2007, Computing in Science and Engineering, 9, 90, \dodoi{10.1109/MCSE.2007.55}

\bibitem[{{Jarrett} {et~al.}(2011){Jarrett}, {Cohen}, {Masci}, {Wright}, {Stern}, {Benford}, {Blain}, {Carey}, {Cutri}, {Eisenhardt}, {Lonsdale}, {Mainzer}, {Marsh}, {Padgett}, {Petty}, {Ressler}, {Skrutskie}, {Stanford}, {Surace}, {Tsai}, {Wheelock}, \& {Yan}}]{jarrett2011}
{Jarrett}, T.~H., {Cohen}, M., {Masci}, F., {et~al.} 2011, ApJ, 735, 112, \dodoi{10.1088/0004-637X/735/2/112}

\bibitem[{{Jones} {et~al.}(2004){Jones}, {Saunders}, {Colless}, {Read}, {Parker}, {Watson}, {Campbell}, {Burkey}, {Mauch}, {Moore}, {Hartley}, {Cass}, {James}, {Russell}, {Fiegert}, {Dawe}, {Huchra}, {Jarrett}, {Lahav}, {Lucey}, {Mamon}, {Proust}, {Sadler}, \& {Wakamatsu}}]{jones2004}
{Jones}, D.~H., {Saunders}, W., {Colless}, M., {et~al.} 2004, MNRAS, 355, 747, \dodoi{10.1111/j.1365-2966.2004.08353.x}

\bibitem[{{Jones} {et~al.}(2009){Jones}, {Read}, {Saunders}, {Colless}, {Jarrett}, {Parker}, {Fairall}, {Mauch}, {Sadler}, {Watson}, {Burton}, {Campbell}, {Cass}, {Croom}, {Dawe}, {Fiegert}, {Frankcombe}, {Hartley}, {Huchra}, {James}, {Kirby}, {Lahav}, {Lucey}, {Mamon}, {Moore}, {Peterson}, {Prior}, {Proust}, {Russell}, {Safouris}, {Wakamatsu}, {Westra}, \& {Williams}}]{jones2009}
{Jones}, D.~H., {Read}, M.~A., {Saunders}, W., {et~al.} 2009, MNRAS, 399, 683, \dodoi{10.1111/j.1365-2966.2009.15338.x}

\bibitem[{{Katgert} {et~al.}(1998){Katgert}, {Mazure}, {den Hartog}, {Adami}, {Biviano}, \& {Perea}}]{katgert1998}
{Katgert}, P., {Mazure}, A., {den Hartog}, R., {et~al.} 1998, A\&AS, 129, 399, \dodoi{10.1051/aas:1998399}

\bibitem[{Klindt(2022)}]{klindt2022}
Klindt, L. 2022, PhD thesis, Durham University.
\newblock \url{http://etheses.dur.ac.uk/14314/}

\bibitem[{{Klindt} {et~al.}(2019){Klindt}, {Alexander}, {Rosario}, {Lusso}, \& {Fotopoulou}}]{klindt2019}
{Klindt}, L., {Alexander}, D.~M., {Rosario}, D.~J., {Lusso}, E., \& {Fotopoulou}, S. 2019, MNRAS, 488, 3109, \dodoi{10.1093/mnras/stz1771}

\bibitem[{{Koss} {et~al.}(2016){Koss}, {Glidden}, {Balokovi{\'c}}, {Stern}, {Lamperti}, {Assef}, {Bauer}, {Ballantyne}, {Boggs}, {Craig}, {Farrah}, {F{\"u}rst}, {Gandhi}, {Gehrels}, {Hailey}, {Harrison}, {Markwardt}, {Masini}, {Ricci}, {Treister}, {Walton}, \& {Zhang}}]{koss2016}
{Koss}, M.~J., {Glidden}, A., {Balokovi{\'c}}, M., {et~al.} 2016, ApJL, 824, L4, \dodoi{10.3847/2041-8205/824/1/L4}

\bibitem[{{Koss} {et~al.}(2022){Koss}, {Trakhtenbrot}, {Ricci}, {Bauer}, {Treister}, {Mushotzky}, {Urry}, {Ananna}, {Balokovi{\'c}}, {den Brok}, {Cenko}, {Harrison}, {Ichikawa}, {Lamperti}, {Lein}, {Mej{\'\i}a-Restrepo}, {Oh}, {Pacucci}, {Pfeifle}, {Powell}, {Privon}, {Ricci}, {Salvato}, {Schawinski}, {Shimizu}, {Smith}, \& {Stern}}]{koss2022}
{Koss}, M.~J., {Trakhtenbrot}, B., {Ricci}, C., {et~al.} 2022, ApJS, 261, 1, \dodoi{10.3847/1538-4365/ac6c8f}

\bibitem[{{Kraft} {et~al.}(1991){Kraft}, {Burrows}, \& {Nousek}}]{kraft1991}
{Kraft}, R.~P., {Burrows}, D.~N., \& {Nousek}, J.~A. 1991, ApJ, 374, 344, \dodoi{10.1086/170124}

\bibitem[{{Lake} {et~al.}(2012){Lake}, {Wright}, {Petty}, {Assef}, {Jarrett}, {Stanford}, {Stern}, \& {Tsai}}]{lake2012}
{Lake}, S.~E., {Wright}, E.~L., {Petty}, S., {et~al.} 2012, AJ, 143, 7.
\newblock \doarXiv{1111.0341}

\bibitem[{{LaMassa} {et~al.}(2019){LaMassa}, {Georgakakis}, {Vivek}, {Salvato}, {Ananna}, {Urry}, {MacLeod}, \& {Ross}}]{lamassa_sdss_2019}
{LaMassa}, S.~M., {Georgakakis}, A., {Vivek}, M., {et~al.} 2019, \apj, 876, 50, \dodoi{10.3847/1538-4357/ab108b}

\bibitem[{{Lansbury} {et~al.}(2015){Lansbury}, {Gandhi}, {Alexander}, {Assef}, {Aird}, {Annuar}, {Ballantyne}, {Balokovi{\'c}}, {Bauer}, {Boggs}, {Brandt}, {Brightman}, {Christensen}, {Civano}, {Comastri}, {Craig}, {Del Moro}, {Grefenstette}, {Hailey}, {Harrison}, {Hickox}, {Koss}, {LaMassa}, {Luo}, {Puccetti}, {Stern}, {Treister}, {Vignali}, {Zappacosta}, \& {Zhang}}]{lansbury2015}
{Lansbury}, G.~B., {Gandhi}, P., {Alexander}, D.~M., {et~al.} 2015, ApJ, 809, 115, \dodoi{10.1088/0004-637X/809/2/115}

\bibitem[{{Lansbury} {et~al.}(2017{\natexlab{a}}){Lansbury}, {Alexander}, {Aird}, {Gandhi}, {Stern}, {Koss}, {Lamperti}, {Ajello}, {Annuar}, {Assef}, {Ballantyne}, {Balokovi{\'c}}, {Bauer}, {Brandt}, {Brightman}, {Chen}, {Civano}, {Comastri}, {Del Moro}, {Fuentes}, {Harrison}, {Marchesi}, {Masini}, {Mullaney}, {Ricci}, {Saez}, {Tomsick}, {Treister}, {Walton}, \& {Zappacosta}}]{lansbury2017}
{Lansbury}, G.~B., {Alexander}, D.~M., {Aird}, J., {et~al.} 2017{\natexlab{a}}, ApJ, 846, 20, \dodoi{10.3847/1538-4357/aa8176}

\bibitem[{{Lansbury} {et~al.}(2017{\natexlab{b}}){Lansbury}, {Stern}, {Aird}, {Alexander}, {Fuentes}, {Harrison}, {Treister}, {Bauer}, {Tomsick}, {Balokovi{\'c}}, {Del Moro}, {Gandhi}, {Ajello}, {Annuar}, {Ballantyne}, {Boggs}, {Brandt}, {Brightman}, {Chen}, {Christensen}, {Civano}, {Comastri}, {Craig}, {Forster}, {Grefenstette}, {Hailey}, {Hickox}, {Jiang}, {Jun}, {Koss}, {Marchesi}, {Melo}, {Mullaney}, {Noirot}, {Schulze}, {Walton}, {Zappacosta}, \& {Zhang}}]{lansbury2017_cat}
{Lansbury}, G.~B., {Stern}, D., {Aird}, J., {et~al.} 2017{\natexlab{b}}, ApJ, 836, 99, \dodoi{10.3847/1538-4357/836/1/99}

\bibitem[{{Lin} {et~al.}(2012){Lin}, {Webb}, \& {Barret}}]{lin2012}
{Lin}, D., {Webb}, N.~A., \& {Barret}, D. 2012, ApJ, 756, 27, \dodoi{10.1088/0004-637X/756/1/27}

\bibitem[{{Luhman} {et~al.}(2010){Luhman}, {Allen}, {Espaillat}, {Hartmann}, \& {Calvet}}]{luhman2010}
{Luhman}, K.~L., {Allen}, P.~R., {Espaillat}, C., {Hartmann}, L., \& {Calvet}, N. 2010, ApJS, 186, 111, \dodoi{10.1088/0067-0049/186/1/111}

\bibitem[{{Luo} {et~al.}(2014){Luo}, {Brandt}, {Alexander}, {Stern}, {Teng}, {Ar{\'e}valo}, {Bauer}, {Boggs}, {Christensen}, {Comastri}, {Craig}, {Farrah}, {Gandhi}, {Hailey}, {Harrison}, {Koss}, {Ogle}, {Puccetti}, {Saez}, {Scott}, {Walton}, \& {Zhang}}]{luo2014}
{Luo}, B., {Brandt}, W.~N., {Alexander}, D.~M., {et~al.} 2014, ApJ, 794, 70, \dodoi{10.1088/0004-637X/794/1/70}

\bibitem[{{Luo} {et~al.}(2017){Luo}, {Brandt}, {Xue}, {Lehmer}, {Alexander}, {Bauer}, {Vito}, {Yang}, {Basu-Zych}, {Comastri}, {Gilli}, {Gu}, {Hornschemeier}, {Koekemoer}, {Liu}, {Mainieri}, {Paolillo}, {Ranalli}, {Rosati}, {Schneider}, {Shemmer}, {Smail}, {Sun}, {Tozzi}, {Vignali}, \& {Wang}}]{luo2017}
{Luo}, B., {Brandt}, W.~N., {Xue}, Y.~Q., {et~al.} 2017, ApJS, 228, 2, \dodoi{10.3847/1538-4365/228/1/2}

\bibitem[{{Lynden-Bell}(1969)}]{lyndenbell1969}
{Lynden-Bell}, D. 1969, Nature, 223, 690

\bibitem[{{Lyu} {et~al.}(2022){Lyu}, {Alberts}, {Rieke}, \& {Rujopakarn}}]{lyu_agn_2022}
{Lyu}, J., {Alberts}, S., {Rieke}, G.~H., \& {Rujopakarn}, W. 2022, \apj, 941, 191, \dodoi{10.3847/1538-4357/ac9e5d}

\bibitem[{{Mainzer} {et~al.}(2011){Mainzer}, {Bauer}, {Grav}, {Masiero}, {Cutri}, {Dailey}, {Eisenhardt}, {McMillan}, {Wright}, {Walker}, {Jedicke}, {Spahr}, {Tholen}, {Alles}, {Beck}, {Brandenburg}, {Conrow}, {Evans}, {Fowler}, {Jarrett}, {Marsh}, {Masci}, {McCallon}, {Wheelock}, {Wittman}, {Wyatt}, {DeBaun}, {Elliott}, {Elsbury}, {Gautier}, {Gomillion}, {Leisawitz}, {Maleszewski}, {Micheli}, \& {Wilkins}}]{mainzer2011}
{Mainzer}, A., {Bauer}, J., {Grav}, T., {et~al.} 2011, ApJ, 731, 53, \dodoi{10.1088/0004-637X/731/1/53}

\bibitem[{{Malizia} {et~al.}(2012){Malizia}, {Bassani}, {Bazzano}, {Bird}, {Masetti}, {Panessa}, {Stephen}, \& {Ubertini}}]{malizia2012}
{Malizia}, A., {Bassani}, L., {Bazzano}, A., {et~al.} 2012, MNRAS, 426, 1750, \dodoi{10.1111/j.1365-2966.2012.21755.x}

\bibitem[{{Marchesi} {et~al.}(2018){Marchesi}, {Ajello}, {Marcotulli}, {Comastri}, {Lanzuisi}, \& {Vignali}}]{marchesi2018}
{Marchesi}, S., {Ajello}, M., {Marcotulli}, L., {et~al.} 2018, ApJ, 854, 49, \dodoi{10.3847/1538-4357/aaa410}

\bibitem[{{Marchesi} {et~al.}(2019){Marchesi}, {Ajello}, {Zhao}, {Marcotulli}, {Balokovi{\'c}}, {Brightman}, {Comastri}, {Cusumano}, {Lanzuisi}, {La Parola}, {Segreto}, \& {Vignali}}]{marchesi2019}
{Marchesi}, S., {Ajello}, M., {Zhao}, X., {et~al.} 2019, ApJ, 872, 8, \dodoi{10.3847/1538-4357/aafbeb}

\bibitem[{{Marocco} {et~al.}(2021){Marocco}, {Eisenhardt}, {Fowler}, {Kirkpatrick}, {Meisner}, {Schlafly}, {Stanford}, {Garcia}, {Caselden}, {Cushing}, {Cutri}, {Faherty}, {Gelino}, {Gonzalez}, {Jarrett}, {Koontz}, {Mainzer}, {Marchese}, {Mobasher}, {Schlegel}, {Stern}, {Teplitz}, \& {Wright}}]{marocco2021}
{Marocco}, F., {Eisenhardt}, P. R.~M., {Fowler}, J.~W., {et~al.} 2021, ApJS, 253, 8, \dodoi{10.3847/1538-4365/abd805}

\bibitem[{{Masini} {et~al.}(2018){Masini}, {Civano}, {Comastri}, {Fornasini}, {Ballantyne}, {Lansbury}, {Treister}, {Alexander}, {Boorman}, {Brandt}, {Farrah}, {Gandhi}, {Harrison}, {Hickox}, {Kocevski}, {Lanz}, {Marchesi}, {Puccetti}, {Ricci}, {Saez}, {Stern}, \& {Zappacosta}}]{masini2018}
{Masini}, A., {Civano}, F., {Comastri}, A., {et~al.} 2018, ApJS, 235, 17, \dodoi{10.3847/1538-4365/aaa83d}

\bibitem[{{Masini} {et~al.}(2020){Masini}, {Hickox}, {Carroll}, {Aird}, {Alexander}, {Assef}, {Bower}, {Brodwin}, {Brown}, {Chatterjee}, {Chen}, {Dey}, {DiPompeo}, {Duncan}, {Eisenhardt}, {Forman}, {Gonzalez}, {Goulding}, {Hainline}, {Jannuzi}, {Jones}, {Kochanek}, {Kraft}, {Lee}, {Miller}, {Mullaney}, {Myers}, {Ptak}, {Stanford}, {Stern}, {Vikhlinin}, {Wake}, \& {Murray}}]{masini2020}
{Masini}, A., {Hickox}, R.~C., {Carroll}, C.~M., {et~al.} 2020, ApJS, 251, 2, \dodoi{10.3847/1538-4365/abb607}

\bibitem[{{Massaro} {et~al.}(2009){Massaro}, {Giommi}, {Leto}, {Marchegiani}, {Maselli}, {Perri}, {Piranomonte}, \& {Sclavi}}]{massaro2009}
{Massaro}, E., {Giommi}, P., {Leto}, C., {et~al.} 2009, VizieR Online Data Catalog, J/A+A/495/691

\bibitem[{Massey {et~al.}(1992)Massey, Valdes, \& Barnes}]{massey1992}
Massey, P., Valdes, F., \& Barnes, J. 1992, A User’s Guide to Reducing Slit Spectra with IRAF (available at: \url{https://www.mn.uio.no/astro/english/services/it/help/visualization/iraf/spect.pdf} accessed 14/01/2016)

\bibitem[{{Mateos} {et~al.}(2013){Mateos}, {Alonso-Herrero}, {Carrera}, {Blain}, {Severgnini}, {Caccianiga}, \& {Ruiz}}]{mateos2013}
{Mateos}, S., {Alonso-Herrero}, A., {Carrera}, F.~J., {et~al.} 2013, MNRAS, 434, 941, \dodoi{10.1093/mnras/stt953}

\bibitem[{{Mateos} {et~al.}(2012){Mateos}, {Alonso-Herrero}, {Carrera}, {Blain}, {Watson}, {Barcons}, {Braito}, {Severgnini}, {Donley}, \& {Stern}}]{mateos2012}
---. 2012, MNRAS, 426, 3271.
\newblock \doarXiv{1208.2530}

\bibitem[{{Merloni} {et~al.}(2012){Merloni}, {Predehl}, {Becker}, {B{\"o}hringer}, {Boller}, {Brunner}, {Brusa}, {Dennerl}, {Freyberg}, {Friedrich}, {Georgakakis}, {Haberl}, {Hasinger}, {Meidinger}, {Mohr}, {Nandra}, {Rau}, {Reiprich}, {Robrade}, {Salvato}, {Santangelo}, {Sasaki}, {Schwope}, {Wilms}, \& {German eROSITA Consortium}}]{merloni_erosita_2012}
{Merloni}, A., {Predehl}, P., {Becker}, W., {et~al.} 2012, arXiv e-prints, arXiv:1209.3114, \dodoi{10.48550/arXiv.1209.3114}

\bibitem[{{Monet} {et~al.}(2003){Monet}, {Levine}, {Canzian}, {Ables}, {Bird}, {Dahn}, {Guetter}, {Harris}, {Henden}, {Leggett}, {Levison}, {Luginbuhl}, {Martini}, {Monet}, {Munn}, {Pier}, {Rhodes}, {Riepe}, {Sell}, {Stone}, {Vrba}, {Walker}, {Westerhout}, {Brucato}, {Reid}, {Schoening}, {Hartley}, {Read}, \& {Tritton}}]{monet2003}
{Monet}, D.~G., {Levine}, S.~E., {Canzian}, B., {et~al.} 2003, AJ, 125, 984, \dodoi{10.1086/345888}

\bibitem[{{Moretti} {et~al.}(2003){Moretti}, {Campana}, {Lazzati}, \& {Tagliaferri}}]{moretti2003}
{Moretti}, A., {Campana}, S., {Lazzati}, D., \& {Tagliaferri}, G. 2003, ApJ, 588, 696, \dodoi{10.1086/374335}

\bibitem[{{Mori} {et~al.}(2015){Mori}, {Hailey}, {Krivonos}, {Hong}, {Ponti}, {Bauer}, {Perez}, {Nynka}, {Zhang}, {Tomsick}, {Alexander}, {Baganoff}, {Barret}, {Barri{\`e}re}, {Boggs}, {Canipe}, {Christensen}, {Craig}, {Forster}, {Giommi}, {Grefenstette}, {Grindlay}, {Harrison}, {Hornstrup}, {Kitaguchi}, {Koglin}, {Luu}, {Madsen}, {Mao}, {Miyasaka}, {Perri}, {Pivovaroff}, {Puccetti}, {Rana}, {Stern}, {Westergaard}, {Zhang}, \& {Zoglauer}}]{mori2015}
{Mori}, K., {Hailey}, C.~J., {Krivonos}, R., {et~al.} 2015, ApJ, 814, 94, \dodoi{10.1088/0004-637X/814/2/94}

\bibitem[{{Mullaney} {et~al.}(2015){Mullaney}, {Del-Moro}, {Aird}, {Alexander}, {Civano}, {Hickox}, {Lansbury}, {Ajello}, {Assef}, {Ballantyne}, {Balokovi{\'c}}, {Bauer}, {Brandt}, {Boggs}, {Brightman}, {Christensen}, {Comastri}, {Craig}, {Elvis}, {Forster}, {Gandhi}, {Grefenstette}, {Hailey}, {Harrison}, {Koss}, {LaMassa}, {Luo}, {Madsen}, {Puccetti}, {Saez}, {Stern}, {Treister}, {Urry}, {Wik}, {Zappacosta}, \& {Zhang}}]{mullaney2015}
{Mullaney}, J.~R., {Del-Moro}, A., {Aird}, J., {et~al.} 2015, ApJ, 808, 184, \dodoi{10.1088/0004-637X/808/2/184}

\bibitem[{{Mushotzky} {et~al.}(2000){Mushotzky}, {Cowie}, {Barger}, \& {Arnaud}}]{mushotzky2000}
{Mushotzky}, R.~F., {Cowie}, L.~L., {Barger}, A.~J., \& {Arnaud}, K.~A. 2000, Nature, 404, 459, \dodoi{10.1038/35006564}

\bibitem[{{Nandra} {et~al.}(2005){Nandra}, {Laird}, {Adelberger}, {Gardner}, {Mushotzky}, {Rhodes}, {Steidel}, {Teplitz}, \& {Arnaud}}]{nandra2005}
{Nandra}, K., {Laird}, E.~S., {Adelberger}, K., {et~al.} 2005, \mnras, 356, 568, \dodoi{10.1111/j.1365-2966.2004.08475.x}

\bibitem[{{Netzer}(2015)}]{netzer2015}
{Netzer}, H. 2015, ARA\&A, 53, 365, \dodoi{10.1146/annurev-astro-082214-122302}

\bibitem[{{Nidever} {et~al.}(2021){Nidever}, {Dey}, {Fasbender}, {Juneau}, {Meisner}, {Wishart}, {Scott}, {Matt}, {Nikutta}, \& {Pucha}}]{nidever2021}
{Nidever}, D.~L., {Dey}, A., {Fasbender}, K., {et~al.} 2021, AJ, 161, 192, \dodoi{10.3847/1538-3881/abd6e1}

\bibitem[{Nikutta {et~al.}(2020)Nikutta, Fitzpatrick, Scott, \& Weaver}]{nikutta_data_2020}
Nikutta, R., Fitzpatrick, M., Scott, A., \& Weaver, B. 2020, Astronomy and Computing, 33, 100411, \dodoi{https://doi.org/10.1016/j.ascom.2020.100411}

\bibitem[{{Nilsson}(1998)}]{nilsson1998}
{Nilsson}, K. 1998, A\&AS, 132, 31, \dodoi{10.1051/aas:1998442}

\bibitem[{{Nishiyama} {et~al.}(2008){Nishiyama}, {Nagata}, {Tamura}, {Kandori}, {Hatano}, {Sato}, \& {Sugitani}}]{nishiyama_interstellar_2008}
{Nishiyama}, S., {Nagata}, T., {Tamura}, M., {et~al.} 2008, \apj, 680, 1174, \dodoi{10.1086/587791}

\bibitem[{{Oh} {et~al.}(2018){Oh}, {Koss}, {Markwardt}, {Schawinski}, {Baumgartner}, {Barthelmy}, {Cenko}, {Gehrels}, {Mushotzky}, {Petulante}, {Ricci}, {Lien}, \& {Trakhtenbrot}}]{oh2018}
{Oh}, K., {Koss}, M., {Markwardt}, C.~B., {et~al.} 2018, ApJS, 235, 4, \dodoi{10.3847/1538-4365/aaa7fd}

\bibitem[{{Parisi} {et~al.}(2014){Parisi}, {Masetti}, {Rojas}, {Jim{\'e}nez-Bail{\'o}n}, {Chavushyan}, {Palazzi}, {Bassani}, {Bazzano}, {Bird}, {Galaz}, {Minniti}, {Morelli}, \& {Ubertini}}]{parisi2014}
{Parisi}, P., {Masetti}, N., {Rojas}, A.~F., {et~al.} 2014, A\&A, 561, A67, \dodoi{10.1051/0004-6361/201322409}

\bibitem[{{Paturel} {et~al.}(2005){Paturel}, {Vauglin}, {Petit}, {Borsenberger}, {Epchtein}, {Fouqu{\'e}}, \& {Mamon}}]{paturel2005}
{Paturel}, G., {Vauglin}, I., {Petit}, C., {et~al.} 2005, A\&A, 430, 751, \dodoi{10.1051/0004-6361:20041162}

\bibitem[{{Penzias} \& {Wilson}(1965)}]{penzias1965}
{Penzias}, A.~A., \& {Wilson}, R.~W. 1965, ApJ, 142, 419, \dodoi{10.1086/148307}

\bibitem[{{Predehl} {et~al.}(2021){Predehl}, {Andritschke}, {Arefiev}, {Babyshkin}, {Batanov}, {Becker}, {B{\"o}hringer}, {Bogomolov}, {Boller}, {Borm}, {Bornemann}, {Br{\"a}uninger}, {Br{\"u}ggen}, {Brunner}, {Brusa}, {Bulbul}, {Buntov}, {Burwitz}, {Burkert}, {Clerc}, {Churazov}, {Coutinho}, {Dauser}, {Dennerl}, {Doroshenko}, {Eder}, {Emberger}, {Eraerds}, {Finoguenov}, {Freyberg}, {Friedrich}, {Friedrich}, {F{\"u}rmetz}, {Georgakakis}, {Gilfanov}, {Granato}, {Grossberger}, {Gueguen}, {Gureev}, {Haberl}, {H{\"a}lker}, {Hartner}, {Hasinger}, {Huber}, {Ji}, {Kienlin}, {Kink}, {Korotkov}, {Kreykenbohm}, {Lamer}, {Lomakin}, {Lapshov}, {Liu}, {Maitra}, {Meidinger}, {Menz}, {Merloni}, {Mernik}, {Mican}, {Mohr}, {M{\"u}ller}, {Nandra}, {Nazarov}, {Pacaud}, {Pavlinsky}, {Perinati}, {Pfeffermann}, {Pietschner}, {Ramos-Ceja}, {Rau}, {Reiffers}, {Reiprich}, {Robrade}, {Salvato}, {Sanders}, {Santangelo}, {Sasaki}, {Scheuerle}, {Schmid}, {Schmitt}, {Schwope}, {Shirshakov}, {Steinmetz}, {Stewart}, {Str{\"u}der},
  {Sunyaev}, {Tenzer}, {Tiedemann}, {Tr{\"u}mper}, {Voron}, {Weber}, {Wilms}, \& {Yaroshenko}}]{predehl_erosita_2021}
{Predehl}, P., {Andritschke}, R., {Arefiev}, V., {et~al.} 2021, \aap, 647, A1, \dodoi{10.1051/0004-6361/202039313}

\bibitem[{{Ricci} {et~al.}(2017){Ricci}, {Trakhtenbrot}, {Koss}, {Ueda}, {Del Vecchio}, {Treister}, {Schawinski}, {Paltani}, {Oh}, {Lamperti}, {Berney}, {Gandhi}, {Ichikawa}, {Bauer}, {Ho}, {Asmus}, {Beckmann}, {Soldi}, {Balokovi{\'c}}, {Gehrels}, \& {Markwardt}}]{ricci2017}
{Ricci}, C., {Trakhtenbrot}, B., {Koss}, M.~J., {et~al.} 2017, ApJS, 233, 17, \dodoi{10.3847/1538-4365/aa96ad}

\bibitem[{{Richards} {et~al.}(2003){Richards}, {Hall}, {Vanden Berk}, {Strauss}, {Schneider}, {Weinstein}, {Reichard}, {York}, {Knapp}, {Fan}, {Ivezi{\'c}}, {Brinkmann}, {Budav{\'a}ri}, {Csabai}, \& {Nichol}}]{richards2003}
{Richards}, G.~T., {Hall}, P.~B., {Vanden Berk}, D.~E., {et~al.} 2003, AJ, 126, 1131

\bibitem[{{Richards} {et~al.}(2009){Richards}, {Myers}, {Gray}, {Riegel}, {Nichol}, {Brunner}, {Szalay}, {Schneider}, \& {Anderson}}]{richards2009}
{Richards}, G.~T., {Myers}, A.~D., {Gray}, A.~G., {et~al.} 2009, ApJS, 180, 67, \dodoi{10.1088/0067-0049/180/1/67}

\bibitem[{{Ross} {et~al.}(2015){Ross}, {Hamann}, {Zakamska}, {Richards}, {Villforth}, {Strauss}, {Greene}, {Alexandroff}, {Brandt}, {Liu}, {Myers}, {P{\^a}ris}, \& {Schneider}}]{ross2015}
{Ross}, N.~P., {Hamann}, F., {Zakamska}, N.~L., {et~al.} 2015, MNRAS, 453, 3932.
\newblock \doarXiv{1405.1047}

\bibitem[{{Salvato} {et~al.}(2009){Salvato}, {Hasinger}, {Ilbert}, {Zamorani}, {Brusa}, {Scoville}, {Rau}, {Capak}, {Arnouts}, {Aussel}, {Bolzonella}, {Buongiorno}, {Cappelluti}, {Caputi}, {Civano}, {Cook}, {Elvis}, {Gilli}, {Jahnke}, {Kartaltepe}, {Impey}, {Lamareille}, {Le Floc'h}, {Lilly}, {Mainieri}, {McCarthy}, {McCracken}, {Mignoli}, {Mobasher}, {Murayama}, {Sasaki}, {Sanders}, {Schiminovich}, {Shioya}, {Shopbell}, {Silverman}, {Smol{\v{c}}i{\'c}}, {Surace}, {Taniguchi}, {Thompson}, {Trump}, {Urry}, \& {Zamojski}}]{salvato2009}
{Salvato}, M., {Hasinger}, G., {Ilbert}, O., {et~al.} 2009, ApJ, 690, 1250, \dodoi{10.1088/0004-637X/690/2/1250}

\bibitem[{{Salvato} {et~al.}(2018){Salvato}, {Buchner}, {Budav{\'a}ri}, {Dwelly}, {Merloni}, {Brusa}, {Rau}, {Fotopoulou}, \& {Nandra}}]{salvato2018}
{Salvato}, M., {Buchner}, J., {Budav{\'a}ri}, T., {et~al.} 2018, MNRAS, 473, 4937, \dodoi{10.1093/mnras/stx2651}

\bibitem[{{Schlegel} {et~al.}(1998){Schlegel}, {Finkbeiner}, \& {Davis}}]{schlegel1998}
{Schlegel}, D.~J., {Finkbeiner}, D.~P., \& {Davis}, M. 1998, ApJ, 500, 525, \dodoi{10.1086/305772}

\bibitem[{{Schmidt} \& {Green}(1983)}]{schmidt1983}
{Schmidt}, M., \& {Green}, R.~F. 1983, ApJ, 269, 352, \dodoi{10.1086/161048}

\bibitem[{{Schwope} {et~al.}(2000){Schwope}, {Hasinger}, {Lehmann}, {Schwarz}, {Brunner}, {Neizvestny}, {Ugryumov}, {Balega}, {Tr{\"u}mper}, \& {Voges}}]{schwope2000}
{Schwope}, A., {Hasinger}, G., {Lehmann}, I., {et~al.} 2000, Astronomische Nachrichten, 321, 1.
\newblock \doarXiv{astro-ph/0003039}

\bibitem[{{Seeberger} {et~al.}(1994){Seeberger}, {Huchtmeier}, \& {Weinberger}}]{seeberger1994}
{Seeberger}, R., {Huchtmeier}, W.~K., \& {Weinberger}, R. 1994, A\&A, 286, 17

\bibitem[{{Shen} {et~al.}(2011){Shen}, {Richards}, {Strauss}, {Hall}, {Schneider}, {Snedden}, {Bizyaev}, {Brewington}, {Malanushenko}, {Malanushenko}, {Oravetz}, {Pan}, \& {Simmons}}]{shen2011}
{Shen}, Y., {Richards}, G.~T., {Strauss}, M.~A., {et~al.} 2011, ApJS, 194, 45.
\newblock \doarXiv{1006.5178}

\bibitem[{{Stern} {et~al.}(2005){Stern}, {Eisenhardt}, {Gorjian}, {Kochanek}, {Caldwell}, {Eisenstein}, {Brodwin}, {Brown}, {Cool}, {Dey}, {Green}, {Jannuzi}, {Murray}, {Pahre}, \& {Willner}}]{stern2005}
{Stern}, D., {Eisenhardt}, P., {Gorjian}, V., {et~al.} 2005, ApJ, 631, 163, \dodoi{10.1086/432523}

\bibitem[{{Stern} {et~al.}(2012){Stern}, {Assef}, {Benford}, {Blain}, {Cutri}, {Dey}, {Eisenhardt}, {Griffith}, {Jarrett}, {Lake}, {Masci}, {Petty}, {Stanford}, {Tsai}, {Wright}, {Yan}, {Harrison}, \& {Madsen}}]{stern2012}
{Stern}, D., {Assef}, R.~J., {Benford}, D.~J., {et~al.} 2012, ApJ, 753, 30.
\newblock \doarXiv{1205.0811}

\bibitem[{{Strauss} {et~al.}(1992){Strauss}, {Huchra}, {Davis}, {Yahil}, {Fisher}, \& {Tonry}}]{strauss1992}
{Strauss}, M.~A., {Huchra}, J.~P., {Davis}, M., {et~al.} 1992, ApJS, 83, 29, \dodoi{10.1086/191730}

\bibitem[{{SubbaRao} {et~al.}(2002){SubbaRao}, {Frieman}, {Bernardi}, {Loveday}, {Nichol}, {Castander}, \& {Meiksin}}]{subbarao2002}
{SubbaRao}, M., {Frieman}, J., {Bernardi}, M., {et~al.} 2002, in Society of Photo-Optical Instrumentation Engineers (SPIE) Conference Series, Vol. 4847, Astronomical Data Analysis II, ed. J.-L. {Starck} \& F.~D. {Murtagh}, 452--460, \dodoi{10.1117/12.461108}

\bibitem[{{Toba} {et~al.}(2014){Toba}, {Oyabu}, {Matsuhara}, {Malkan}, {Gandhi}, {Nakagawa}, {Isobe}, {Shirahata}, {Oi}, {Ohyama}, {Takita}, {Yamauchi}, \& {Yano}}]{toba2014}
{Toba}, Y., {Oyabu}, S., {Matsuhara}, H., {et~al.} 2014, ApJ, 788, 45, \dodoi{10.1088/0004-637X/788/1/45}

\bibitem[{{Tomsick} {et~al.}(2017){Tomsick}, {Lansbury}, {Rahoui}, {Clavel}, {Fornasini}, {Hong}, {Aird}, {Alexander}, {Bodaghee}, {Chiu}, {Grindlay}, {Hailey}, {Harrison}, {Krivonos}, {Mori}, \& {Stern}}]{tomsick2017}
{Tomsick}, J.~A., {Lansbury}, G.~B., {Rahoui}, F., {et~al.} 2017, \apjs, 230, 25, \dodoi{10.3847/1538-4365/aa7517}

\bibitem[{{Tomsick} {et~al.}(2018){Tomsick}, {Lansbury}, {Rahoui}, {Aird}, {Alexander}, {Clavel}, {Cuturilo}, {Fornasini}, {Hong}, {Klindt}, \& {Stern}}]{tomsick2018}
---. 2018, ApJ, 869, 171, \dodoi{10.3847/1538-4357/aaf007}

\bibitem[{{Torres-Alb{\`a}} {et~al.}(2021{\natexlab{a}}){Torres-Alb{\`a}}, {Marchesi}, {Zhao}, {Ajello}, {Silver}, {Ananna}, {Balokovi{\'c}}, {Boorman}, {Comastri}, {Gilli}, {Lanzuisi}, {Murphy}, {Urry}, \& {Vignali}}]{torres_alba2021}
{Torres-Alb{\`a}}, N., {Marchesi}, S., {Zhao}, X., {et~al.} 2021{\natexlab{a}}, ApJ, 922, 252, \dodoi{10.3847/1538-4357/ac1c73}

\bibitem[{{Torres-Alb{\`a}} {et~al.}(2021{\natexlab{b}}){Torres-Alb{\`a}}, {Marchesi}, {Zhao}, {Ajello}, {Silver}, {Ananna}, {Balokovi{\'c}}, {Boorman}, {Comastri}, {Gilli}, {Lanzuisi}, {Murphy}, {Urry}, \& {Vignali}}]{torres2021}
---. 2021{\natexlab{b}}, ApJ, 922, 252, \dodoi{10.3847/1538-4357/ac1c73}

\bibitem[{{Traulsen} {et~al.}(2020){Traulsen}, {Schwope}, {Lamer}, {Ballet}, {Carrera}, {Ceballos}, {Coriat}, {Freyberg}, {Koliopanos}, {Kurpas}, {Michel}, {Motch}, {Page}, {Watson}, \& {Webb}}]{trauslen2020}
{Traulsen}, I., {Schwope}, A.~D., {Lamer}, G., {et~al.} 2020, A\&A, 641, A137, \dodoi{10.1051/0004-6361/202037706}

\bibitem[{{V{\'e}ron-Cetty} \& {V{\'e}ron}(2006)}]{veroncetty2006}
{V{\'e}ron-Cetty}, M.~P., \& {V{\'e}ron}, P. 2006, A\&A, 455, 773, \dodoi{10.1051/0004-6361:20065177}

\bibitem[{{V{\'e}ron-Cetty} \& {V{\'e}ron}(2010)}]{veroncetty2010}
---. 2010, A\&A, 518, A10, \dodoi{10.1051/0004-6361/201014188}

\bibitem[{Virtanen {et~al.}(2020)Virtanen, Gommers, Oliphant, Haberland, Reddy, Cournapeau, Burovski, Peterson, Weckesser, Bright, {van der Walt}, Brett, Wilson, Millman, Mayorov, Nelson, Jones, Kern, Larson, Carey, Polat, Feng, Moore, {VanderPlas}, Laxalde, Perktold, Cimrman, Henriksen, Quintero, Harris, Archibald, Ribeiro, Pedregosa, {van Mulbregt}, \& {SciPy 1.0 Contributors}}]{virtanen_scipy_2020}
Virtanen, P., Gommers, R., Oliphant, T.~E., {et~al.} 2020, Nature Methods, 17, 261, \dodoi{10.1038/s41592-019-0686-2}

\bibitem[{{Webb} {et~al.}(2020){Webb}, {Coriat}, {Traulsen}, {Ballet}, {Motch}, {Carrera}, {Koliopanos}, {Authier}, {de la Calle}, {Ceballos}, {Colomo}, {Chuard}, {Freyberg}, {Garcia}, {Kolehmainen}, {Lamer}, {Lin}, {Maggi}, {Michel}, {Page}, {Page}, {Perea-Calderon}, {Pineau}, {Rodriguez}, {Rosen}, {Santos Lleo}, {Saxton}, {Schwope}, {Tom{\'a}s}, {Watson}, \& {Zakardjian}}]{webb2020}
{Webb}, N.~A., {Coriat}, M., {Traulsen}, I., {et~al.} 2020, A\&A, 641, A136, \dodoi{10.1051/0004-6361/201937353}

\bibitem[{{Wik} {et~al.}(2014){Wik}, {Hornstrup}, {Molendi}, {Madejski}, {Harrison}, {Zoglauer}, {Grefenstette}, {Gastaldello}, {Madsen}, {Westergaard}, {Ferreira}, {Kitaguchi}, {Pedersen}, {Boggs}, {Christensen}, {Craig}, {Hailey}, {Stern}, \& {Zhang}}]{wik2014}
{Wik}, D.~R., {Hornstrup}, A., {Molendi}, S., {et~al.} 2014, ApJ, 792, 48, \dodoi{10.1088/0004-637X/792/1/48}

\bibitem[{{Winkler} \& {Long}(1997)}]{winkler1997}
{Winkler}, P.~F., \& {Long}, K.~S. 1997, ApJ, 486, L137, \dodoi{10.1086/310850}

\bibitem[{{Worsley} {et~al.}(2005){Worsley}, {Fabian}, {Bauer}, {Alexander}, {Hasinger}, {Mateos}, {Brunner}, {Brandt}, \& {Schneider}}]{worsley2005}
{Worsley}, M.~A., {Fabian}, A.~C., {Bauer}, F.~E., {et~al.} 2005, MNRAS, 357, 1281, \dodoi{10.1111/j.1365-2966.2005.08731.x}

\bibitem[{{Wright} {et~al.}(2010){Wright}, {Eisenhardt}, {Mainzer}, {Ressler}, {Cutri}, {Jarrett}, {Kirkpatrick}, {Padgett}, {McMillan}, {Skrutskie}, {Stanford}, {Cohen}, {Walker}, {Mather}, {Leisawitz}, {Gautier}, {McLean}, {Benford}, {Lonsdale}, {Blain}, {Mendez}, {Irace}, {Duval}, {Liu}, {Royer}, {Heinrichsen}, {Howard}, {Shannon}, {Kendall}, {Walsh}, {Larsen}, {Cardon}, {Schick}, {Schwalm}, {Abid}, {Fabinsky}, {Naes}, \& {Tsai}}]{wright2010}
{Wright}, E.~L., {Eisenhardt}, P.~R.~M., {Mainzer}, A.~K., {et~al.} 2010, AJ, 140, 1868.
\newblock \doarXiv{1008.0031}

\bibitem[{{Yan} {et~al.}(2019){Yan}, {Hickox}, {Hainline}, {Stern}, {Lansbury}, {Alexander}, {Hviding}, {Assef}, {Ballantyne}, {Dipompeo}, {Lanz}, {Carroll}, {Koss}, {Lamperti}, {Civano}, {Del Moro}, {Gandhi}, \& {Myers}}]{yan2019}
{Yan}, W., {Hickox}, R.~C., {Hainline}, K.~N., {et~al.} 2019, ApJ, 870, 33, \dodoi{10.3847/1538-4357/aaeed4}

\bibitem[{{Yuan} {et~al.}(2015){Yuan}, {Lidman}, {Davis}, {Childress}, {Abdalla}, {Banerji}, {Buckley-Geer}, {Carnero Rosell}, {Carollo}, {Castander}, {D'Andrea}, {Diehl}, {Cunha}, {Foley}, {Frieman}, {Glazebrook}, {Gschwend}, {Hinton}, {Jouvel}, {Kessler}, {Kim}, {King}, {Kuehn}, {Kuhlmann}, {Lewis}, {Lin}, {Martini}, {McMahon}, {Mould}, {Nichol}, {Norris}, {O'Neill}, {Ostrovski}, {Papadopoulos}, {Parkinson}, {Reed}, {Romer}, {Rooney}, {Rozo}, {Rykoff}, {Sako}, {Scalzo}, {Schmidt}, {Scolnic}, {Seymour}, {Sharp}, {Sobreira}, {Sullivan}, {Thomas}, {Tucker}, {Uddin}, {Wechsler}, {Wester}, {Wilcox}, {Zhang}, {Abbott}, {Allam}, {Bauer}, {Benoit-L{\'e}vy}, {Bertin}, {Brooks}, {Burke}, {Carrasco Kind}, {Covarrubias}, {Crocce}, {da Costa}, {DePoy}, {Desai}, {Doel}, {Eifler}, {Evrard}, {Fausti Neto}, {Flaugher}, {Fosalba}, {Gaztanaga}, {Gerdes}, {Gruen}, {Gruendl}, {Honscheid}, {James}, {Kuropatkin}, {Lahav}, {Li}, {Maia}, {Makler}, {Marshall}, {Miller}, {Miquel}, {Ogando}, {Plazas}, {Roodman}, {Sanchez}, {Scarpine},
  {Schubnell}, {Sevilla-Noarbe}, {Smith}, {Soares-Santos}, {Suchyta}, {Swanson}, {Tarle}, {Thaler}, \& {Walker}}]{yuan2015}
{Yuan}, F., {Lidman}, C., {Davis}, T.~M., {et~al.} 2015, MNRAS, 452, 3047, \dodoi{10.1093/mnras/stv1507}

\bibitem[{{Yuan} {et~al.}(2016){Yuan}, {Strauss}, \& {Zakamska}}]{yuan_spectroscopic_2016}
{Yuan}, S., {Strauss}, M.~A., \& {Zakamska}, N.~L. 2016, \mnras, 462, 1603, \dodoi{10.1093/mnras/stw1747}

\bibitem[{{Zhao} {et~al.}(2021{\natexlab{a}}){Zhao}, {Marchesi}, {Ajello}, {Cole}, {Hu}, {Silver}, \& {Torres-Alb{\`a}}}]{zhao2021_obscured}
{Zhao}, X., {Marchesi}, S., {Ajello}, M., {et~al.} 2021{\natexlab{a}}, A\&A, 650, A57, \dodoi{10.1051/0004-6361/202140297}

\bibitem[{{Zhao} {et~al.}(2021{\natexlab{b}}){Zhao}, {Civano}, {Fornasini}, {Alexander}, {Cappelluti}, {Chen}, {Cohen}, {Elvis}, {Gandhi}, {Grogin}, {Hickox}, {Jansen}, {Koekemoer}, {Lanzuisi}, {Maksym}, {Masini}, {Rosario}, {Ward}, {Willmer}, \& {Windhorst}}]{zhao2021}
{Zhao}, X., {Civano}, F., {Fornasini}, F.~M., {et~al.} 2021{\natexlab{b}}, MNRAS, 508, 5176, \dodoi{10.1093/mnras/stab2885}

\end{thebibliography}
\bibliographystyle{aasjournal}



\end{document}